# US HL-LHC Accelerator Upgrade Project

## MQXFA Final Design Report

**Prepared by:**
Giorgio Ambrosio, US HL-LHC AUP 302.2 Magnets L2 Manager, FNAL
Kathleen Amm, US HL-LHC AUP BNL lab representative, BNL
Mike Anerella, US HL-LHC AUP 302.2.06 Deputy Manager, BNL
Giorgio Apollinari, US HL-LHC AUP Project Manager, FNAL
Maria Baldini, US HL-LHC AUP Magnets L2 Deputy Manager, FNAL
Anis Ben Yahia, US HL-LHC AUP 302.4.01 Manager, BNL
James Blowers, US HL-LHC AUP QA Manager, FNAL
Ruben Carcagno, US HL-LHC AUP Deputy Project Manager, FNAL
Daniel Cheng, US HL-LHC AUP 302.2.07 Deputy Manager, LBNL
Guram Chlachidze US HL-LHC AUP 302.4.04 Manager, FNAL
Lance Cooley, US HL-LHC AUP 302.2.02 Manager, FNAL
Sandor Feher, US HL-LHC AUP 302.4 Cryo-assembly L2 Manager, FNAL
Paolo Ferracin, US HL-LHC AUP Magnets L2 Deputy Manager, LBNL
Henry Hocker, US HL-LHC AUP QA Manager at BNL, BNL
Susana Izquierdo Bermudez, HL-LHC WP3 Project Engineer, CERN
Piyush Joshi, US HL-LHC AUP 302.4.01 Deputy Manager, BNL
Vito Lombardo, US HL-LHC AUP 302.2.02 Deputy Manager, FNAL
Vittorio Marinozzi, US HL-LHC AUP 302.2.01 Scientist, FNAL
Joseph Muratore, US HL-LHC AUP 302.4.01 Former Manager, BNL
Michael Naus, US HL-LHC AUP 302.2.03 Deputy Manager, LBNL
Fred Nobrega, US HL-LHC AUP 302.2.05 Manager, FNAL
Heng Pan, US HL-LHC AUP 302.2.07 Engineer, LBNL
Marcellus Parker, US HL-LHC AUP 302.2.05 Engineer, FNAL
Ian Pong, US HL-LHC AUP 302.2.03 Manager, LBNL
Soren Prestemon, US HL-LHC AUP 302.2.07 Manager and LBNL lab representative, LBNL
Emmanuele Ravaioli, HL-LHC WP3 Scientist, CERN
Katherine L. Ray, US HL-LHC AUP QA Manager at LBNL, LBNL
GianLuca Sabbi, US HL-LHC AUP 302.2.01 Scientist, LBNL
Jesse Schmalzle, US HL-LHC AUP 302.2.06 Manager, BNL
Ezio Todesco, HL-LHC WP3 Manager, CERN
Melanie Turenne, US HL-LHC AUP 302.2.01 Scientist, FNAL
Miao Yu, US HL-LHC AUP 302.2.04 Manager, FNAL

**Approved by:**
Giorgio Ambrosio, US HL-LHC AUP Magnets L2 Manager, FNAL
Ruben Carcagno, US HL-LHC AUP Deputy Project Manager, FNAL
Giorgio Apollinari, US HL-LHC AUP Project Manager, FNAL

**Revision History**

| Revision | Date | Section No. | Revision Description |
|---|---|---|---|
| V1 | 5/15/18 | All | Initial Release |
| V2 | 5/16/18 | 5.1, 6.2, 7 | Updates in coil fabrication tolerances, assembly specifications, magnet test |
| V3 | 12/31/19 | 4.3, 4.6, 5.1, 5.5 | Updates in structural design, quench protection, coil technical specifications, handling and shipping |
| | 4/19/20 | 5.1 | Added “Specifications for series coils are in QXFA Series Coil Production Specification, US-HiLumi-doc-2986. The specifications presented below are only for reference”; adjusted coil length from 4532 mm to 4529 mm and midplane tolerance from +/- 0.05 mm to +0.075, -0.125 mm. |
| | 7/6/20 | 1, 2, 4.3, 5, 6, 7 | 1: Added reference to Memorandum of Understanding (MOU) between CERN and FNAL on AUP, US HiLumi doc 2511.<br>2: Added text about interfaces and added references: [3] Q1/Q3 Cryostat Assembly and Horizontal Test Final Design Report, US-HiLumi-doc-2882. [4] Interface Control Document WBS 302.2.07-302.4.02, US-HiLumi-doc-216. [5] MQXFA Magnet Interface Specification, US-HiLumi-doc-1674 (EDMS 2031177).<br>4.3 and 6: Updated structure design based on DCR 2679, DCR 2943, and Tech Board Decision for MQXFA03 Preload (doc-2496).<br>5: Updated coil fabrication chapter adding reference to Series Coils Production Specification (doc-2986).<br>7: updated prototype info and added MQXFA03 training history; updated table 7.1 and reference [1] (doc-1109).<br>All sections: checked and updated references and links. |
| | 10/23/20 | 3, 4, 5, 6, 7 | 3.2: Updated HT specification reference.<br>4.1: Changed Table 4.2 adding columns for the updated values of nominal (16.23 kA) and ultimate current (17.49 kA)<br>4, 5, 6, 7: changed references to 16.47 kA from “nominal current” to “nominal current plus margin” and changed references to 17.89 kA from “ultimate current” to “ultimate current plus margin”.<br>4.4: Updated figures and references |
| | 3/10/22 | 4 | In Table 4.9 added $\sigma_y$ = 1120 MPa for Nitronic 50;<br>In paragraph 4.6.4 added plan for meeting the requirement that the MQXFA magnet voltage to ground must be lower or equal to 353 V [3]. |
| | | | |

# Contents

# 1 Introduction

The MQXFA Quadrupole magnets will be installed in HL LHC to form the Q1 and Q3 inner triplet optical elements in front of the interaction points 1 (ATLAS) and 5 (CMS). A pair of MQXFA units is assembled in a stainless steel helium vessel, including the end domes, to make the Q1 Cold Mass or the Q3 Cold Mass. The US HL LHC Accelerator Upgrade Project is responsible for the design, manufacturing and test of the Q1/Q3 Cold Masses and the complete MQXFA magnets [1]. CERN provides the cryostat components and is responsible for integration and installation in HL LHC [2].

The MQXFA quadrupoles have 150 mm aperture, 4.2 m magnetic length, nominal gradient of 132.2 T/m, and coil peak field of 11.3 T. They use $Nb_3Sn$ conductor and a support structure made of segmented aluminum shells pre-loaded by using bladders and keys. The Q2 optical elements, to be fabricated by CERN, will be based on the same quadrupole design with a longer magnetic length of 7.15 m. These units are called MQXFB and MQXF is used as a general reference to both MQXFA and MQXFB.

The design criteria for MQXFA magnets are documented in [3-5]. These criteria were developed based on experience from the US LHC Accelerator Research Program (LARP) and other high field magnet R&D programs in the US and Europe [6]. In particular, the 120 mm aperture High-gradient Quadrupole by LARP provided the main design reference for MQXF.

A 1$^{st}$ generation short model (MQXFS1) [7] was developed by LARP and CERN and tested in 2016. This magnet exceeded ultimate current, demonstrated excellent memory and large temperature margin. Additional feedback from conductor development and magnet analysis was implemented in the 2$^{nd}$ generation (final) design [8]. The final design was tested through a series of 1.5 m short models (MQXFS3/4/5/6) developed in close collaboration by LARP and CERN, and it was implemented in the MQXFA long prototypes fabricated and tested by LARP. The first prototype (MQXFA1) had coils of 4 m magnetic length in a full-length structure. The MQXFA2 prototype had the final 4.2 m magnetic length. MQXFA03 [9], the first pre-series MQXFA magnet, was tested in winter 2019 and met all requirements up to nominal current.

# 2 Requirements and Interfaces

The Functional Requirements Specification for the MQXFA magnets can be found in [1]. The MQXFA functional requirements are classified into two groups: “Threshold” requirements and “Objective” requirements. Threshold requirements are requirements that contain at least one parameter that the project must achieve, and Objective requirements are requirements that the project should achieve and will strive to achieve. A summary of the MQXFA threshold and objective requirements can be found at the end of [1].

Criteria for acceptance of MQXFA magnets are presented in [2].

MQXFA magnets are installed in LMQXFA cold masses [3]. There are two MQXFA magnets in each LMQXFA cold mass. The LMQXFA cold mass, when surrounded by the QQXFA or QQXFC cryostat shields, piping, and vacuum vessel, is then the LQXFA cryo-assembly for Q1 and the LQXFB cryo-assembly for Q3 [3].

Interfaces between the MQXFA magnet and the LMQXFA cold mass are controlled by Interface Control Document WBS 302.2.07 - 302.4.02 [4] and are described in MQXFA Magnet Interface Specification [5].

# 3 Superconductor

The Rutherford cable for the MQXF magnets is fabricated using 40 $Nb_3Sn$ strands of diameter 0.85 mm. The baseline strand is a Rod-Restack-Process RRP® strand manufactured by Oxford-Instrument Superconducting Technology (OST), a company of Bruker Energy & Supercon Technologies (BEST).

## 3.1 Strand

Initial acceptance tests of several billets at OST and within LARP for a wire diameter of 0.778 mm (this diameter wire was used for the 120 mm aperture HQ magnets) showed that, although the wires meet the critical current requirement, the residual resistance ratio (RRR) was in many cases barely above the minimum requirements when reacted using the standard reaction schedule of 210°C/72 h + 400°C/48 h + 650°C/48 hr. The reaction used by OST to qualify the strand before delivery of strand to LARP is 210°C/48 h + 400°C/48 h + 650°C/50 h.

To increase the manufacturing margin and increase the likelihood of RRR exceeding 150 in the round wire, the tin content in the sub-element core was reduced by 5% from the standard amount. The wires from these "Reduced-Sn" billets showed a marked increase in RRR to values over 300, demonstrating that MQXF conductor requirements can be met by RRP conductor. The protection offered by the "Reduced-Sn" modification also allowed the final heat treatment temperature to be more aggressive for increasing critical current at 15 T, where the reaction used by OST changed the final stage to 665°C/75 h. The duration was subsequently reduced to 50 h. The resulting statistical distribution of properties exhibited acceptable margins above both the RRR and critical current.

The number of superconducting filaments shall be equal to or larger than 108, and the mean copper/non-copper ratio shall be 1.2. These requirements imply that the sub-element diameter by design is ~ 55 µm. The US HL-LHC-AUP strand specification and requirements [1] are based on the LARP strand specification and requirements used to procure strands for the MQXF prototypes [2]. The main characteristics of the strand are summarized in Table 3.1.

The QXF coil heat-treatment schedule for cables using this RRP strand is the following: 210°C/48h + 400°C/48h + 665°C/50h. This is based on supplier recommendation, and is consistent with past observations by LARP experts, showing that the reduction of the final stage duration, compared to the duration previously applied by the supplier, improved RRR for extracted strands at locations of the kinked edges from cabling, with negligible reduction in critical current.

Table 3.1 : Main Parameters of the HL-LHC AUP Strand.

| Parameter or characteristic | Value | Unit |
|---|---|---|
| Superconductor composition | Ti-alloyed $Nb_3Sn$ | |
| Strand Diameter | 0.850 ± 0.003 | mm |
| Critical current at 4.2 K and 12 T | > 632 | A |
| Critical current at 4.2 K and 15 T | > 331 | A |
| *n*-value at 15 T | > 30 | |
| Count of sub-elements<br>(Equivalent sub-element diameter) | ≥ 108<br>(≤ 55) | <br>(μm) |
| Cu : Non-Cu volume Ratio<br>Variation around mean | ≥ 1.2<br>± 0.1 | |
| Residual Resistance Ratio *RRR*<br>for reacted final-size strand | ≥ 150 | |
| Magnetization at 3 T, 4.2 K<br>(reported for information) | < 256<br>(< 320) | kA $m^{-1}$<br>(mT) |
| Twist Pitch | 19.0 ± 3.0 | mm |
| Twist Direction | Right-hand screw | |
| Strand Spring Back | < 720 | arc degrees |
| Minimum piece length | 500 | m |
| High temperature HT duration | ≥ 40 | Hours |
| Total heat treatment duration from start of ramp to power off and furnace cool | ≤ 240 | Hours |
| Heat treatment heating ramp rate | ≤ 50 | °C per hour |
| Rolled strand (0.7225 mm thick) critical current at 4.2 K and 12 T | > 600 | A |
| Rolled strand critical current at 4.2 K and 15 T | > 314 | A |
| Rolled strand *RRR* after reaction | > 100 | |

## 3.2 Cable

The MQXFA cables have a reduced keystone angle of 0.40° (cf. first generation: 0.55°). They will have a minimum length of 455 m, respooled from 500 m of strands. The strand maps will be drawn up by LBNL to optimize wire usage with a reasonable billet blending. The strands will be respooled as-received (i.e. without pre-cable annealing). An in-line dual-axis optical micrometer will be used during respooling to verify diameter. The cables will be fabricated according to the latest “US-HiLumi Cable Specification” [3], which is based on the LARP specification used for the cables of the MQXF prototype coils fabricated by LARP. The salient cable parameters are listed below in Table 3.2.

Table 3.2 : MQXFA Cable Parameters.

| | |
|---|---|
| Number of Wires in Cable | 40 |
| Cable Mid-Thickness | 1.525 ± 0.010 mm |
| Cable Width | 18.15 ± 0.05 mm |
| Cable Keystone Angle | 0.40° ± 0.1° |
| Cable Lay Direction | Left |
| Cable Lay Pitch | 109 ± 3 mm |
| Cable Core Material | 316 L Stainless Steel |
| Cable Core Width | 12 mm |
| Cable Core Thickness | 0.025 mm |
| Cable Core Position | Biased towards the major edge |
| Maximum Cable Residual Twist | 150°/m |

The cables will be identified using the following scheme, which is compatible with the CERN system (Figure 3.1 and Table 3.3).

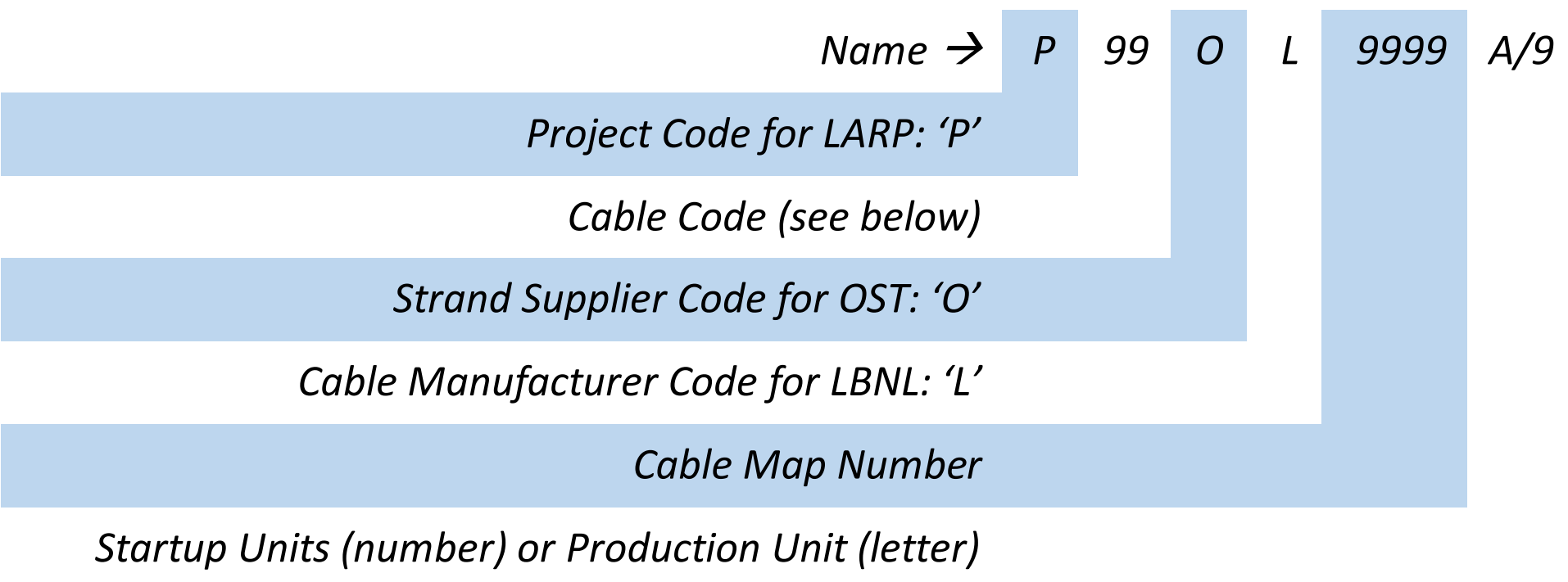

Figure 3.1: Cable ID Scheme based on US HiLumi DocDB # 41.

Table 3.3: Cable codes relevant to MQXF 2nd generation cable designs used during R&D and production.

| Cable code | 2nd Generation Cable Description |
|---|---|
| 43 | QXF R&D w/ Core, using un-annealed 108/127 wires |
| 45 | QXF R&D w/ Core, using un-annealed 132/169 wires |
| 47 | QXF R&D w/ Core, using un-annealed 144/169 wires |

Each cable unit length will be accompanied by a Release Note (a.k.a. Cable Report), including a cable summary, the Respool Log (identifying the strand-spool-brake-fork relationship), charts of the cable dimension parameters (keystone angle, width, and mid-thickness) from the in-line measurements, optical micrograph (for assessing strand damage), and a report of any non-conformity. It will be formatted to allow exporting the required data to the Vector Database.

Strands representative of the billet blend will be extracted according to the Respool Log for $I_C$ and RRR measurements. The heat treatment schedule will be that specified in the specification for quadrupole magnet conductor [1].

## 3.3 Cable Insulation

The MQXFA cables are insulated according to the HL-LHC AUP specification [5] based on LARP specification [6] using a high-strength glass yarn (S-2 Glass® direct sized yarn) made of 9 µm continuous glass strands twisted together and treated with an inorganic sizing (“933”), supplied by AGY (2558 Wagener Road, Aiken, South Carolina, USA 29801). The single-ply yarn is then made into a 2-ply yarn with a twist pitch of 3 inches (i.e. 0.34 twist per inch) by an undisclosed subcontractor of New England Wire Technologies (NEWT, 130 North Main Street, Lisbon, NH 03585), who braids the yarns of fiberglass onto our superconducting cables. The yarn is identified by HL-LHC-AUP and its vendors as “SCG75 1/0 0.7Z fiber with 933 sizing”. For completeness, Table 3.4 translates the US- and EU-style nomenclature.

Table 3.4: US- and EU-style nomenclature.

| US style | S | C | G | 75 | 1/0 | 0.7Z |
|---|---|---|---|---|---|---|
| EU style | S | C | 9 | 66 | 1x0 | Z28 |
| Meaning | High-strength glass | Continuous filament, aka strand | Filament diameter = 9 µm | Yield is 7500 yd/lb = 66 g/km (TEX) | 1 strand twisted into continuous filament ends, no piling | Right hand twist at 0.7 turns per inch = 28 turns per meter |

The braiding is done using a 48 bobbin-carrier braiding machine at NEWT, set to achieve 18 picks per inch in a vertical orientation. The bobbin payoff tension is between 1/4 to 3/8 lbs, and the cable payoff and take-up spool tensions are 2 and 12 lbs, respectively. The insulation specification thickness is 0.145 ± 0.005 mm, common between CERN and HL-LHC-AUP. The insulation thickness is checked at the start of the insulation run using a 5-foot sample according to a 10-stack measurement procedure agreed between CERN and HL-LHC-AUP. The insulated 10-stack cable is measured at 5 MPa during three loading cycles, repeated using the same stack with the insulation removed. The per-side insulation thickness (calculated by subtracting the bare-stack average from the insulated-stack average and then divided by two) measured by the vendor is communicated to HL-LHC-AUP and must be within the tolerance before commencing further work. A second 5-foot sample is sent to HL-LHC-AUP for verification measurements.

# 4 Magnet Design

## 4.1 2D Magnetic Design

### 4.1.1 Coil main parameters

The cross-section of the MQXFA is based on the cos2$\theta$-layout with two conductor blocks in each layer and similar pole angle in both layers. These features are based on the 120 mm HQ design in order to achieve the following design objectives:

- Minimize the peak stress by optimizing the balance of Lorentz force and pre-load among the two layers.
- High conductor packing to maximize the engineering current density and increase the operating margin
- Minimize field errors, in particular the high order harmonic components, by optimizing the position and angle of each conductor block
- Incorporate to the extent possible the HQ experience in tooling design and coil fabrication.

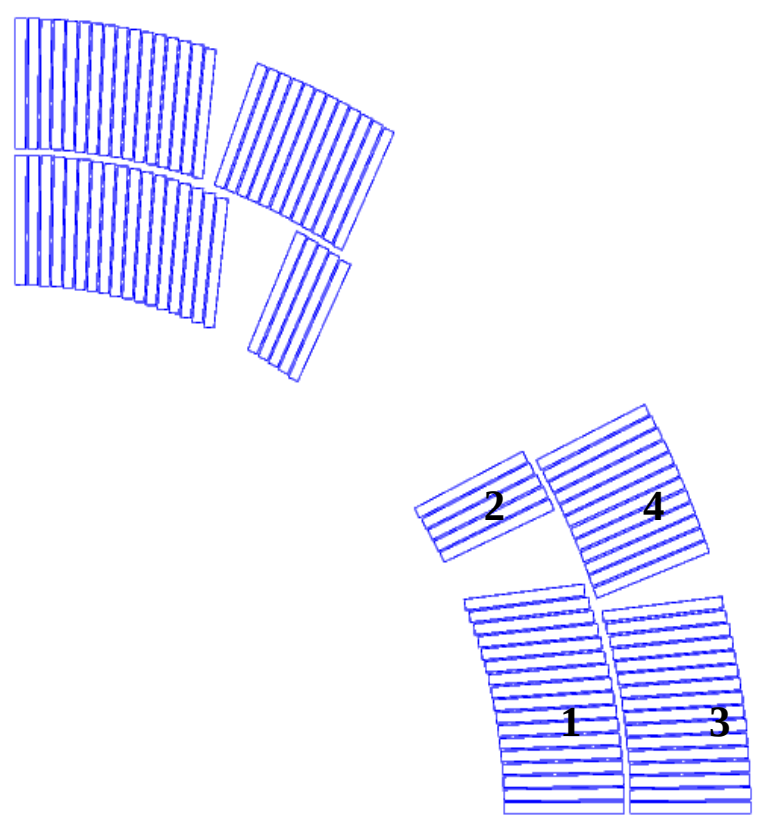

Figure 4.1: Cross-section of the MQXF coil.

Figure 4.1 shows the cross-section design for the second generation coil (MQXF_v2) implementing various improvements with respect to the original design [1], which was used for the first practice coils and short model (MQXFS1). In order to minimize the impact on coil fabrication and tooling the guidelines for coil re-optimization were to 1) to keep same number of conductors per block; 2) to keep the pole turns of the inner and outer layer aligned so as to have the same concept of layer jump (only hard way bending); 3) to keep the same coil inner and outer diameters.

The main coil parameters are summarized in Table 4.1. Based on the measurements performed to characterize the conductor dimensional change during heat treatment [2], and in order to find the best compromise between performance and field quality, for the second generation design it was decided to: 1) reduce the radial space in the tooling to accommodate for a cable width expansion to 1.2% instead of 2%; 2) keep the same azimuthal space, corresponding to a thickness expansion of 4.5%. The reduction of the nominal insulation thickness from 150 µm to 145 µm without changing the actual insulation thickness will help to assure a better azimuthal position of the coil turns. Dimensions of the cable before and after reaction are both given in Table 4.1. Design calculations are based on the dimension of the reacted cable.

The additional radial space due to the decreased cable width after reaction is partially absorbed by the increased inter-layer insulation (from 0.500 mm to 0.660 mm) and the outer layer of S2-glass that is installed in the outer coil diameter before impregnation (which increases from 0.150 mm to 0.310 mm). For the second generation design, we also consider a thicker mid-plane and pole insulation to allow fine tuning of field quality. The insulation between the mid-plane and the first insulated conductor increases from 0.250 mm to 0.375 mm, and from 0.350 mm to 0.500 mm between the pole and the insulated conductor. More information about the modifications implemented in the second generation design can be found in [3].

Table 4.1: Nominal parameters of the MQXF coil.

| | unit | |
|---|---|---|
| Coil aperture radius | mm | 75.000 |
| Layer 1 outer radius | mm | 93.653 |
| Inter-layer thickness | mm | 0.660 |
| Outer layer inner radius | mm | 94.313 |
| Outer layer outer radius | mm | 112.966 |
| Mid-plane shim thickness (per coil) | mm | 0.375 |
| Number of turns in block 1 | | 17 |
| Number of turns in block 2 | | 5 |
| Number of turns in block 3 | | 16 |
| Number of turns in block 4 | | 12 |
| Bare unreacted/**reacted** conductor width | mm | 18.150/**18.363** |
| Bare unreacted/**reacted** conductor thickness | mm | 1.525/**1.594** |
| Nominal keystone angle | deg | 0.40 |
| Nominal insulation thickness | mm | 0.145 |

The relative position of the block number 2 and 4 (pole block of layer 1 and pole block of layer 2) is such that the broad face of the layer jump turn (orange conductor) is parallel to the broad face of the top conductor of the upper block outer layer (see Figure 4.2). This way a shim with a uniform thickness (~0.25 mm) can be used. Fabrication tolerances for cable and insulation are provided in the conductor section.

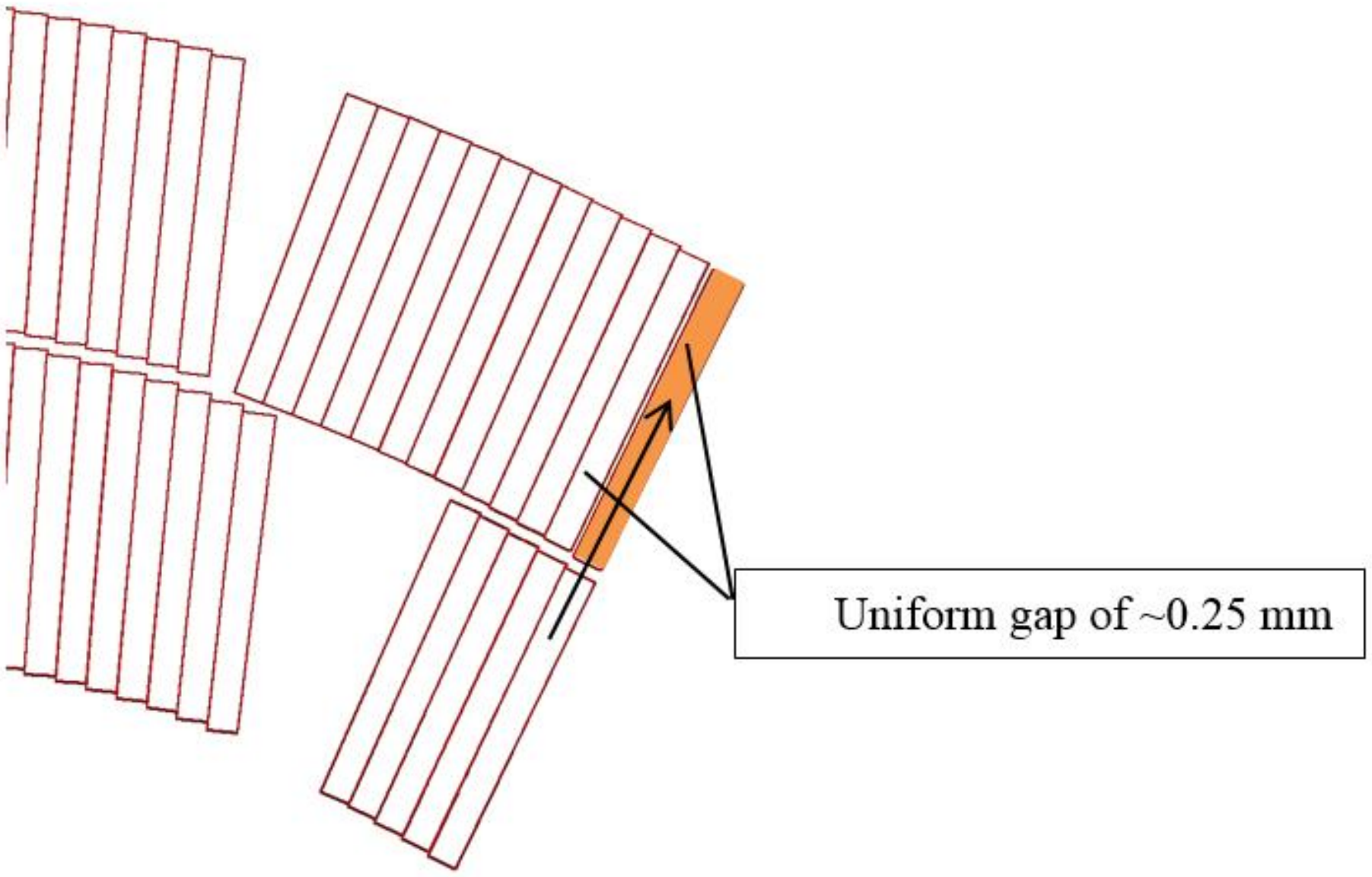

Figure 4.2: Coil layer jump.

Figure 4.3 compares the turn position for the first and second generation coil design.

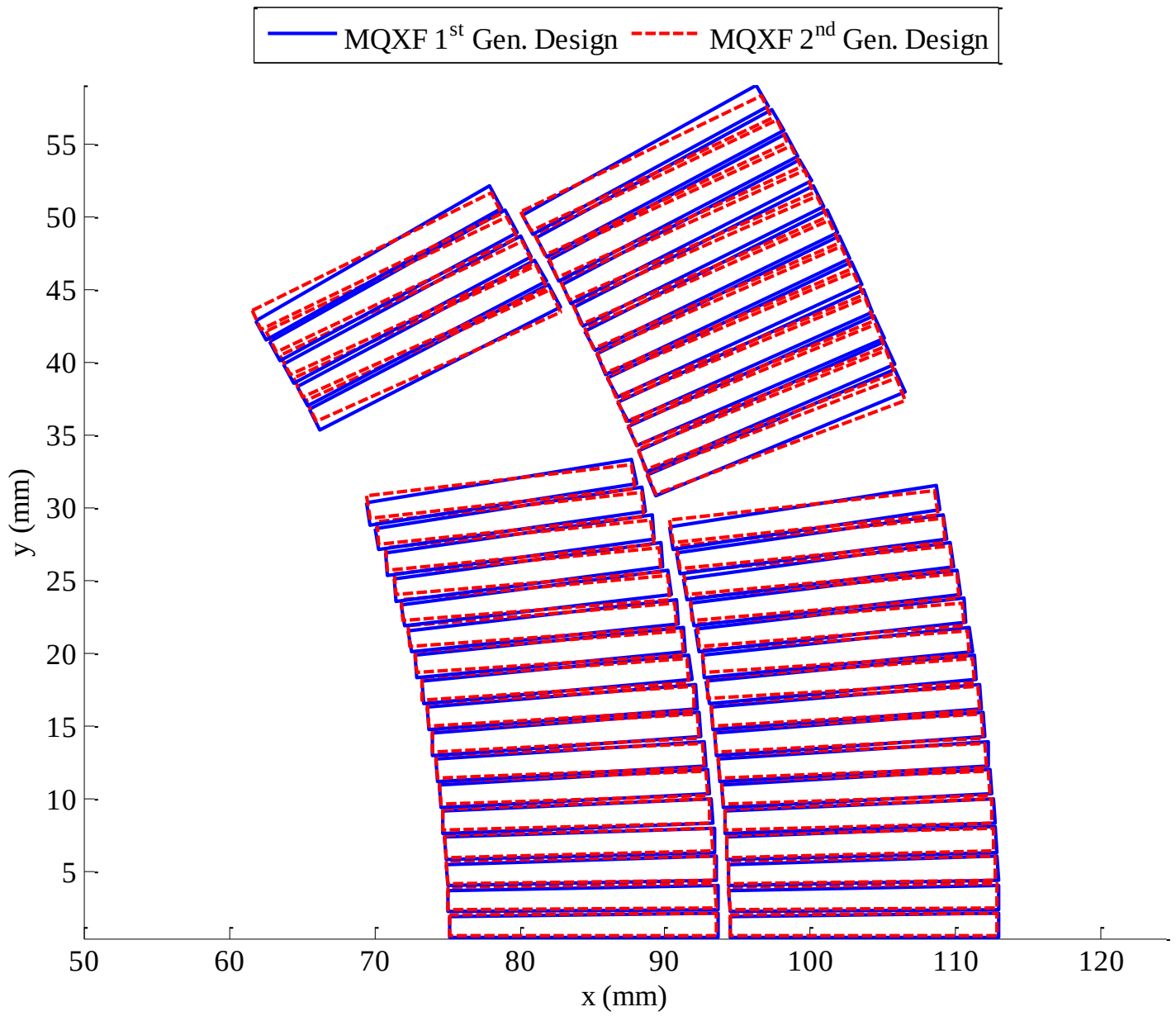

Figure 4.3: Cross-section of the MQXF second generation coil (red) superimposed to the cross-section of the first generation coil (blue).

### 4.1.2 Roxie model

The computation of the magnetic field was performed with the Roxie software [4]. In Figure 4.4 it is given the 2D data table as implemented in Roxie. The option "alignment of the conductor on the coil OD" was selected (ODFAC = 1) since it provides the best results based on the experience from previous magnets.

Block Data 2D

| No | Type | NCab | X | Y | α | Current | Cable name | N1 | N2 | Imag | Turn |
|---|---|---|---|---|---|---|---|---|---|---|---|
| 1 | Cos | 17 | 75 | 0.28648 | 0 | 16471 | XF145HTH5 | 2 | 20 | 0 | 0 |
| 2 | Cos | 5 | 75 | 28.5 | 24 | 16471 | XF145HTH5 | 2 | 20 | 0 | 0 |
| 3 | Cos | 16 | 94.313 | 0.22782 | 0 | 16471 | XF145HTH5 | 2 | 20 | 0 | 0 |
| 4 | Cos | 12 | 94.313 | 19 | 20.8049 | 16471 | XF145HTH5 | 2 | 20 | 0 | 0 |

Figure 4.4: Input-file of the MQXF coil used in ROXIE.

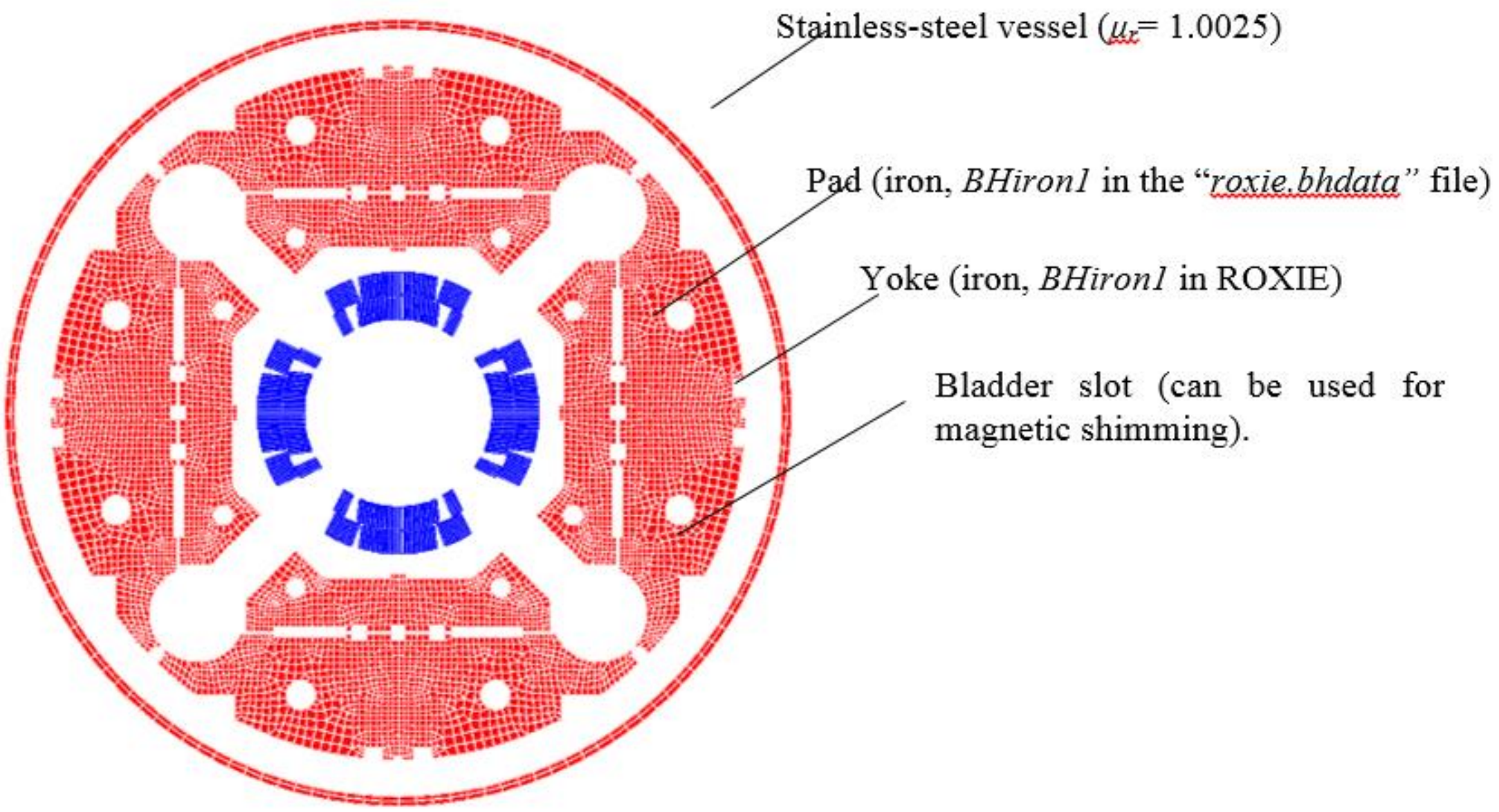

Figure 4.5: MQXF magnet model used in ROXIE.

In the MQXF assembly the superconducting coil is surrounded by iron pads and yokes which reinforce the field in the aperture and reduce the stray field outside the magnet. These elements are also integral components of the shell-based support structure. As a consequence the yoke and the pad implement slots for the insertion of the bladders and grooves for the keys. The 2D FEM model used for the computation of the magnetic field is shown in Figure 4.5. The *BH* characteristic used for the iron components is defined as "BHiron1" in the *roxie.bhdata* file. This *BH* curve assumes a filling factor of 1 (full body). Note that no thermal contraction factor was used for the computation of the harmonics, *i.e.*, the coil is assumed to be at room temperature. The impact of cool-down and mechanical deformation on field quality will be discussed later.

### 4.1.3 Magnet performance

Minimal requirements for wire manufacturing set by CERN and HL-LHC-AUP require a critical current larger than 632 A and 331 A in respectively 12 T and 15 T applied field and at a temperature of 4.2 K. For the computation of the magnet short sample current 5% degradation on the current due to cabling is assumed and a correction factor of 0.429 T/kA is used to take into account the strand self-field. It corresponds to the magnetic field produced at 89% of the radius of a straight wire. For the characterization of the critical surface the scaling law developed in [5] is used. Table 4.2 presents the magnetic parameters of the MQXF magnet when powered at short sample ($I_{ss}$), nominal ($I_{nom}$) and ultimate-operation ($I_{ult}$) currents. The official values for nominal and ultimate operation current are reported in [6]. Table 4.2 shows main parameters also at 16.47 and 17.89 kA. The first value (16.47 kA = nominal current plus 240 A margin) has been used for many computations presented in this report and it will be referred to as $I_{nom+margin}$. The MQXF load line is shown in Figure 4.6. The operational temperature is 1.9 K.

In Figure 4.7 the magnetic field density in the coil (left) and in the yoke (right) at $I_{nom+margin}$ are plotted. At I = $I_{nom+margin}$ the peak field in the coil reaches 11.41 T. The peak field is located in the pole block of the inner layer (block 2). The proximity of the yoke to the coil results in a highly saturated iron yoke that translate in a ~9% reduction in the transfer function from injection to nominal current (Figure 4.8-left). For the same reason the differential inductance Ld is non-linear (Figure 4.8-right). In spite of being highly saturated the iron yoke still contribute to enhancing the magnet gradient by ~ 8% at $I_{nom+margin}$ (from 121.7 T/m to 132.6 T/m).

Table 4.2: Main magnetic parameters of the MQXF cross-section considering an operational temperature of 1.9 K.

| Parameters | Units | $I_{ss}$ | | $I_{ult}$ | $I_{nom+margin}$ | $I_{nom}$ |
|---|---|---|---|---|---|---|
| Current | kA | 21.24 | 17.89 | 17.49 | 16.47 | 16.23 |
| $I/I_{ss}$ | % | 100 | 84 | 82 | 78 | 76 |
| Gradient | T/m | 168.1 | 143.2 | 142.1 | 132.6 | 132.2 |
| Coil peak field | T | 14.5 | 12.3 | 12.1 | 11.4 | 11.3 |
| Stored energy | MJ/m | 1.89 | 1.38 | 1.32 | 1.18 | 1.15 |
| Current sharing temperature | K | 1.9 | 5.8 | 6.0 | 6.8 | 7.0 |
| Differential inductance | mH/m | 8.13 | 8.18 | 8.23 | 8.21 | 8.26 |
| Superconductor current density ($j_{sc}$) | A/mm$^2$ | 2059 | 1734 | 1695 | 1596 | 1573 |
| Engineering current density ($j_{eng}$) | A/mm$^2$ | 936 | 788 | 771 | 726 | 715 |
| Forces x | MN/m | 3.83 | 2.85 | 2.74 | 2.47 | 2.41 |
| Forces y | MN/m | -5.68 | -4.08 | -3.94 | -3.48 | -3.41 |

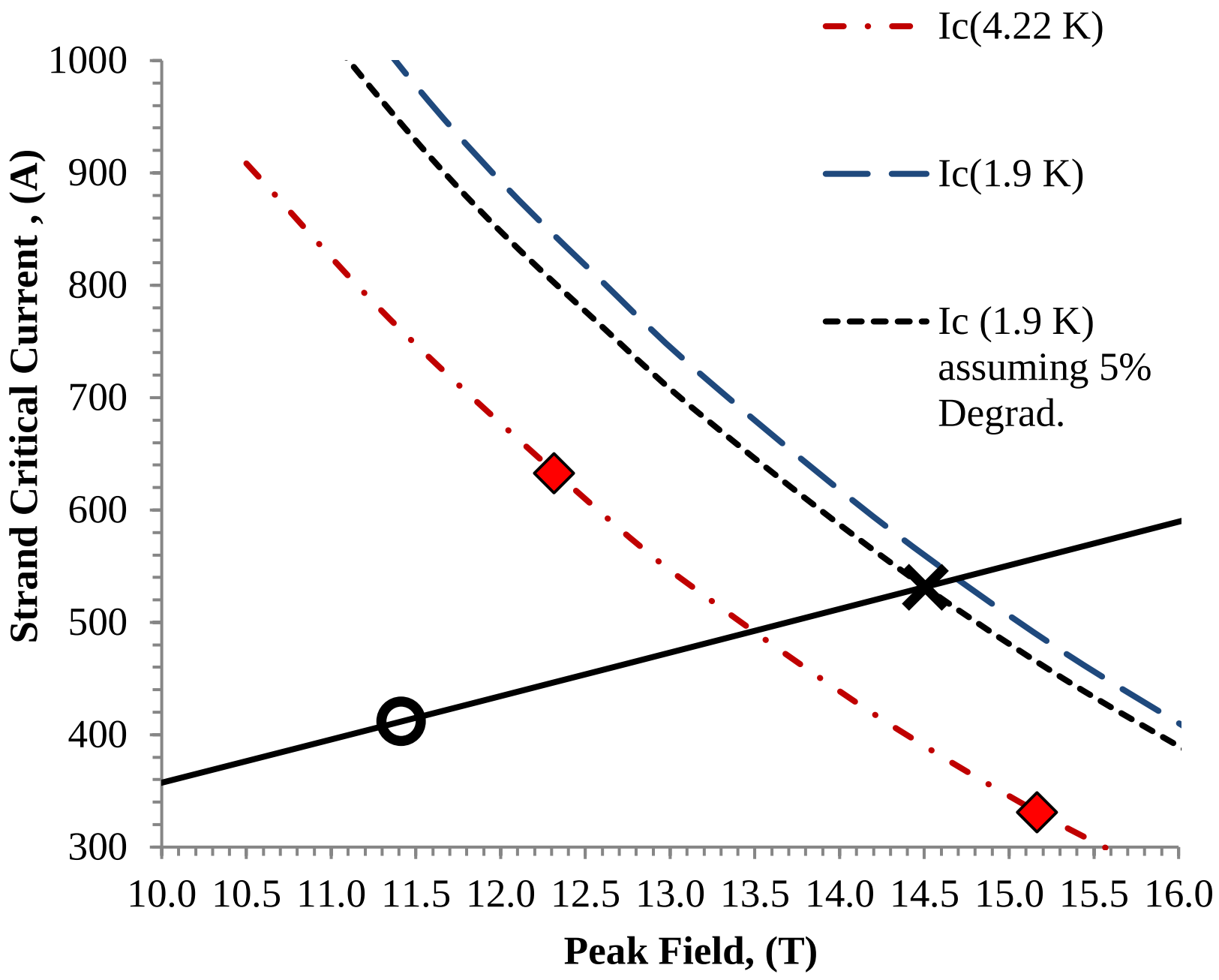

Figure 4.6: MQXF quadrupole load line (current in the cable versus peak field in the coil). The peak field is always located in block 2. The load line has been obtained by gradually increasing the current in the coil. This way the non-linearity of the load line is taken into account.

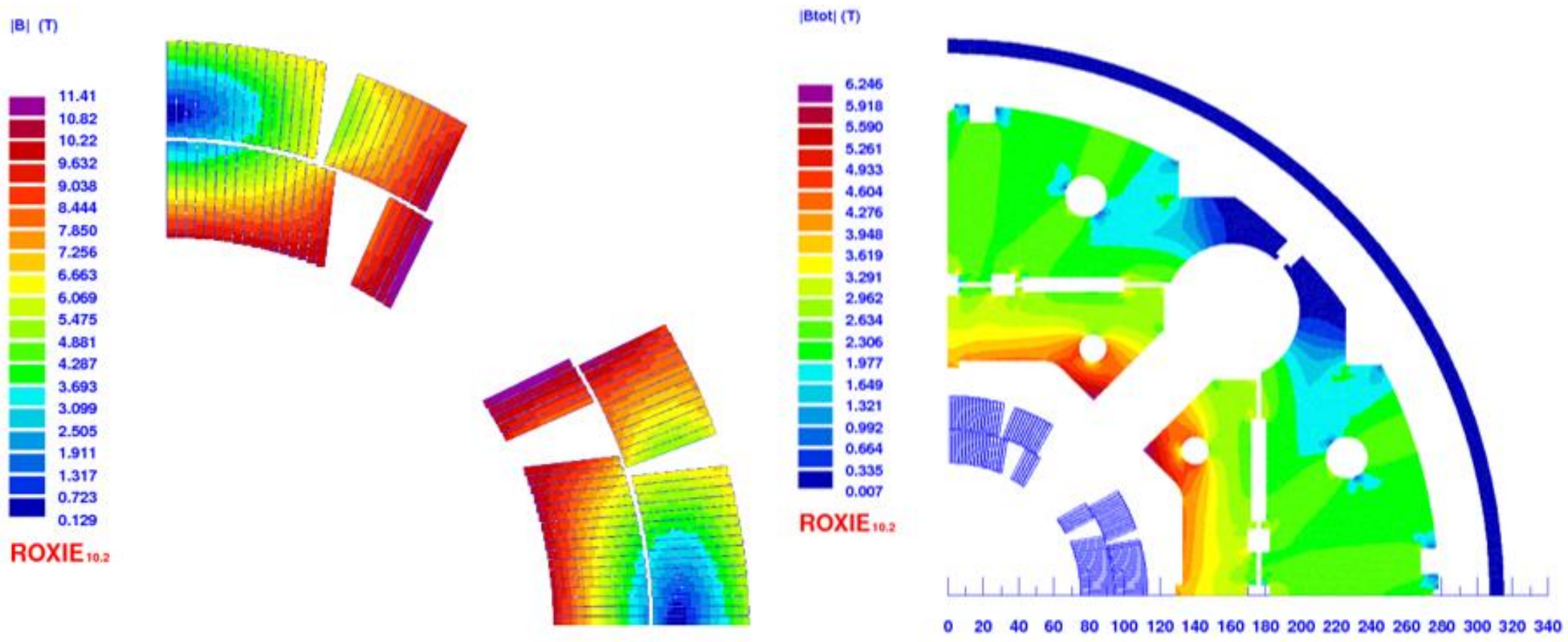

Figure 4.7: Magnetic flux density in the coil (left) and in the yoke (right) at nominal current + 240 A ($I_{nom+margin}$).

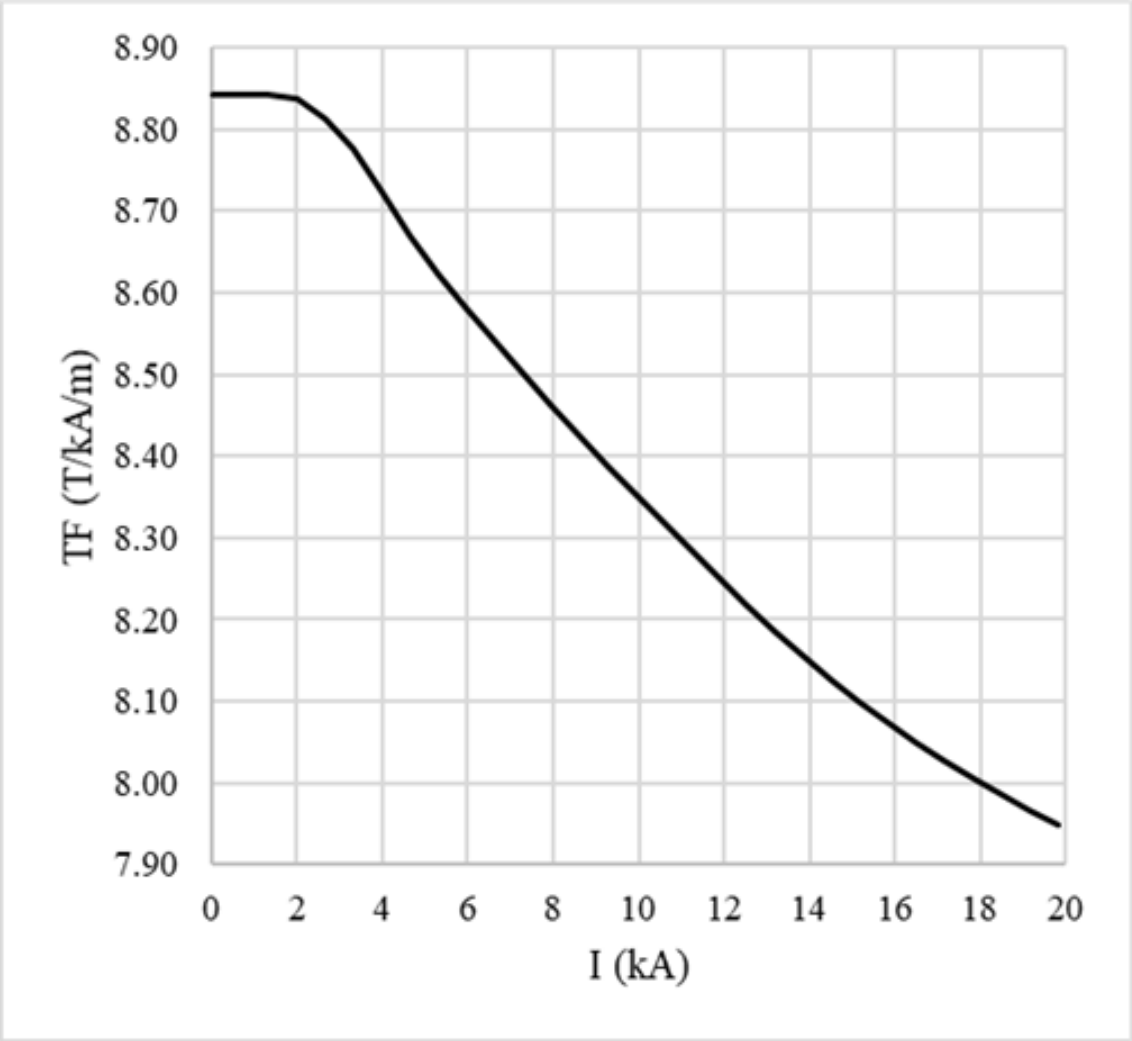

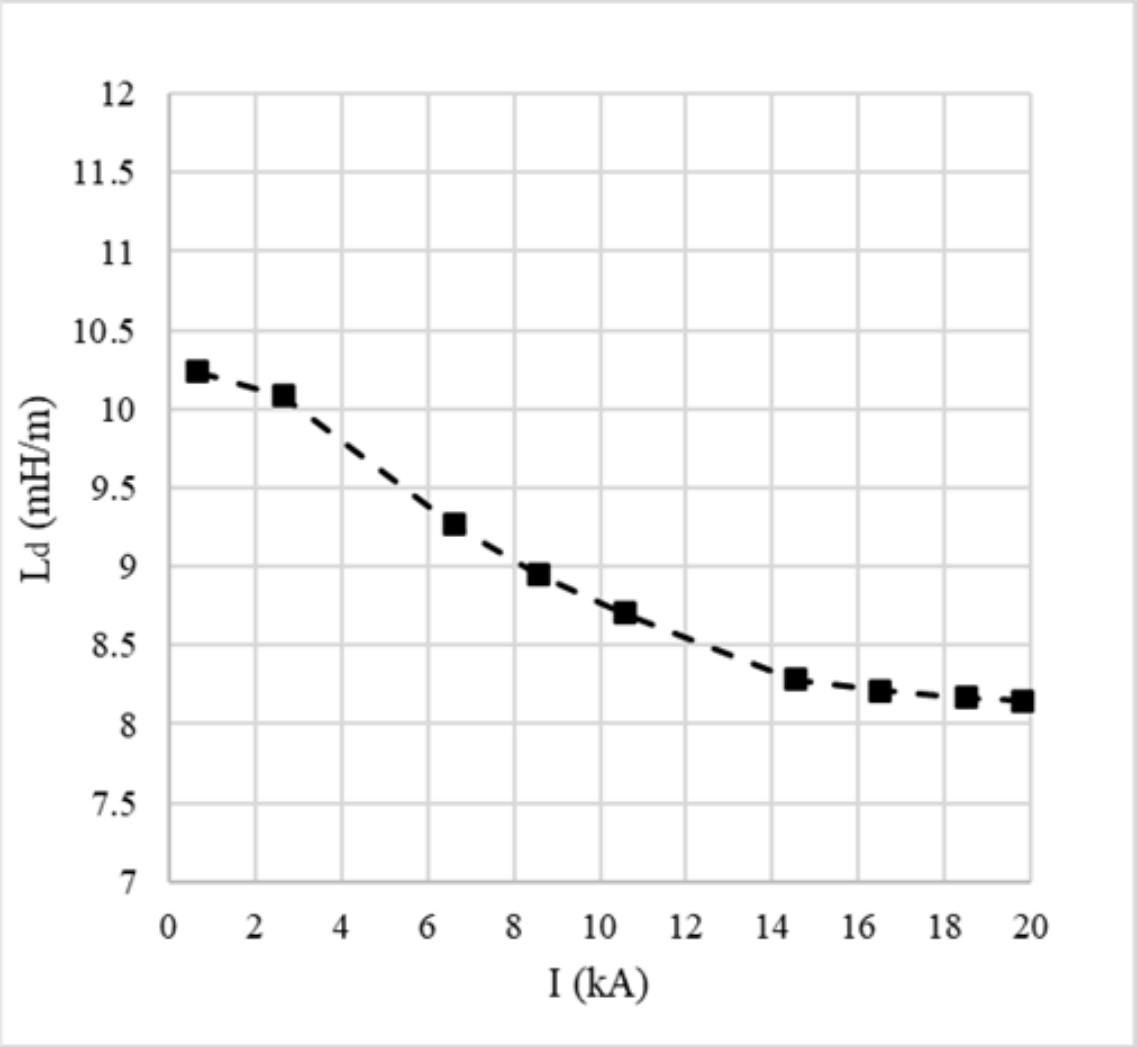

Figure 4.8: Left: Transfer function defined as the ratio between the gradient and the current and expressed in [T/m/kA] plotted versus current. Right: differential inductance Ld in mH/m. Roxie was used for the computations.

### 4.1.4 Field quality

Due to the large beam size and orbit displacement in the final focusing triplet, these magnets have challenging targets for field quality requirements at nominal operating current. The coil cross section is optimized such that all allowed harmonics are within one unit. The following convention for the definition of the multipoles is used, taking as reference radius 2/3 of the coil aperture radius ($R_{ref}$ = 50 mm).

$$B_y + iB_x = 10^{-4} B_2 \sum_{n=1}^{\infty} (b_n + ia_n) \frac{(x+iy)^{n-1}}{R_{ref}^{n-1}}$$

For the second generation design, coil cross section has been re-optimized to account for the effect of coil deformation on field quality [7] and the contribution of the splices and connection leads [8] to the integral field quality. The impact of coil deformation is an offset of +0.9 units on $b_6$, mostly caused by the azimuthal coil deformation during cool down as it can be observed in Figure 4.9. The evaluation of the impact of coil deformation on field quality was carried out by importing the coil displacement map extracted from the ANSYS solution into the 2D magnetic model implemented in Roxie. The displacement map corresponds to the state of the coil after the room temperature pre-load, the cool-down and the excitation to $I_{nom+margin}$, and it is estimated with respect to the design coil geometry at room temperature without any pre-load. The computed displacements were applied to every strand of the magnetic model and a harmonic analysis was performed with the displaced strand distribution. The deformation of the iron yoke was not taken into account during this analysis. The results of the mechanical analysis indicate a radial displacement of the blocks of -0.3 to -0.4 mm and an azimuthal displacement of -0.04 to -0.05 mm. The output of the numerical magnetic model showed a change in the normalized $b_6$ harmonic of about 0.9 units and a negligible change of other allowed harmonics such as $b_{10}$ and $b_{14}$. The offset of the $b_6$ is mostly caused by the azimuthal coil deformation, which results from the pre-load applied to the structure during the assembly and the structure contraction during the cool-down phase. As expected for quadrupole magnets, the deformations resulting from the electro-magnetic forces have a negligible effect on the $b_6$, as shown in Figure 4.9.

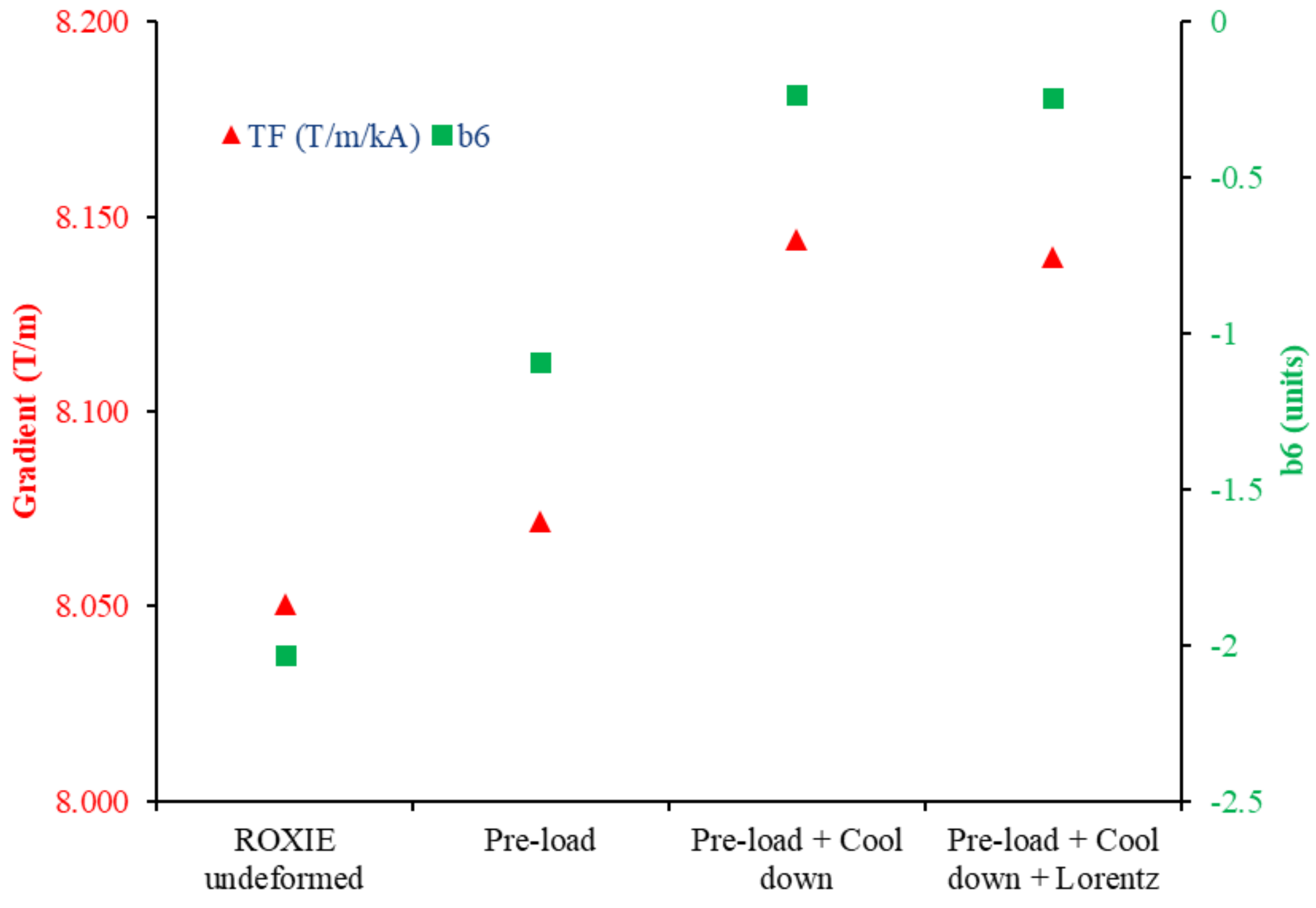

Figure 4.9: Impact of coil deformations due to cool-down, pre-loading, and energizing on the gradient and b6 at 7 TeV. In order to retain only the effect of the mechanical deformation, the magnet current is the same for the different steps to have the same contribution from the iron.

The contribution of the $Nb_3Sn$-NbTi splices and connection leads is an offset of +0.5 units on $b_6$ as it will be described in 3D section. The change on the rest of the harmonics is negligible. Figure 4.10 shows the evolution of $b_6$ as a function of the current including the mechanical deformations and 3D effects. As it can be seen from the plot, the main contribution to $b_6$ is coming from the iron saturation. Even if the difference between injection and nominal current is less than one unit, the maximum variation of $b_6$ for the full range of current is 4 units.

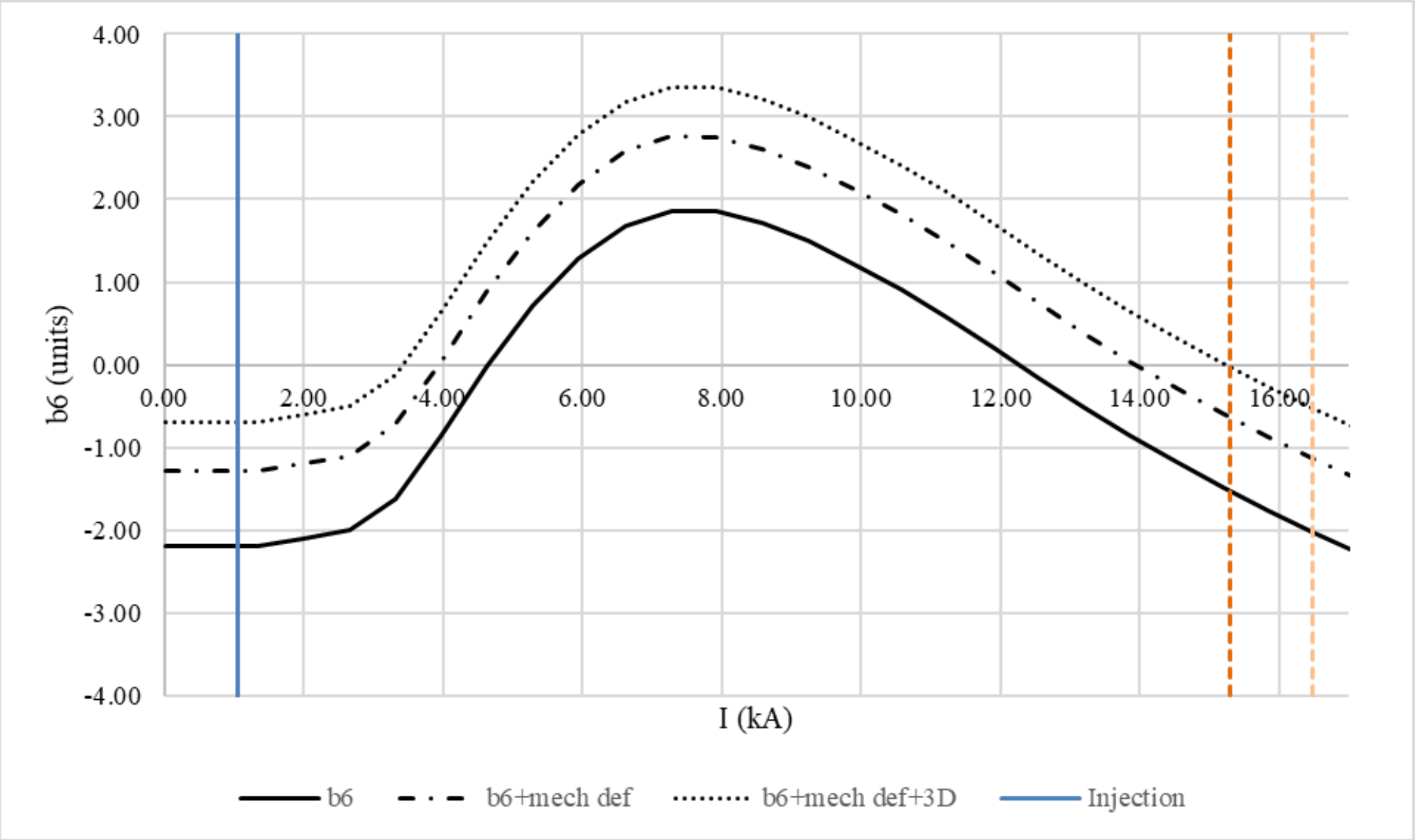

Figure 4.10: First allowed harmonic ($b_6$) plotted versus the current, including the contribution of the mechanical deformation and 3D effects. Harmonics are expressed in units. Dashed lines show current for 6.5 TeV and 7 TeV.

The evolution of the first four allowed harmonics ($b_6$, $b_{10}$, $b_{14}$ and $b_{18}$) with the current is plotted in Figure 4.11. Numerical values of the field components calculated for 7 TeV, and 6.5 TeV are given in Table 4.3.

Measurements of the field quality of short models MQXFS3, MQXFS4 and MQXFS5, together with warm field quality measurements of AUP prototype MQXFAP2 indicated a systematic b6 far from the target of 0 units at nominal current. Therefore, decision was made by HL-LHC WP3 leader to apply a "b6 correction" in coil manufacturing. This change consists in adding 0.125 mm insulation thickness on the coil pole and in removing 0.125 mm from the coil midplane insulation during the coil manufacturing phase (before reaction and impregnation). The expected impact on magnet straight section is b6 increase by 5.3 units. It was also decided that the geometry of the end spacers shall undergo no modifications. This change introduced a discontinuity of 0.125 mm in the cross-section that was judged to be well inside the assembly tolerances. More details about this decision are in ref [9].

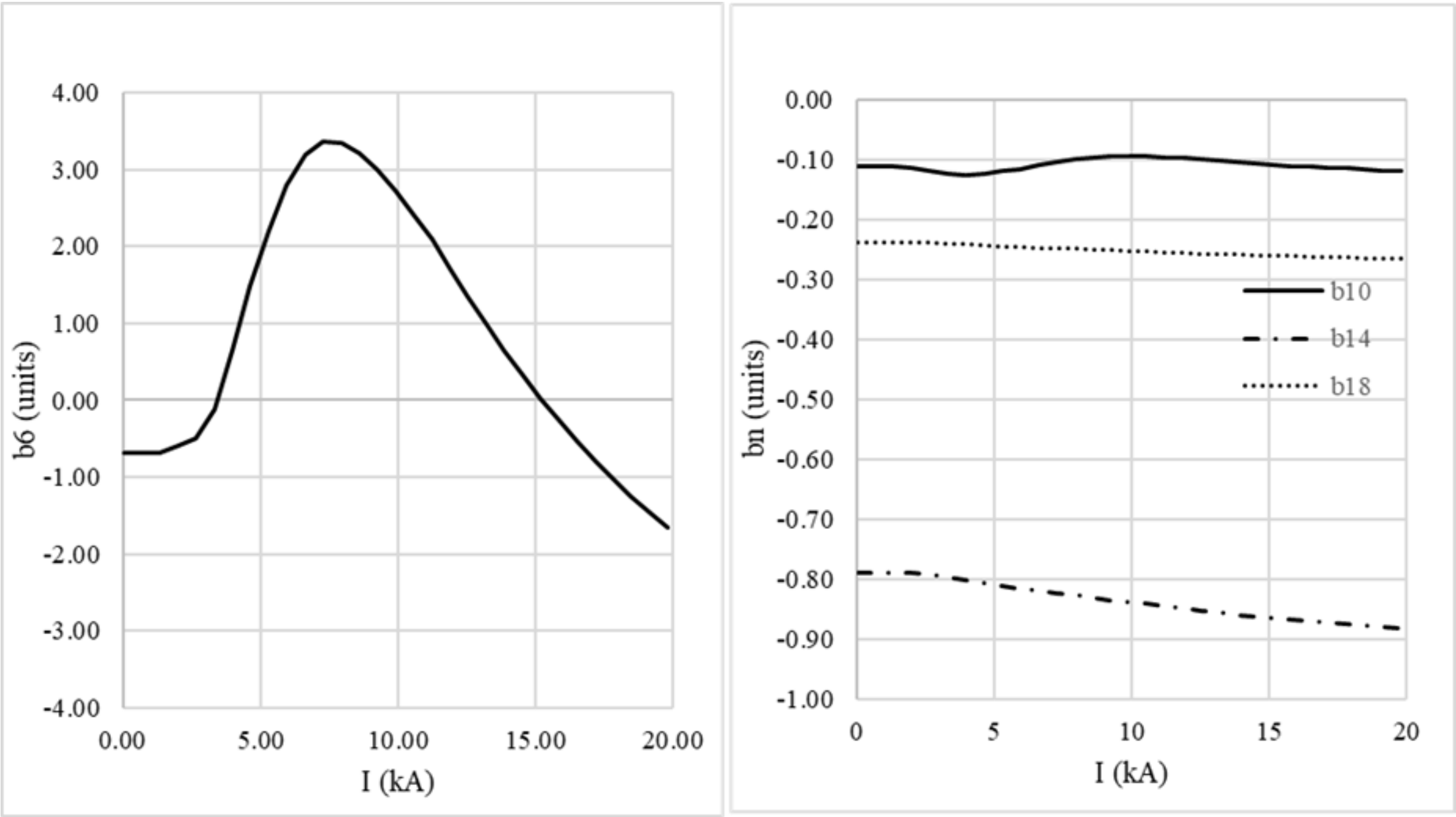

Figure 4.11: First four allowed harmonics ($b_6$, $b_{10}$, $b_{14}$ and $b_{18}$) plotted versus current. Harmonics are expressed in units.

Table 4.3: Field Harmonic components calculated for ~7 TeV (Nominal + margin) and 6.5 TeV.

| | unit | *7 TeV* | *6.5 TeV* |
|---|---|---|---|
| Current | kA | 16.47 | 15.29 |
| Gradient | T/m | 132.6 | 123.8 |
| $b_6$ | unit | -0.69 | -0.03 |
| $b_{10}$ | unit | -0.11 | -0.11 |
| $b_{14}$ | unit | -0.26 | -0.26 |
| $b_{18}$ | unit | -0.86 | -0.86 |

### 4.1.5 Systematic field error

Field calculations are performed assuming that conductors are aligned on their OD based on the experience from previous (NbTi) magnet. However, it is clear that one does not totally control the position of the cable in the impregnation cavity due to the necessity to allow space for the cable to grow during heat treatment. Here we estimate the field errors due to various defects in tooling, coil size and asymmetry, and turn shifts.

Certain coil variances are observed from 1st generation coil cross section and CMM. At the winding and block level there are azimuthal and radial shifts and rotations. The position of turns near the pole and near the midplane have very little displacement as observed in coil cross sections.

Therefore, shifts mostly affect the middle turns of each level near the wedge. The variability along the magnetic length alone is the subject of this analysis. A demonstration of each turn shift is shown in Figure 4.12 at greatly exaggerated magnitudes. Coil size and asymmetry is also included in this analysis based on 1st generation coil CMM data. Field errors due to tooling defects were calculated in the first generation MQXF design report and are included in this analysis.

A Monte-Carlo code was written in Java in conjunction with COMSOL Multiphysics to calculate the harmonics based on 40 line currents uniformly distributed within each cable. The code neglects effects from iron. Random turn/block/coil shifts have a normal distribution with a standard deviation of 50 µm. Errors larger than 1 standard deviation are not included.

The normal and skew harmonic standard deviations are presented in Table 4.4. All random harmonics assume a 50 µm independent RMS displacement. This level of RMS aberration is based on initial coil cross section analysis and coil CMM variance. Machining tolerance for coil parts is consistent with this choice of RMS aberration. Radial variation seems to have the largest effect on harmonics. This is partly an artifact that the entire block is shifted for radial variation as seen in coil cross sections while rotational and azimuthal variation principally is a shift of only the central turns between the midplane and pole. Regardless this analysis indicates that attention should be paid on the materials and thickness of radial insulation.

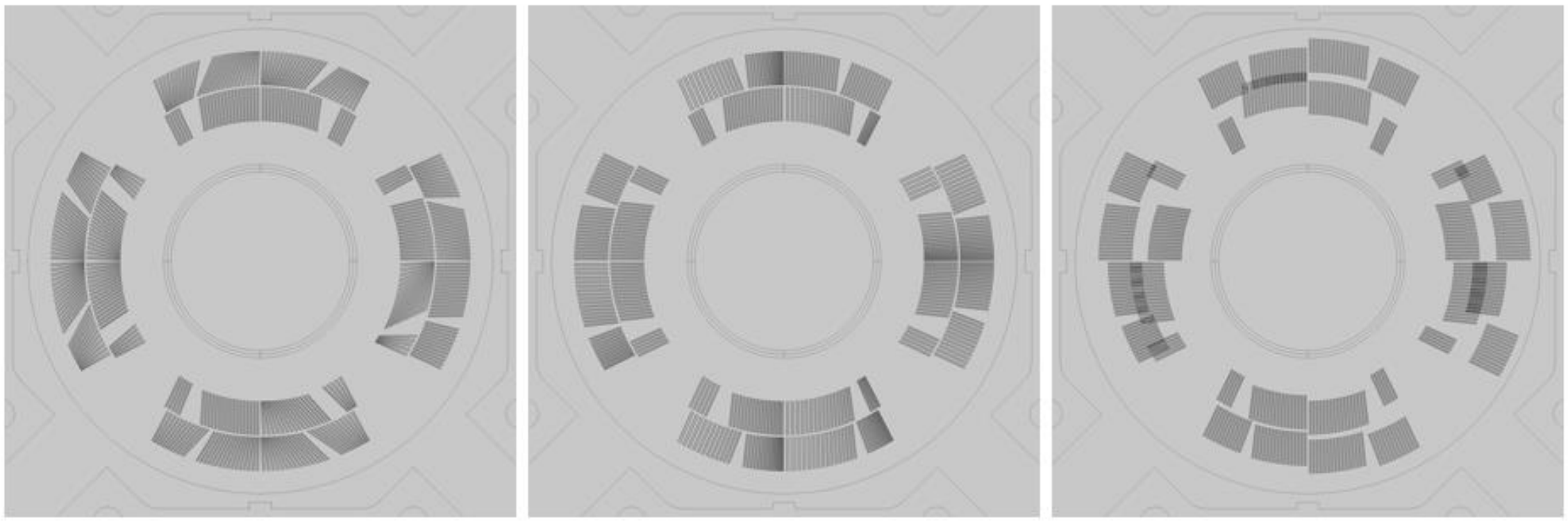

Figure 4.12: On the left a magnet cross section as displayed in COMSOL with random block rotations. The middle image has random azimuthal shifts. The right image has random radial shifts. Each type of shift has a standard deviation and peak greatly exaggerating displacements for demonstration purposes only.

Table 4.4: Random Harmonics based on 50 µm RMS variation in tooling defects, coil CMM, and turn shifts.

| | b3 | b4 | b5 | b6 | b7 | b8 | b9 | b10 | b11 | b12 | b13 | b14 |
|---|---|---|---|---|---|---|---|---|---|---|---|---|
| **Total Random Normal** | **1.560** | **1.070** | **0.680** | **1.880** | **0.270** | **0.160** | **0.100** | **0.220** | **0.041** | **0.024** | **0.015** | **0.032** |
| Coil Size and Asymmetry | 0.859 | 0.331 | 0.134 | 0.002 | 0.027 | 0.005 | 0.005 | 0.000 | 0.000 | 0.001 | 0.001 | 0.001 |
| Tooling Defects | 0.829 | 0.623 | 0.412 | 0.301 | 0.159 | 0.121 | 0.061 | 0.040 | 0.028 | 0.018 | 0.011 | 0.007 |
| Radial Shifts | 0.761 | 0.564 | 0.418 | 0.195 | 0.173 | 0.029 | 0.068 | 0.029 | 0.024 | 0.016 | 0.008 | 0.003 |
| Azimuthal Shifts | 0.650 | 0.566 | 0.317 | 0.287 | 0.128 | 0.096 | 0.047 | 0.022 | 0.015 | 0.004 | 0.005 | 0.000 |
| Rotational Shifts | 0.093 | 0.105 | 0.072 | 0.074 | 0.038 | 0.031 | 0.016 | 0.008 | 0.006 | 0.001 | 0.002 | 0.000 |

| | a3 | a4 | a5 | a6 | a7 | a8 | a9 | a10 | a11 | a12 | a13 | a14 |
|---|---|---|---|---|---|---|---|---|---|---|---|---|
| **Total Random Skew** | **1.550** | **0.960** | **0.680** | **0.390** | **0.280** | **0.180** | **0.100** | **0.066** | **0.036** | **0.021** | **0.010** | **0.008** |
| Coil Size and Asymmetry | 0.846 | 0.302 | 0.142 | 0.083 | 0.028 | 0.016 | 0.004 | 0.001 | 0.000 | 0.001 | 0.001 | 0.000 |
| Tooling Defects | 0.849 | 0.585 | 0.406 | 0.192 | 0.161 | 0.069 | 0.060 | 0.030 | 0.020 | 0.012 | 0.008 | 0.005 |
| Radial Shifts | 0.784 | 0.628 | 0.421 | 0.328 | 0.181 | 0.155 | 0.069 | 0.050 | 0.025 | 0.011 | 0.008 | 0.006 |
| Azimuthal Shifts | 0.594 | 0.311 | 0.305 | 0.007 | 0.124 | 0.052 | 0.043 | 0.030 | 0.015 | 0.012 | 0.005 | 0.004 |
| Rotational Shifts | 0.098 | 0.063 | 0.070 | 0.001 | 0.037 | 0.017 | 0.016 | 0.011 | 0.006 | 0.005 | 0.002 | 0.002 |

Coil size and asymmetry harmonics decay much quicker than other block or winding level aberrations as a function of harmonic order. This is expected because the aberration is applied over an entire coil rather than a single turn or block. Thus, the coil size and asymmetry has seemingly negligible effect on high order harmonics. This partly explains the larger than expected $a_3$ and $b_3$ harmonic seen in MQXFS01 assembly test for example.

The 2$^{nd}$ generation MQXF design has increased radial and azimuthal insulation and has reduced the free space by roughly half the 1$^{st}$ generation design. It is expected therefore that 2$^{nd}$ generation MQXF should likewise have reduced longitudinal variability of turns.

## 4.2 3D Magnetic Design

### 4.2.1 Design objectives and process

From the magnetic point of view, the design objectives for coil ends optimization are:

- to limit the peak field enhancement in the ends;
- to keep the coil end as compact as possible in order to increase the magnetic length for a given coil length;
- to minimize the multipole content of the integrated field.

Magnetic and mechanical optimization of the coil ends for the first generation coil design is described in [8]. Following the positive feedback from winding and destructive inspection of the first generation practice coils, the overall shape of the coil ends was not modified. Only a fine tuning was needed to adapt to the new cable geometry and optimize field quality. In order to compensate the non-negligible positive contribution of the coil layer jump and $Nb_3Sn$/NbTi splice to $b_6$ [8], the following actions were taken: 1) The magnet longitudinal loading system has been moved from the connection side to the non-connection side of the magnet to minimize the length of the current leads; 2) Re-optimization of the longitudinal position of the coil blocks at the ends. Figure 4.13 compares the conductor longitudinal position for the first and second generation design; 3) Coil cross section has been optimized aiming to a $b_6$ close to -0.5 units in the straight section to minimize the $b_6$ integrated over the entire magnet length.

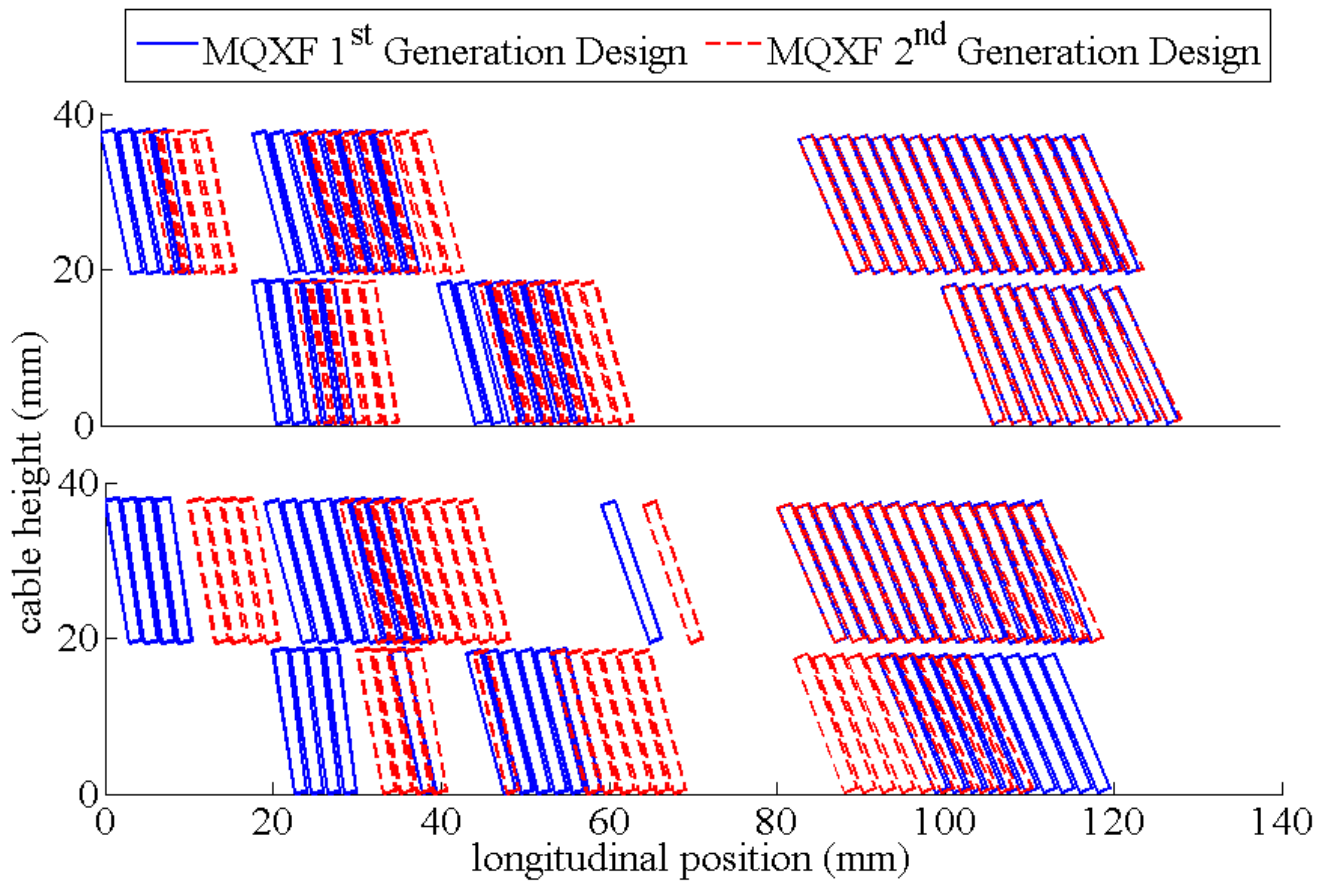

Figure 4.13: Comparison of the conductor position on coil ends for the first and second generation coil design.

Figure 4.14 to Figure 4.17 provide the 3D data table as implemented in ROXIE for each coil end.

Block Data 2D

| No | Type | NCab | R | φ | α | Current | Cable name | N1 | N2 | Imag | Turn | Ne |
|---|---|---|---|---|---|---|---|---|---|---|---|---|
| 1 | Cos | 10 | 75 | 0.28648 | 0 | 15310 | XF145HTH5 | 2 | 20 | 0 | 0 | 1 |
| 2 | Cos | 7 | 75 | 14.2547 | 3.9938 | 15310 | XF145HTH5 | 2 | 20 | 0 | 0 | 2 |
| 3 | Cos | 5 | 75 | 28.5 | 24 | 15310 | XF145HTH5 | 2 | 20 | 0 | 0 | 3 |

More options :

| No | String | N/a | N/a |
|---|---|---|---|
| | | | |
| | | | |
| | | | |

Block Data 3D

| Ne | Type | β | Bo | zo | Wi | Wo | Hwed | Horder |
|---|---|---|---|---|---|---|---|---|
| 1 | Diff. Geometry f | 71 | 2 | 266 | 0.35 | 0.02 | 18.5 | 3.2 |
| 2 | Diff. Geometry f | 76 | 1.9 | 211 | 0.17 | 0.02 | 18.5 | 3 |
| 3 | Diff. Geometry f | 81 | 2.23 | 189 | 0.15 | 0.02 | 18.5 | 2.8 |

Figure 4.14: ROXIE input-file of the MQXF_V2 return end, inner layer.

Block Data 2D

| No | Type | NCab | X | Y | α | Current | Cable name | N1 | N2 | Imag | Turn | Ne |
|---|---|---|---|---|---|---|---|---|---|---|---|---|
| 1 | Cos | 10 | 75 | 1.6762 | 0.3994 | -15310 | XF145HTH5 | 1 | 1 | 0 | 0 | 1 |
| 2 | Cos | 6 | 75 | 15.6684 | 4.3932 | -15310 | XF145HTH5 | 1 | 1 | 0 | 0 | 2 |
| 3 | Cos | 1 | 75 | 28.5 | 24 | -15310 | XF145HTH5 | 1 | 1 | 0 | 0 | 3 |
| 4 | Cos | 4 | 75 | 29.8946 | 24.3994 | -15310 | XF145HTH5 | 1 | 1 | 0 | 0 | 4 |
| 5 | Cos | 10 | 75 | 0.28648 | 0 | 15310 | XF145HTH5 | 1 | 1 | 1 | 90 | 1 |
| 6 | Cos | 6 | 75 | 14.255 | 3.9938 | 15310 | XF145HTH5 | 1 | 1 | 1 | 90 | 2 |
| 7 | Cos | 1 | 75 | 22.8184 | 6.3901 | 15310 | XF145HTH5 | 1 | 1 | 1 | 90 | 3 |
| 8 | Cos | 4 | 75 | 28.5 | 24 | 15310 | XF145HTH5 | 1 | 1 | 1 | 90 | 5 |

More options :

| No | String | N/a | N/a |
|---|---|---|---|
| | | | |
| | | | |
| | | | |

Block Data 3D

| No | Type | β | Bo | zo | Wi | Wo | Hwed | Horder |
|---|---|---|---|---|---|---|---|---|
| 1 | Diff. Geometry f | 71 | 2 | 382 | 0.35 | 0.02 | 18.5 | 3.2 |
| 2 | Diff. Geometry f | 76 | 1.9 | 353 | 0.17 | 0.02 | 18.5 | 2.8 |
| 3 | Diff. Geometry f | 78 | 1.85 | 344 | 0.17 | 0.02 | 18.5 | 2.5 |
| 4 | Diff. Geometry f | 80 | 2.5528 | 330 | 0.15 | 0.02 | 18.5 | 2.8 |
| 5 | Diff. Geometry f | 80 | 2.2305 | 330 | 0.15 | 0.02 | 18.5 | 2.5 |

Figure 4.15: ROXIE input-file of the MQXF_V2 lead end, inner layer.

Block Data 2D

| No | Type | NCab | R | φ | α | Current | Cable name | N1 | N2 | Imag | Turn | Ne |
|---|---|---|---|---|---|---|---|---|---|---|---|---|
| 1 | Cos | 16 | 94.313 | 0.22782 | 0 | 15310 | XF145HTH5 | 2 | 20 | 0 | 0 | 1 |
| 2 | Cos | 8 | 94.313 | 19 | 20.8049 | 15310 | XF145HTH5 | 2 | 20 | 0 | 0 | 2 |
| 3 | Cos | 4 | 94.313 | 27.8448 | 23.9999 | 15310 | XF145HTH5 | 2 | 20 | 0 | 0 | 3 |

More options :

| No | String | N/a | N/a |
|---|---|---|---|
| | | | |
| | | | |
| | | | |

Block Data 3D

| Ne | Type | β | Bo | zo | Wi | Wo | Hwed | Horder |
|---|---|---|---|---|---|---|---|---|
| 1 | Diff. Geometry f | 68.5 | 2 | 249 | 0.2 | 0.02 | 18.5 | 3 |
| 2 | Diff. Geometry f | 76 | 1.9 | 189 | 0.15 | 0.02 | 18.5 | 2.8 |
| 3 | Diff. Geometry f | 79 | 1.804 | 171 | 0.15 | 0.02 | 18.5 | 2.3 |

Figure 4.16: ROXIE input-file of the MQXF_V2 return end, outer layer.

Block Data 2D

| No | Type | NCab | R | φ | α | Current | Cable name | N1 | N2 | Imag | Turn | Ne |
|---|---|---|---|---|---|---|---|---|---|---|---|---|
| 1 | Cos | 15 | 94.313 | 0.22782 | 0 | -15310 | XF145HTH5 | 1 | 1 | 0 | 0 | 1 |
| 2 | Cos | 1 | 94.313 | 16.9004 | 5.9907 | -15310 | XF145HTH5 | 1 | 1 | 0 | 0 | 2 |
| 3 | Cos | 8 | 94.313 | 19 | 20.8049 | -15310 | XF145HTH5 | 1 | 1 | 0 | 0 | 3 |
| 4 | Cos | 4 | 94.313 | 27.8448 | 23.9999 | -15310 | XF145HTH5 | 1 | 1 | 0 | 0 | 4 |
| 5 | Cos | 15 | 94.313 | 1.3329 | 0.3994 | 15310 | XF145HTH5 | 1 | 1 | 1 | 90 | 1 |
| 6 | Cos | 1 | 94.313 | 19 | 20.8049 | 15310 | XF145HTH5 | 1 | 1 | 1 | 90 | 2 |
| 7 | Cos | 8 | 94.313 | 20.1054 | 21.2043 | 15310 | XF145HTH5 | 1 | 1 | 1 | 90 | 3 |
| 8 | Cos | 3 | 94.313 | 28.9526 | 24.3993 | 15310 | XF145HTH5 | 1 | 1 | 1 | 90 | 6 |
| 9 | Cos | 1 | 94.313 | 32.4347 | 25.5974 | 15310 | XF145HTH5 | 1 | 1 | 1 | 90 | 5 |

More options :

| No | String | N/a | N/a |
|---|---|---|---|
| | | | |
| | | | |
| | | | |

Block Data 3D

| Ne | Type | β | Bo | zo | Wi | Wo | Hwed | Horder |
|---|---|---|---|---|---|---|---|---|
| 1 | Diff. Geometry f | 68.5 | 2 | 380 | 0.2 | 0.02 | 18.5 | 3 |
| 2 | Diff. Geometry f | 72 | 1.95 | 364 | 0.15 | 0.02 | 18.5 | 2.9 |
| 3 | Diff. Geometry f | 76 | 1.9 | 328 | 0.15 | 0.02 | 18.5 | 2.8 |
| 4 | Diff. Geometry f | 80 | 1.788 | 310 | 0.15 | 0.02 | 18.5 | 2.3 |
| 5 | Diff. Geometry f | 80 | 2.042 | 310 | 0.15 | 0.02 | 18.5 | 2.5 |
| 6 | Diff. Geometry f | 79.9968 | 2.042 | 311.996 | 0.15 | 0.02 | 18.5 | 2.5 |

Figure 4.17: ROXIE input-file of the MQXF_V2 lead end, outer layer.

### 4.2.2 Study of Integrated Field Harmonics

The objective is to have an integrated multipole content lower than the random components, defining the integrated multipole content as

$$\bar{b}_n = \frac{\int_{-\infty}^{+\infty} B_n(I,z)dz}{B_2^{ss} l_{mag}(I)}$$

where $B_n$ follows the convention

$$B_y + iB_x = \sum (B_n + iA_n)(x+iy)^{n-1},$$

$B_2^{ss}$ is the main field in the straight section and $l_{mag}$ is the magnetic length defined as

$$l_{mag}(I) = \frac{1}{B_2^{ss}(I)} \int_{-\infty}^{+\infty} B_2(I,z)dz.$$

Integration limits are $\pm\infty$ when providing the total integral of the harmonic content. The contribution of each magnet end is also provided in a separate column in Table 4.5. As it can be observed, even if the integral of $b_6$ over the connection side of 400 mm length is close to 9 units, the total integral is 0.32 units for Q1/Q3 and -0.07 units for Q2a/b. The rest of the harmonics are also summarized in the table, providing the local contribution on the magnet connection side (RE), non-connection side (LE) and the total integral. Only the harmonics where the end contribution is larger than 0.1 units are included in the table.

Table 4.5: Field Harmonics ($R_{ref}$ = 50mm).

| | | Ends | | Integral | |
|---|---|---|---|---|---|
| | Straight part | RE | LE | Q1/Q3 | Q2a/Q2b |
| Magnetic length | -- | 0.400 | 0.341 | 4.2 | 7.15 |
| $b_6$ | -0.640 | 8.943 | -0.025 | 0.323 | -0.075 |
| $b_{10}$ | -0.110 | -0.189 | -0.821 | -0.175 | -0.148 |
| $a_2$ | 0.000 | -31.342 | 0.000 | -2.985 | -1.753 |
| $a_6$ | 0.000 | 2.209 | 0.000 | 0.210 | 0.124 |

### 4.2.3 Magnetic and Physical Lengths

In order to minimize the impact on beam dynamics of the reduction of the nominal gradient from 140 T/m to 132.6 T/m (subsequently changed to 132.2 T/m), the magnetic length has been increased by 200 mm for Q1/Q3 and by 350 mm for Q2a/b. Table 4.6 summarizes the magnetic and physical lengths of the coil, pad and yoke for MQXFS, for the 4.2-m length magnet (Q1/Q3) and for the 7.15-m length magnet (Q2a/Q2b).

Table 4.6: Magnetic length and physical lengths

| Parameters | Units | MQXFS | Q1/Q3 | Q2 |
|---|---|---|---|---|
| Magnetic length at 1.9 K | mm | 1196 | 4200 | 7150 |
| Magnetic length at RT | mm | 1200 | 4213 | 7172 |
| Overall coil length at RT (including splice extension) | mm | 1510 | 4532 | 7482 |
| Magnetic yoke extension at RT | mm | 1552 | 4565 | 7524 |
| Magnetic pad extension at RT | mm | 975 | 3988 | 6947 |
| Cable length per coil | m | 126 | 431 | 721 |
| Cable unit length (including winding margin) | m | 150 | 455 | 750 |

Changing the number of iron yoke laminations on the magnet's ends, the magnetic length will be affected.
The maximum increase of magnetic length we could achieve is around 18 mm (peak field enhancement on the coil ends of 0.4 T). The maximum decrease of magnetic length is around 23 mm (Figure 4.18).

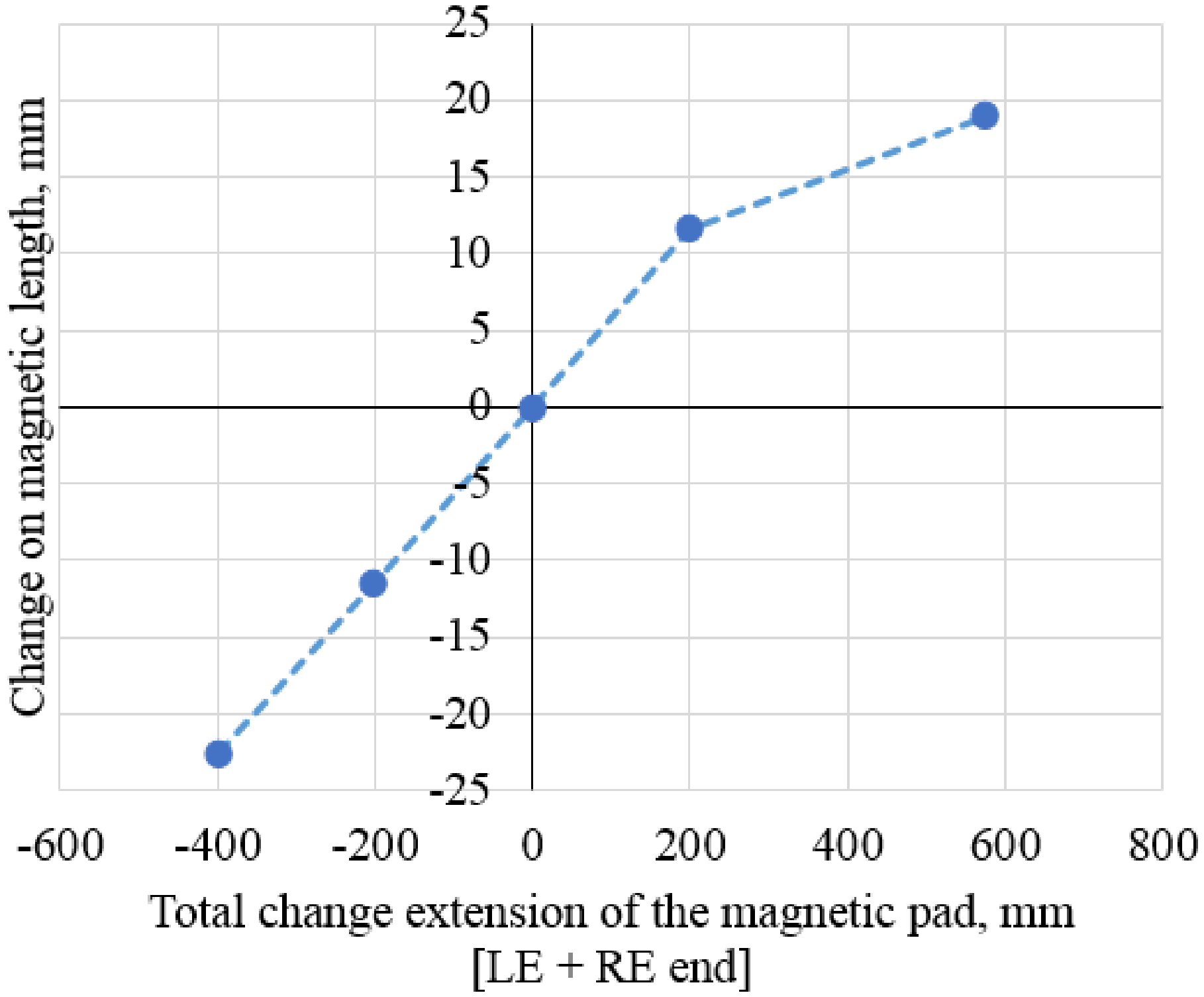

Figure 4.18: Magnetic length change respect to the extension of the magnetic pad change.

### 4.2.4 Fringe Field

The maximum field in a circular path 20 mm from the outer surface of the cryostat will be 0.0068T (Figure 4.19).

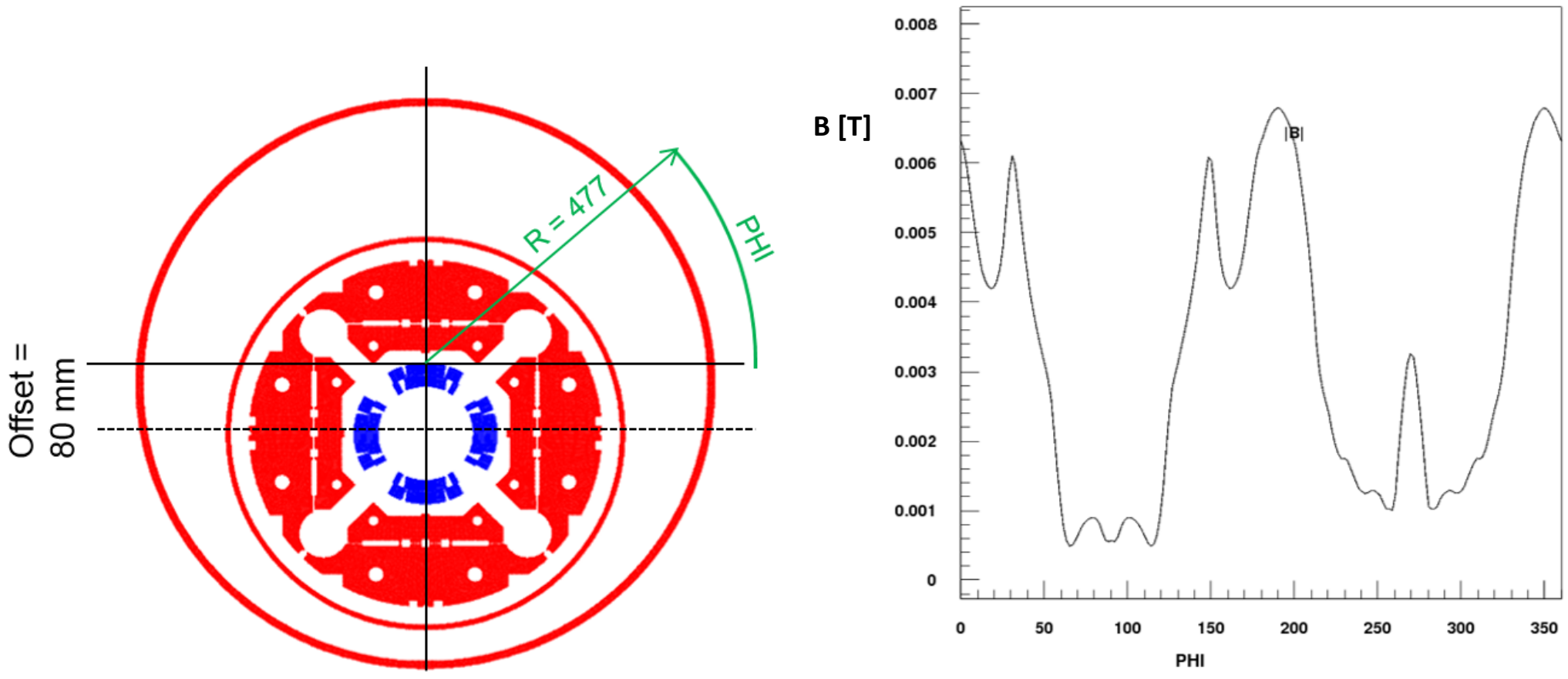

Figure 4.19: Magnetic field around the cryostat at 20 mm (function of the angle).

## 4.3 Structural Design

### 4.3.1 Design features and goals

The support structure of MQXF relies on an aluminum shell pre-stressed at room temperature with bladders and interference keys (i.e. bladder and key technology), which has been demonstrated in the previous successful series of LARP magnets such as HQ. The cross section of the structure of MQXF is a direct scale-up from the HQ models which featured a 120 mm aperture [1]. As shown Figure 4.20, the MQXF quadrupoles feature an aperture of 150 mm and provides a nominal field gradient of 132.2 T/m by utilizing $Nb_3Sn$ superconductor over a magnetic length of 4.2 m (MQXFA) and 7.15 m (MQXFB) at cold [2].

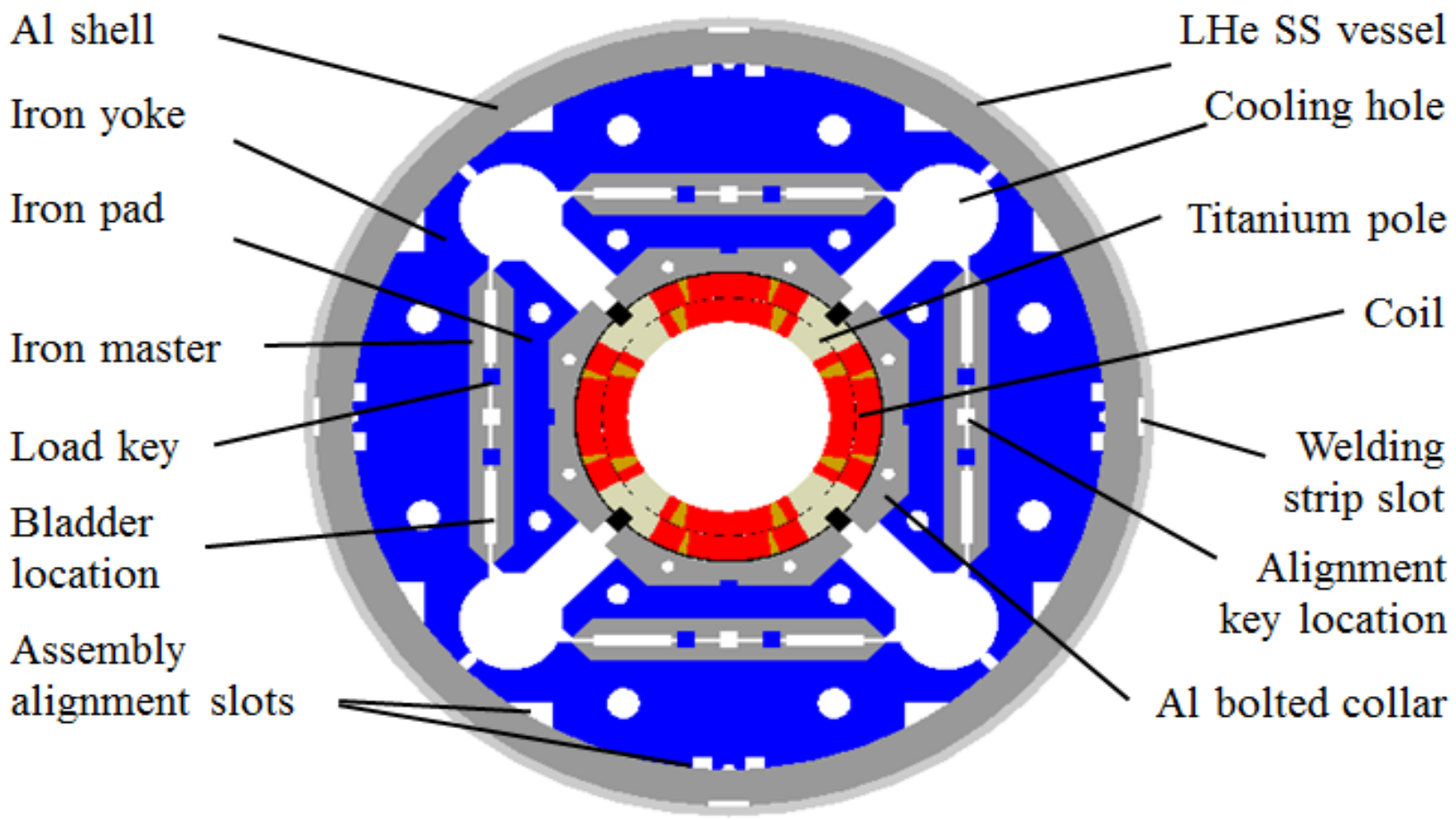

Figure 4.20: Cross section of the MQXF.

The design of the structure comprises an iron yoke assembly surrounded by a 29 mm thick aluminum shell, four iron pads, and the coil-collar subassembly which consists of four aluminum collars bolted around the coils and the G11 pole alignment keys. The yoke, pads and collars are made of thin laminations assembled with tie rods. Between each pad and yoke, the master package on each quadrant contains two interference keys to balance the azimuthal tension in the outer shell with the azimuthal compression in the inner coils.

Maintaining contact between the coils and poles pieces at all stages is achieved by azimuthal load applied on the shell: it relies on a system of water-pressurized bladders and keys to pre-compress the coil-pack and to pre-tension to aluminum shell at room temperature. During the pre-load operation, the pressurized bladders open up the master package and allow inserting the load keys with shims of the designed thickness, thus creating an interference between the coil-pack and the shell-yoke sub-assembly. The final pre-load is achieved during the cool-down phase, when the tensioned aluminum shell increases its stress because of its high thermal contraction.

In operational conditions the MQXFA magnet will experience a total Lorentz force in the axial direction (Z direction, parallel to the magnet's bore) of the order of 1.17 MN at 16.47 kA (nominal current plus margin). Axial pre-stress is therefore designed to withstand the total axial forces generated by the coil ends. Four tensioned steel rods within the pads' gaps are connected to endplates to provide the axial pre-stress. Similarly to the azimuthal pre-stress, the room temperature axial pre-stress is tuned to counteract after cool-down the axial Lorentz force (see Sect. 4.3.3.3 3D

Results).

To summarize, the main features of the MQXF structural design are:

(1) Shell-based support structure relying on “bladder and key” technology to perform azimuthal pre-loads, which allows reversible assembly process and tunable preload;
(2) G11 alignment keys inserted into the pole pieces provide coil alignment by assuring azimuthal contact between coil and collars after cool-down;
(3) The aluminum shell provides additional pre-load to the coil during cool-down;
(4) Axial pre-load is provided by four SS rods and two end-plates.

### 4.3.2 Graded Approach of Structural Analysis

Structural failure can occur via one of the followings:

a) Plastic collapse typically associated with "tough" materials that yield in a smooth manner under the influence of large loads;

b) Linear elastic fracture, typically associated with brittle materials under significant loads coupled with stress concentration factors such as defects or voids;

c) Ductile tearing, i.e. materials subjected to a combination of the elements above.

MQXFA structural designs account the failures above by referencing ASME FFS-1 as a standard to accept use of nominally ‘brittle’ materials with assumed flaws. A graded approach [3] which is expected to yield structural designs that are safe for operation in the Large Hadron Collider is developed and used in MQXFA designs, shown in Figure 4.21, wherein the design criteria are evaluated using consecutively more advanced and detailed analysis as the components and load cases are found to result in reduced margin with respect to relevant mechanical figures of merit.

Due to the complexity of the magnet design and the various load conditions encountered during fabrication, assembly, and operation, most structural elements of the MQXFA magnet are subjected to 2D and/or 3D FEA. At each grade level criteria are defined which, if FEA results exceeded, will trigger the next level of analysis and/or require modifications to the design.

The graded procedure consists of four levels analysis from ‘hand calculations’ to FEA, advance FEA to LEFM—in order of complexity. MQXFA structural design uses FEA for every structural component; In the grade I and II analysis, the FEA results will be evaluated with the material properties, parts could be reported to be satisfactory when the results meet Von Mises criteria; Grade III will be required when the results show stress singularity or concentration, or any other cases that exceed the Von Mises criteria. Grade IV analysis will be triggered when the material is brittle and indicates high stress in the structure. Associate with grade III and IV analysis, mitigations were performed as well to ensure the design is deemed appropriate.

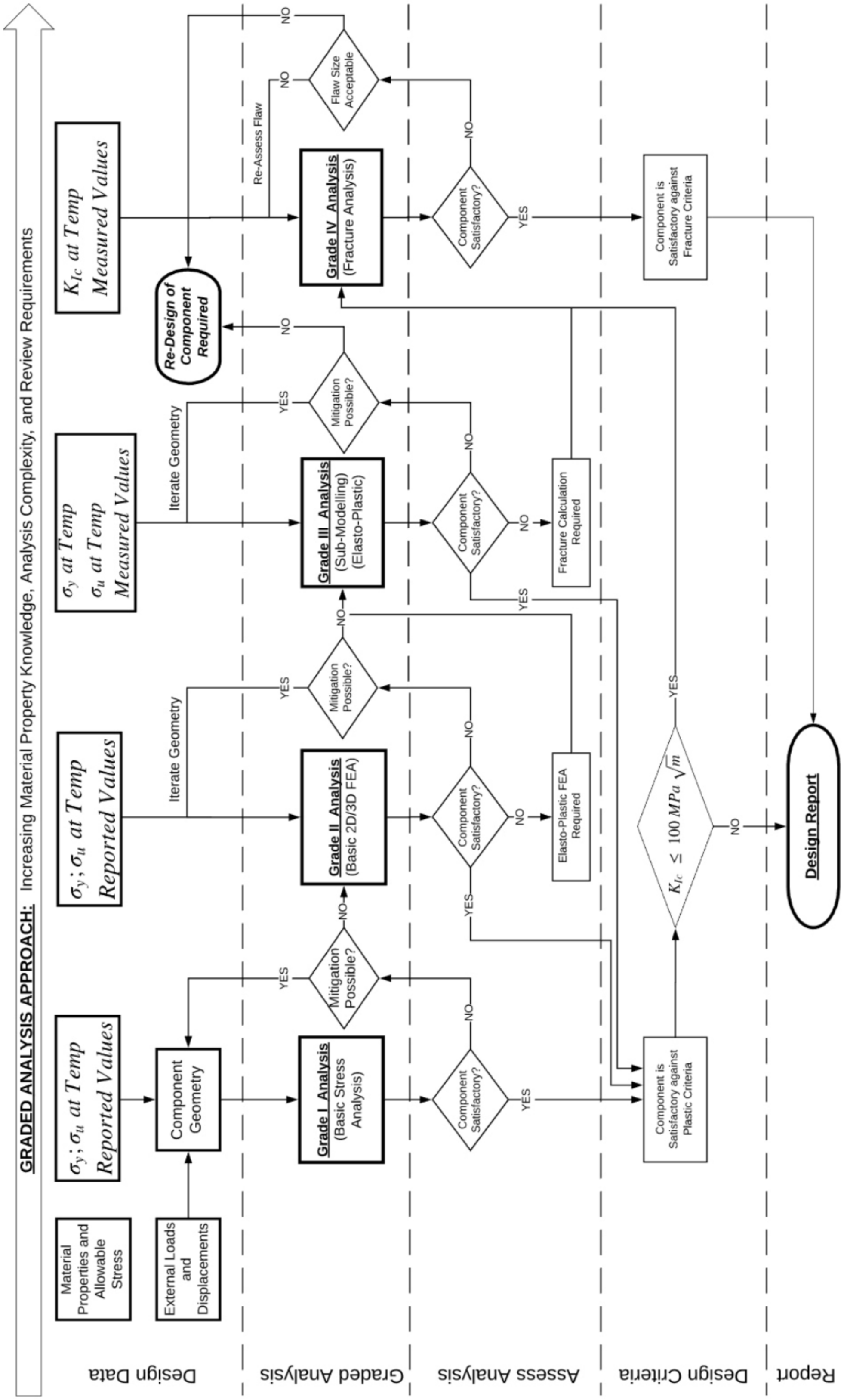

Figure 4.21: Schematic of Graded Approach to Analysis Procedures.

### 4.3.3 Grade II Analysis

#### 4.3.3.1 Model Description

Standard ANSYS 2D/3D finite element models are used in grade II analysis. The ANSYS models' parameters are defined based on the experiences of the previous MQXF magnets. The models are octant symmetric azimuthally. Each coil block, including cables, epoxy resin and fiberglass, is modeled as a homogenous object with average properties determined in [2]. The 2D model, as shown in Figure 4.22, employs high order 8-node element (plane183) for the main structures; uses elements of CONTA172 and TARGE169 to model the contact interfaces. The half-length 3D model (Figure 4.23) uses higher order 20-node elements (SOLID186) for all volumes. Similarly, 3D contact elements CONTA174 and TARGE170 with similar contact parameters are used in the 3D model. Every part inside a coil, in either 2D or 3D models, is bonded at all steps; frictional contact with frictional coefficient 0.2 was used at each interface of all the other components.

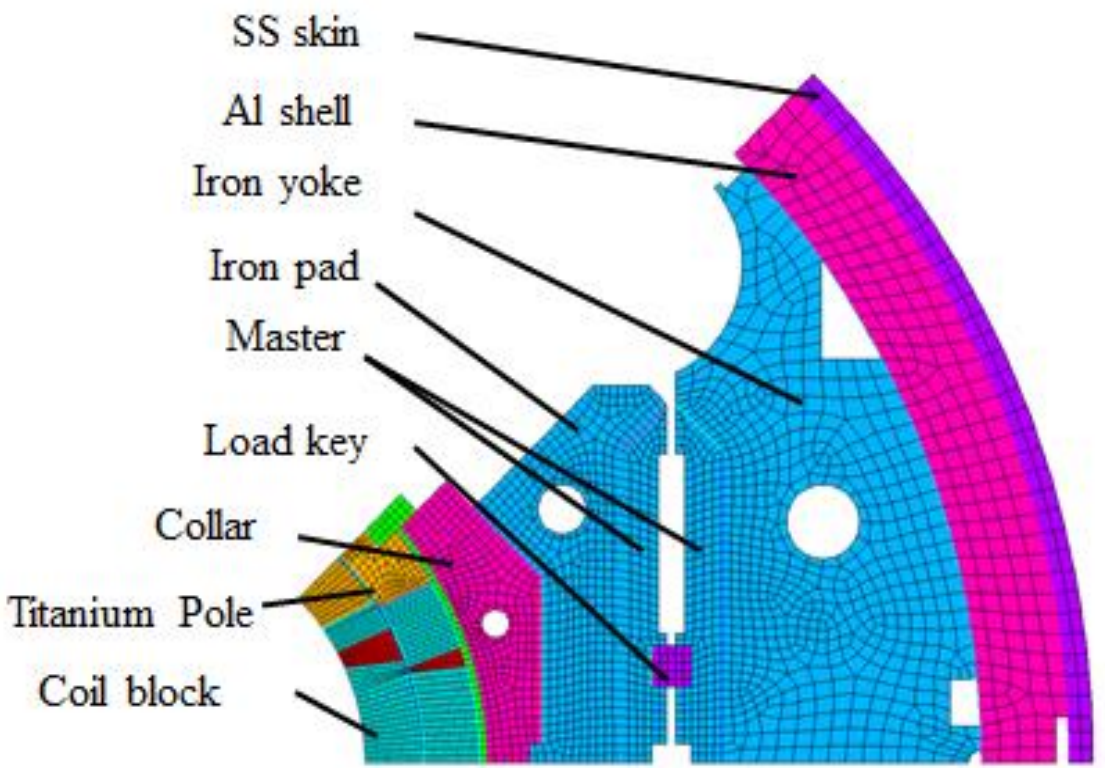

Figure 4.22: Mesh view of the 2D octant model (materials are displayed by colors).

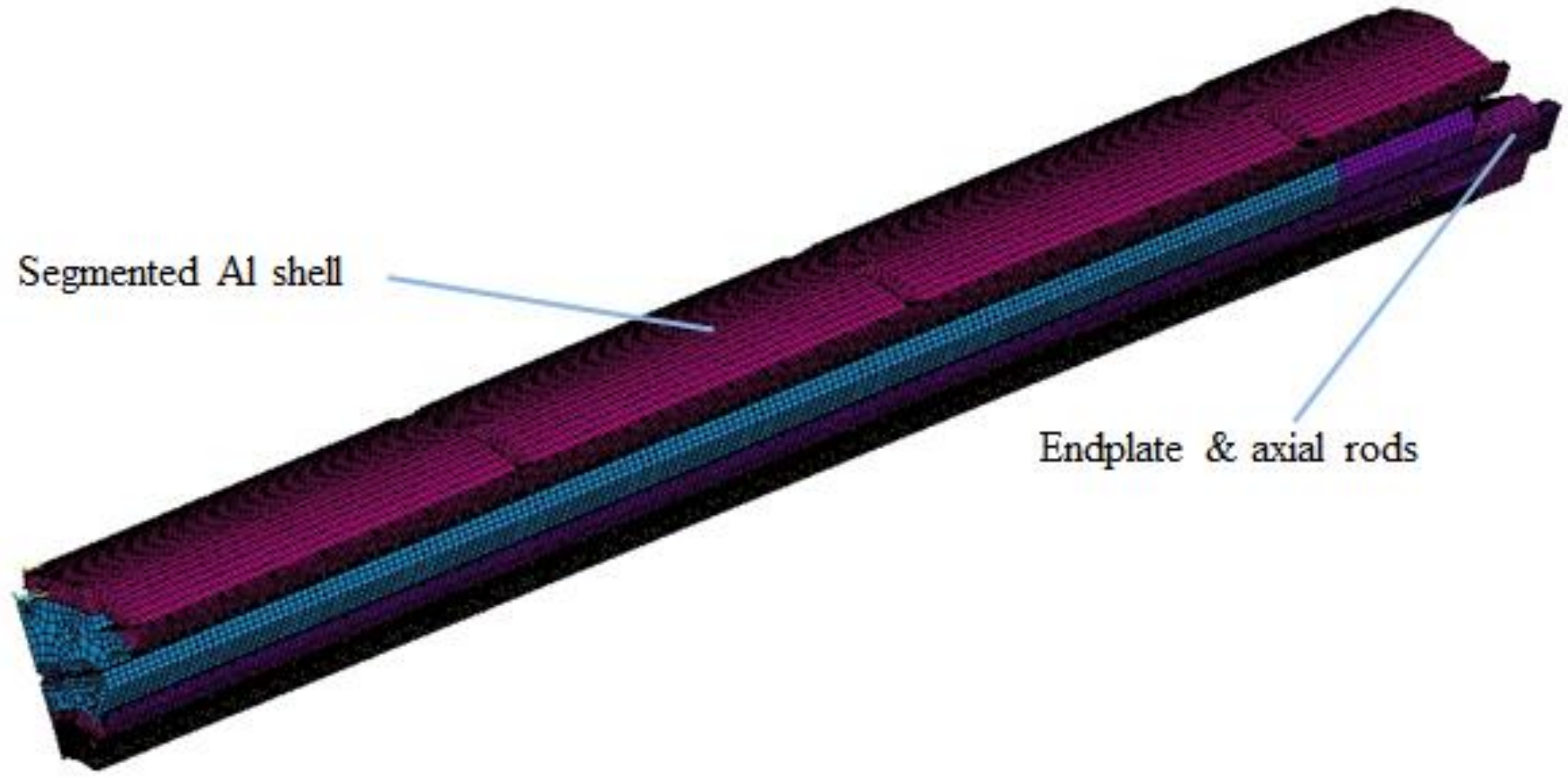

Figure 4.23: Mesh view of the 3D model (materials are displayed by colors).

Material properties used in the MQXFA design are shown in Table 4.7. The operation process was simulated by the following four steps as defined in [3] for MQXFAP1 magnet:

A. Room-temperature load (load 1b) --- Azimuthal and axial pre-load at room temperature; The models utilize contact offset between load key / pad-master contact interface to simulate the azimuthal preload shims, and 3D model applies displacement on rod to simulate axial preload.

B. Cool-down to 1.9 K: the temperature of all solids was changed from 300 K to 1.9 K.
C. Operation (nominal current plus margin: 16.47 kA): Import the coil from ROXIE to opera, and then compute forces in opera and import from opera to ANSYS.
D. Operation (ultimate current plus margin: 17.89 kA): scale the imported Lorentz force to the level of ultimate current plus margin.

Table 4.7 Material Properties used in FE models

| Materials | E at 293K | E at 4.2K | Poisson Ratio | Integrated thermal contraction (293K-4.2K) |
|---|---|---|---|---|
| | GPa | GPa | | |
| Coil | 20 | 20 | 0.3 | 3.88E-03 |
| Aluminum bronze | 110 | 120 | 0.28 | 3.24E-03 |
| Aluminum 7075T6 | 70 | 79 | 0.34 | 4.20E-03 |
| Iron | 213 | 224 | 0.28 | 2.00E-03 |
| G11 (normal direction) | 10 | 10 | 0.3 | 7.30E-03 |
| G11 (layer direction) | 15 | 15 | 0.3 | 2.44E-03 |
| Nitronic 50 | 210 | 225 | 0.28 | 2.60E-03 |
| Stainless Steel | 193 | 210 | 0.28 | 2.90E-03 |
| Titanium | 115 | 125 | 0.3 | 1.70E-03 |

The graded analyses shown here use the MQXFAP1 design conditions. In design of MQXFAP1 the azimuthal interference is 640 µm, and the pre-tension on each rod is 580 µε. (MQXFAP2 used 750 µm based on the test performance of MQXFAP1).

The computations in both the 2D and 3D (discussed in next section) models were carried out in light of the structural design criteria [3] and superconducting elements design criteria [4] (***detailed definition of the design procedures and calculation approaches can be found in [3]and [4], this report mainly focuses on the results computed in the work frame defined by the design criteria***):

1) Limit the coil peak stress to 120 MPa at room temperature and 200 MPa at 1.9 K;
2) Coil axial stain remains below 0.2% with axial pre-load >50% of the full axial Lorentz force.
3) The coil pre-load is selected to maintaining contact between the coils and poles pieces up to the operating gradient. Localized tension up to 20 MPa at the pole-coil interface is acceptable provided that average compression is maintained across the pole width.
4) Maintain the stress in the support structure components within the material limits (discussed in next section).

#### 4.3.3.2 2D Results

Stresses in the coils have been computed and displayed in a cylindrical coordinate system. Azimuthal stress distribution is checked in the coil as shown in Figure 4.24. The coils are compressed from the bladder operation to cool-down; as mentioned above, the inner and outer layers at the coil/pole interface show an average tension of about 22 MPa when energized to the 16.47 kA (nominal current plus margin). The maximum $\sigma_\theta$ in coil is -94 MPa at room temperature and -158 MPa after cool-down. Coil stress from the 2D computation is within the limit; more detailed coil stress will be further discussed in next section (*4.3.3.3 3D* Results).

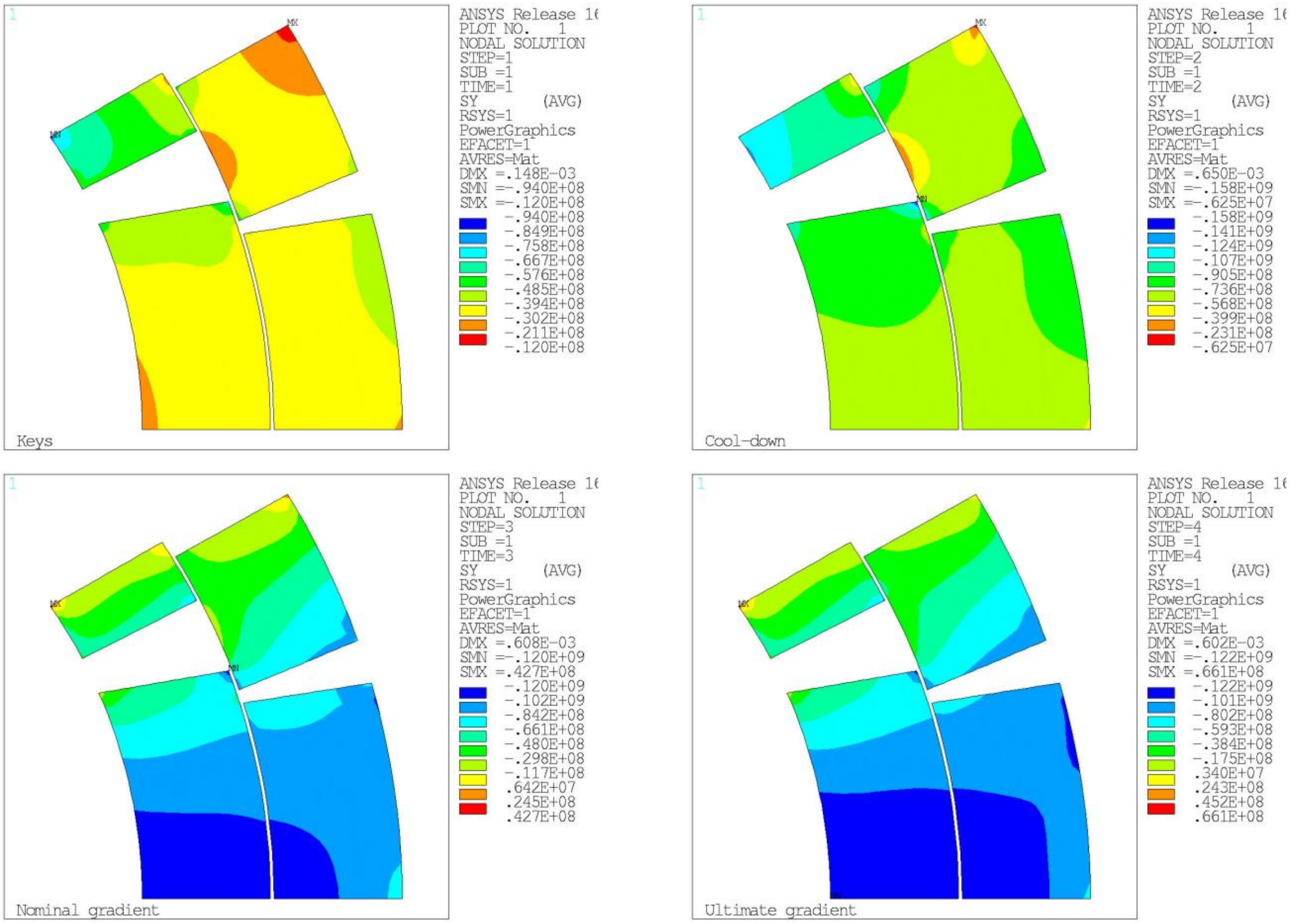

Figure 4.24: Azimuthal stress distribution of MQXFAP1.

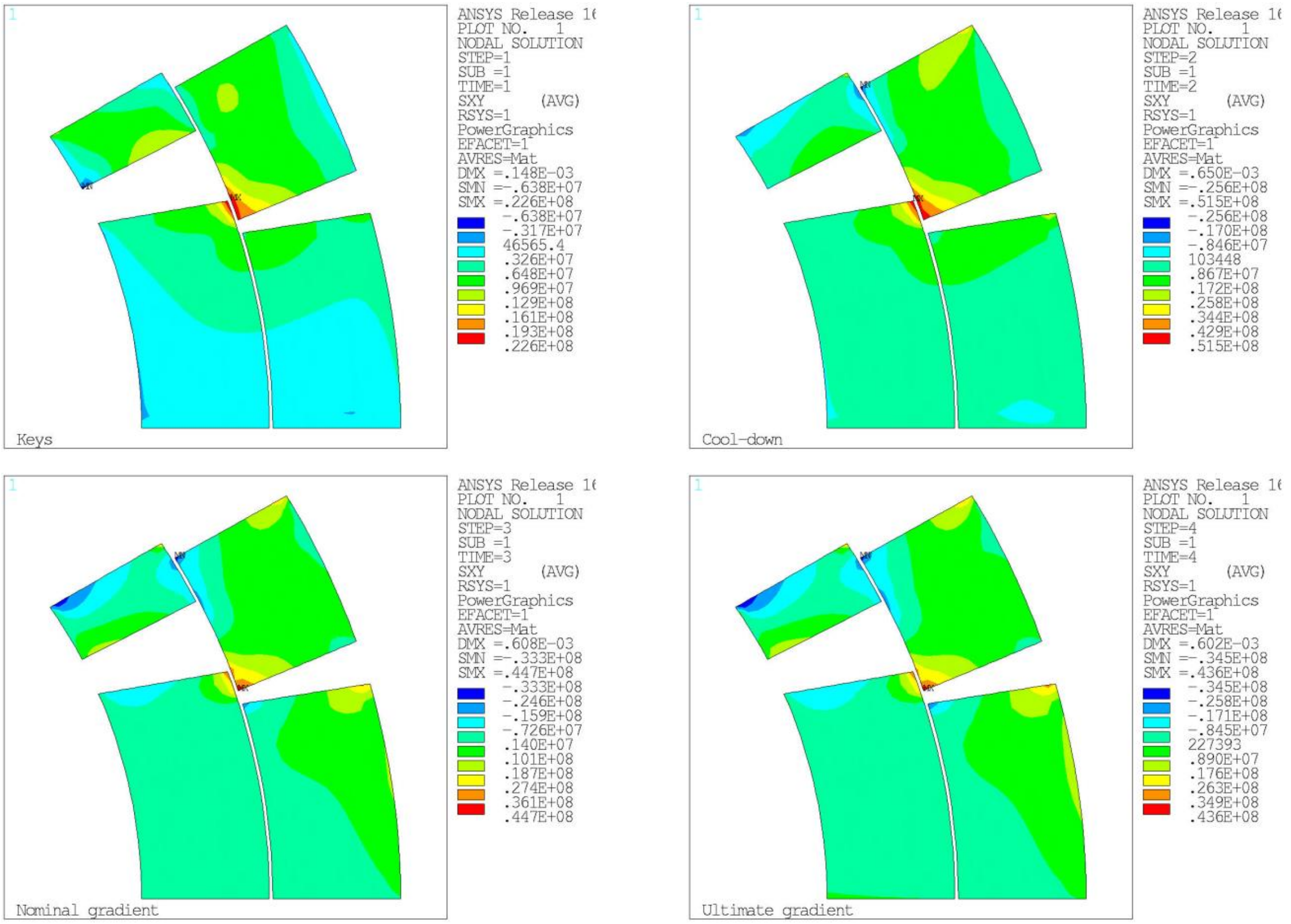

Figure 4.25: Shear stress distribution of MQXFAP1.

Shear stress in the coils after cooldown varies in the ± 25 ~ 35 MPa range with the exception of a few corners (the outer corner of layer 1 pole turn and the inner corner of layer 2 pole turn, (Figure 4.25). A peak shear stress of 51 MPa is observed after cool-down at the corners on the coil/wedge interface. According to measurements performed at 77 K [5], the shear stress inside coils is in the acceptable range.

Detailed analysis on the support structure (metallic parts) was performed with 3D analysis.

#### 4.3.3.3 3D Results

The 3D model applies the same azimuthal interference of 640 μm as the baseline. Axial preload is provided by pre-tensioning to 580 με on the axial steel rods.

The coil azimuthal stress over load steps is shown in Figure 4.26. Peak coil stress usually appears on the ends of the coil straight section. Compared with the 2D solution, the coil azimuthal stress reaches the maximum compression of −102 MPa at room temperature after the bladder operations; and −165 MPa at 1.9 K. When the coils are energized, the location of the peak coil stress moves from the pole-turn to the mid-plane, and the magnitude reduces to about -125 MPa.

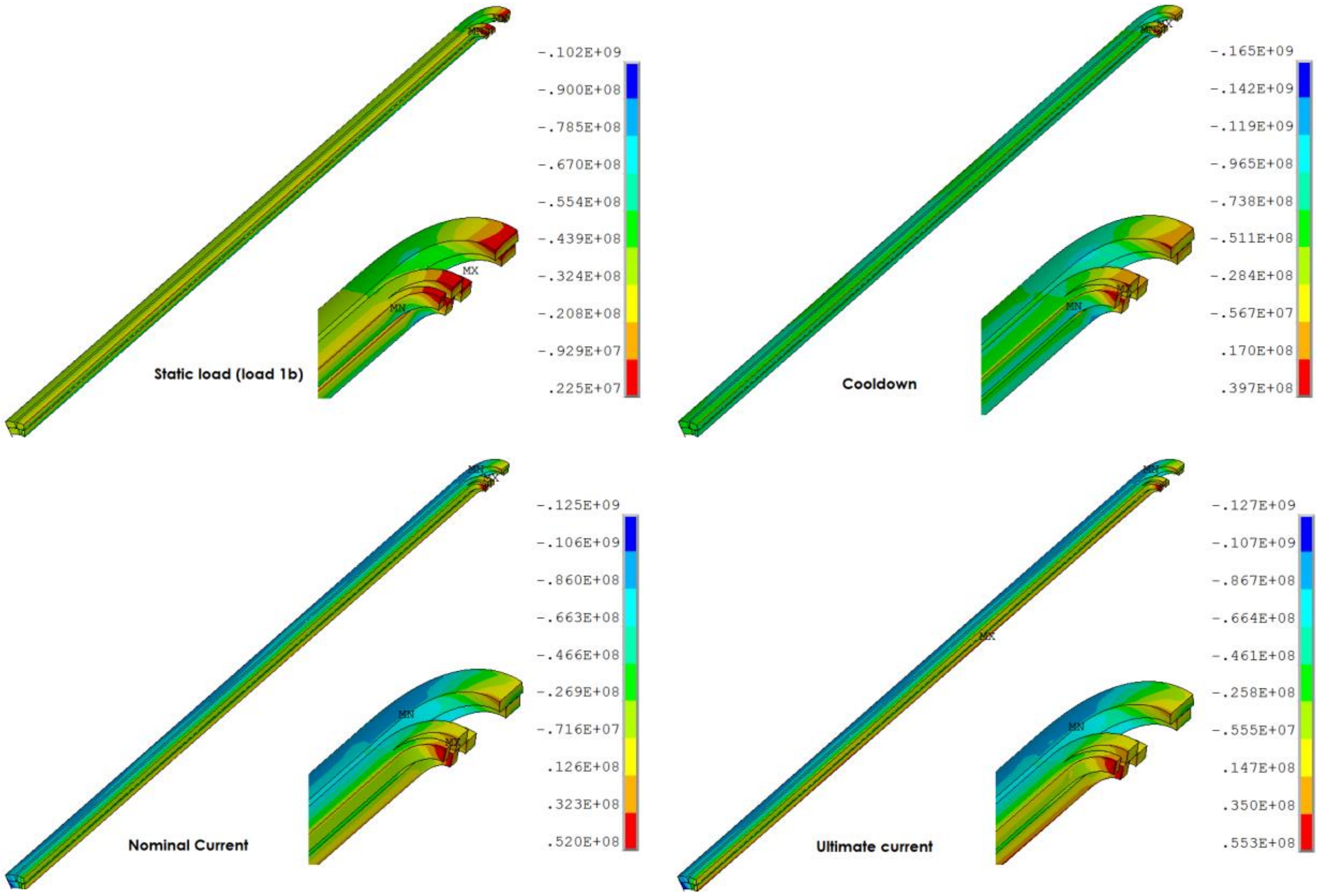

Figure 4.26: Coil azimuthal stress distributions (Unit: Pa)

The axial load that counteracts the axial Lorentz force of 1.17 MN on the coils is provided by four pre-tensioned 32 mm diameter steel rods and two Nitronic 50 end plates, which prevent the coils from detaching from the pole or end-spacers. Although the rods and plates were sized to accommodate the full axial load (Fz), in the simulation the magnet was axially preloaded to only ~54% of this value following previous experience on the LARP HQ series.

The longitudinal strain in the coil straight section is listed in Table 4.8. It’s dominantly affected by the difference of thermal contraction rate of coil blocks and Titanium poles. The coil stress and longitudinal strain of the baseline case are within the design criteria defined in [4].

Table 4.8 Longitudinal strain on coil straight section

| Load steps | Point 1 | Point 2 | Point 3 |
|---|---|---|---|
| Cooldown | 0.180% | 0.170% | -0.002% |
| Nominal Current + margin (16.47 kA) | 0.183% | 0.176% | 0.162% |
| Ultimate Current + margin (17.89 kA) | 0.184% | 0.177% | 0.191% |

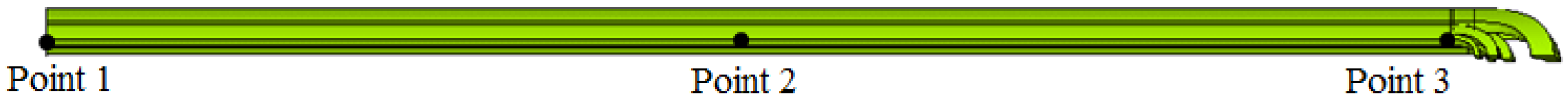

The peak stresses in the metallic parts are listed in Table 4.9. Please note that ARMCO steel is ultra-low carbon steel with very low impurities produced by AK steel, and it was tested at CERN at both room temperature and 4 K. It shows purely brittle behavior at 4 K, thus only principal stress is considered.

From the Table 4.9 most of the structural components are within the yield limits; therefore, the stresses in those parts can be accepted after a successful Grade II analysis is completed. Shell and yoke clearly exceed their yield limits at either room temperature or 1.9 K. The peak stresses in the end shell and yoke are located at the cut-out inner corners, as shown in Figure 4.27, where stress concentrations occur due to the sharp corners in the model.

Table 4.9 Peak stresses in the metallic parts

| Part | Material | Principal Stress (MPa) | | Von Mises Stress (MPa) | | $\sigma_y$ (MPa) | |
|---|---|---|---|---|---|---|---|
| | | 293 K | 1.9 K | 293 K | 1.9 K | 293 K | 1.9 K |
| Collar | Al 7075 [6] | - | - | 121 | 273 | 420 | 550 |
| SS Pad | SS 316 [6] | - | - | 82 | 277 | 289 | 375 |
| Iron Pad | ARMCO [7] | 98 | 152 | - | - | 223 | - |
| Yoke | ARMCO | 246 | 306 | - | - | 223 | - |
| Shell | Al 7075 | 280 | 610 | 320 | 573 | 420 | 550 |
| Endplate | Nitronic 50 | - | - | 137 | 333 | 517 | 1120 |

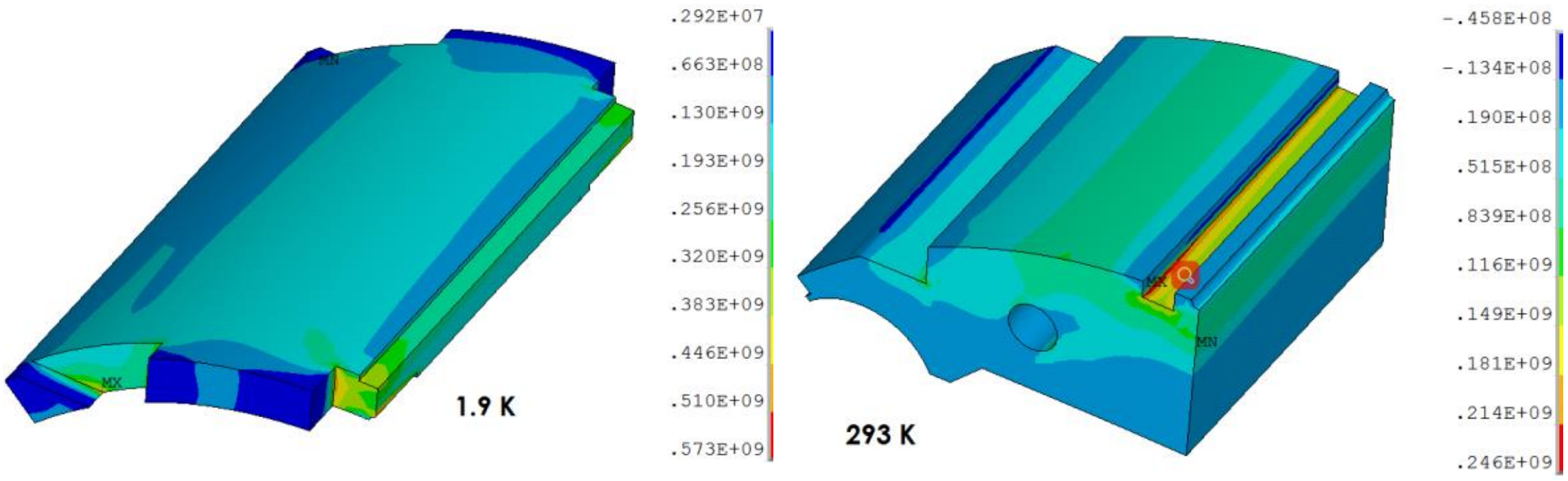

Figure 4.27: Peak Von Mises stress (Pa) in the end shell (Left) and peak principal stress (Pa) in yoke (Right)

The stress concentration cannot easily be resolved by increasing the local mesh density. This usually means that the model needs to include the actual geometric features such as fillets or chamfers. On the other hand, Al 7075 and ARMCO steel are considered as brittle material due to low fracture toughness. Therefore, the end shell and yoke need to be analyzed in Grade III with detailed features and mesh refinements; then, also Grade IV analysis is carried out for fracture assessment.

### 4.3.4 Grade III Analysis

For components exhibiting stress concentrations that cannot be readily resolved via routine mesh refinement studies in the primary FEA model, sub-modeling was employed to evaluate the stress distribution around the particular concerning area.

Sub-model is a separate finite element model of the local region of interest, which imposes displacements on the cutting boundaries from the original model. The basis of a sub-model is St. Venant's principle when picking the cutting boundary locations from the original model.

Use the end shell for instance: Figure 4.28 illuminates the verification of the end shell sub-model. The Von Mises stress on the picked locations of the cutting boundary has compared with the stress on the same locations of the original model. The agreement of the two sets of stress indicates that the cutting boundaries are remote enough to meet St. Venant's principle.

Same verification process has been performed on the yoke sub-model.

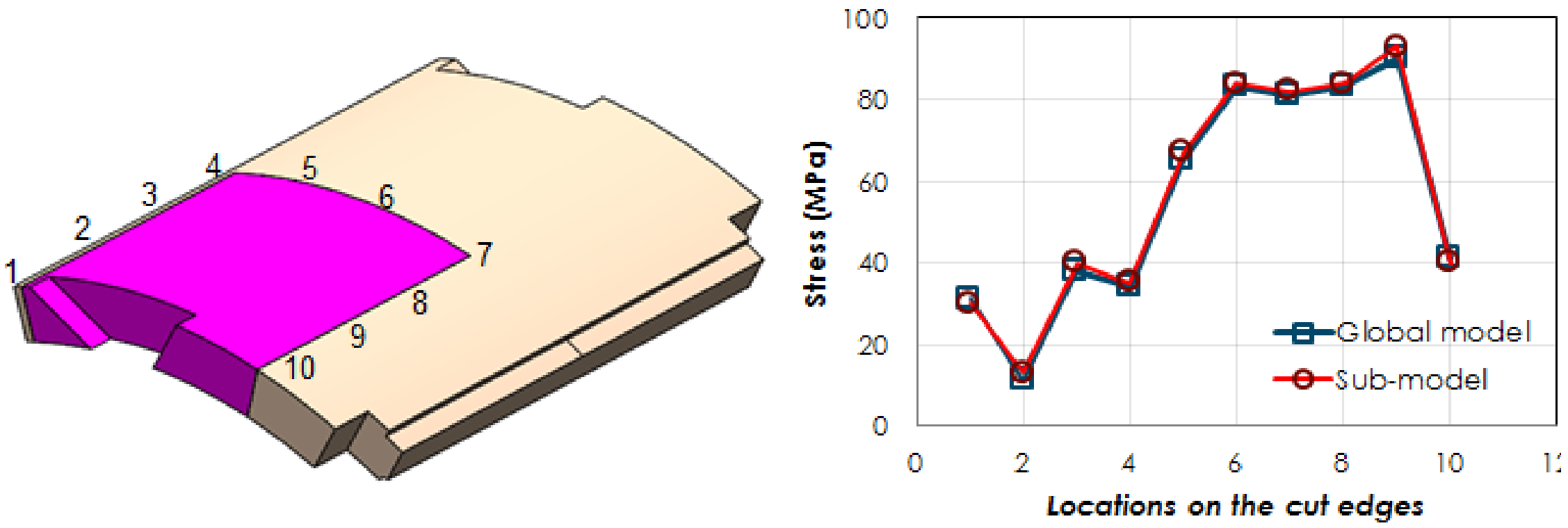

Figure 4.28: End shell sub-model verification

#### 4.3.4.1 4End shell sub-model results

Once the sub-model has been verified, mesh refinements have been performed on the sub-model. There are additional cases with different fillet sizes were performed for the end shell as well.

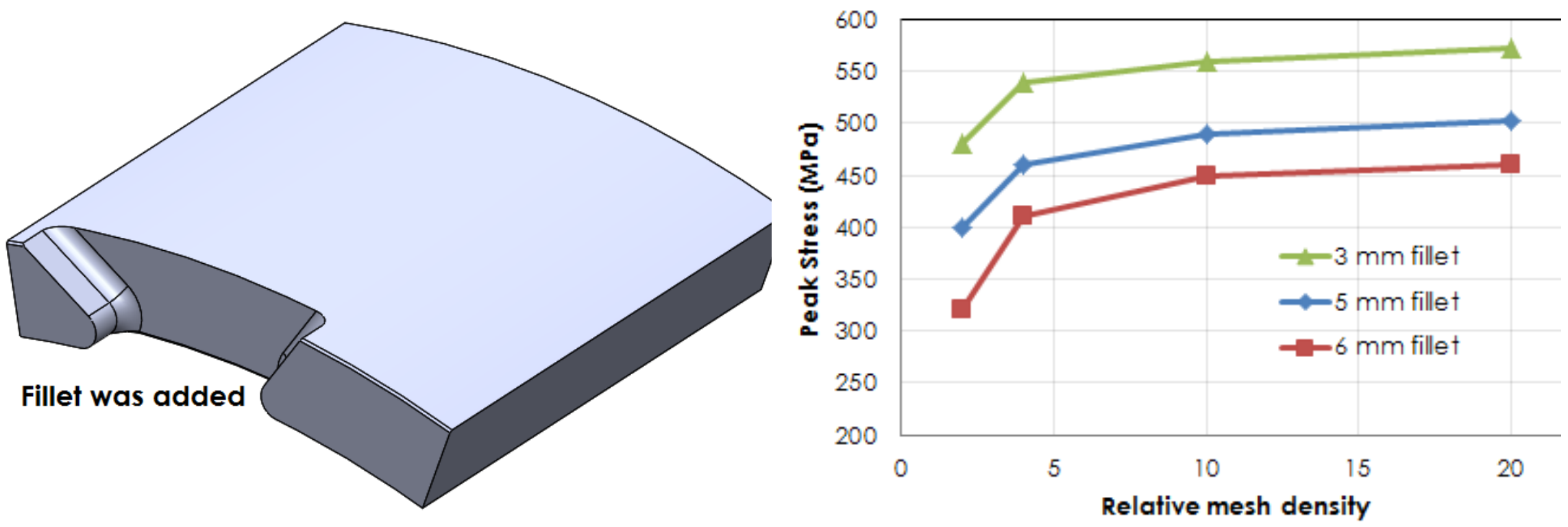

Figure 4.29: End shell sub-model (left) and Peak Von Mises stress at 1.9 K vs. relative mesh density (right)

The initial elements size in the original model was 8 mm. The mesh refinement studies (Figure 4.29) show the peak stress in end shell increases and converges finally with increased the mesh density. Thus, the following studies then used relative mesh density of 10 (element size 0.8 mm). Elasto-plastic study has been performed for the end shell sub-model at 1.9 K to obtain the plastic zone size.

Fillet certainly reduces stress concentrations at the local corner; the peak stress decreases to 490 MPa with 5 mm fillet at 1.9 K, which dropped within the yield limits of 550 MPa for Aluminum 7075.

Although there was no plastic deformation occurred with 5 or 6 mm fillet, Grade IV analysis will be performed due to low fracture toughness of Al 7075.

#### 4.3.4.2 Yoke sub-model results

The yoke is considered brittle at room temperature and 1.9 K according to the test results conducted by CERN [7]. But it still shows yielding point at room temperature. Sub-model was used to determine the plastic deformation and the stress on the likely path if part-through a crack presents itself. Similarly, the yoke sub-model also includes different fillet radiuses (Figure 4.30).

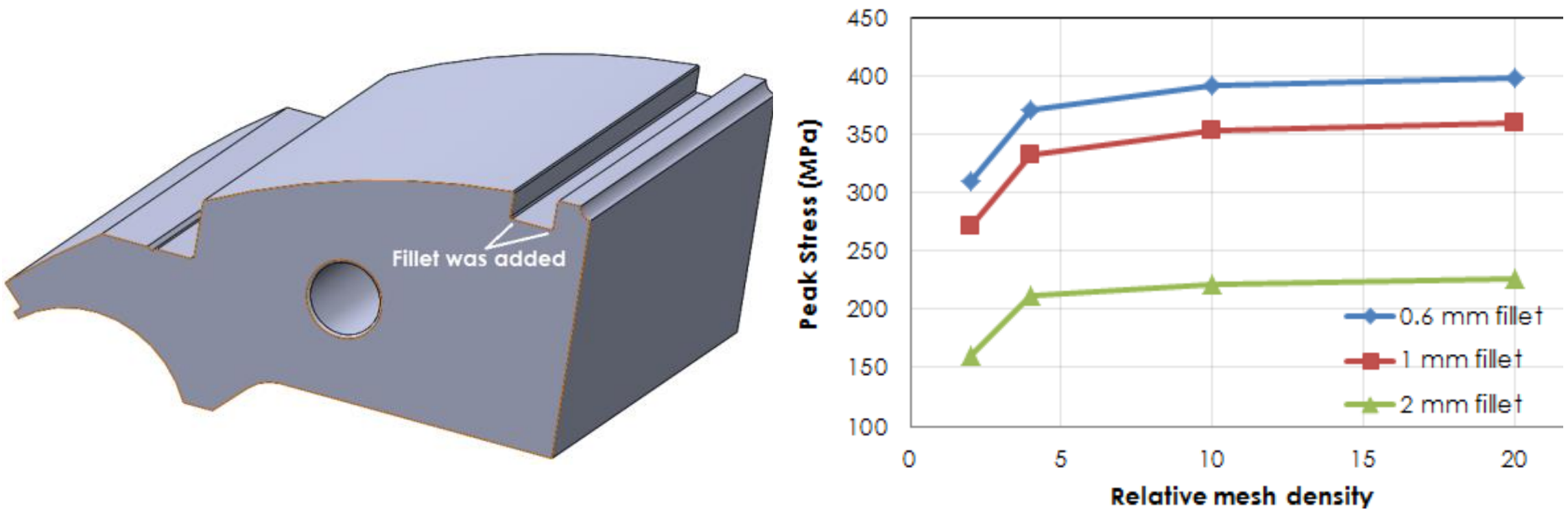

Figure 4.30: Yoke sub-model (left). Peak principle stress at room temperature vs. relative mesh density (right)

The initial elements size of the yoke was 6 mm. As shown in Figure 4.30, the mesh refinement studies show that the peak stress at room temperature in the yoke has similar trends as seen in end shell sub-model. It converges finally once the mesh density larger than 10. Thus, the following studies then used relative mesh density of 10 (element size 0.6 mm). Elasto-plastic study has been performed for the yoke sub-model at room temperature.

Adding a fillet lowers the peak stress at the corner, too; however, plastic deformation still occurs at room temperature. As a fracture dominated material, the yoke will be analyzed in Grade IV to determine the plastic deformation and load factor with fillets.

### 4.3.5 Grade IV Analysis

Grade III analyses for the end shell and yoke show that fillets around the corners are recommended. As defined in the work flow (Figure 4.31), end shell and yoke will be analyzed in Grade IV.

Grade IV analysis will determine if their load factor meet the design criteria with different fillet sizes. The approach in this grade relies on the R6 FAD (Failure Assessment Diagram) [8], captures failure by LEFM (elastic fracture), and plastic collapse simultaneously.

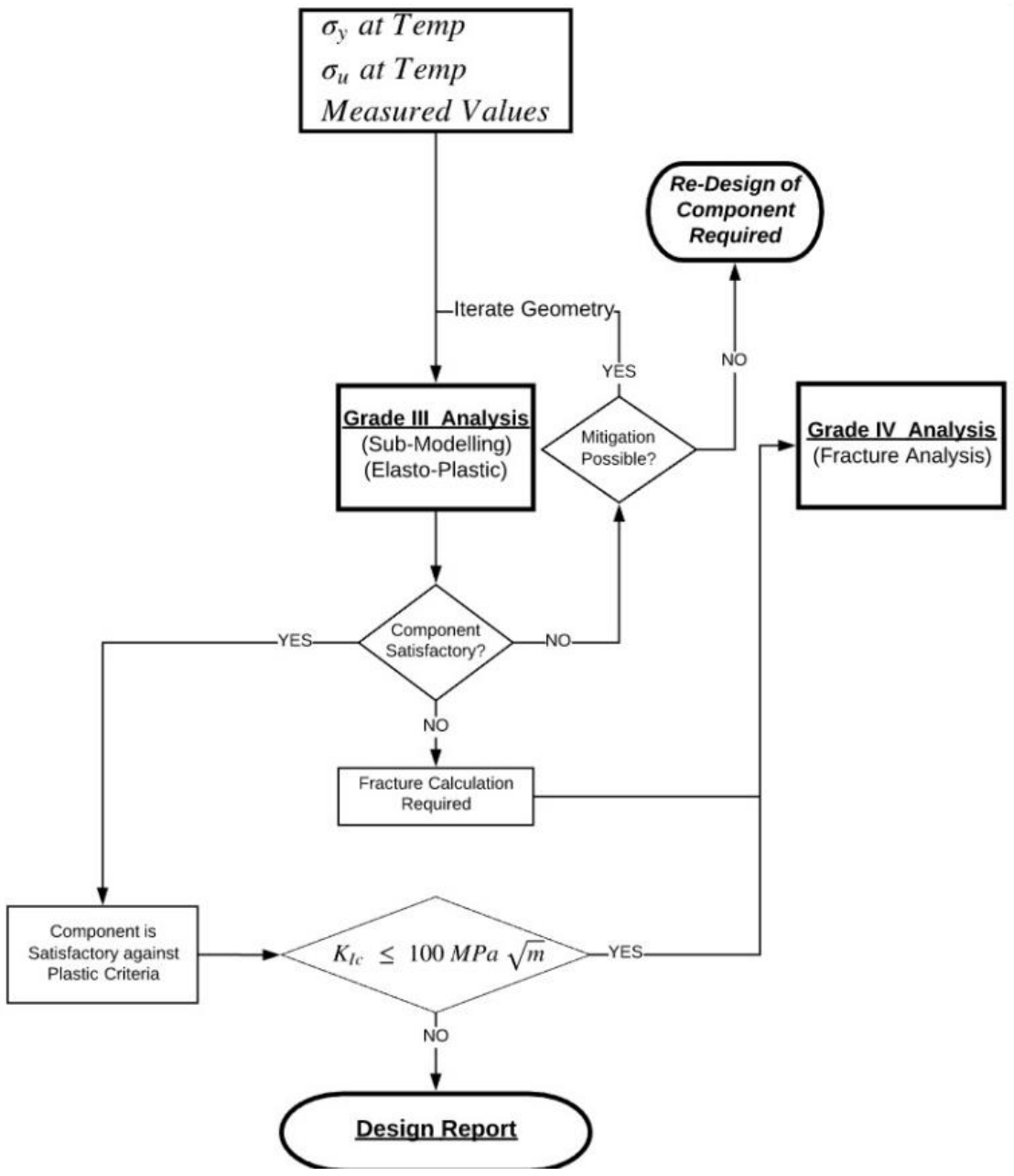

Figure 4.31: Flow chart describing procedure from Grade III to Grade IV

For the purposes of design, semi-elliptic part-through cracks are assumed with flaw features intersecting and centered on the components surface as these typically have the highest stress intensities. The major process of fracture analysis is to determine the applied stress intensity $K_I$. For part-through cracks subject to primary stresses, $K_I$ can be written in the following form:

$$K_I = F\sigma(x)\sqrt{\frac{\pi a}{Q}} \tag{4-1}$$

$$Q = 1 + 1.464\left(\frac{a}{c}\right)^{1.65} \tag{4-2}$$

where $a$ and $c$ are the elliptical radius of a crack, $F$ is geometry constants that can be obtained from FEA or published data. $\sigma(x)$ can be approximated as a cubic expansion of a load profile extracted from an unflawed elastic analysis in the direction of assumed crack propagation through part thickness $x = a$ direction. $F\sigma(x)$ can be approximated as:

$$F\sigma(x) = \sum_{i=0}^{3} G_i A_i x^i \tag{4-3}$$

where $G_i$ is shown in [10] appendix A.2; $A_i$ is the curve-fitting coefficient. The stress on the path of assumed crack propagation has been defined in the end shell and yoke sub-models. The path direction of a crack growth is normal to the max. principal stress and in the most energetic direction (the direction of lowest stress gradient).

### 4.3.5.1 End shell fracture analysis at 1.9 K

The total Von Mises stresses of the end shell with different fillets at 1.9 K , considered on the path as described above, are shown in Figure 4.32. Since the end shell does not yield at room temperature, the analysis focuses on the stress at cold.

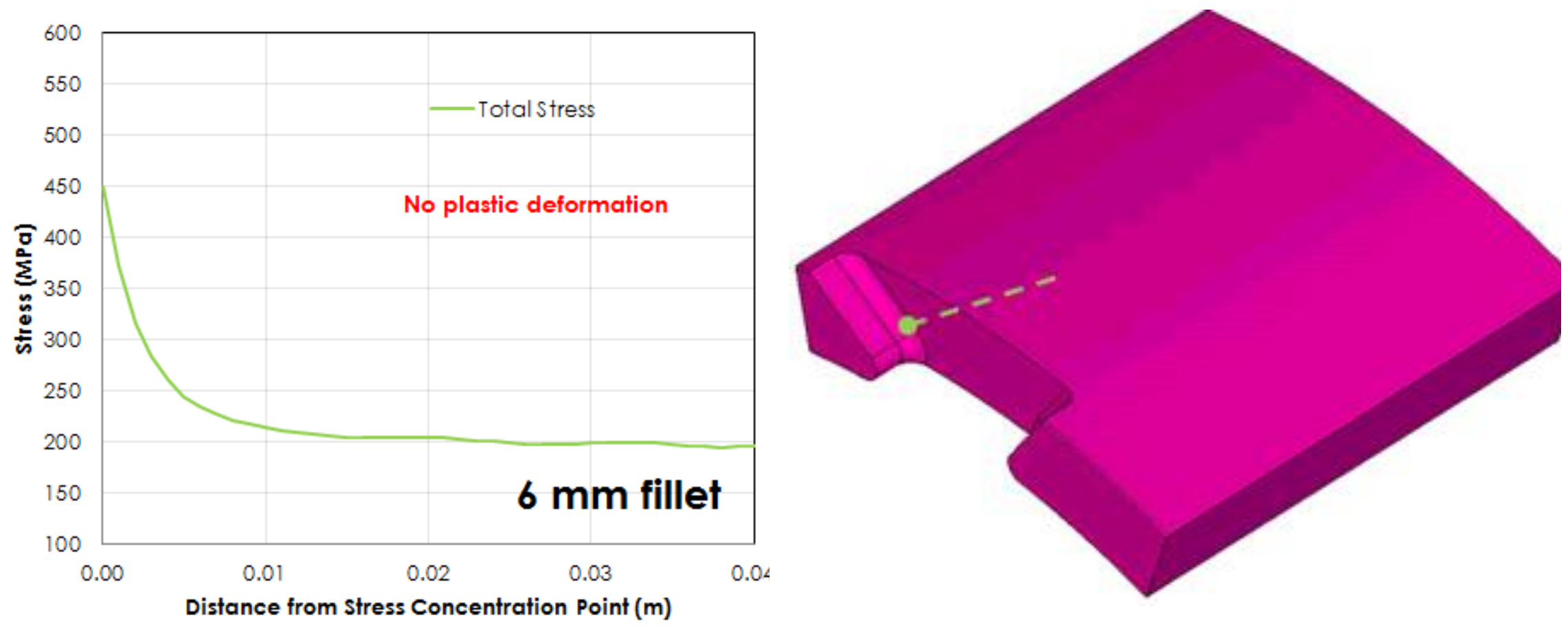

Figure 4.32: Total stress on the path where a crack is most likely to propagate.

The local stress concentration results in local plastic deformations that limit the effective stress state. The fully elastic analysis will coincide with the elasto-plastic model at some distance from the structural discontinuity. The distance to the coincidence is considered as the plastic depth on the defined path. With 3 mm fillet, plastic deformation occurred around the stress concentration area, with a 0.6 mm plastic depth along the path; the peak stress dropped when increasing the fillet radius.

Considering the total stress illustrated in Figure 4.32 and a given crack length, the “load point” of each case can be determined in FAD. Load points inside of the FAD curve are safe from failure, load points falling outside or on the curve constitute a risk of failure. Each load point determines a “load line” by connecting the original point. The load factor in the plot is defined as:

$$Load\ factor = \frac{L}{L'} \tag{4-4}$$

where *L* is the length from the original point to the load point; *L'* is the length from the original point to the "projected load point", which is the intersection of load line and the FAD curve. A flaw is considered critical in size if its Load Factor is equal to one.

The critical flaw sizes at 1.9 K for 5 mm fillet is 2 mm. Accordingly, critical flaw sizes are 1.6 mm and 2.5 mm for 3 mm and 6 mm fillet end shell, respectively. Assuming 2 mm crack started at the stress concentration spot, the load points in FAD of the stress concentration area of different cases can be seen in Figure 4.33.

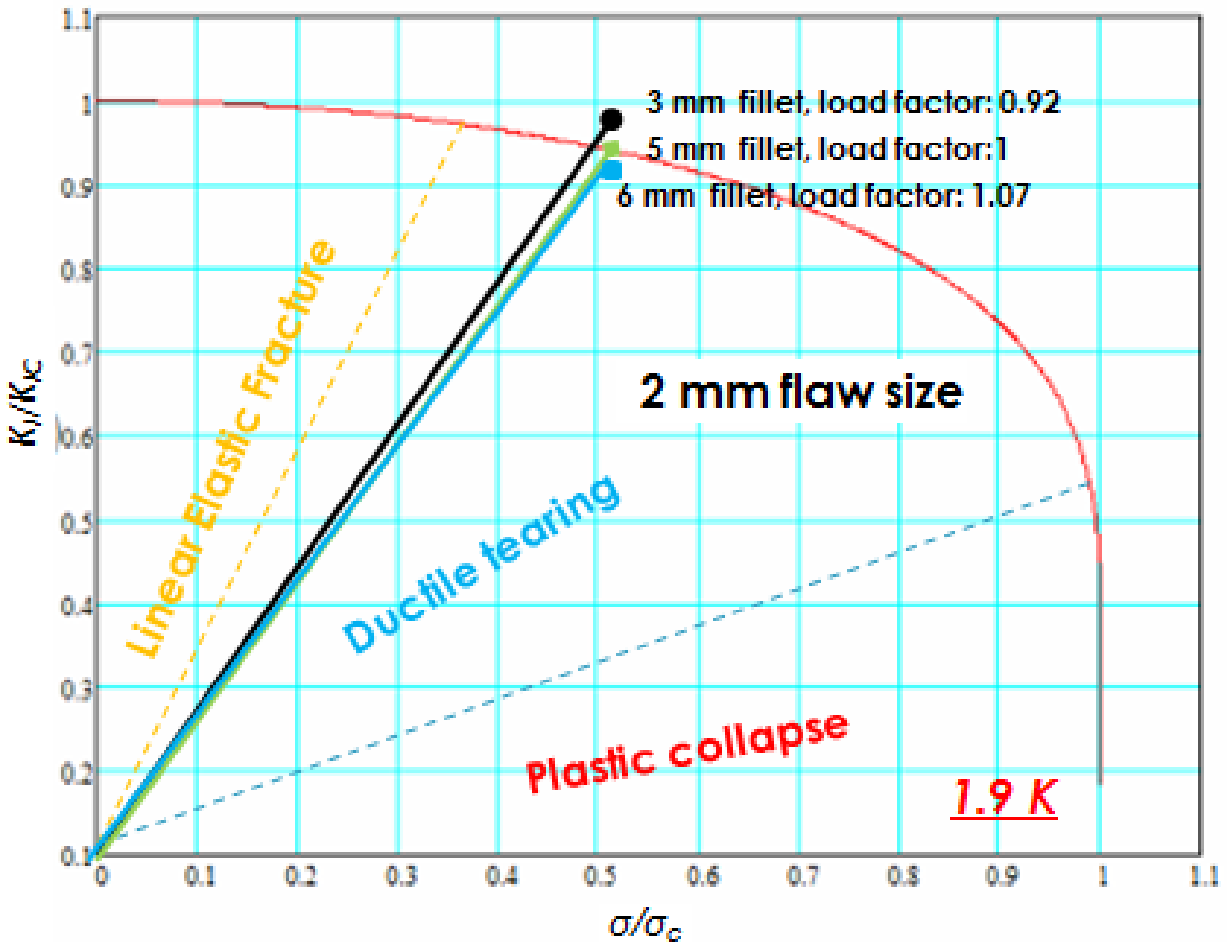

Figure 4.33: FAD for 2 mm crack in stress concentration area of end shell with different fillet sizes

The plastic deformation depth in the case of 3 mm fillet is less than the critical flaw size of 1.6 mm, thus the plastic deformation will not affect the crack propagation. However, according to the flaw sizes correlated to Inspection Grades for Aluminum Forgings [11], if a flaw is 2 mm, an inspection Class of "AA" is required; a higher Class of "AAA" is required if a flaw is less than 1.77 mm.

Based on the calculations above, in pre-series and series magnets the end shell cut-out corners were rounded up to 10 mm fillet radius on triangular slots and 15 mm fillet radius on square slots, and each batch of the forged material will be inspected in the Class of "AA". Please note that the Structural Design Criteria [3] require load factor of 1.2 or larger, and the load factors in all shell cut-outs are larger than 1.4.

#### 4.3.5.2 Yoke fracture analysis results

The yoke yields at the notch corner at room temperature in the cases of fillet radius from 0.6 mm to 2 mm. The path of a likely crack propagation along the yoke is shown in Figure 4.34, pointing to the tie-rod hole.

Figure 4.35 shows the total stress on the path at room temperature. The yield depths at room temperature of the cases are 2.1 mm and 0.8 mm for 1 mm and 2 mm fillet radius, respectively.

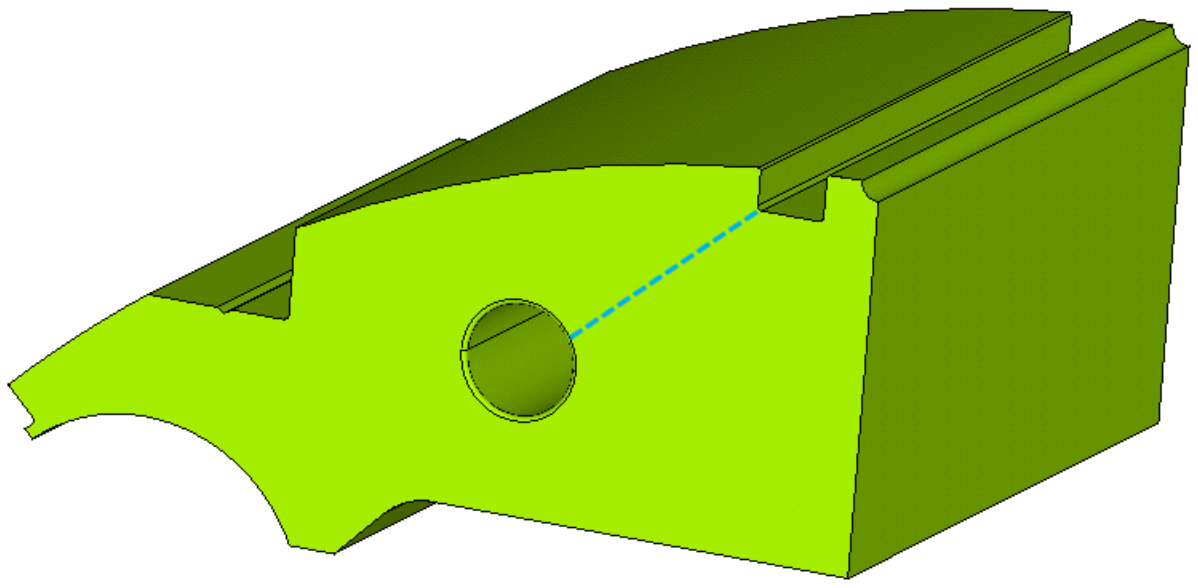

Figure 4.34: Path of a crack is most likely to propagate on yoke

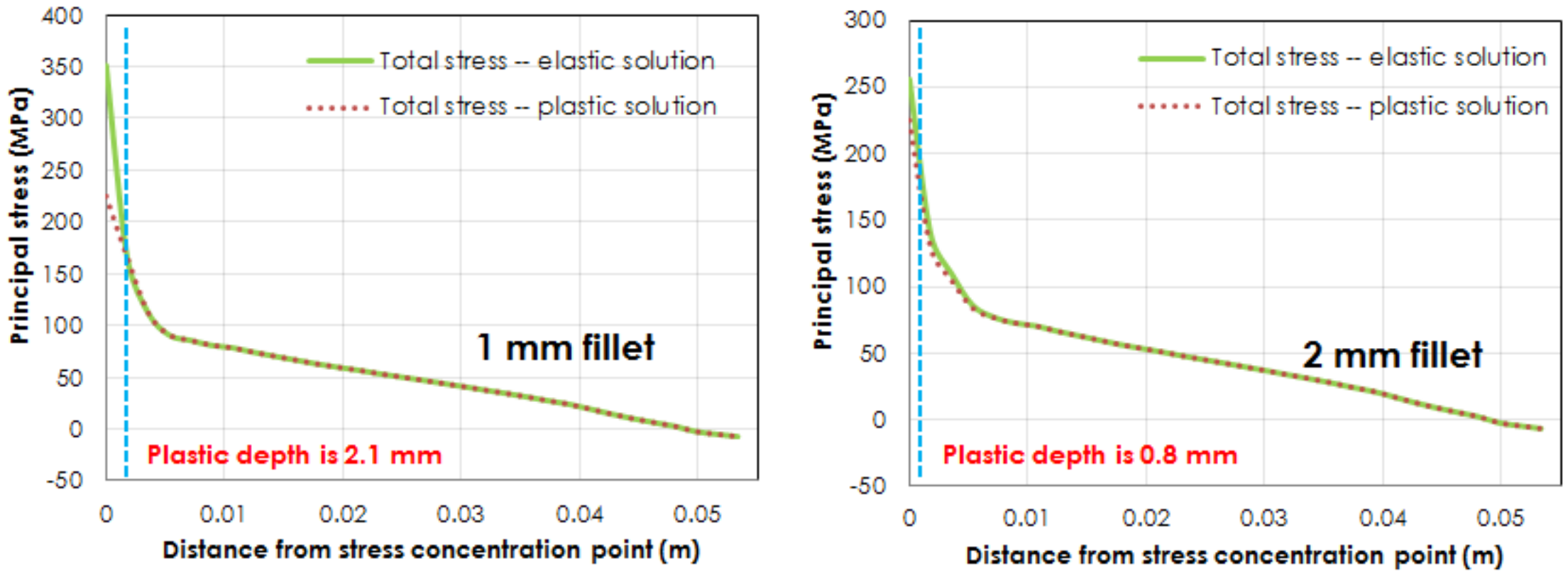

Figure 4.35: Total stress on the path at room temperature

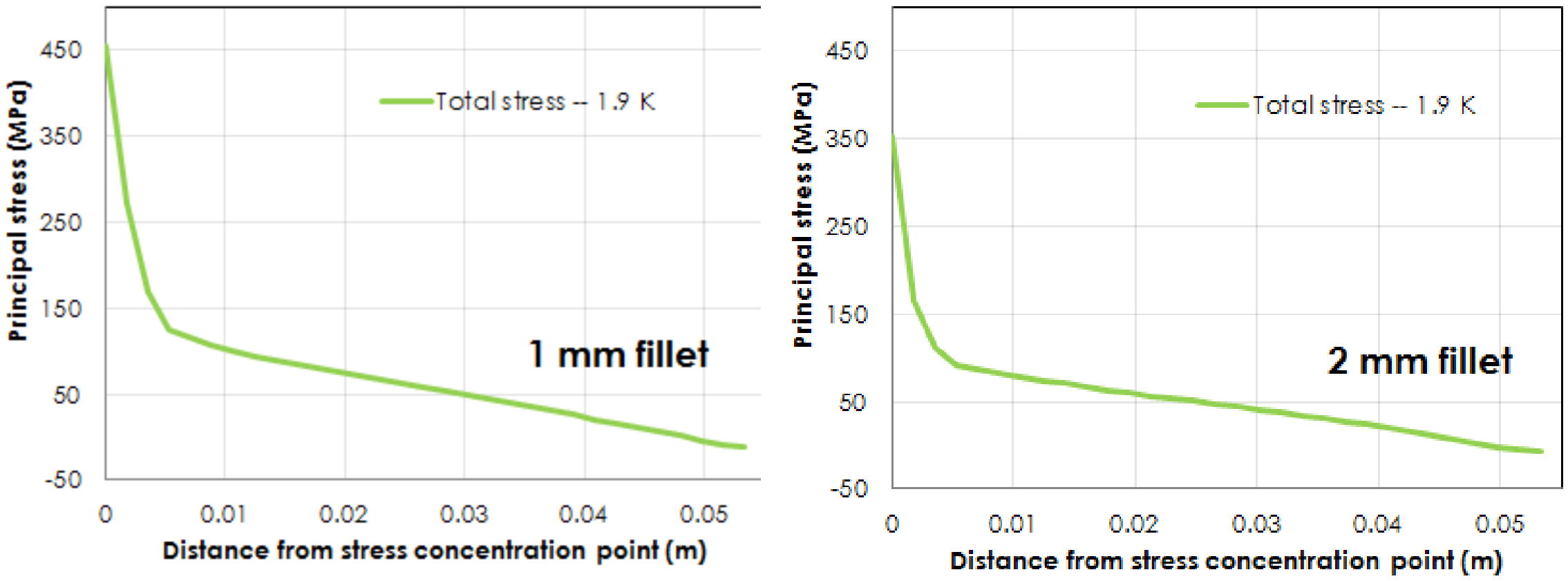

Figure 4.36: Total stress on the path at 1.9 K

The critical flaw sizes for 1 mm fillet case are calculated as 8.6 mm and 5.5 mm at room temperature and 1.9 K, using the total stress illustrated in Figure 4.35 and Figure 4.36. The plastic zone seen on yoke is much smaller than the critical flaw size.

The FAD of yoke is shown in Figure 4.37 when a 5 mm flaw size is considered. For all the cases of 1 mm and 2 mm fillets, the load points for the yoke are within the envelope of the FAD, which indicates that 2 mm fillet will be adequate for the yoke at both room temperature and 1.9 K.

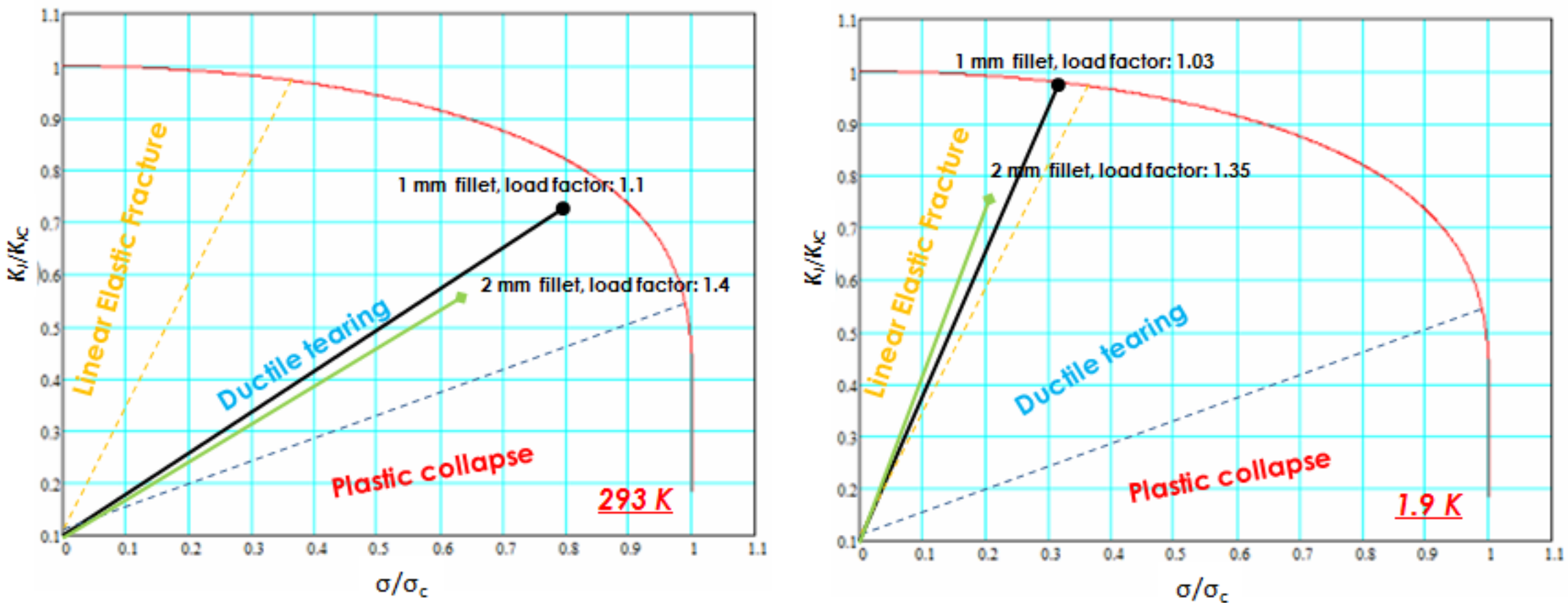

Figure 4.37: FAD for 5 mm thumbnail crack in end shell with different fillet radiuses

With 2 mm fillet radius, the load factor at both room temperature and 1.9 K are about 1.4.

### 4.3.6 Peak Stresses in prototype and preseries magnets

MQXFAP2 has an azimuthal preload target of 750 µm interference at room temperature, which corresponds to an increase of 17.2% compared with MQXFAP1. The axial pre-load of MQXFAP2 remains the same as the one applied in MQXFAP1. MQXFA03, the first pre-series magnet, incorporates lessons learned from the protypes, resulting in reduced stresses, particularly in the shells. All production magnets will be assembled and pre-loaded as MQXFA03.

Table 4.10 lists the peak azimuthal stress of coils computed with the models of MQXFAP1 and MQXFAP2. The alignment G11 pole key used in MQXFAP1 is only for the coil straight section; the coil stress can be lowered by extending the pole to the same length of the entire coil. Therefore, MQXFAP2 intends to use the full length pole key. The coil peak stress predicted by the MQXFAP2 model is higher than that from MQXFAP1, but it's within the design criteria of superconducting elements.

Table 4.10 Peak coil azimuthal stress of MQXFAP1, MQXFAP2 and the pre-series magnet MQXFA03

| | MQXFAP1 | MQXFAP1b | MQXFAP2 | MQXFA03 |
|---|---|---|---|---|
| 293 K | -104 | -89 | -108 | -82 |
| 1.9 K (on entire coil, keys for straight sections) | -160 | -146 | N/A | -132 |
| 1.9 K (on entire coil, full length key) | N/A | N/A | -167 | N/A |

Table 4.11 lists the peak stresses in metallic parts computed for MQXFAP1 and MQXFAP2. All the parts are still within the yielding limits. Higher peak stresses are foreseen in the shell and yoke. The peak stresses increase by 13% maximally in the shell and yoke.

Based on the high stresses in the prototype shells, modifications were made to the shell cutouts to reduce stress concentrations in preparation for the pre-series magnets. Furthermore, the pole key gap was increased, thereby reducing force interception by the collars and furthermore reducing the shell stresses. The stress values on metal components for the pre-series magnets are shown in Table 4.12.

Table 4.11 Peak stresses in the metallic parts of MQXFAP1 and MQXFAP2

| Part | Principal Stress (MPa) | | | | | | Von Mises Stress (MPa) | | | | | |
|---|---|---|---|---|---|---|---|---|---|---|---|---|
| | 293 K | | | 1.9 K | | | 293 K | | | 1.9 K | | |
| | MQXFAP1a | AP1b | AP2 | MQXFAP1a | AP1b | AP2 | MQXFAP1a | AP1b | AP2 | MQXFAP1a | AP1b | AP2 |
| Collar | - | | - | - | | - | 90 | 78 | 110 | 273 | 270 | 276 |
| SS Pad | - | | - | - | | - | 73 | 65 | 82 | 258 | 254 | 280 |
| Iron Pad | 82 | 72 | 99 | 146 | 122 | 150 | - | - | - | - | - | - |
| Yoke | 208 | 200 | 220 | 305 | 295 | 320 | - | - | - | - | - | - |
| Shell | 320 | 290 | 365 | 650 | 520 | 690 | 342 | 320 | 410 | 635 | 536 | 714 |
| Endplate | - | | - | - | | - | 165 | 160 | 169 | 330 | 326 | 335 |
| | | | | | | | | | | | | |

Table 4.12 Peak stresses in the metallic parts of pre-series magnet MQXFA03

| Part | Principal Stress (MPa) | | Von Mises Stress (MPa) | |
|---|---|---|---|---|
| | 293 K | 1.9 K | 293 K | 1.9 K |
| | MQXFA03 | MQXFA03 | MQXFA03 | MQXFA03 |
| Collar | - | - | 73 | 263 |
| SS Pad | - | - | 60 | 250 |
| Iron Pad | 54 | 110 | - | - |
| Yoke | 140 | 232 | - | - |
| Shell | 203 | 460 | 232 | 512 |
| Endplate | - | - | 250 | 408 |

### 4.3.7 **Conclusion**

The MQXFA03 magnet all production magnets have been designed according to the structural design criteria [3],[4]. The coil stress and strain have been evaluated in both 2D and 3D finite element models, and the results meet the requirements defined in [4].

The structural metallic parts have been designed following a graded approach described in [3]. Most of the structural components meet the design criteria at grade II level; End shell and yoke presented high stress concentration in Grade II analyses, therefore Grade III and Grade IV then were performed. The analysis pointed out the need of adding fillets on the cut-out corners of these components. The load factors of the shells and yokes are at least 1.4 after the fillets were added.

### 4.4 Radiation Effects

Proton-proton collisions at 7 TeV generate on average about 120 secondary particles per interaction. Many of these particles, and their decay products, are intercepted by the detector components. However, those emitted at small angles with respect to the beam direction exit the detector region through the TAS, a 60 mm aperture, 1.8-m long absorber located at 20 m from the IP. While less than 10% of the secondary particles cross the TAS, they carry a large fraction of the total energy, resulting in 3.8 kW of power on each side of IP1 and IP5 at the nominal HL-LHC luminosity of $5\times10^{34}$cm$^{-2}$s$^{-1}$. This has a critical impact on both thermal loads and radiation dose to the IR quadrupoles.

The protection strategy of the IT magnets is centered on a tungsten absorber placed in the magnet bore between the beam screen and the coil. The absorber is mounted on the beam screen, which is made of stainless steel and has an octagonal shape. This assembly allows both shielding the superconducting coils and collecting a significant amount of debris energy at a higher temperature, reducing the heat load on the 1.9 K system.

Detailed simulations of the radiation load for HL-LHC were carried out using FLUKA at CERN [1-2] and MARS at Fermilab [3-4]. Both codes have been extensively used and validated in past studies for LHC and many other applications. Both FLUKA and MARS simulations used identical models of the HL-LHC machine geometry and materials, and the results turned out to be in close agreement [5]. A comprehensive analysis of the radiation effects for the High Luminosity LHC is reported in [6] and references therein. Details of the IR quadrupole model are shown in Figure 4.38.

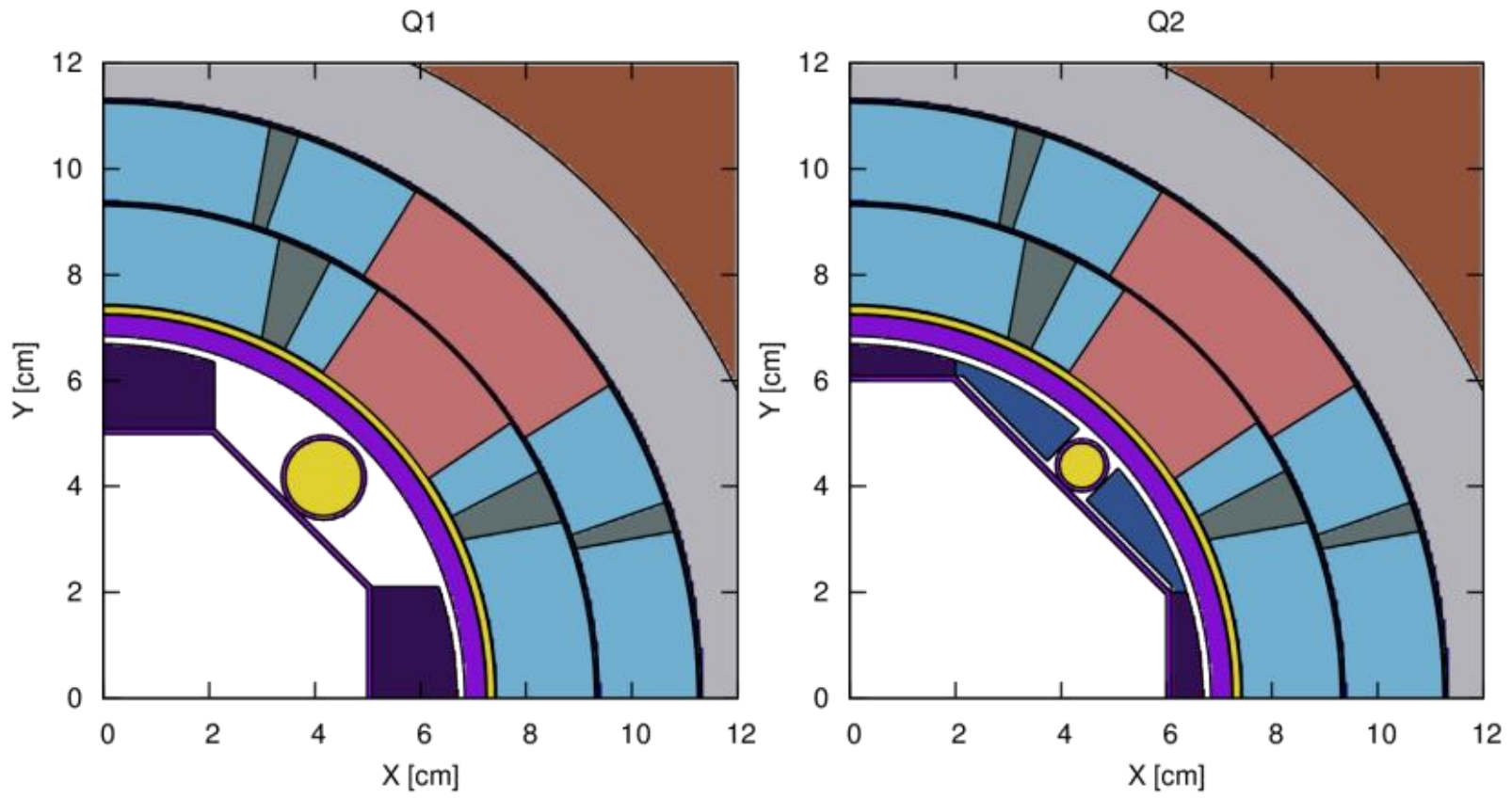

Figure 4.38: Details of the model used in FLUKA for the inner region of the IR quadrupole Q1 (left) and Q2-Q3 (right). [6]

The power density and dynamic heat load are normalized to a luminosity of 5×1034cm−2s−1. The absorbed dose, neutron fluence and DPA are normalized to an integrated luminosity of 3000 fb−1. Figure 4.39 shows the longitudinal profile of the peak power density on the inner coils of the IT magnets. The peak value in the Nb3Sn quadrupoles is below 2 mW/cm3, considerably lower than the quench limit at the operating point. For the total length of the cold mass, the average dynamic heat load on it is ~12 W/m. This is within a design range of 10–15 W/m used for the LHC and assumed for the HL-LHC.

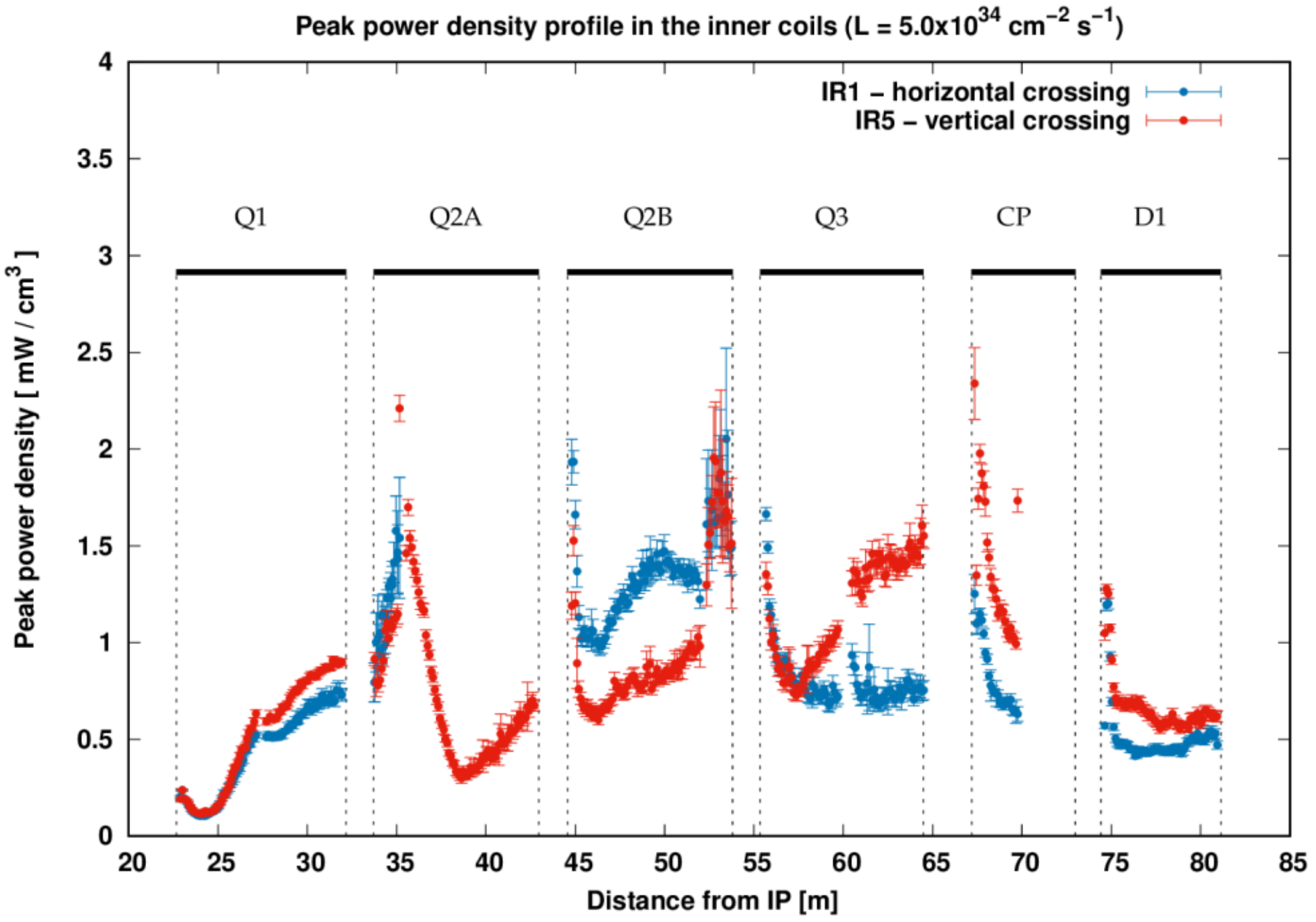

Figure 4.39: Longitudinal peak power density profile on the inner coils of the IT magnets, for both vertical and horizontal crossing. Values are averaged over the full radial thickness of the coil.

Figure 4.40 shows the longitudinal profile of peak dose on the inner coils and insulating materials. Thanks to the increased aperture and tungsten absorbers, the dose in the HL-LHC inner triplet at 3000 $fb^{-1}$ is comparable to that of the present LHC at 300 $fb^{-1}$. The peak dose in the insulation reaches 20 MGy for the main quadrupoles. The results of a detailed characterization of the radiation resistance of the ceramic epoxy used for MQXF are reported in [7]. The materials have been irradiated with 50 MGy, using 4 MeV electron beam at 77 K which well represents the electromagnetic radiation impact relevant to the IR magnet case. The electrical strength of the irradiated samples was found to be between 45 kV/mm and 65 kV/mm, which is still significantly higher than the required 5 kV/mm. This provides adequate margin for the HL-LHC integrated luminosity target. Further studies of the beam screen and absorber configuration are underway to guide the engineering design incorporating all details required for fabrication and installation of these components.

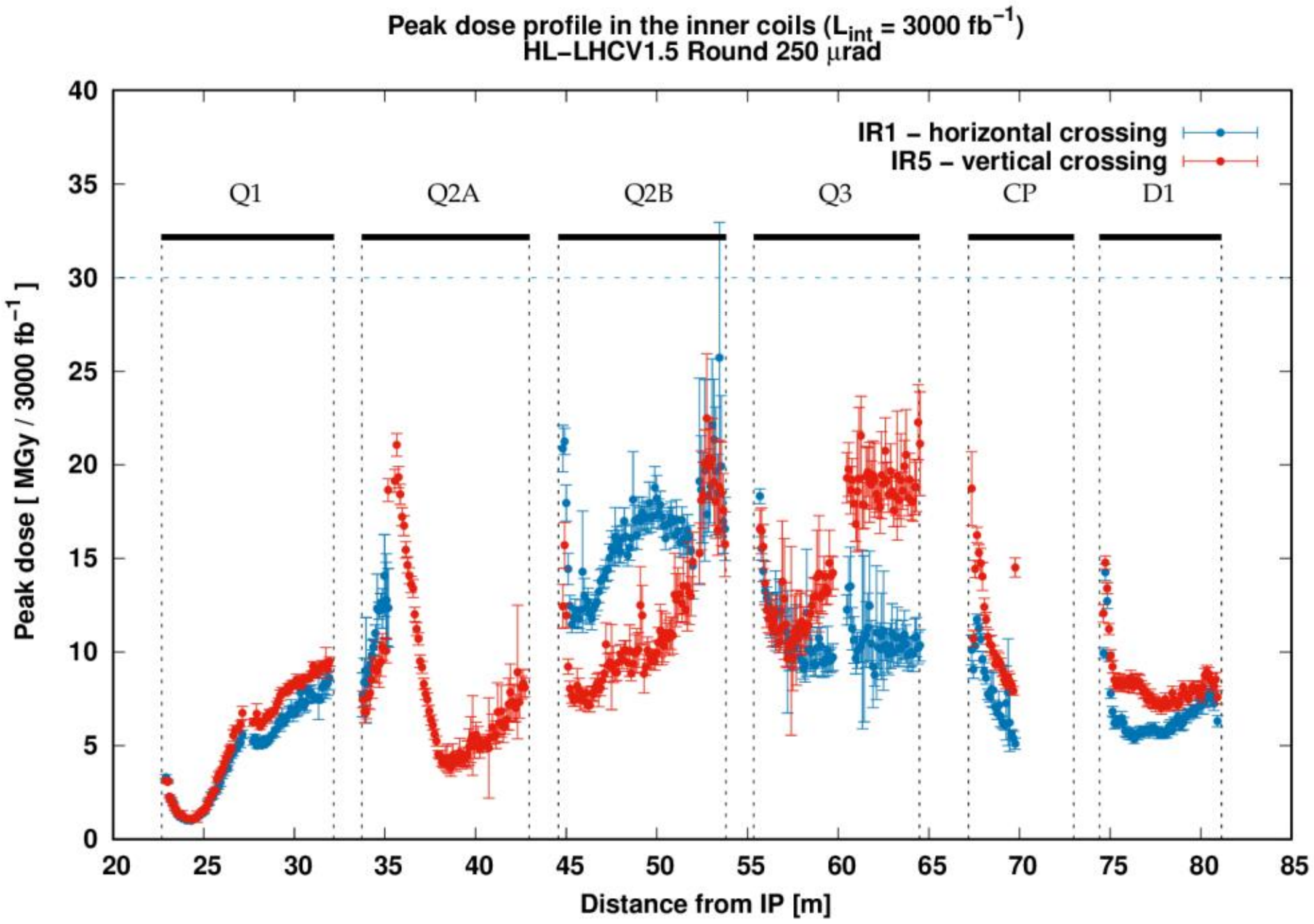

Figure 4.40: Longitudinal peak dose profile on inner coils and embedded insulators, for both vertical and horizontal crossing. Values refer to a 3 mm radial resolution.

The radiation effect on $Nb_3Sn$ superconductor, copper stabilizer and structural components is best characterized by integrated neutron fluence and DPA over the expected magnet lifetime. Figure 4.41 shows the corresponding profiles on the quadrupole inner coils, where the peaks are located. The highest DPA is about $2\times10^{-4}$ DPA per 3000 fb−1 integrated luminosity, which is acceptable for the superconductor. A similar conclusion can be reached by comparing the neutron fluence in the coils with the known limits. In the quadrupole coils, the peak fluence is $\sim2\times10^{17}cm^{-2}$ which is substantially lower than the $3\times10^{18}cm^{-2}$ limit used for the $Nb_3Sn$ superconductor. The integrated DPA in the magnet mechanical structures is 0.003 to 0.01 in the steel beam screen and tungsten absorber, $\sim10^{-4}$ in the collar and yoke, and noticeably less outside. These are to be compared to a ~10 DPA limit for mechanical properties of these materials. Neutron fluence in the IT mechanical structures range from $3\times10^{16}$ $cm^{-2}$ to $3\times10^{17}$ $cm^{-2}$ compared to the $10^{21}$ $cm^{-2}$ to $7\times10^{22}$ $cm^{-2}$ limits.

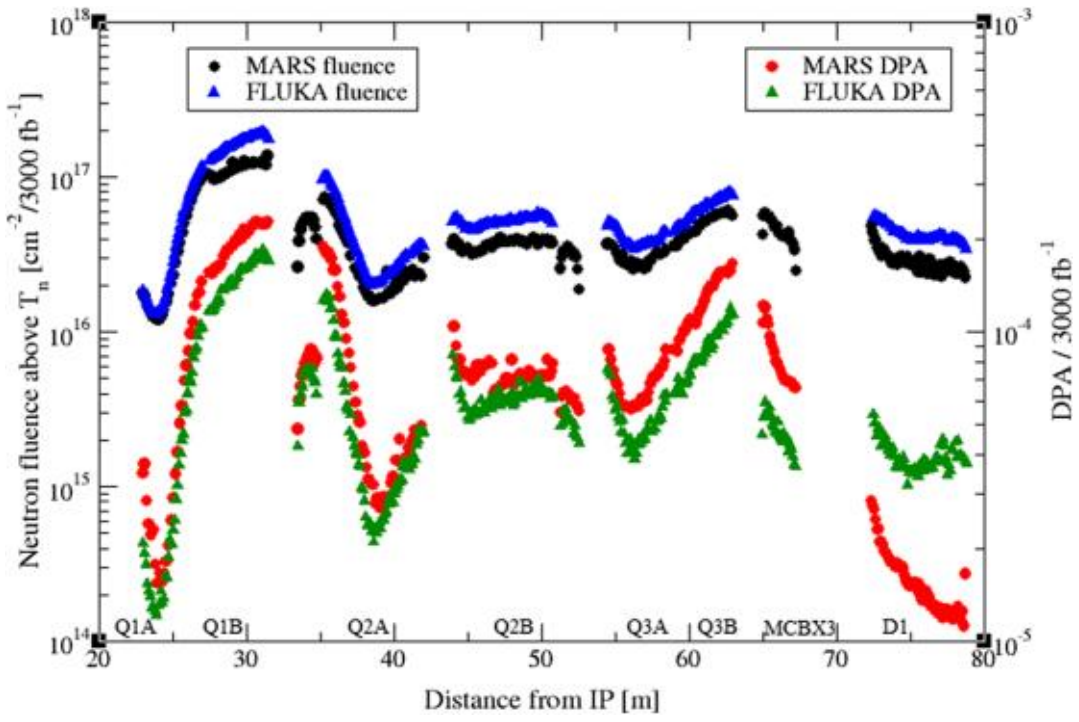

Figure 4.41: Longitudinal peak neutron fluence and peak DPA profiles along the hottest regions in the IT magnet coils [5].

## 4.5 Thermal Design

The final focusing magnets will receive a heat load due to debris coming from the adjacent particle interaction points: the computed peak heat deposition in the coil is up to 4 $mW/cm^3$ on the inner edge (over a radial interval of 3 mm) at the ultimate luminosity of $7.5\times10^{34}$ $cm^{-2}$ $s^{-1}$.

The cooling method is based on the one used in the present LHC, and the overall cryogenic infrastructure needed is similar to the existing configuration. The cooling performance is evaluated in terms of temperature margin of the magnets under full steady state heat load conditions, and in terms of local maximum sustainable load.

An annular gap, filled with He, separates the cold bore from the superconducting coil, which is surrounded by the mechanical support structure to form the cold mass. The cold-mass is enclosed in a vessel, to be kept at the chosen operating temperature and supported in a vacuum insulated cryostat. This section is focused on the cooling configuration specific to the cold mass [1].

The heat loads due to debris from the adjacent interaction point are intercepted at two distinct magnet locations and temperature levels. A first heat intercept is on tungsten absorbers which are placed inside the beam pipe vacuum and which will be cooled in the 40 K to 60 K range. The remaining heat load will fall on the cold mass volume comprised of the yoke, collars and coils. For the purpose of the evaluation of the cooling, the heat load to the cold masses is 930 W for the ultimate luminosity of $7.5\times10^{34}$ $cm^{-2}$ $s^{-1}$.

### 4.5.1 Cooling requirements for magnet cold masses

The cooling principle, depicted in Figure 4.42 is an evolution of the one proposed for the LHC-Phase-I Upgrade [2]. The cold masses will be cooled in a pressurized static superfluid helium bath at 1.3 bar and at a temperature of about 1.9 K. The heat generated in the magnets will be extracted by vaporization of superfluid helium which travels as a low pressure two-phase flow in two parallel bayonet heat exchangers (HX) protruding the magnet yokes (depicted as one bold line in Figure 4.42). The low vapor pressure inside the heat exchanger is maintained by a cold compressor system, with a suction pressure of 15 mbar, corresponding to the saturation temperature of 1.776 K.

The flow diagrams for the four site implementations (Left of IP1, Right of IP1, Left of IP5, and Right of IP5), are very similar. They differ in orientation left-right, so Q1 will always face the IP, and have slope dependence for the He II two-phase flow in the HX, so the flow is always down-stream. Figure 4.43 shows the flow diagram version applicable to the right side of IP5. On top are depicted the supply headers for cryogens, with two jumper connections between it and the magnet cryostats. All cryogenic valves are on the supply header-side.

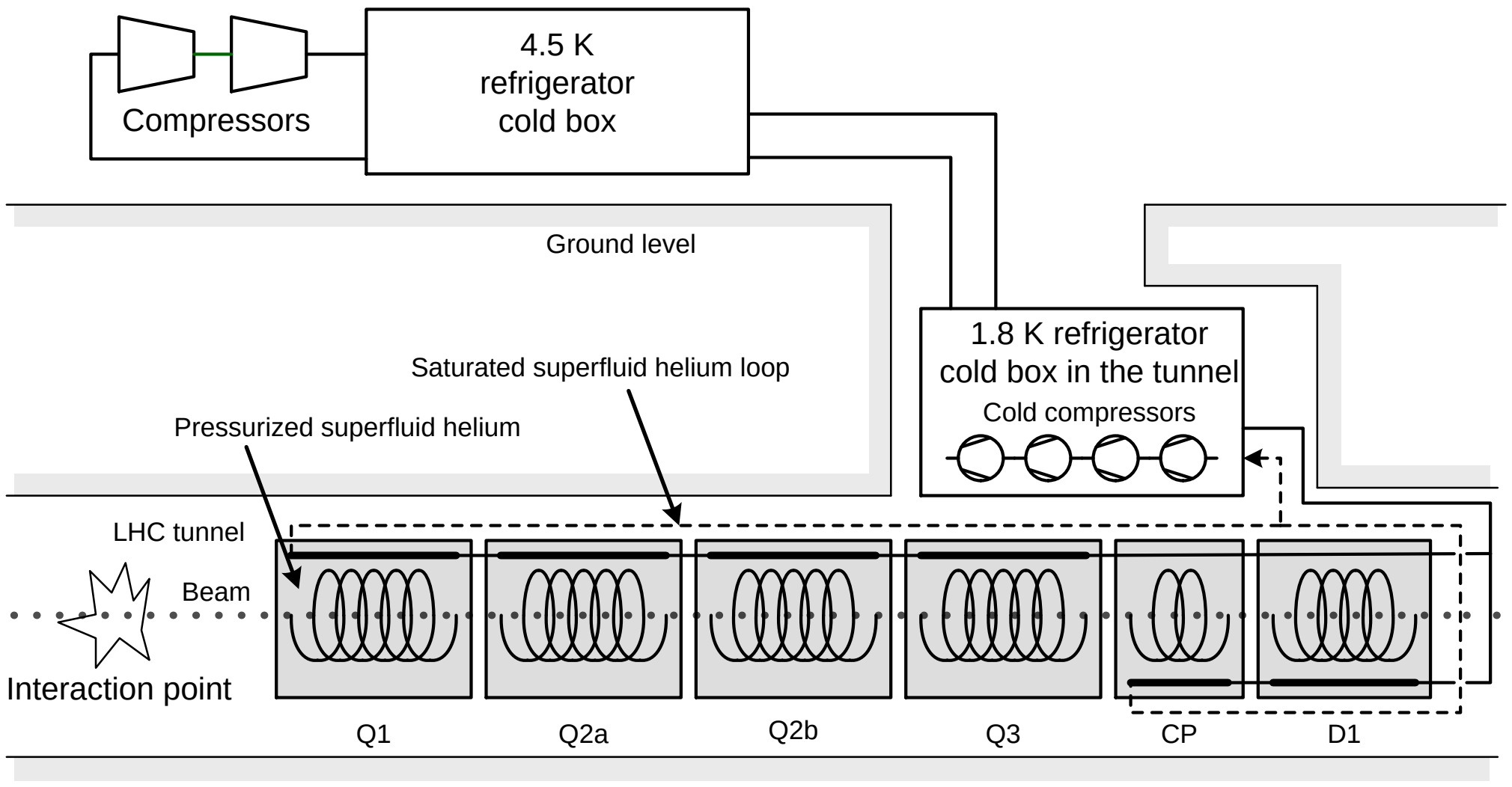

Figure 4.42: Architecture of the cooling by using superfluid helium.

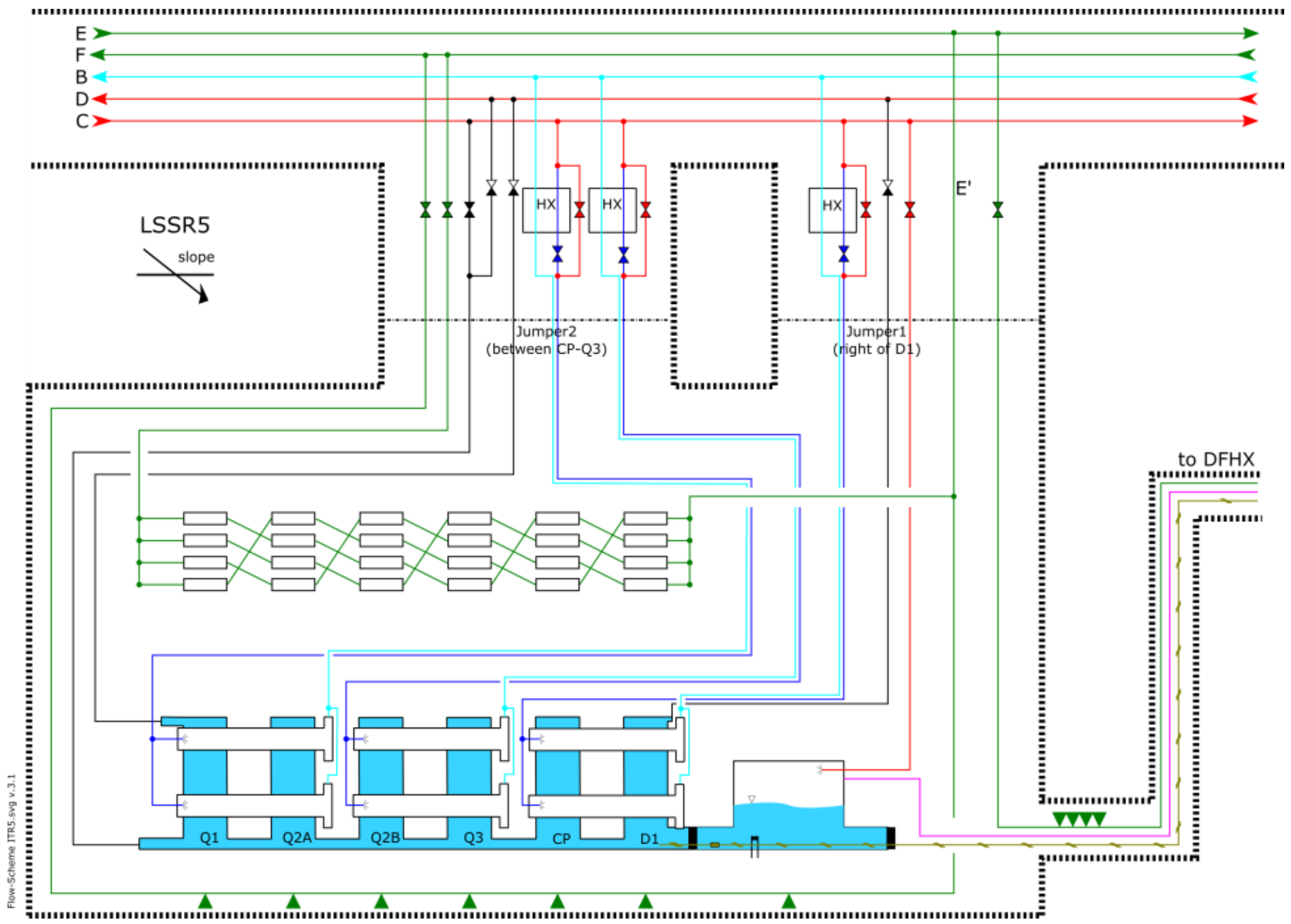

Figure 4.43: flow diagram for inner triplets, right of IP5

Due to constraints on magnet design, the bayonet heat exchangers cannot be continuous all the way from the quadrupole magnets through the corrector package and D1 dipole. The size and number of heat exchangers is determined by the maximum vapor velocity of 7 m/s above which the heat exchangers do not function anymore and the total available heat exchange area, when they are wetted over their full length. For the quadrupole heat exchangers, the vapor velocity limit is the more stringent condition and is met if one uses two heat exchangers in parallel with inner diameters greater than 68 mm and lengths limited to two quadrupoles in series. As a consequence, the flow diagram exhibits three sets of bayonet heat exchanger cooling-units (D1+CP, Q3+Q2b, Q2a+Q1), each with their own supply and low vapor pressure return. These supplies and return lines are forcibly proper to the cooling loops such as to keep the counter flow heat exchangers, situated on the cryogenic supply headers' side, balanced. Each of these cooling loops integrates a phase separator at its end. These have to absorb the liquid present in the bayonet heat exchangers in case magnet quenches would drive it out. The liquid volumes to accumulate, without obstructing the low-vapor pressure return inlet, are estimated to be 12.5 ℓ for each of the quadrupole cooling loops and 5.5 ℓ for the D1+CP cooling loop.

The heat exchangers themselves are to be made of copper to assure proper heat conduction across the walls. A wall thickness of about 3 mm is required to sustain the external design pressures of 20 bar. With, in addition, an annular space of 1.5 mm between the HX and the yoke to allow contact area of the pressurized superfluid helium on the coil-side, the yoke-hole size required is 77 mm minimum. With this configuration about 800 W can be safely extracted. The 77 mm yoke-hole size is compatible with the mechanical design of the magnet, but should not be increased otherwise one would need to increase as well the overall diameter of the cold mass. Coping with the remaining 250 W is to be done via active cooling of the D1 and CP. Two parallel bayonet heat exchangers of 51 mm ID through D1 and CP are foreseen, requiring yoke holes of 60 mm diameter.

For correct functioning of the two-phase heat exchanger configuration, heat must be given some freedom to redistribute along the length of the cold-masses. This is no hard criterion, and a free longitudinal area of $\geq 150\ cm^2$ through the Q1, Q2a, Q2b, Q3, and their interconnections and $\geq 100\ cm^2$ through D1, CP and their interconnections are deemed to be sufficient.

Two pressure relief devices are foreseen as safety in case of sudden energy release to the cold mass helium due to either magnet quenches or catastrophic loss of insulation vacuum. The quench energies released are substantial for the MQXF – quadrupoles and D1 – dipole only. The energy per local helium volume for the magnets in the corrector package are so low that they can be absorbed by the surrounding helium without consequences for the neighboring magnets

### 4.5.2 Radial heat extraction in IR quadrupoles

The $Nb_3Sn$ quadrupole coils are fully impregnated, without any helium penetration. The heat loads from the coils and the beam-pipe area can only evacuate to the two heat exchangers by means of the pressurized He II. To this end, the cold mass design needs to incorporate the necessary radial helium passages.

Figure 4.44 shows the typical heat flow path: out from the coil areas, through the annular spacing between cold bore and inner coil-block, and subsequently via free passages through the titanium pole pieces and G11-alignment key, and around the axial rods towards the cooling channels where the two-phase flow bayonet heat exchangers will be inserted. They will occupy the two-upper yoke-holes marked "Cooling channel". Since only two of the four possible cooling channels will house bayonet heat exchangers, free helium paths interconnecting these four cooling channel holes shall be implemented in the cold mass design. Doing so allows for equilibrating the heat flows and increasing the heat extraction margins as a whole.

The annular space between cold bore and inner coil block is set at 1.5 mm and the free passage needed through the titanium insert and G11-alignment key should be of the order of 8 mm holes repeated every 40 mm – 50 mm along the length of the magnet. Magnet design is presently integrating 8 mm diameter holes every 50 mm. This value and repetition rate will be used in the temperature margin evaluation in subsequent paragraphs. Around the axial rods a free passage of 1.5 mm has to be guaranteed.

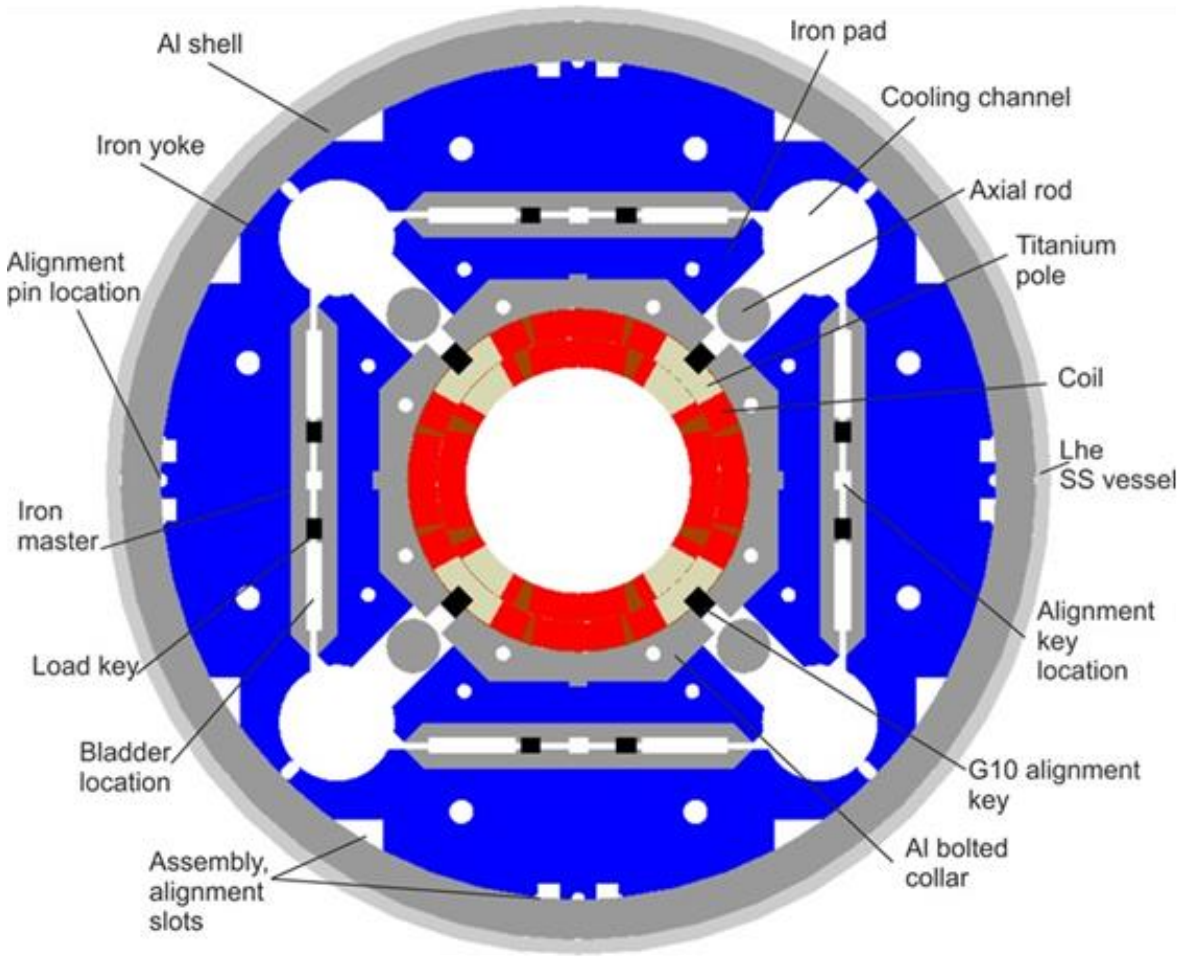

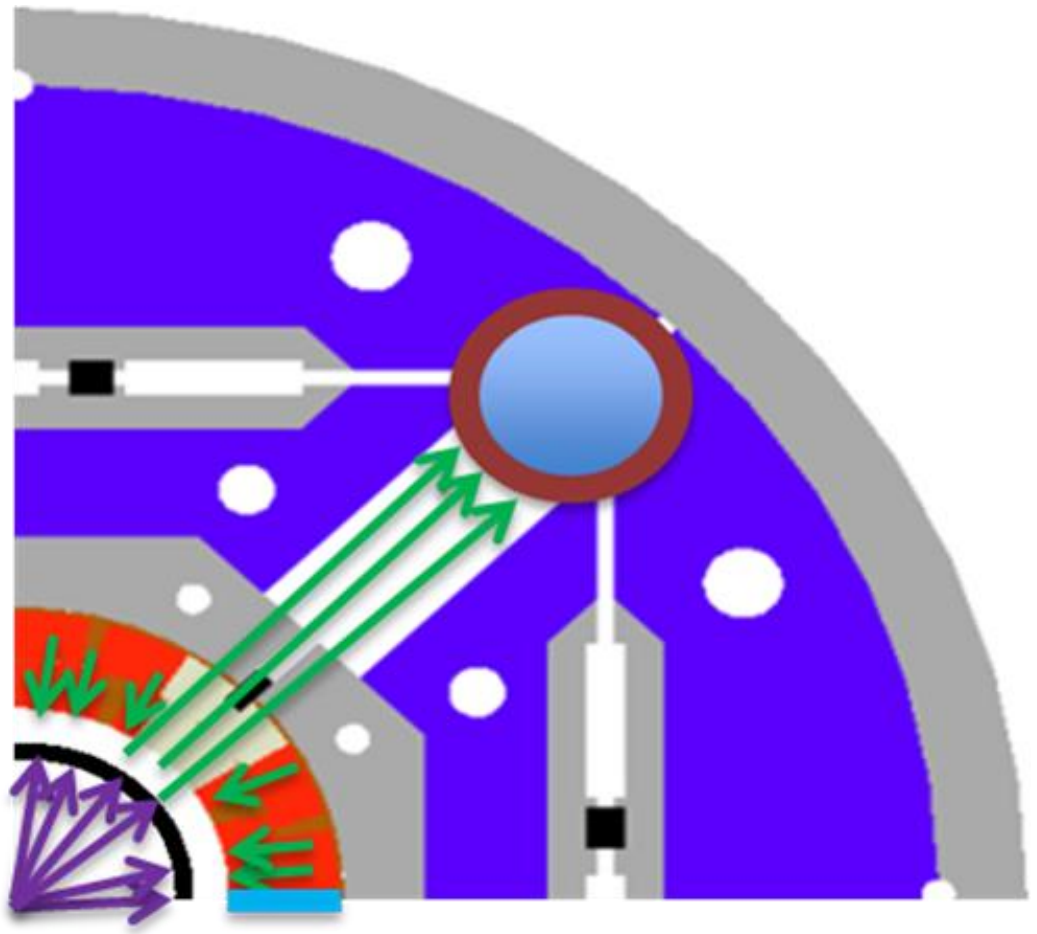

Figure 4.44: Heat flow paths from coil to one of the two-phase heat exchangers located in the upper right quadrant

### 4.5.3 Thermal performance evaluation

The heat load distribution at the most unfavorable location is used as reference (Figure 4.45) [2]. The peak power densities reached in the quadrupoles are close to 7 mW/cm$^3$.

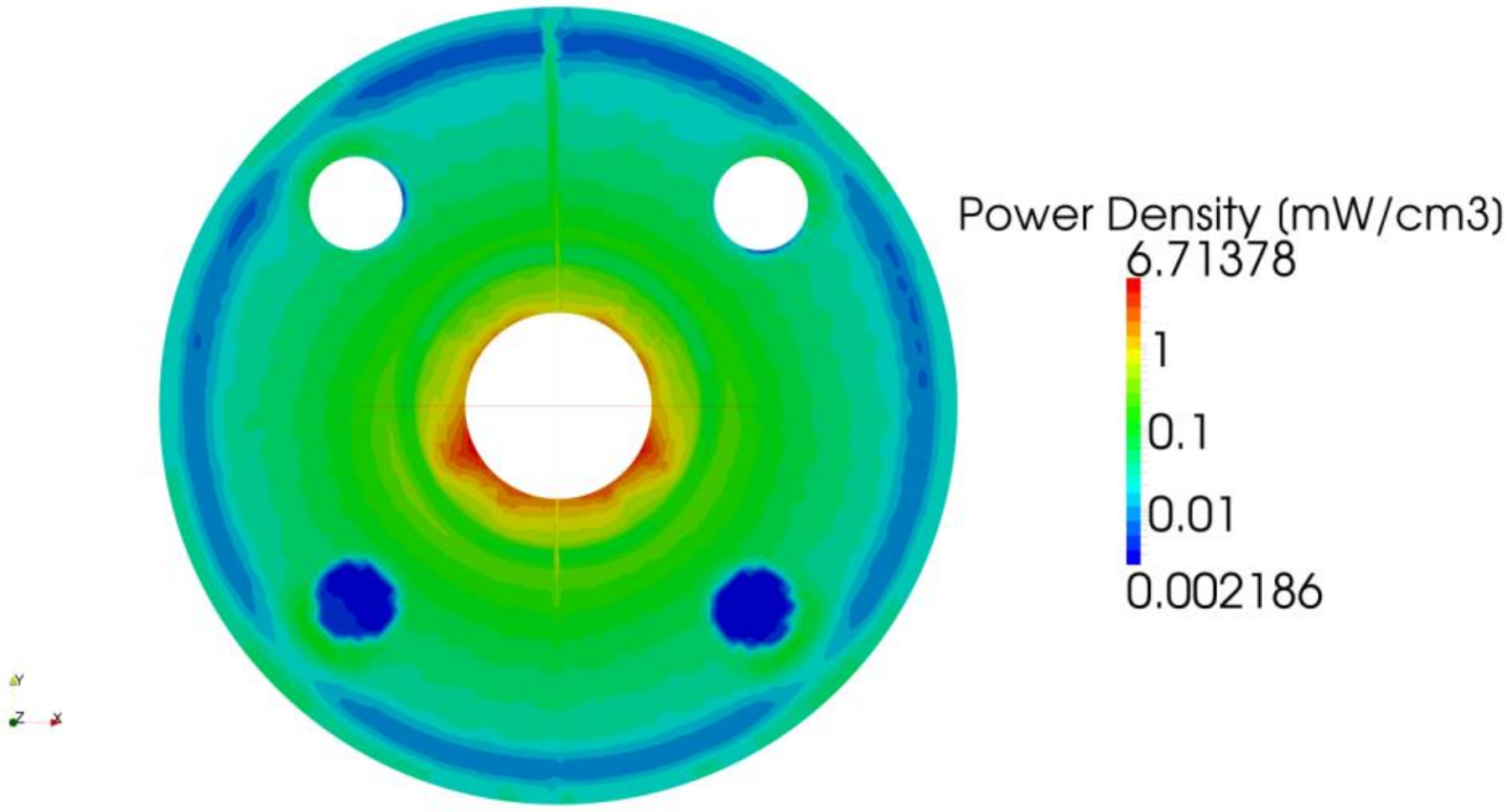

Figure 4.45: Power deposition map for Q3 at ultimate peak luminosity.

These power deposition maps are used as input to produce temperature distribution maps of the cold masses. The results are then converted into temperature margin maps by comparison with the respective current sharing maps, calculated on an assumed homogeneous temperature distribution of 1.9 K (Figure 4.46).

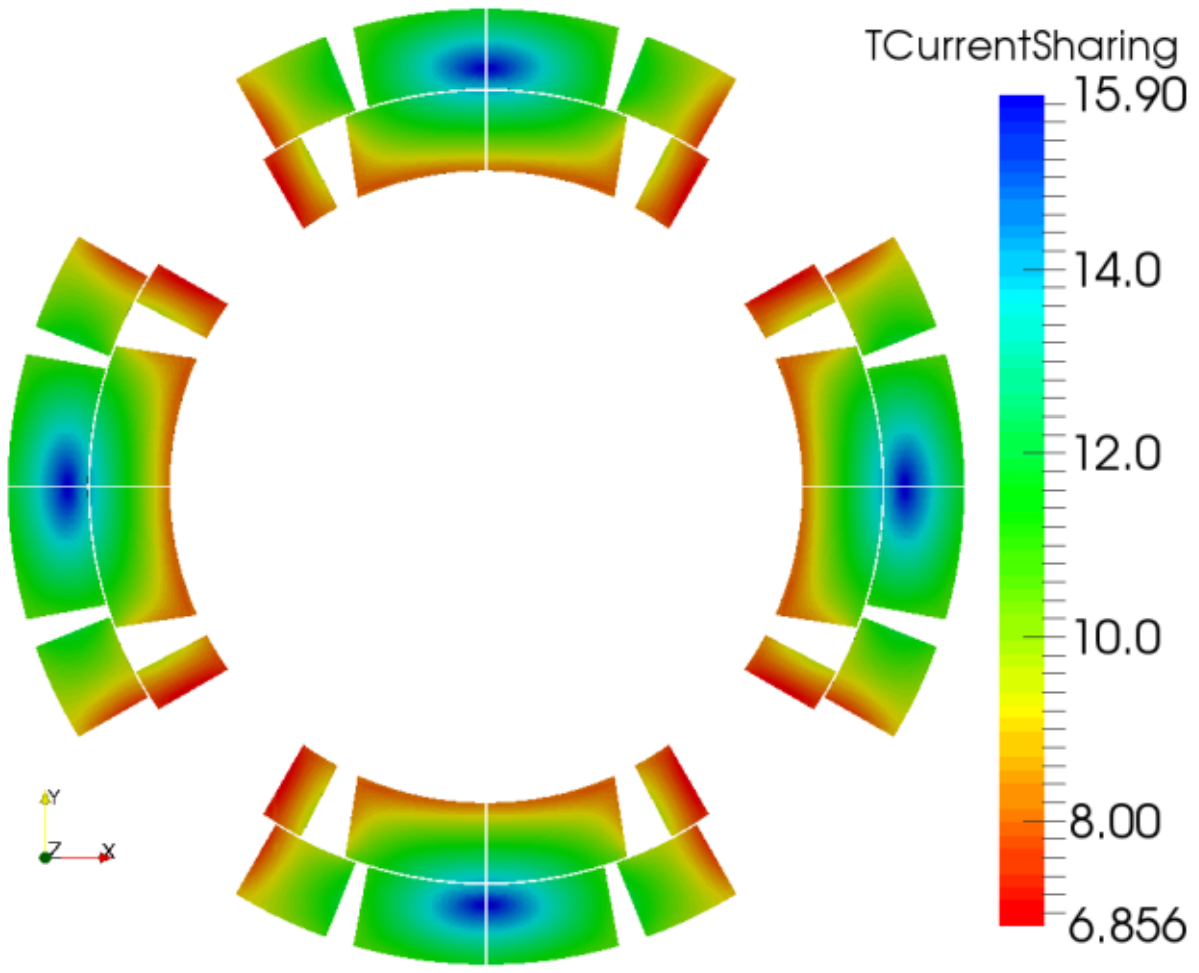

Figure 4.46: Current sharing map of MQXF quadrupole.

The MQXF-quadrupole coil materials were implemented with the coil pack details as shown in Figure 4.47. “Porous” quench heater traces were assumed to be placed on the MQXF inner coil layer (Figure 4.48) to perform the analysis. Since inner layer quench heater traces have been later removed from the quench protection scheme, these results are conservative (being the traces right in the heat extraction path).

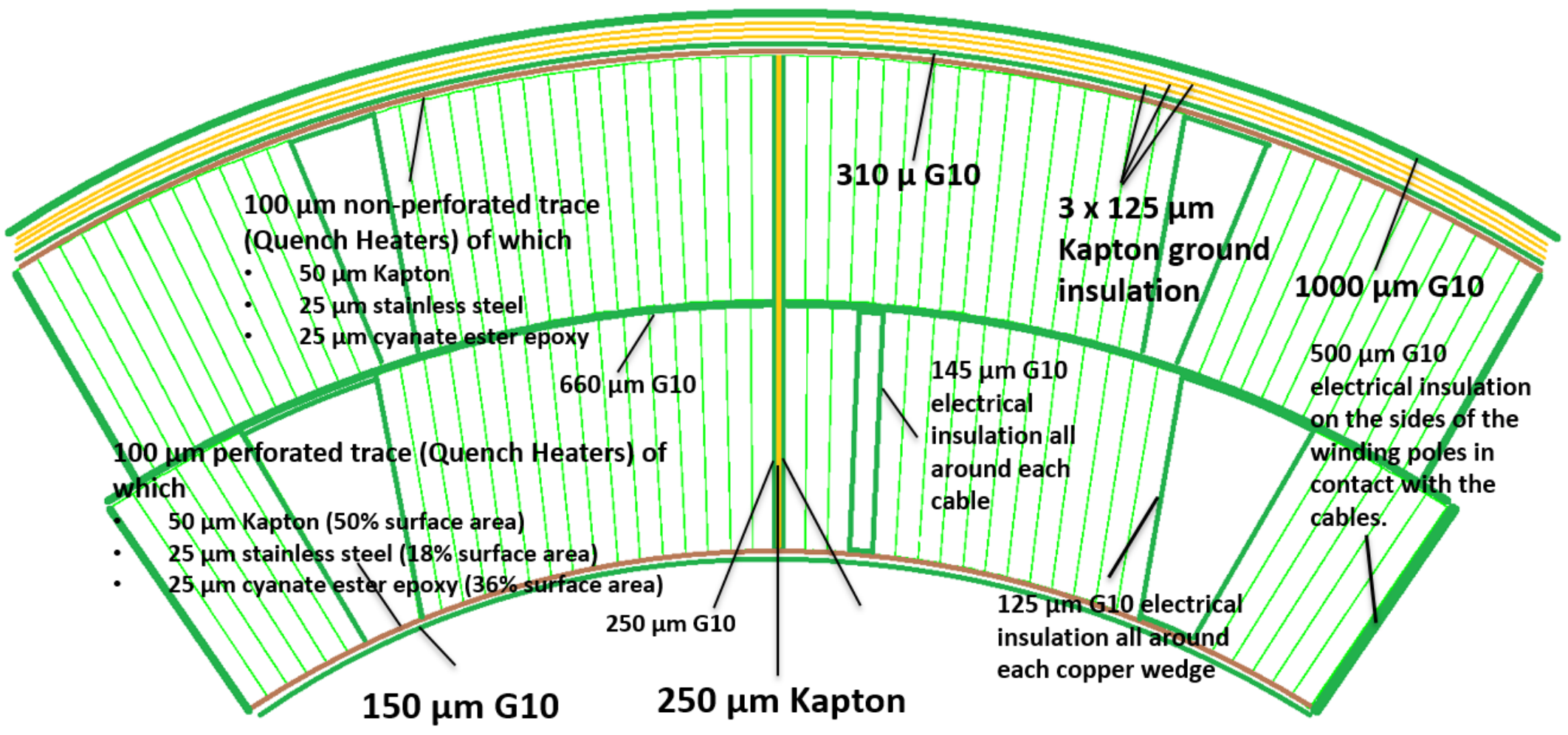

Figure 4.47: Quadrupole $Nb_3Sn$ coil materials assumed for the calculations.

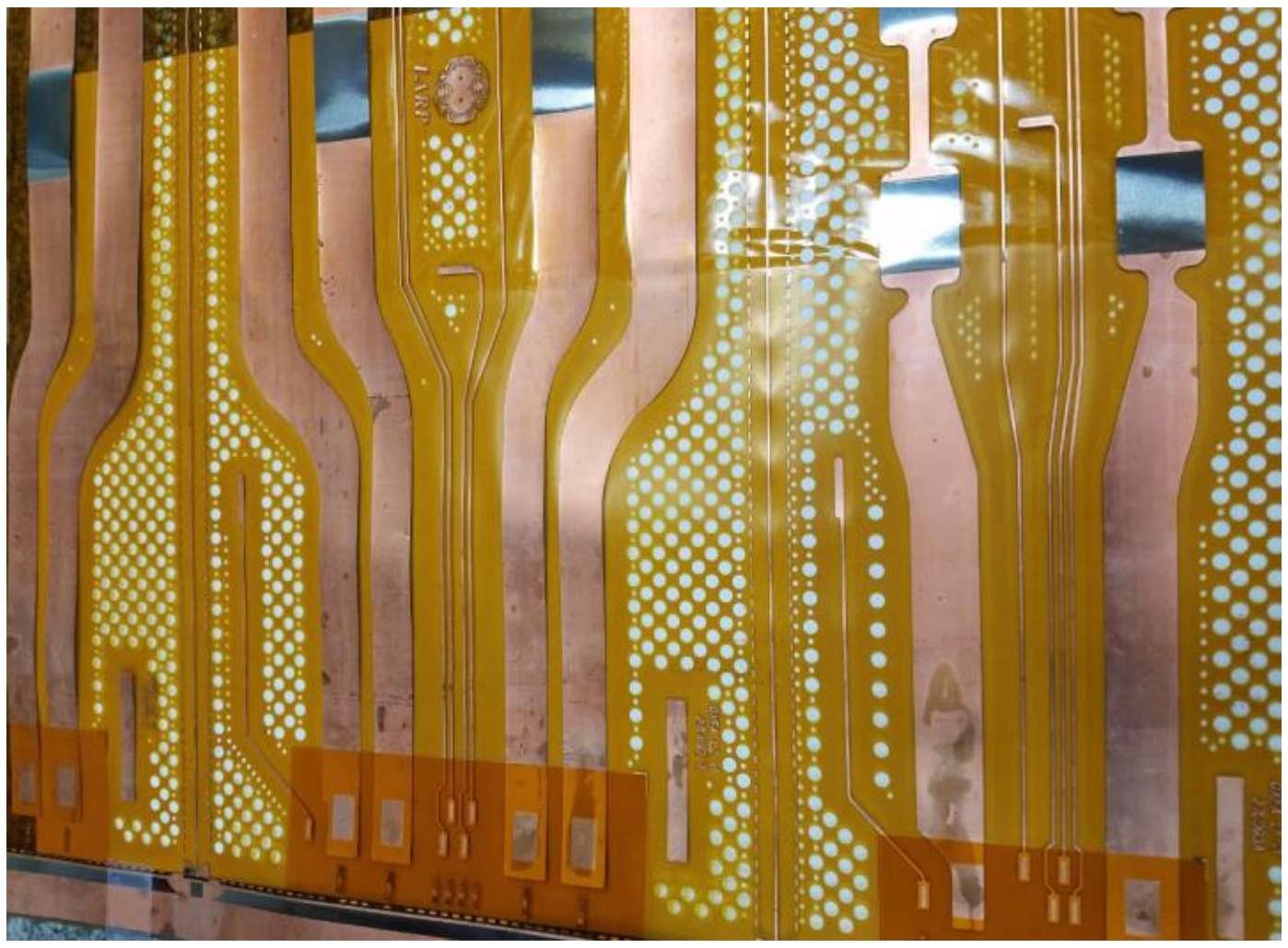

Figure 4.48: Perforated Kapton quench heater traces.

### 4.5.4 Steady state temperature, T margin and local maximum-sustainable load

Figure 4.49, Figure 4.50, Figure 4.51 and Figure 4.52 show the temperature and temperature-margin maps with their respective zooms. These were all calculated assuming a 1.9 K cold source (bayonet heat exchanger) temperature. Although the highest temperatures of about 3.15 K are reached in the outer coil layer, when combined with the current sharing maps, we find the most critical zones at the inner layer pole edges. There, the temperature margin goes down to about 4.1 K. Note that the value of 4.1 K is reached in a small fraction of the coil (Figure 4.52), and further optimization of the beam screen W absorber could remove this singularity and bring the temperature margin above 5 K all over the coil.

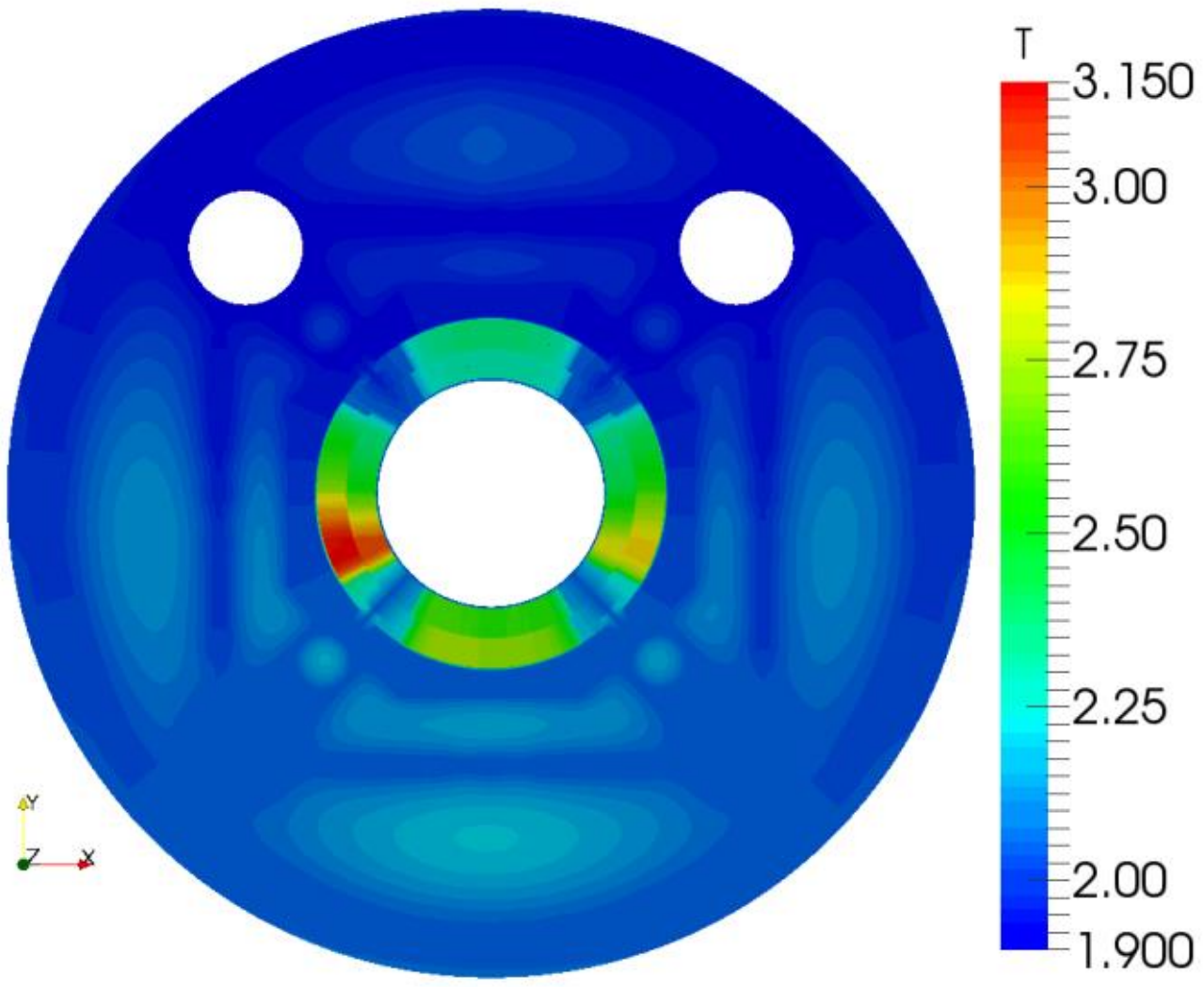

Figure 4.49: MQXF quadrupole temperature map; full section.

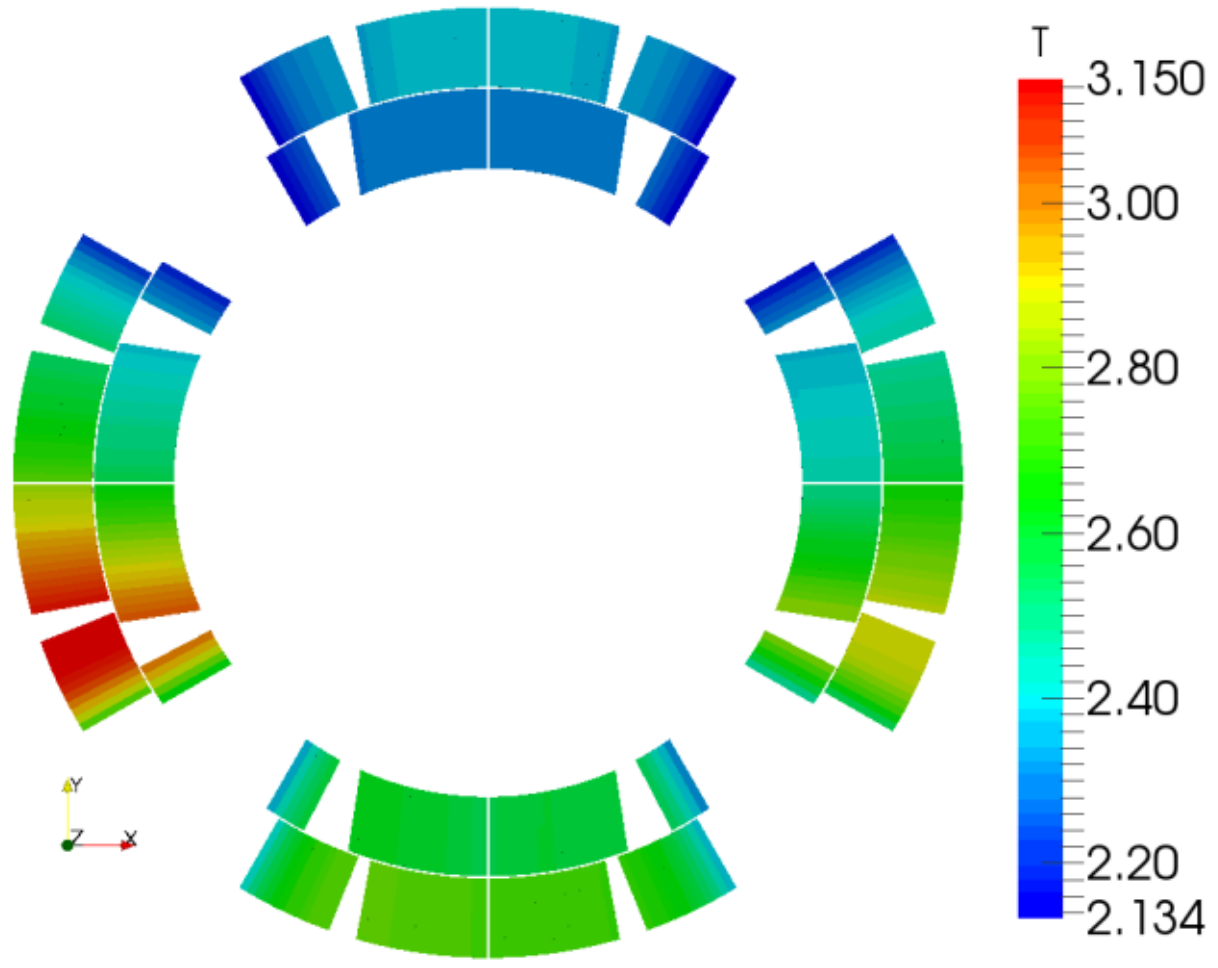

Figure 4.50: MQXF quadrupole temperature map; coil section.

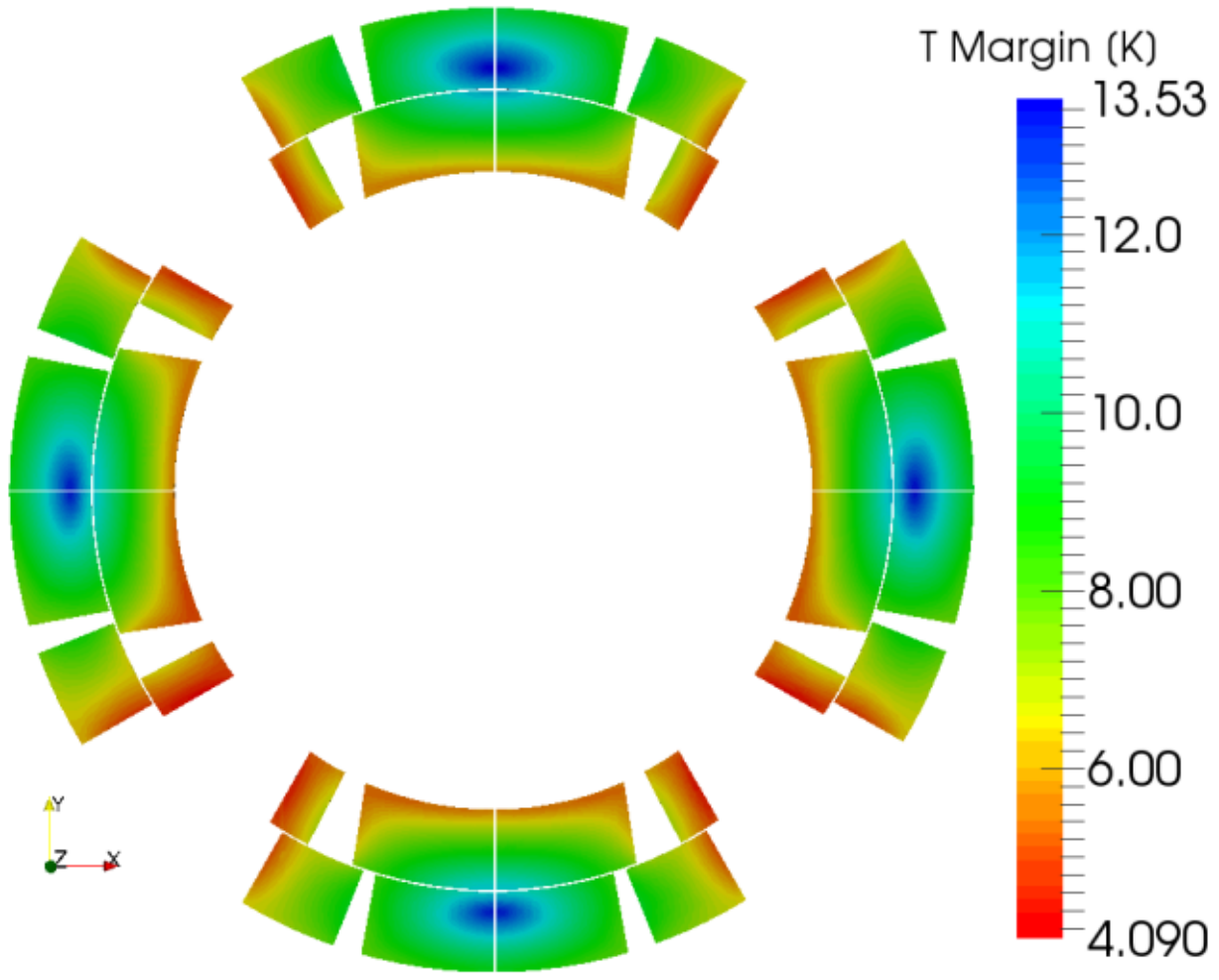

Figure 4.51: MQXF quadrupole temperature-margin map.

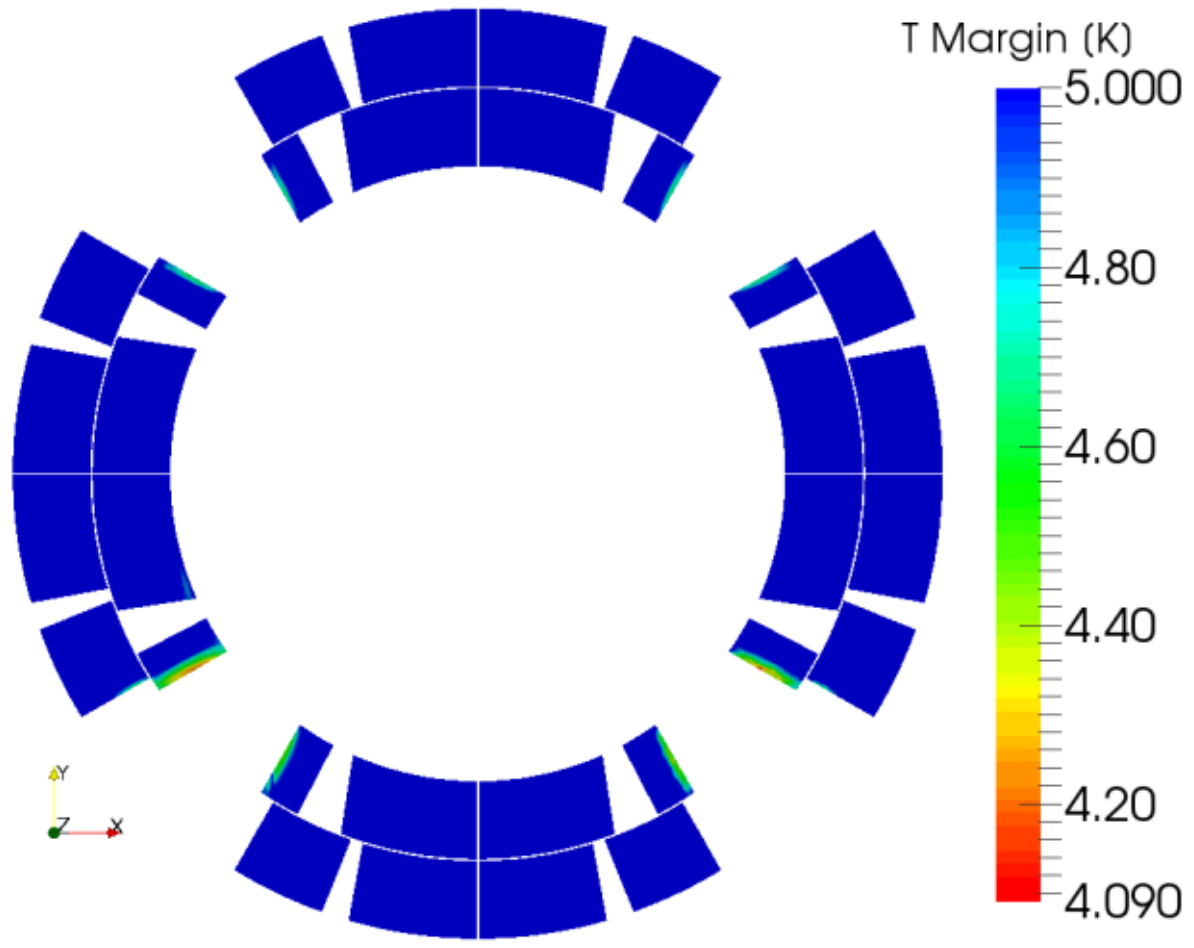

Figure 4.52: MQXF quadrupole temperature-margin map, capped off at 5 K to expose the critical zones.

The robustness of the MQXF thermal design is addressed by steady state local power deposit values that the coil can withstand without either the cooling breaking down or the cable reaching a temperature margin of 0 K. We found that in these steady state cases the local cooling break-down occurs first. Figure 4.53 shows that locally we can sustain powers from 56 mW/cm$^3$ at 1.9 K down to 19 mW/cm$^3$ at 2.1 K bayonet heat exchanger temperature. This constitutes a factor 8 at 1.9 K down to 3 at 2.1 K with respect to the expected peak load of 6.7 mW/cm$^3$ at ultimate luminosity.

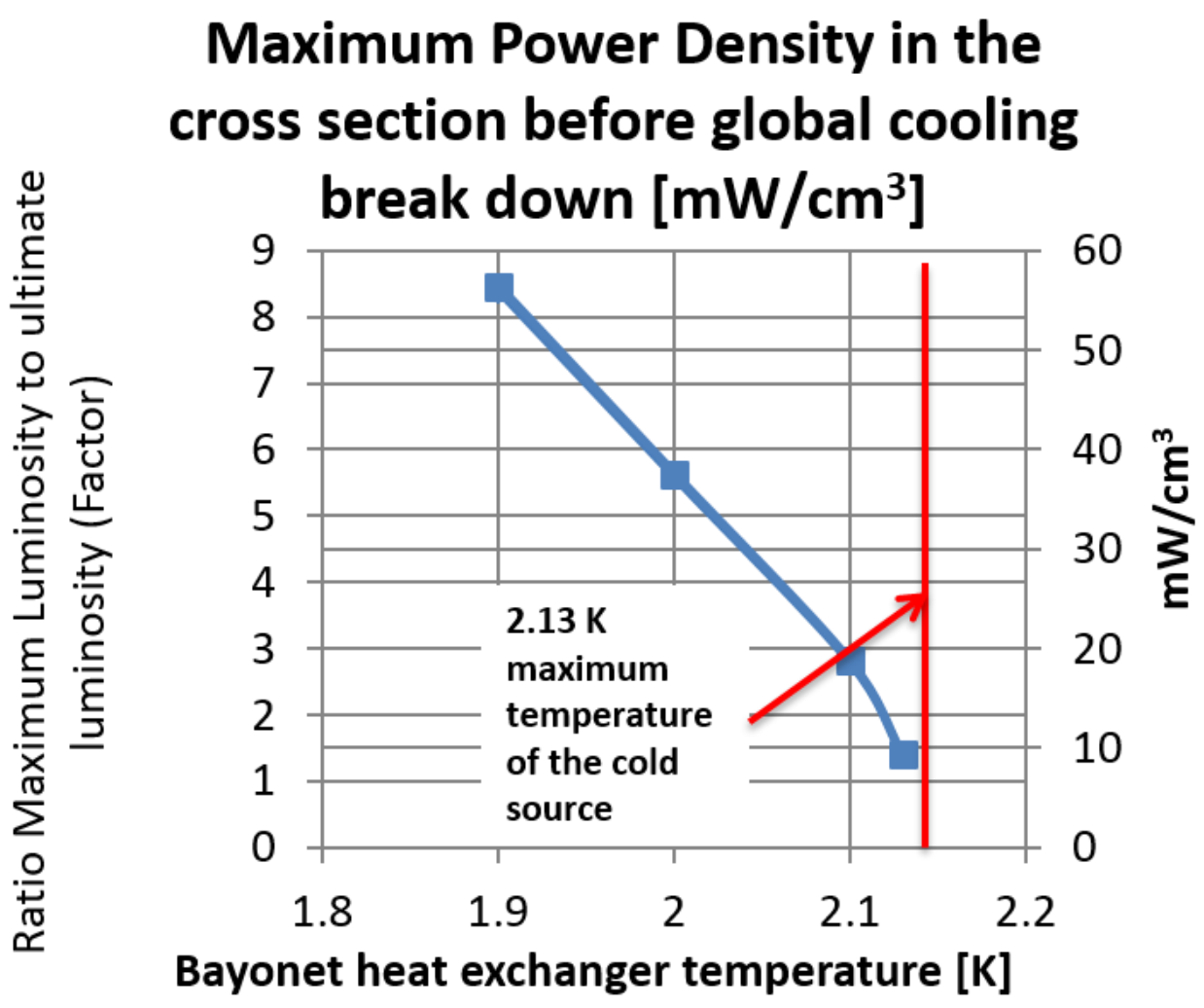

Figure 4.53: MQXF Quadrupole steady state power margin.

## 4.6 Quench Protection

The requirements for MQXFA magnets protection in operating conditions are: hot-spot temperature < 350 K [1,2] coil-ground voltages < 670 V [3] and, turn-to-turn voltages < 160 V [4].

The MQXFA is protected with a combination of quench heaters (QH) and Coupling-Loss Induced Quench (CLIQ) system [5-8]. Upon quench detection, both QH and CLIQ units are triggered simultaneously. The QH units introduce a current through the QH strips attached to the coil and heat up the conductor by heat diffusion through a thin 50 µm insulation layer. The CLIQ units introduce oscillations in the magnet transport currents. The resulting local magnetic-field change introduces high inter-filament and inter-strand coupling loss, which heats the copper matrix of the strands. The two systems together constitute an effective synergy since they are based on different heating mechanisms and they deposit energy mainly in different parts of the winding pack.

Following a detailed study and evaluation of both Inner Layer (IL) and Outer Layer (OL) quench heaters, the configuration selected for the production magnets includes only the outer layer quench heaters [9]. Each magnet is equipped with 16 quench heater strips glued to the outer layers, and with two CLIQ terminals. This system has been optimized by assessing the performance of different types of quench heater strips, of alternative CLIQ configurations, and of combinations of these. Dedicated experimental studies have confirmed that adequate quench protection of the MQXF magnets in the LHC tunnel can be achieved in both nominal conditions and failure scenarios

### 4.6.1 Quench heater strips

Four quench heater strips are glued to the outer layer of each pole:

- Two strips (one on each side of the pole) in the low-field region near the magnetic mid-plane
- Two strips (one on each side of the pole) in the high-field region near the magnetic pole (B02, B03)

The quench heaters are made of stainless steel strips with copper plated sections to lower the resistance between heating stations, as shown in Figure 4.54. Their position in the coil cross-section and naming conventions are shown in Figure 4.55. Note that the strips of each pole are numbered starting from the strip closest to the lead end.

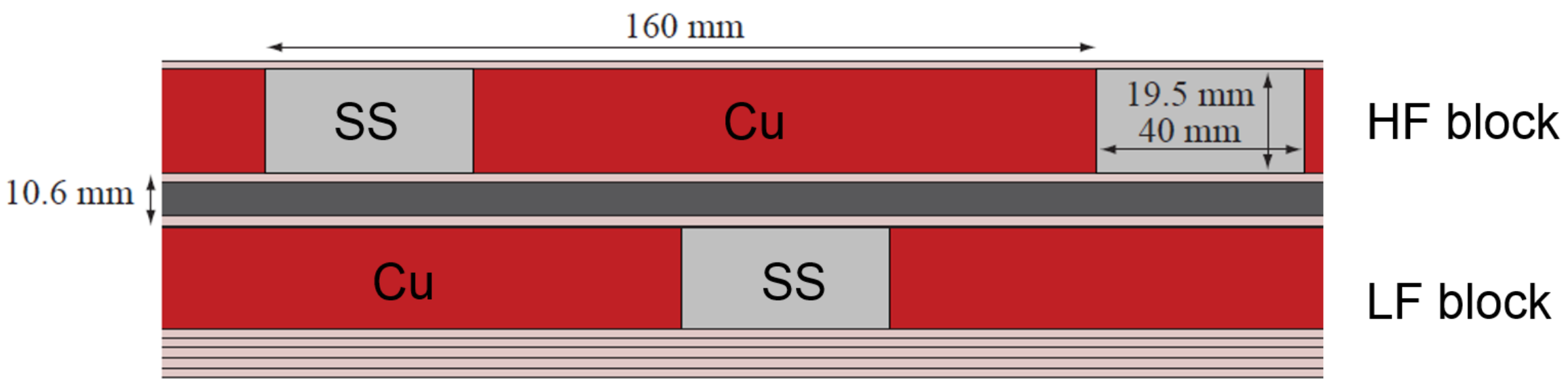

Figure 4.54: Design of the MQXFA outer layer quench heaters.

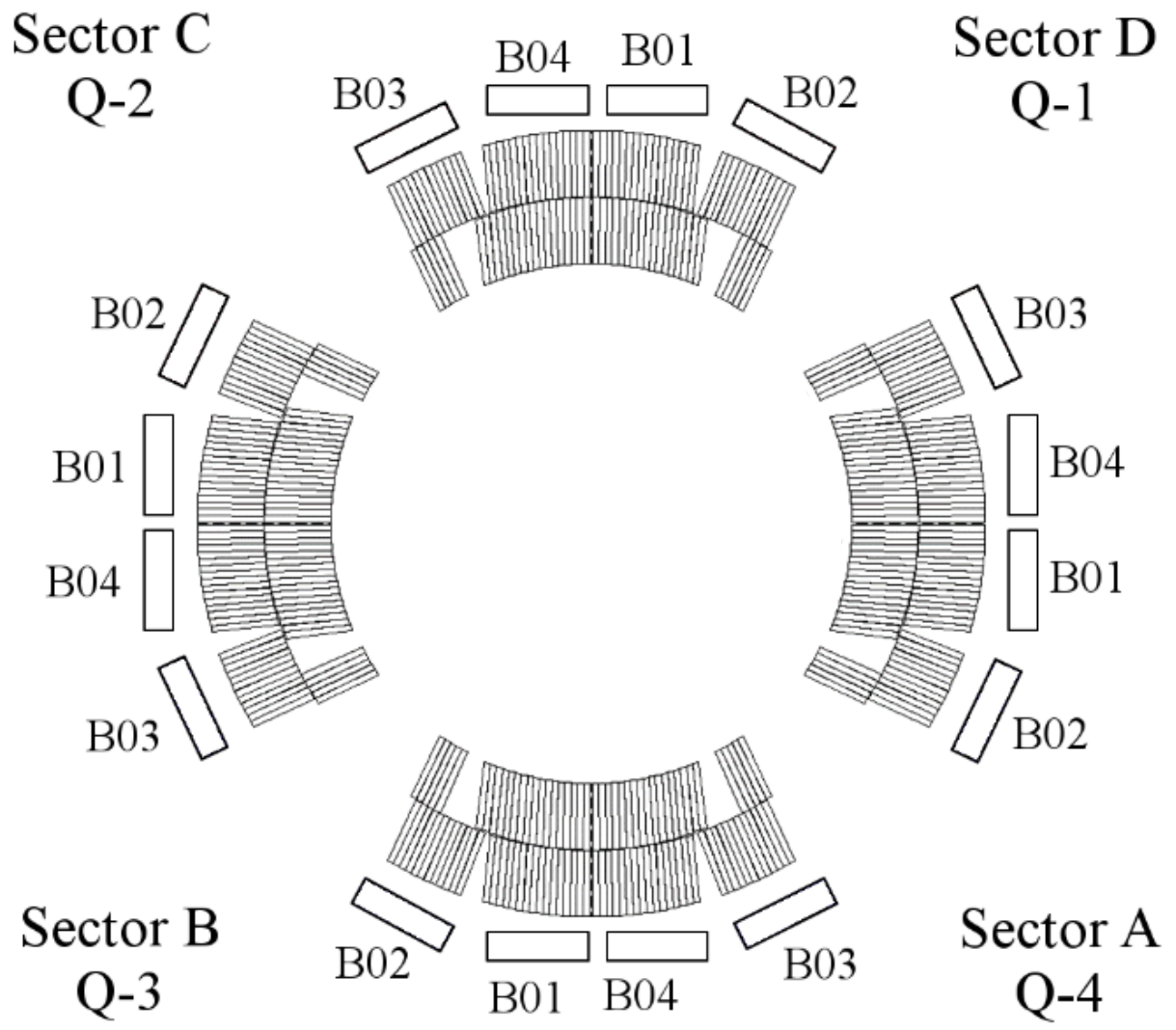

Figure 4.55: Position of the heater strips attached to the MQXFA magnets, viewed from the lead side.

The main parameters of the two types of heater strips and their estimated resistances at cryogenic and room temperature are summarized in Table 4.. For the resistance calculations, the applied magnetic field is considered nihil, the resistivity of stainless steel is assumed to be 5.45E-7 and 7.31E-7 Ωm at 10 and 293 K, respectively (RRR=1.34), and that of copper 6.00E-10 and 1.76E-8 Ωm, respectively (RRR=29).

Table 4.13: Parameters of the quench heater strips attached to the MQXFA magnet

| QH strip location | High-field outer layer | Low-field outer layer |
|---|---|---|
| QH design type [10] | Cu-plated 2 OL | Cu-plated 2 OL |
| Strip length [m] | 4 | 4 |
| Heater SS width [mm] | 20 | 20 |
| Heater Cu width [mm] | 20 | 20 |
| Heater SS thickness [mm] | 0.025 | 0.025 |
| Heater Cu thickness [mm] | 0.01 | 0.01 |
| Station length [mm] | 40 | 40 |
| Station period [mm] | 160 | 160 |
| Number of stations | 26 | 26 |
| Strip resistance @ 10 K [Ω] | 1.10 | 1.14 |
| Strip resistance @ 293 K [Ω] | 1.73 | 1.80 |

Each pair of QH strips is connected in series to an LHC standard QH power supply, featuring a capacitor bank of 7.05 mF charged to 900 V, with a series resistance that will reduce the voltage across the heater circuit to ~600 V. In order to reduce the peak heater-to-coil voltages , the pairs are composed of strips glued to poles that are adjacent to each other (P1-P2 and P3-P4), as shown in Figure 4.55. The resistance of the copper leads connected to the strips is assumed to be 0.6 Ω per circuit, the same as the present LHC QH circuits. The expected performances of the QH circuits are summarized in Table 4.14.

Table 4.14: Expected performances of the QH circuits protecting MQXFA

| QH strip location | High-field outer layer | Low-field outer layer |
|---|---|---|
| Number of strips in series | 2 | 2 |
| Strip resistance @ 10 K [Ω] | 1.10 | 1.14 |
| Resistance of the warm leads [Ω] | 0.60 | 0.60 |
| QH circuit resistance @ 10 K [Ω] | 2.80 | 2.89 |
| QH supply charging voltage [V] | 900 | 900 |
| QH supply capacitance [mF] | 7.05 | 7.05 |
| Expected peak QH voltage [V] | 600 | 600 |
| Expected peak QH current [A] | 214 | 208 |
| Expected QH discharge time constant [ms] | 32 | 32 |
| Peak QH power density [W/cm$^2$] | 250 | 235 |
| QH energy density [J/cm$^2$] | 2.47 | 2.40 |

### 4.6.2 CLIQ terminals and leads

Each cold mass (Q1/Q3) is composed of two MQXFA magnets, powered in series. The MQXF electrical circuit is presented in [11]. The presence of parallel elements (thyristors, diodes) across each cold mass allows analyzing the discharge of each cold mass separately. In Figure 4.56, the part of the circuit corresponding to one cold mass is schematized. Each two MQXFA magnets are protected by two CLIQ units connected as shown in Figure 4.56. Two CLIQ leads are attached to each magnet, located at the joint between two electrically connected poles (see taps 'a' and 'b' shown in Figure 4.56). Their terminals are connected at the coil ends in the “pizza box” in Figure 8.10. The electrical order of the four poles is as follows: Q-1, Q-4, Q-2, Q-3 [12]. The resulting configuration, called “Crossed-Poles”, allows introducing opposite current changes in poles which are physically adjacent (P1-P3 and P2-P4), which is the most effective option for this magnet [5,8].

Each CLIQ unit features a 40 mF capacitor bank charged to 600 V.

The parameters of the conductor used for the CLIQ leads are summarized in Table 4.. The copper cross-section of the leads is dimensioned to avoid overheating during the CLIQ discharge and to not limit the CLIQ performance due to an excessive electrical resistance. The temperature in the leads is expected to remain well below 300 K.

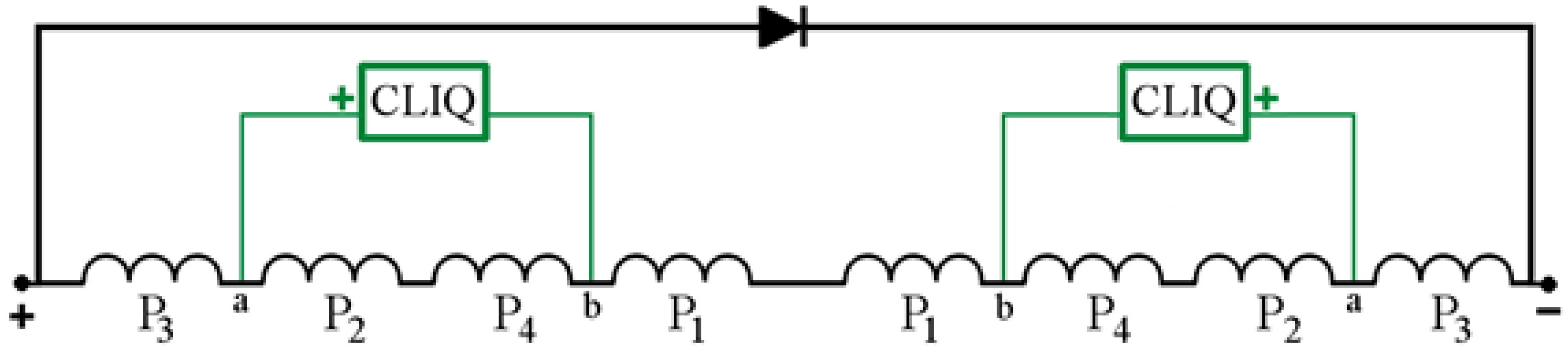

Figure 4.56: MQXFA electrical order of the poles and position of the CLIQ terminals.

Table 4.15: Parameters of the conductor used for the CLIQ leads of the MQXFA magnet

| Parameter | Value |
|---|---|
| Material | Silver Plated Copper |
| Construction | 133 filaments |
| Filament diameter | 0.287 mm |
| Strand diameter | 4.09 mm |
| Conductor area | 8.60 mm$^2$ |
| Conductor resistance per unit length | 0.217 Ω/100m |
| Insulation material | Extruded polyimide |
| Strand diameter with insulation | 4.95 mm |
| RRR | >100 |

### 4.6.3 Expected quench protection performance

The detailed study of the MQXF quench protection and redundancy is presented in [11]. This analysis actually shows the performance of the protection system in the case of the 7.15 m long Q2a/Q2b magnets, which have the same cross-section and conductor parameters but different length. However, the same limits and safety factors on the coil's hot-spot temperature and peak voltage to ground are assumed for the longer and shorter magnets. In particular, the electrical quality assurance of the 4.2 m long MQXFA magnet will be performed at the same voltages as the 7.15 m long magnet.

Two options for the quench protection system are analyzed in the above-mentioned report:

- QH attached to the coil's outer layers (O-QH)
- QH attached to the coil's outer layers and CLIQ (O-QH + CLIQ).

The main quench protection simulation results, at nominal current plus margin and at ultimate current plus margin, are summarized in Table 4.16. Sensitivity to strand parameters and initial quench position are included in the analysis. It is shown that a protection system including only quench heaters attached to the coil's outer layers does not offer the required level of redundancy and barely maintains the coil's hot-spot temperature below 350 K, which is considered a conservative limit against permanent degradation. On the contrary, the other protection option effectively protects the magnet and limit the hot-spot temperature well below 350 K.

The consequences of QH circuit failures are studied, and the results reported in

Table 4.. The worst-case considered in the analysis is the failure of two independent QH circuits. Before the publication of the report [11], the worst-case $U_{g,peak}$ reference value was 520 V, calculated in the case of O-QH+CLIQ [12]. The increase with respect to this value comes from the improvement in the model accuracy and from the detailed analysis of the effect of the initial hot-spot position. However, it is recommended that no correction of the test values during electrical quality be asked, considering that prudent safety margins were applied.

We assume that the total detection plus validation time is 15 ms [11]. This number is compatible with a voltage threshold of 100 mV, and a validation time of 10 ms.

Table 4.16: Simulated coil hot-spot temperature $T_{hot}$, peak voltage to ground $U_{g,peak}$ and peak turn to turn voltage $U_{t,peak}$ obtained after a quench at nominal+margin and at ultimate+margin current, for varying fraction of copper in the conductor, RRR and strand diameter. Uncertainty ranges also include the effect of different quench locations.

| Configuration | $T_{hot}$ [K] | $U_{g,peak}$ [V] | $U_{t,peak}$ [V] |
|---|---|---|---|
| Nominal current plus margin (16.47 kA) | | | |
| O-QH | 293-364 | 304-619 | 62-123 |
| O-QH + CLIQ | 215-248 | 521-658 | 49-90 |
| Ultimate current plus margin (17.89 kA) | | | |
| O-QH | 312-389 | 362-860 | 72-145 |
| O-QH + CLIQ | 237-273 | 664-924 | 61-109 |

Table 4.17: Failure case analysis. Simulated hot-spot temperature, peak voltage to ground and peak turn to turn voltage obtained for one failure or two simultaneous failures of QH circuits, at nominal+margin and at ultimate+margin current. Uncertainty ranges are due to the different locations of the initial quench and of the failing QH circuits.

| Configuration | $T_{hot}$ [K] | | | $U_{g,peak}$ [V] | | | $U_{t,peak}$ [V] | | |
|---|---|---|---|---|---|---|---|---|---|
| | No f | 1 | 2 | No f | 1 | 2 | No f | 1 | 2 |
| Nominal current plus margin (16.47 kA) | | | | | | | | | |
| O-QH | 330-345 | 345-362 | 363-384 | 577 | 702 | 868 | 113 | 122 | 132 |
| O-QH + CLIQ | 236-237 | 238-240 | 239-242 | 641 | 668 | 666 | 83 | 84 | 86 |
| Ultimate current plus margin (17.89 kA) | | | | | | | | | |
| O-QH | 352-369 | 364-385 | 379-406 | 808 | 916 | 1068 | 133 | 141 | 152 |
| O-QH + CLIQ | 260-262 | 261-264 | 262-267 | 874 | 910 | 909 | 101 | 103 | 105 |

### 4.6.4 **Peak voltages in MQXFA**

As already stated, MQXFA electrical test criteria [3] are based on MQXFB peak voltages, which have been reported in the previous section. For reference, we report also MQXFA peak voltages in this section.

The expected peak coil-ground voltage during a quench in MQXFA is

- MQXFA peak voltage coil-to-ground: 340 V

This peak voltage is computed at 16.47 kA (nominal current plus margin), in case of two QH failures. The peak voltage occurs immediately after the quench protection system is triggered, when the magnet is still cold (1.9 K). This voltage is lower than in MQXFB due to the lower CLIQ voltage in MQXFA (600 V) than in MQXFB (1000 V).

However, due to variations of RRR and Cu/NCu of coils (with values within conductor acceptance limits), the peak voltage to ground can be larger than nominal value. Indeed, due to RRR and Cu/NCu variations, coils have different resistance during a quench, leading to voltage unbalances inside the magnet. The Electrical Design Criteria document [3] states that the MQXFA magnet voltage to ground must be lower or equal to 353 V. In order to meet this requirement we proved that, even with the most extreme (and therefore unlikely) combination of coils RRR and Cu/NCu within accepted conductor limits, it is always possible to find some coil electrical orderings with peak voltage lower than 353 V [14]. This means that, once coils are produced and assigned to a magnet, it always possible to find some coil orderings that meet the Electrical Design Criteria.

Therefore, after 5 coils (including a spare coil) are assigned to a magnet, the following plan shall be executed:

- Average RRR and Cu/NCu are estimated for each coil using conductor data. RRR are estimated using rolled strand, minor and major edge extracted strands and witness sample value. Analysis of magnet test data showed that the best estimates of coil RRR are obtained using vendor rolled strand data and minor edge extracted strand data, as discussed in the MQXFA03 and 04 test reports [15].
- Peak voltages are computed for each possible coil electrical ordering, using both minor edge and rolled strand RRR data.
- All orderings, which have peak voltage lower than 353 V using both assumptions, are selected as candidates.
- The L3 or deputy for Magnet assembly selects the coil ordering among the list of candidates based on coil shimming and stress criteria.
- The selected coil ordering is reviewed during the Structure and Shims Review of each magnet.

### 4.6.5 Failure analysis of the coil-to-heater insulation

The insulation between coils and quench heaters is constituted by 150 µm of cable insulation (S2-glass impregnated with epoxy resin), plus 50 µm of polyimide layer (with an additional 50 µm polyimide layer in the ends). In case of damage to the polyimide insulation, the electrical robustness of the coil may become marginal because of the possible presence of epoxy cracks due to thermal contractions and electromagnetic forces. Indeed, the cracks may be filled by superfluid helium that evaporates during a quench. The breakdown voltage of helium gas decreases significantly as temperature increases and it could allow a short between heater and coil.

A complete analysis of a failure of the insulation between quench heaters and coil in MQXF magnet was performed [13]. Peak voltages between coils and quench heaters were computed during a quench to verify if they are still sustainable by the magnets when a damage in the insulation is present. The failure analysis was carried out under the following assumptions: (i) the polyimide insulation under an MQXF quench heater has some damages or cuts, (ii) the heater is not disconnected from the Heater Firing Unit, (iii) the location of a cut coincides with the location of a crack in the cable insulation (made of fiber-glass filled epoxy), (iv) they generate the worst-case scenario of a 200 um direct path cable-to-heater filled with helium, and (v) this path is located where the maximum coil-to-heater voltage is developed during a quench.

Under these assumptions, the coil-to-heater voltages at intermediate temperatures (353 V at ~100 K) during quench is comparable to the helium gas breakdown voltage at 1 bar pressure. However, the pressure of helium gas in the crack is expected to increase significantly with temperature, and the helium voltage breakdown value increases with pressure. A significant helium pressure increase is expected in the coil outer layer cracks where the polyimide ground-insulation, compressed between coil and structure, acts as seal. In MQXFA magnets the contact pressure between coil and structure is always larger than 100 bar. Assuming that the seal can sustain at least 10% of this pressure, the pressure of helium in the crack should be equal to or higher than 10 bar. The helium voltage breakdown at 100 K and 10 bar is higher than 1 kV for a 0.2 mm path. Therefore, it provides ≥ 1200 V margin in MQXFA magnets.

The case of the helium, close to the heaters, warming up immediately after quench heaters are fired was also taken into account. Since heaters warm-up and cool-down very rapidly with respect to coils, the scenario to be considered is the one immediately (~20 ms) after heaters firing. At this time the helium close to the heaters is warm and the coil-to-heater voltages can reach high values. On the other hand, the coil is still cold at this time. Assuming that half of the helium is “cold” (10

K), and half of the helium is "warm" (150 K), helium alone can withstand up to 1.5 kV, that is larger than the expected worst peak coil-to-heater voltage (~353 V in MQXFA) immediately after heater discharge

This failure analysis [13] shows that under those assumptions, MQXFA magnets are expected not to suffer coil-to-heater voltage breakdown even in the worst-case scenario of a direct coil-to-heater path.

### 4.6.6 Quench protection during single magnet tests

During the test of a single MQXFA magnet, one 40 mF, 600 V CLIQ unit is connected to the magnet. The resulting configuration, shown in Figure 4.57, allows introducing in the magnet poles the same oscillating currents as in the baseline LHC configuration.

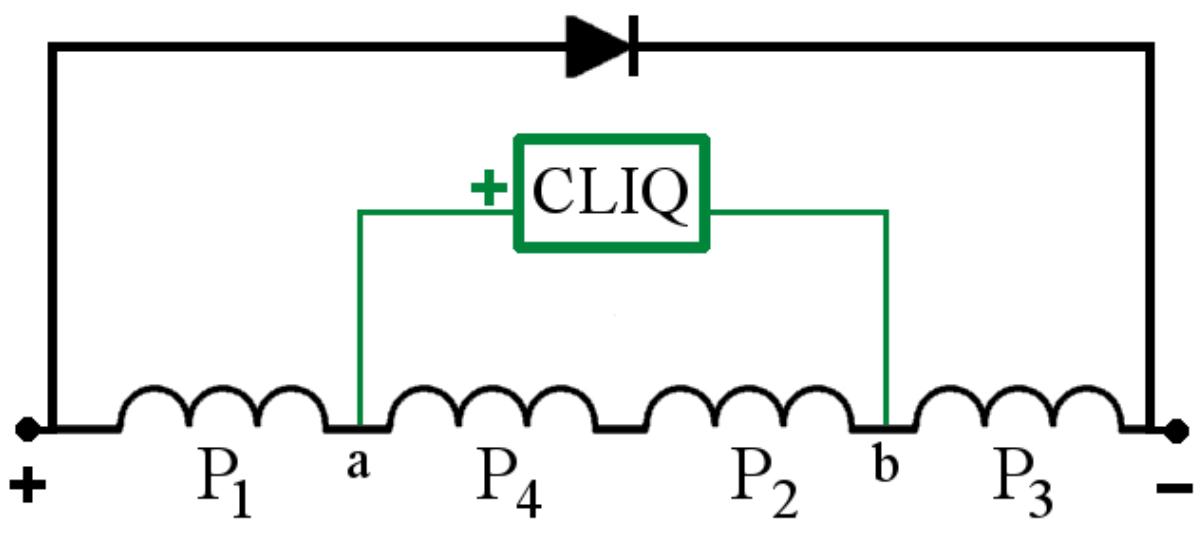

Figure 4.57: MQXFA electrical order of the poles and position of the CLIQ terminals.

### 4.6.7 Quench protection during cold mass tests

During the test of an entire cold mass, composed of two MQXFA magnets in series, the CLIQ configuration is identical to the final LHC machine case. Accordingly, the CLIQ connection scheme is presented in Figure 4.56.

# 5 MQXFA Coils

## 5.1 Coil Design and QA/QC Plan

The QXFA coil has a magnetic length of 4.2 m at the magnet operating temperature and is 4.529 m long at room temperature. It consists of $Nb_3Sn$ cable, stainless steel end parts, copper wedges, titanium pole pieces, and fiberglass insulation all encapsulated by epoxy. Specifications for series coils are in *QXFA Series Coil Production Specification* [1]. The specifications presented below are only for information only.

### 5.1.1 Coil Design

The QXFA coil is a two-layer cos-2$\theta$ coil with saddle-shaped ends. The two-layer coil is wound continuously, without a splice at transition between the inner and outer layers, using the double-pancake technique successfully used in all LARP coils and by several other $Nb_3Sn$ magnets. The cross-section for coil design is shown in Table 5.1. For the inner layer, there are 5 turns in the first conductor group, 17 turns for the second conductor group, and inner layer wedges in between the two conductor groups on both coil sides. For the outer layer, there are 12 turns in the first conductor group, 16 turns in the second conductor group, and outer layer wedges in between the two conductor groups on both coil sides.

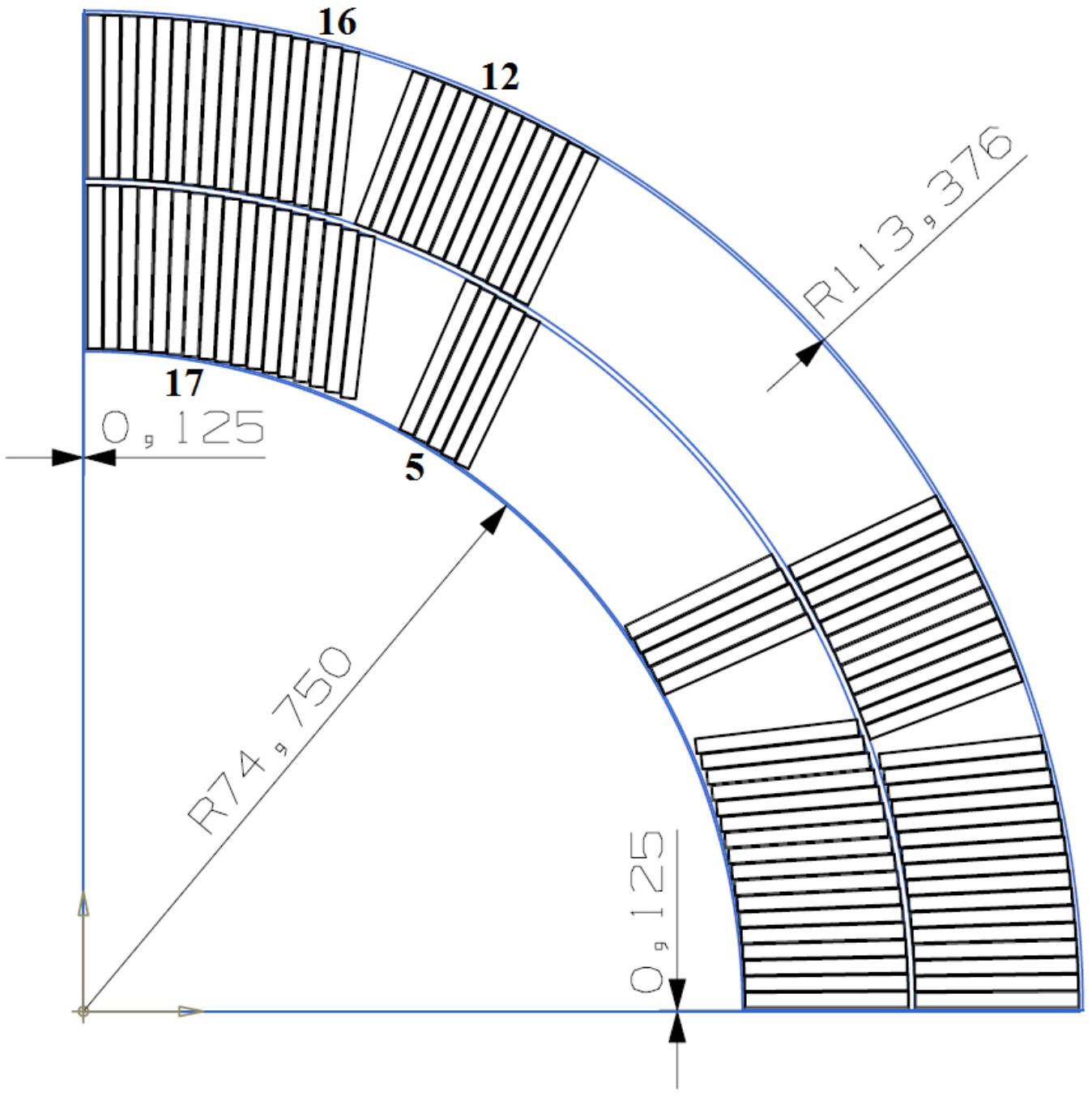

Figure 5.1: Coil Cross-section.

#### 5.1.1.1 Coil dimensions

Table 5.1 lists the coil dimensional specifications at room temperature. It should be noted that positive tolerances specify excessive material and negative tolerances specify less-than-nominal material. To meet the requirement of a magnetic length of 4.2 m, the overall coil length is 4.529 m. Reference drawing is QXFA Coil Assembly Drawing [2], and coil main components are shown in Fig 5.2.

Table 5.1 Coil Dimensional Specification (see text for note about tolerances)

| | Dimension (mm) | Tolerance (mm) |
|---|---|---|
| Inner radius profile at pole | 74.750 | + 0.0, - 0.250 |
| Inner radius profile except pole | 74.750 | + 0.150, - 0.250 |
| Outer radius profile | 113.376 | + 0.200, - 0.250 |
| Coil to midplane gap | 0.125 | + 0.075, - 0.125 |
| Coil length | 4529.0 | ± 5.0 |
| Pole Keyway azimuthal position | Centered with respect to midplanes | ± 0.25 |

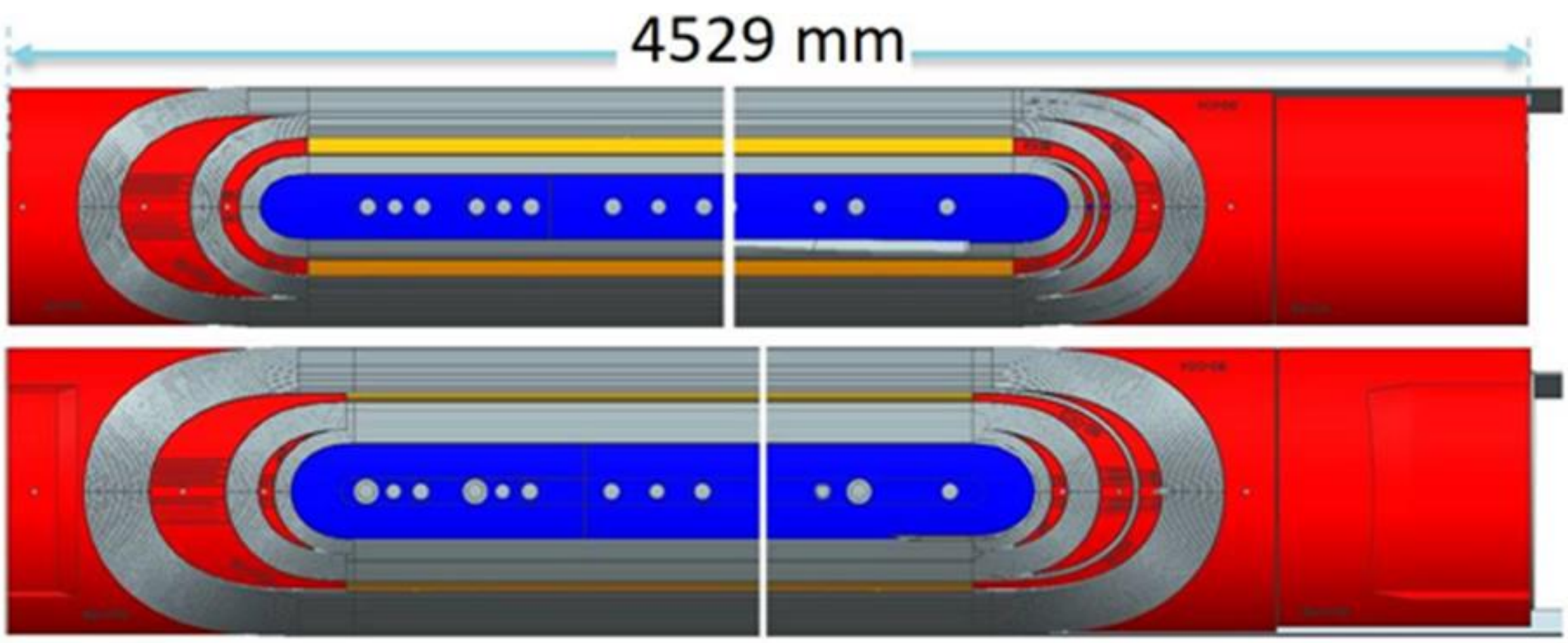

Figure 5.2: QXFA Coil inner layer (top) and outer layer (bottom) showing main components: $Nb_3Sn$ cable (in grey), plasma coated stainless steel end parts (in red), copper wedges (in orange), titanium pole pieces (in blue).

#### 5.1.1.2 Coil Heat Treatment

Coil heat treatment takes place in a gas tight oven using an automatic program. The reaction fixture must be sufficiently gas tight to assure internal overpressure.

The coil heat treatment cycle is shown in Table 5.2. Coil heat treatment specifications are in [1]. Coil heat treatment cycle is measured by thermocouples set in thermal contact with the reaction fixture. The durations and ramp rates for the oven control may be varied in order to achieve the target temperature for the reaction fixture. Witness samples are the main tool used to verify quality of the heat treatment cycle. Witness samples are tested shortly after each coil heat treatment and shall demonstrate critical current and RRR above requirement values (section 3.1). The thermocouples set on the reaction fixture are used to verify the uniformity of the heat treatment on the whole coil.

Table 5.2: QXFA Coil Reaction Cycle.

| | Step | Description | Average Rate | Time |
|---|---|---|---|---|
| Step 1 | Ramp | from 20 °C to 210 °C | 25 C/hour | |
| Step 2 | Soak | 210 °C | | 48 hours |
| Step 3 | Ramp | from 210 °C to 395 °C | 50 C/hour | |
| Step 4 | Soak | 395 °C | | 48 hours |
| Step 5 | Ramp | from 395 °C to 665 °C | 50 C/hour | |
| Step 6 | Soak | 665 °C | | 50 hours |
| Step 7 | Ramp | from 665 °C to 20 °C | | ~ 80 hours |

In case a coil heat treatment is stopped (for instance by a power outage) the oven may be restarted after checking its functionality, the data acquisition and the argon flow. Restart procedures are described in [1].

#### 5.1.1.3 Coil Epoxy Impregnation

Each coil is impregnated with CTD-101K epoxy. Coil epoxy impregnation specifications are in [1]. The cycle for coil epoxy curing is presented in Table 5.3.

Table 5.3: QXFA Coil Epoxy Curing Cycle.

| Step | Description | Time / Rate |
|---|---|---|
| Ramp | from 55°C to 110°C | 10 C/hour |
| Soak | 110°C | 5-7 hours |
| Ramp | from 110°C to 125°C | 10 C/hour |
| Soak | 125°C | 16-18 hours |

### 5.1.2 Coil QA/QC Plan

#### a. Coil Electrical Measurements

The coil QA/QC plan consists of electrical measurements made throughout the fabrication process. Electrical continuity check between the coil and the ground, and between the coil and the coil parts is performed during coil winding and curing, and during preparation for coil reaction and impregnation. Electrical continuity check between the coil and the quench heater is performed after coil has been epoxy impregnated. When the coil is transferred to the shipping fixture, coil resistance, Ls and Q are measured, and Table 5.4 shows the measurement ranges for acceptance. Hipot test and impulse test are also performed when the coil is ready to be shipped. The specifications for these tests are written in Coil Fabrication Electrical QA [3], based on MQXFA Electrical design criteria [4].

Table 5.4 Coil Electrical Measurement Ranges

| **Measurement** | **Ref. Min-Max** |
|---|---|
| *Coil Resistance @ 1A* | 590.00-610.00 mV |
| *Ls @ 20 Hz* | 4.80-5.10 mH |
| *Q @ 20 Hz* | 0.80-0.90 |
| *Ls @ 100 Hz* | 3.20-3.50 mH |
| *Q @ 100 Hz* | 1.50-1.70 |
| *Ls @ 1 kHz* | 1.80-2.00 mH |
| *Q @ 1 kHz* | 1.90-2.10 |

#### b. Coil Mechanical Measurements

The coil QA/QC plan includes mechanical measurements made throughout the fabrication process. After impregnation, the coil is measured at 14 cross sections with a coordinate measurement machine, the coil CMM locations are presented in [5].

## 5.2 Coil Fabrication Components

QXFA coil, as well as its components, is designed and ready for the series coil production. There are two coil assembly drawings, F10115641 Rev. B and F10115848 Rev. B, that depict the coil dimension and the components at different coil fabrication stages. Fig. 5.3 shows the impregnated coil drawing with its dimensions and tolerances. The coil dimensional specifications are listed in QXFA coil production specification [1]. The drawings and the material of the coil fabrication components, including insulated $Nb_3Sn$ cable, poles, end parts, wedges, quench heater trace, NbTi cable lead, coil insulation materials and other materials, are described in QXFA coil production specification [1].

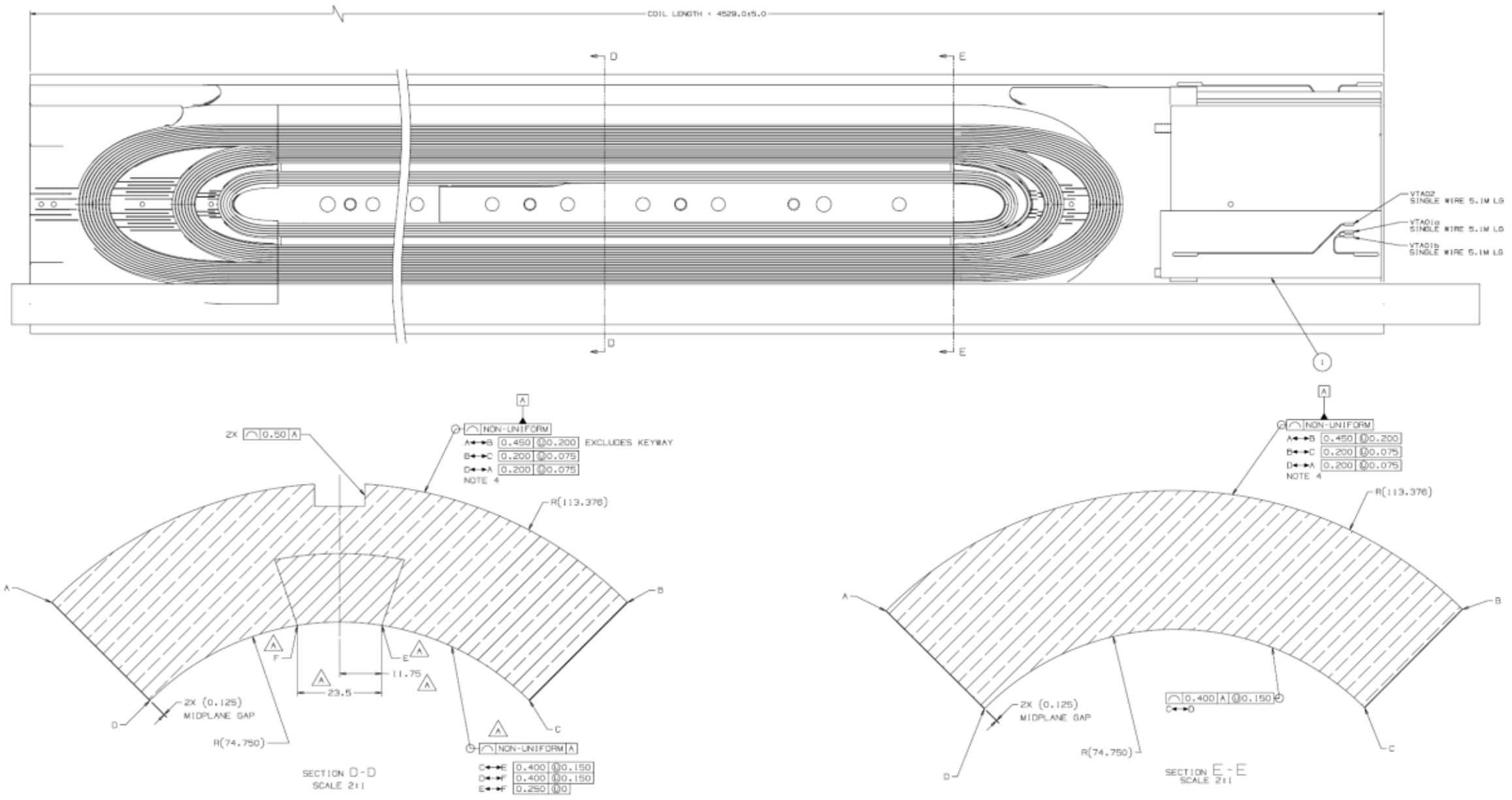

Fig. 5.3 Coil Dimension

The fabrication components for each coil, the part kit, is collected at FNAL and provided to both FNAL and BNL coil fabrication teams. Fig. 5.4 shows the template of the part kit list, with a specified coil ID and the fabrication drawings. In the column of Routing Form number, the QC tracking information, the assigned cable ID and the assigned quench heater trace ID are located.

| | |
|---|---|
| Coil Serial No. | QXFA 215 |
| Kit submitted by: | Miao Yu |
| Date submitted: | 05-Feb-20 |
| Shipping date: | 14-Feb-20 |
| Deliver to: | BNL Bldg 902E - J Schmalzle |

Coil ID

Coil fabrication drawings

| | | | | |
|---|---|---|---|---|
| Drawing Title: | QXFA Coil Assembly_WCR | | QXFA Coil Assembly_Impregnation | |
| Drawing No. | F10115641 | Rev. B | F10115848 | Rev. B |

| | Description | Drawing Number | Rev | Unit | Quantity Required | Location | Quantity Issued | E&F Verified | Routing Form Number | Spares Number / Comment | Step |
|---|---|---|---|---|---|---|---|---|---|---|---|
| 1 | L1 Pole | F10039545 | C | EA | 9 | R6,L2 | 9 | | 104256 | | Wind L1 |
| 2 | L1 Pole RE | F10005738 | C | EA | 1 | R6,L2 | 1 | | 104177 | | Wind L1 |
| 3 | L1 Pole LE | F10039540 | E | EA | 1 | R6,L2 | 1 | | 105856 | | |
| 4 | L2 Pole LE | F10039541 | E | EA | 1 | R6,L2 | 1 | | 105857 | | |
| 5 | L2 Pole | F10039546 | C | EA | 9 | R6,L2 | 9 | | 104237 | | Wind L2 |
| 6 | L2 Pole RE | F10005750 | C | EA | 1 | R6,L2 | 1 | | 104179 | | Wind L2 |
| 7 | QXFA COIL L1 WEDGE_Long | F10068150 | C | EA | 10 | R12,2 | 10 | | 102483 | | Wind Prep |
| 8 | QXFA COIL L1 WEDGE_Short | F10068151 | A | EA | 2 | R12,2 | 2 | | 103151 | | Wind Prep |
| 9 | QXFA COIL L2 WEDGE_Long | F10068148 | C | EA | 10 | R12,2 | 10 | | 102484 | | Wind Prep |
| 10 | QXFA COIL L2 WEDGE_Short | F10068149 | A | EA | 2 | R12,2 | 2 | | 103150 | | Wind Prep |
| 11 | 8 mm dia. 24 mm LG SS316 dowel pin | FC0050264 | - | EA | 22 | R2,S3,L3 | 22 | | 105808 | | |
| 12 | SPACER-1, LEAD END LAYER 1 B2-IC1 | F10088257 | - | EA | 1 | R12,2 | 1 | | 103969 | | |
| 25 | SPACER-1, RETURN END LAYER 2 B2-OR1 | F10042029 | - | EA | 1 | R12,2 | 1 | | 103977 | | |
| 26 | SPACER-2, RETURN END LAYER 2 B2-OR2 | F10042030 | - | EA | 1 | R6,L2 | 1 | | 105821 | | |
| 27 | SADDLE, RETURN END LAYER 2 B2-OR3 | F10042028 | C | EA | 1 | R6,L2 | 1 | | 104077 | | |
| 28 | Ramp turn shim | F10039954 | B | EA | 1 | IB3 | 1 | | PO#627617 | | |
| 29 | BNL QXFA Interlayer Insulation LE | F10130030 | A | EA | 1 | IB3 | 1 | | PRN217935 + PO#632999 | Require pre-cure with binder at BNL | Wind Prep |
| 30 | BNL QXFA Interlayer Insulation RE | F10130031 | A | EA | 1 | IB3 | 1 | | PRN217935 + PO#632999 | Require pre-cure with binder at BNL | Wind Prep |
| 31 | Interlayer Insulation LE Spliceblock | F10105529 | A | EA | 1 | IB3 | 1 | | PRN217935 + PO#632999 | | Impreg |
| 32 | QXFA Reaction Cloth LE | F10130345 | - | EA | 1 | IB3 | 1 | | PO#632999 | | React |
| 33 | QXFA Reaction Cloth RE | F10130346 | - | EA | 1 | IB3 | 1 | | PO#632999 | | React |
| 34 | Impregnation OD Cloth | F10058890 | B | EA | 1 | IB3 | 1 | | PO#632999 | | Impreg |
| 35 | Impregnation ID Cloth | F10058894 | B | EA | 1 | IB3 | 1 | | PRN217935 | | |
| 36 | QXFA OL Trace | su-1009-8330A | A | EA | 1 | IB3 | 1 | | 221-49-02-1 | | |
| 37 | QXFA IL Trace Splice | F10058887 | A | EA | 1 | IB3 | 1 | | | | Impreg |
| 38 | L1 LE Splice Block Pocket Filler | F10105542 | B | EA | 1 | IB3 | 1 | | PRN217935 | | Impreg |
| 43 | L2 RE Saddle Pocket Glass Filler | F10105556 | B | EA | 1 | IB3 | 1 | | 103905 | | Impreg |
| 44 | L2 RE Saddle Pocket Teflon Filler | F10105557 | B | EA | 1 | IB3 | 1 | | | | Impreg |
| 45 | G11 Midplane shim | F10108832 | - | EA | 2 | IB3 | 2 | | PO#650907 | | Impreg |
| 46 | IL Coil End Conductor Group Filler | F10110210 | - | EA | 1 | IB3 | 1 | | PO#632999 | | Impreg |
| 47 | OL Coil End Conductor Group Filler | F10105539 | - | EA | 1 | IB3 | 1 | | PRN217935 | | Impreg |
| 48 | S2 glass powder for paste | FC0066104 | A | Bag | 1 | IB3 | 0 | | PRN228445 | | Wind L2 |
| 49 | L2 LE trace Kapton insulation for Series QH Trace | F10104831 | - | EA | 2 | IB3 | 2 | | | | Impreg |
| 50 | L2 RE trace Kapton insulation | F10104911 | - | EA | 2 | IB3 | 2 | | | | Impreg |
| 51 | Voltage tap wires HH2619_twisted | FC0067203 | - | m | 22 | IB3 | 0 | | 103994 | | Impreg |
| 52 | Quench heater wires HH1819_twisted | FC0067205 | - | m | 12.2 | IB3 | 0 | | 104028 | | Impreg |
| 53 | Wedge Insulation_Heat cleaned | F10132315 | - | m | 20 | IB3 | 0 | | 102639 | | Wind Prep |
| 54 | Pole Insulation_heat cleaned | F10133503 | - | spool | 1 | IB3 | 1 | | PO#650689 | | Wind Prep |
| 55 | End Part Insulation_heat cleaned | F10133814 | - | m | 8 | IB3 | 0 | | PO#614544 | | Wind Prep |
| 56 | NbTi Lead | F10082715 | - | m | 8.55 | IB3 | 0 | | | | Impreg |
| 57 | $Nb_3Sn$ Cable | F10043195 | D | Spool | 1 | IB3 | 0 | | P43OL1129 | | Wind Prep |
| 58 | Witness Samples | - | - | Barrel | 6 | IB3A | 0 | | P43OL1129 | | React |
| 59 | G11 Splice Filler | F10108761 | - | EA | 2 | IB3 | 2 | | MSV-4083 | | Impreg |

RF# or PO#
To track the procurement and QC info.

Laser cut S2 glass

Template hand cut

Assigned QH trace ID

Assigned Nb3Sn cable and its witness samples

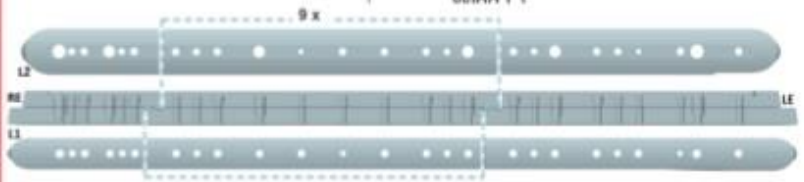

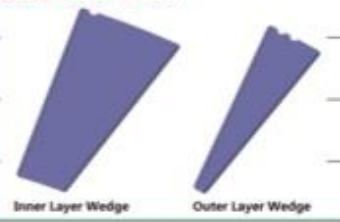

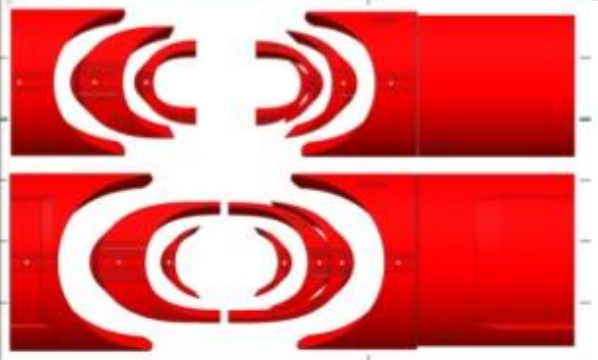

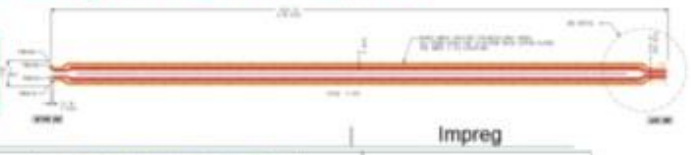

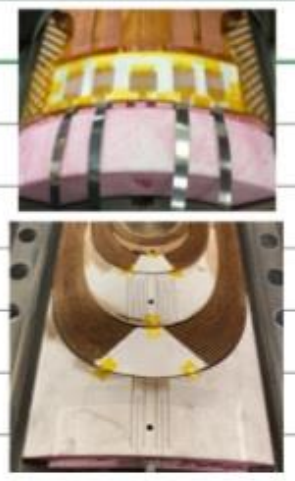

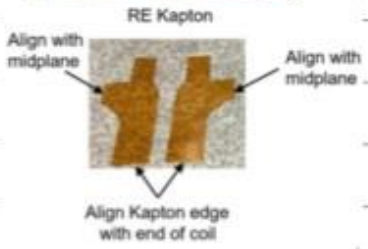

Fig. 5.4 QXFA Coil Part Kit List

## 5.3 Coil Fabrication at FNAL

### 5.3.1 Winding and Curing

The 2-layer QXFA coil is wound without a cable splice between the layers on the SELVA winding machine shown in figure 5.5. The winding machine bridge is on a track and has a boom attached that supports the spool of cable being used during winding. Within the boom is the tension mechanism that maintains the cable tension at the prescribed set point.

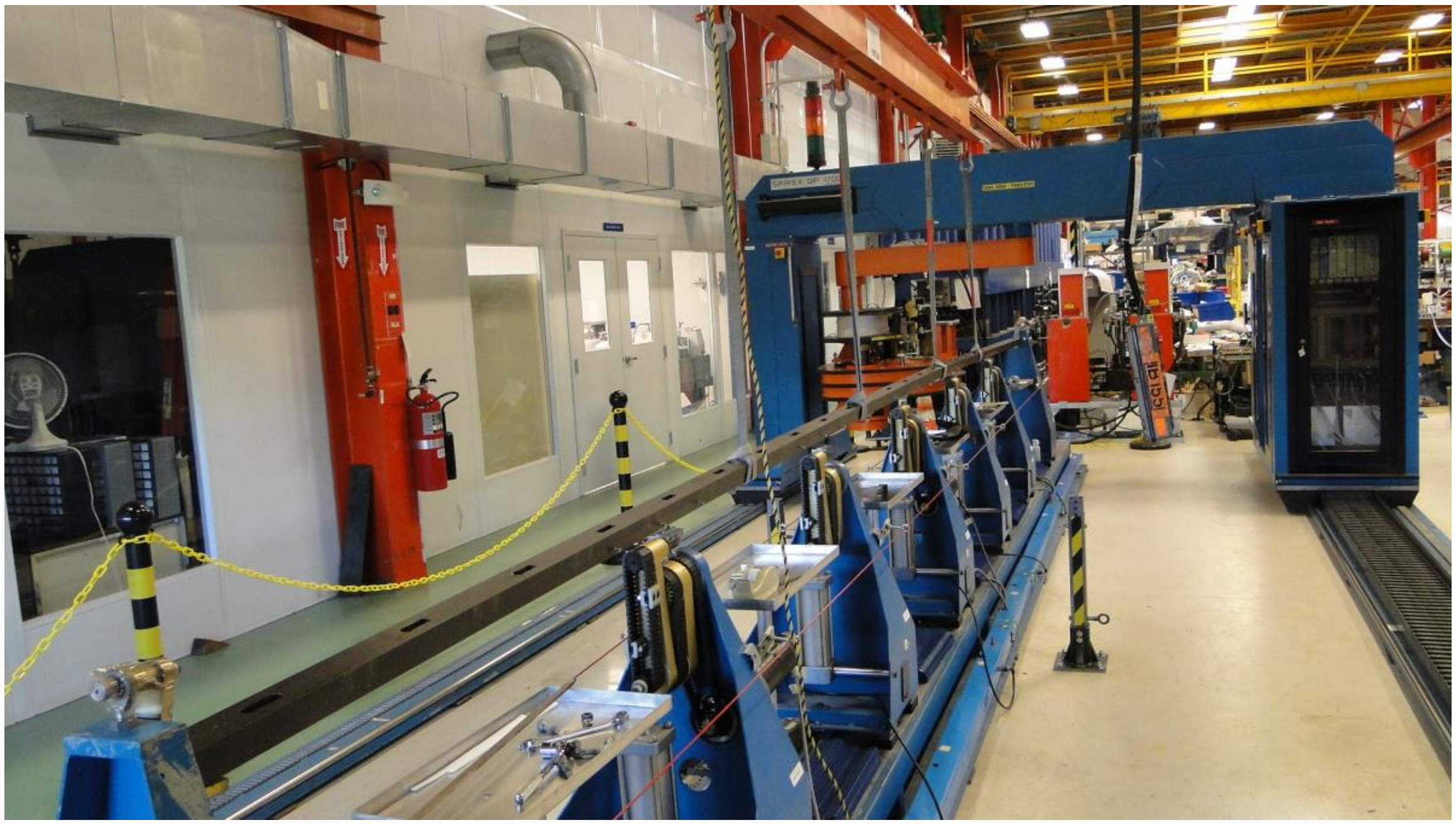

Figure 5.5: SELVA Winder.

The insulated cable is measured to verify the overall length and then split into two spools for the inner layer and the outer layer.

Before winding, the inner pole parts are prepared and setup on the winding mandrel with 5 layers of 0.125 mm thick S2 glass insulation wrapped around. To accommodate the coil length shrinking due to the tension release after curing and thermal effects during reaction, ten pole gaps totaling 12.7 mm are used. The total gap distributed along the wedge segments is 22.75 mm and is required to be no less than 2 mm/m longer than the pole gap, due to the thermal shrinkage difference between the wedge and the coil.

The coil winding and curing process is shown in Figure 5.6. The inner layer is wound first. The wound inner coil is uniformly painted with 324 grams of binder (CTD-1202) at a rate of 1.7 g/m per cable to all but the 3 turns adjacent to the pole. The coil is packaged with the curing tooling and shims, as shown in Figure 5.7. It is transferred to the curing mold through the rollover fixture. The inner layer coil is cured in the curing mold at 150° C for 1 hr and 45 mins, and after curing it is transferred back to SELVA through the rollover fixture. The 0.66 mm pre-cured interlayer insulation is placed on top of the inner layer coil, and the outer layer pole parts are set up on top of it with 4 layers of 0.125 mm thick S2 glass insulation wrapped around continuously from the inner layer. When the outer layer winding is completed, the outer layer is uniformly painted with 450 grams of binder (CTD-1202) at a rate of 1.7 g/m per cable to all but the 3 turns adjacent to the pole The coil is packaged as shown in Figure 5.7. It is transferred through the rollover fixture to the curing mold for 1 hr and 45 mins curing at 150°C. Each layer is cured under pressure in a precise closed cavity mold at 150 °C in air. While the coil is inverted relative to winding orientation, the pole is radially compressed with mandrel cylinders. Subsequently the coil is azimuthally compressed with the platen cylinders until closure which typically occurs at 13 MPa coil pressure. Curing is performed on the coil to set the coil size for reaction, as well as allow the coils to be easily handled, facilitating insertion into the reaction fixture without damage. Spacers simulating the outer layer are used during the curing of the inner layer. The outer layer is cured on the top of the inner layer. Therefore, the same mandrel and mold are used to cure both the inner and the outer layer.

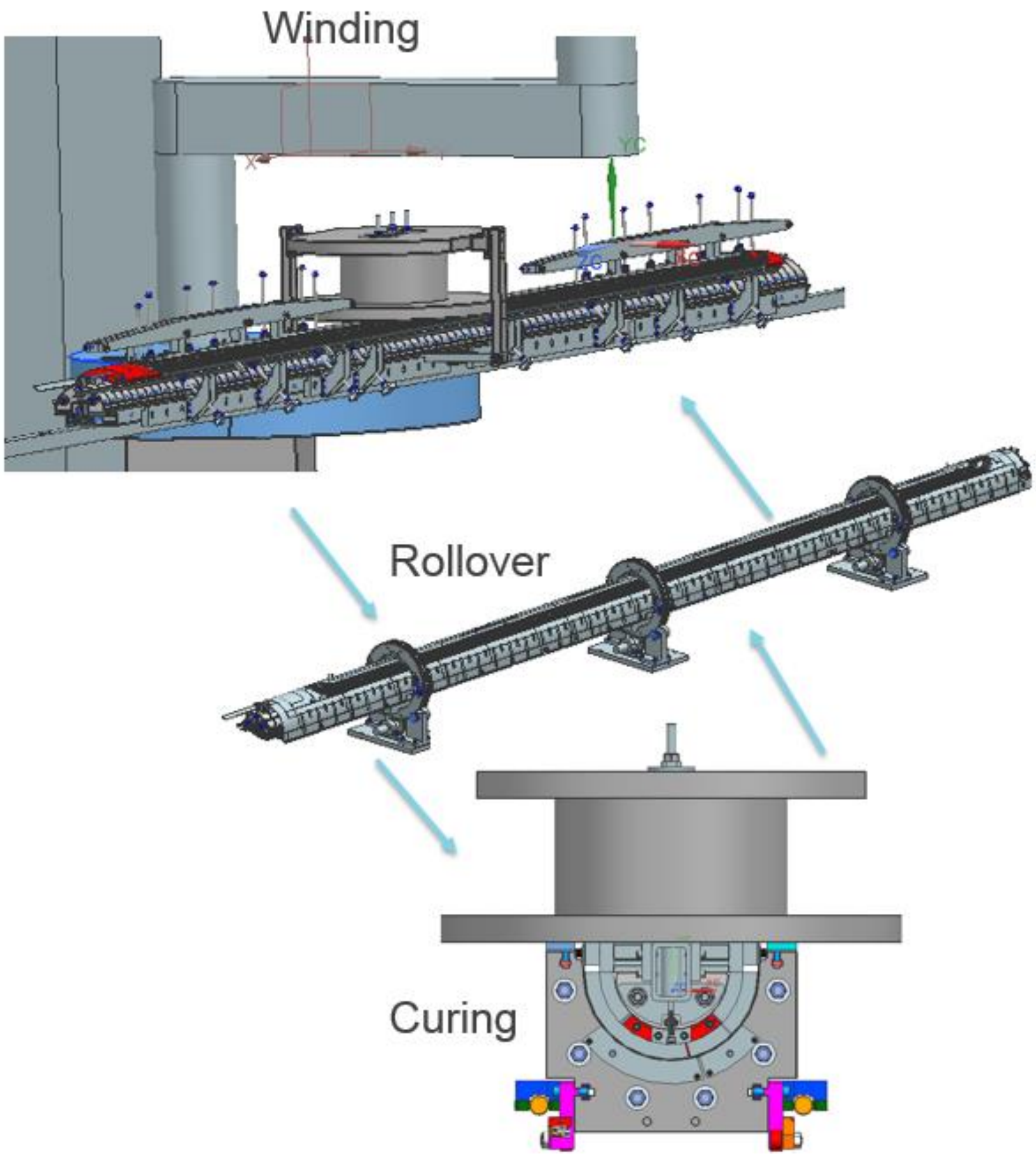

Figure 5.6: Winding and Curing Process.

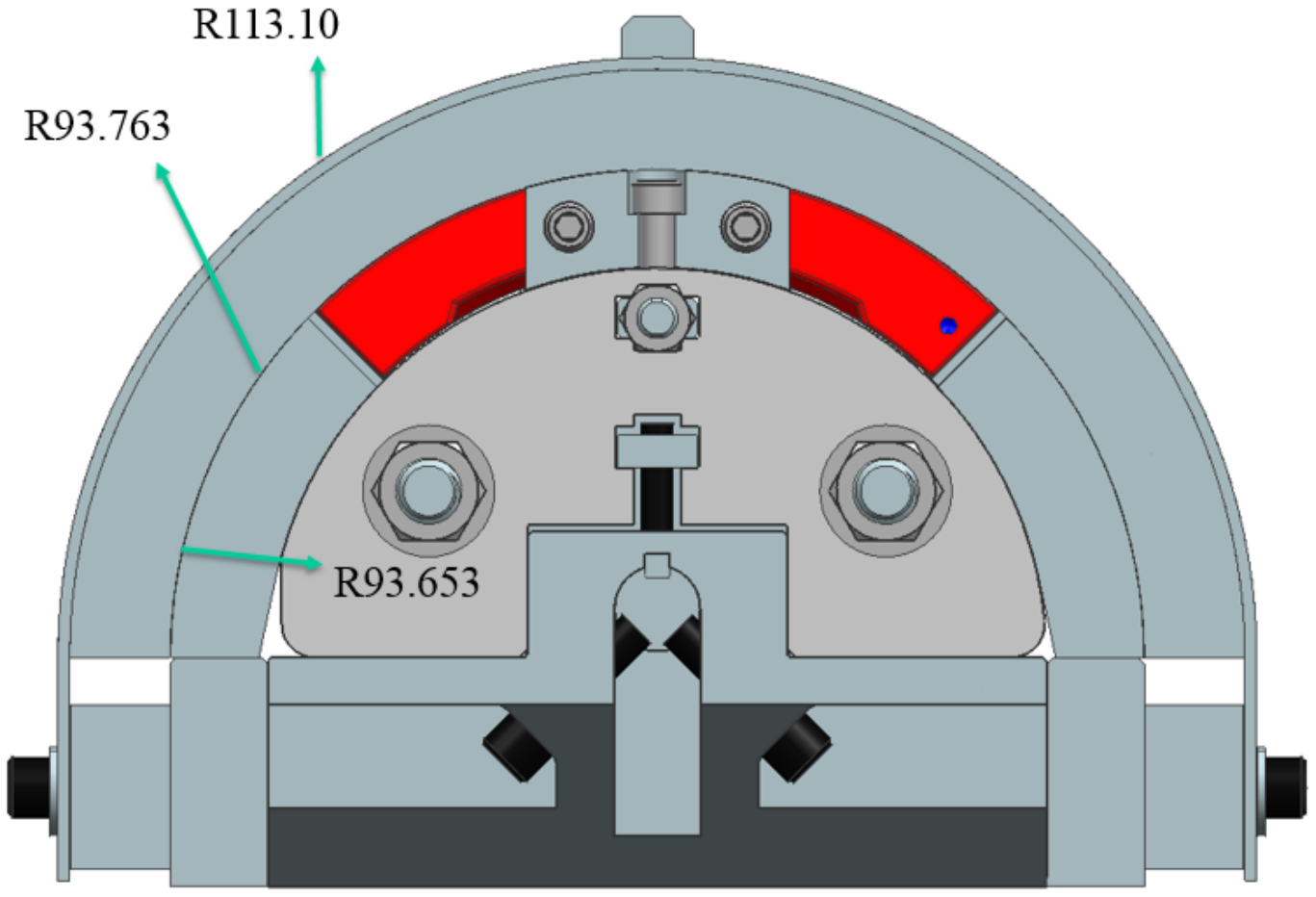

| Tooling/ Shims | R or t (mm) | + tol. (mm) | - tol. (mm) | Max. (mm) | Min. (mm) |
|---|---|---|---|---|---|
| Mandrel | 74.873 | 0.01 | 0.01 | | |
| Kapton | 0.127 | 0.01 | 0.01 | | |
| | 75 | 0.02 | 0.02 | | |
| L1 Sector pusher | 75 | 0.08 | 0 | 0.1 | -0.02 |
| Coil OD | 93.653 | 0 | 0.05 | | |
| Mylar | 0.1 | 0.01 | 0.01 | | |
| | 93.753 | 0.01 | 0.06 | | |
| Curing spacer | 93.763 | 0.08 | 0 | 0.15 | 0 |
| | 113.10 | 0 | 0.08 | | |
| Curing retainer | 113.10 | 0.05 | 0 | 0.13 | 0 |
| 90mil stock | 2.286 | 0.014 | 0.127 | | |
| | 115.386 | 0.014 | 0.127 | | |
| Curing mold | 115.4 | 0.01 | 0 | 0.151 | 0 |

Figure 5.7: Inner Layer Packaging for Curing.

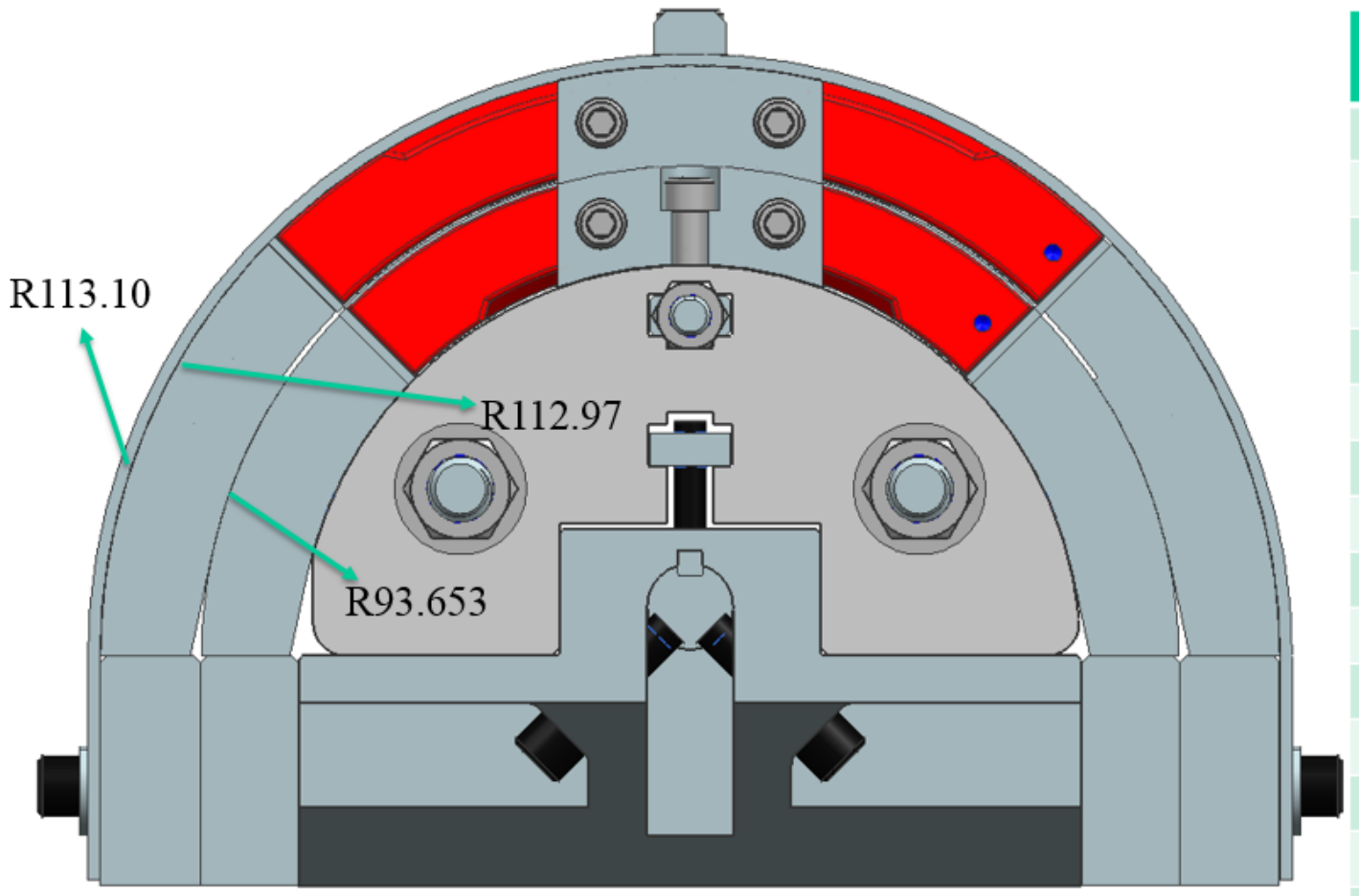

| Tooling/ Shims | R or t (mm) | + tol. (mm) | - tol. (mm) | Max. (mm) | Min. (mm) |
|---|---|---|---|---|---|
| Mandrel | 74.873 | 0.01 | 0.01 | | |
| Kapton | 0.127 | 0.01 | 0.01 | | |
| | 75 | 0.02 | 0.02 | | |
| L1 pusher | 75 | 0.08 | 0 | 0.1 | -0.02 |
| | 93.653 | 0 | 0.08 | | |
| Interlayer | 0.660 | 0.019 | 0.048 | | |
| | 94.313 | 0.019 | 0.128 | | |
| L2 pusher | 93.653 | 0.08 | 0 | 0.16 | 0 |
| Cutout | 94.653 | 0.08 | 0 | 0.321 | 0.548 |
| Coil OD | 112.97 | 0 | 0.05 | | |
| Mylar | 0.1 | 0.01 | 0.01 | | |
| | 113.07 | 0.01 | 0.06 | | |
| Curing retainer | 113.10 | 0.05 | 0 | 0.14 | 0.02 |
| 90 mil stock | 2.286 | 0.014 | 0.127 | | |
| | 115.386 | 0.014 | 0.127 | | |
| Curing mold | 115.4 | 0.01 | 0 | 0.151 | 0 |

Figure 5.8: Outer Layer Packaging for Curing.

The 8.8 m long curing press is shown in Figure 5.9, and the press load to cure a QXFA coil is shown in Table 5.5 The results of the stress analysis are shown in Figure 5.10. The maximum stress under normal operation is less than 100 MPa, lower than 1/3 of steel 1050’s yield strength (580 MPa). A detailed version of the procedures can be found online at [1].

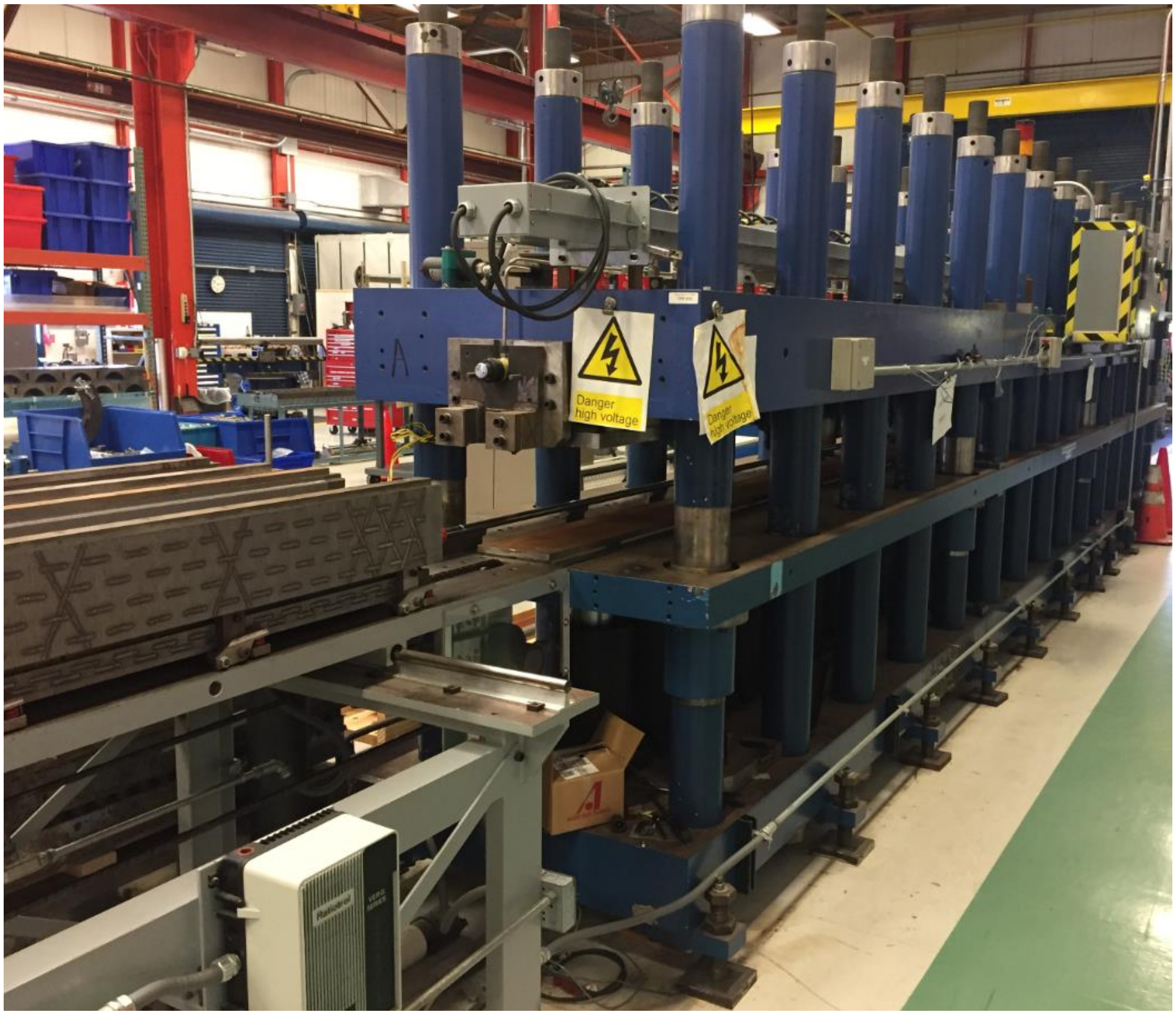

Figure 5.9: Curing Press

Table 5.5: Curing Press Load for Curing QXFA Coil.

| | Capacity (pump psi) | Max. force/cylinder kN (ton) | Spacing cm(inch) | Unit Force kN/m/psi (lb/in/psi) |
|---|---|---|---|---|
| Main Cylinders | 10000 | 1780 (200) | 30 (12) | 0.6 (3.34) |
| Mandrel Cylinders | 10000 | 134 (15) | 15 (6) | 0.09 (0.5) |

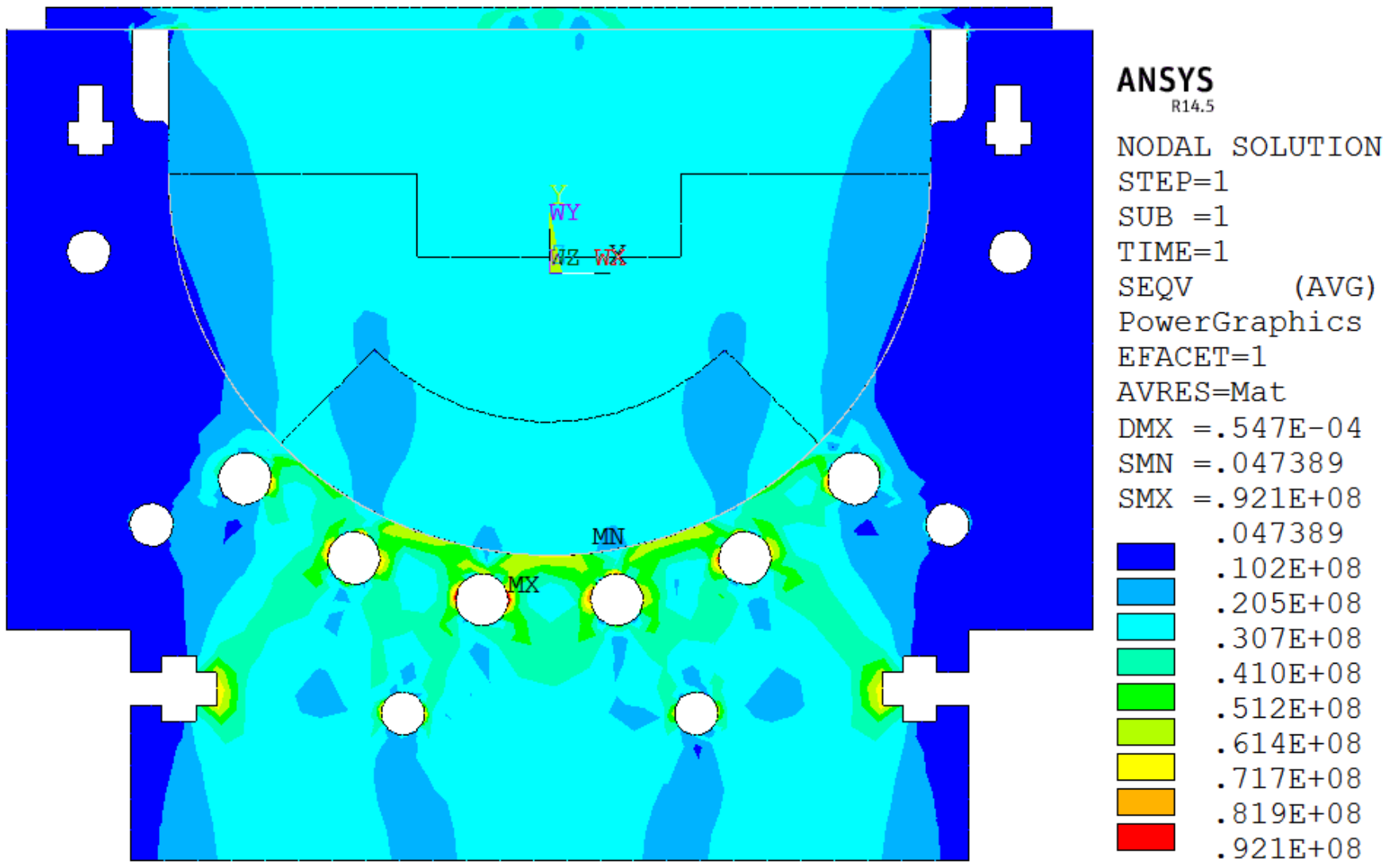

Figure 5.10: QXFA coil curing mold analysis.

To reduce the production risk, the rotation table is prepared as the backup winder for coil winding, as shown in Figure 5.11, which was used for LARP LQ and LHQ coil winding.

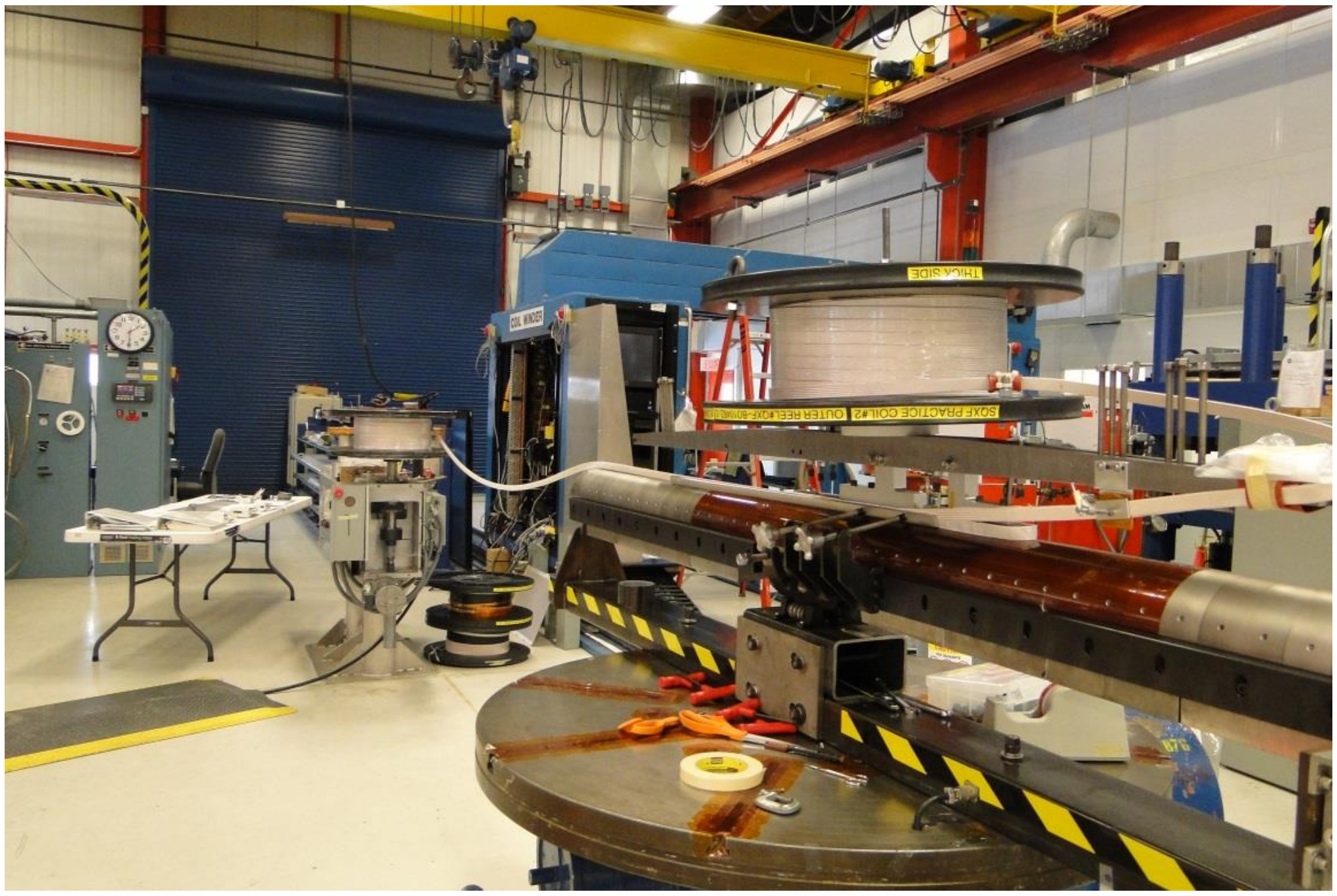

Figure 5.11: Rotation Table.

The coil cross section during winding and curing is shown in Figure 5.12.

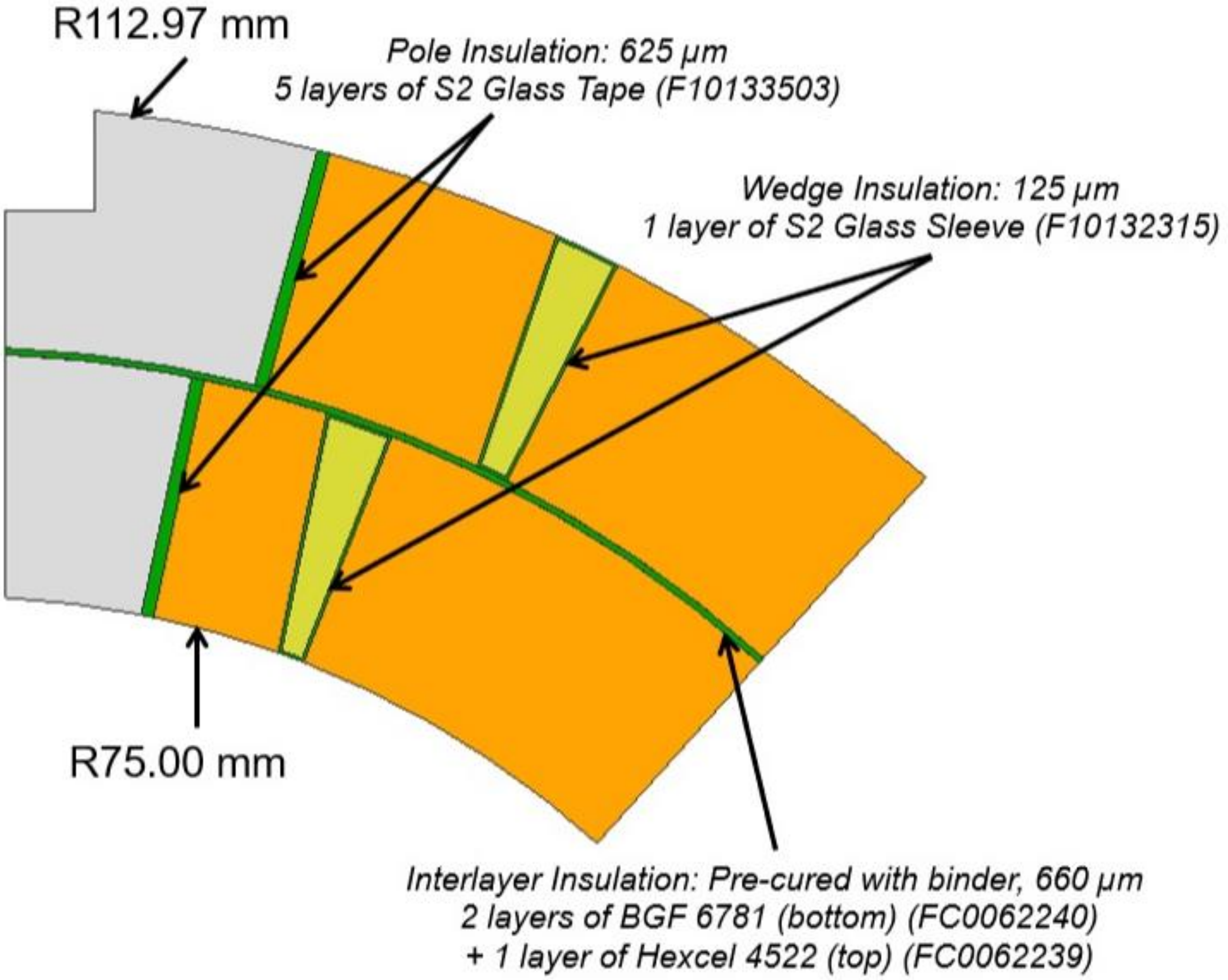

Figure 5.12: Coil cross-section during winding and curing.

Materials:

- Pole insulation – S2 glass tape.
- Wedge insulation – S2 glass sleeve.
- Interlayer insulation – BGF 6781 and Hexcel 4522.

### 5.3.2 Reaction

The QXFA coil reaction/impregnation tooling design is based on the designs used for QXFS, HQ and LHQ coils. A cross section of the reaction fixture is shown in Figure 5.13. The closed cavity mold defines the coil size precisely and alignment pins are used to position the pole pieces during reaction and impregnation. The nominal fixture dimensions and tolerances are shown in Table 5.6. Slender links connect the end saddles to the coil pole to keep the saddles in contact with the cable turns during reaction. A 0.6 mm radial filler is included to allow for the possibility of a small adjustment to the coil outer diameter. The fixture temperature during reaction is monitored continuously by thermocouples bolted to the outside of the reaction fixture.

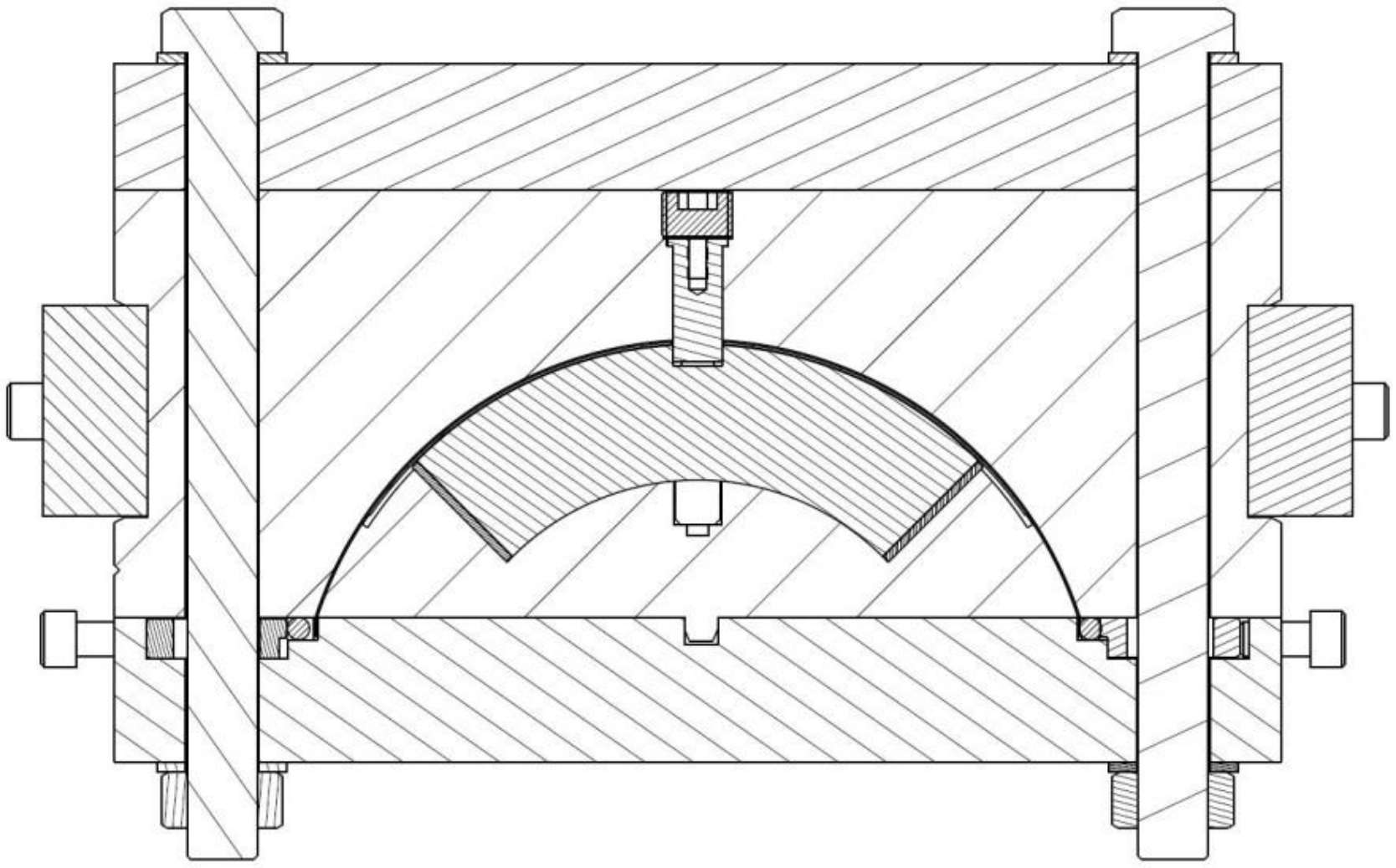

Figure 5.13: Reaction fixture cross section

Table 5.6: QXF Reaction/Impregnation Fixture Dimensions.

| Description | Dimension | Tolerance |
|---|---|---|
| Inner radius | 74.750 mm | +0<br>-0.05 mm |
| Outer radius | 113.630 mm | +0.08<br>-0.03 |
| Midplane offset | 0 mm | |

Note:

Dimensions are with liner, radial filler and midplane shims installed.

Reaction takes place in a gas tight oven using an automatic program. The specification for the coil heat treatment cycle is shown in QXFA Series Coil Production Specification [1]. The durations and ramp rates for the oven control may be varied in order to achieve the target temperature for the reaction fixture. The reaction fixture is sealed and a continuously flowing argon atmosphere is used to carry away any contaminates that are released during the reaction cycle. Argon flow is supplied independently to the oven at 50 SCFH and to the reaction fixture within the oven at 25 SCFH. Flow rates are sufficient to maintain the oven at a small positive pressure relative to the external atmosphere and the reaction fixture at a positive pressure relative to the internal oven atmosphere. Argon flow is maintained throughout the entire cycle, until the fixture has cooled to a temperature below 100°C. Temperature uniformity within the furnace volume is maintained to within ±7°C at the 210°C and 400°C plateau's and ±5°C at the 665°C plateau.

Mica sheets are set around the coil in preparation for the reaction to reduce friction between materials with different thermal expansions. A sketch of the layers of material installed for reaction is shown in Figure 5.14.

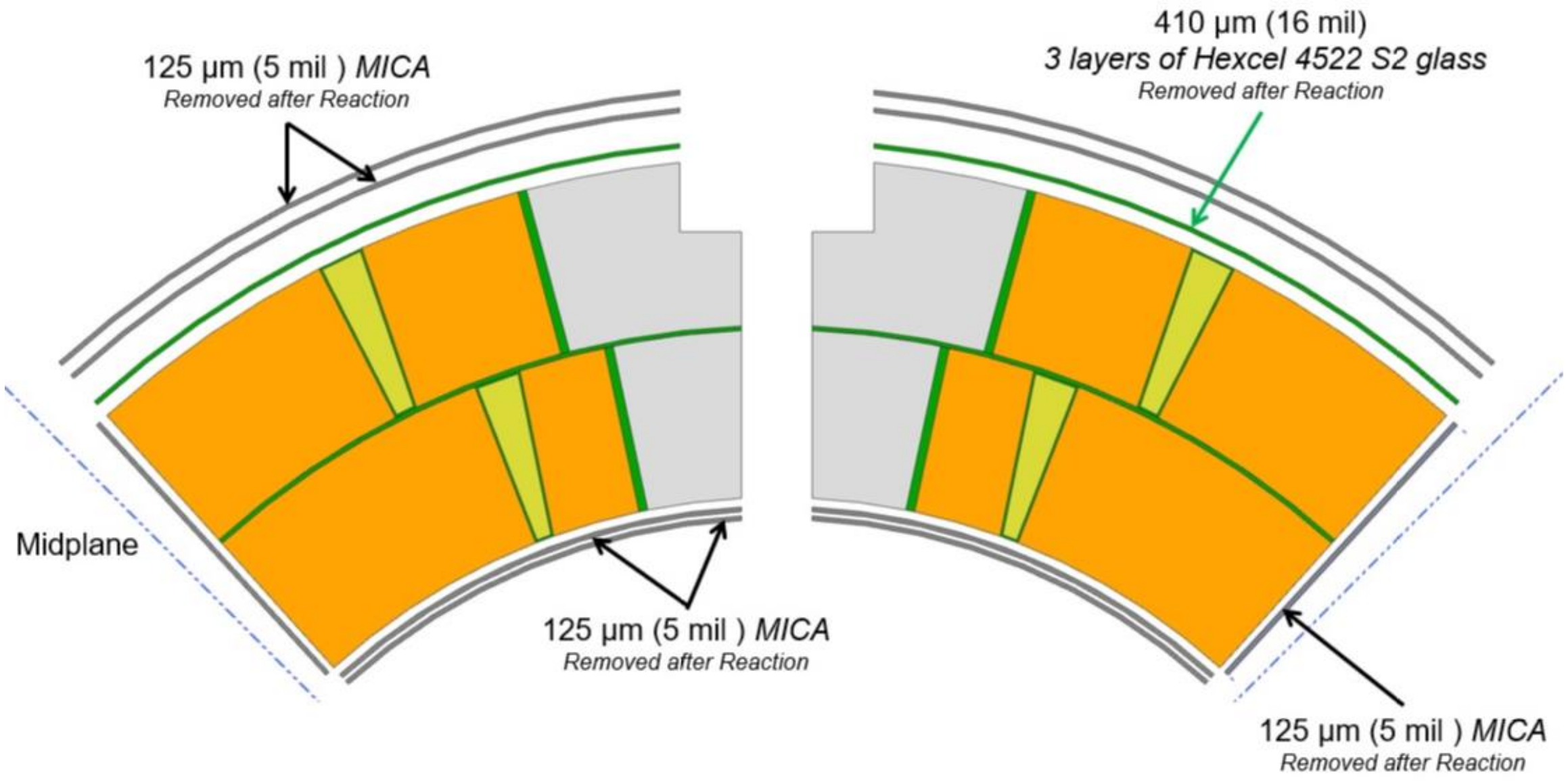

Figure 5.14: Coil cross section during Reaction.

Materials:

- Mica – Cogebi Cogemica Hi-Temp 710-1.
- Fiberglass – Hexcel 4522.

### 5.3.3 Impregnation

After reaction, coil lead splices are made. The $Nb_3Sn$ coil leads are soldered to a pair of NbTi cables using MOB39 flux and 96/4 tin/silver solder. Next the outer layer protection heater trace is installed made of copper plated stainless steel foil glued to a layer of polyimide film. Fiberglass cloth and G11 shims are then added to the coil. The fixture is closed and rotated along its long axis to install fiberglass on the coil ID. A sketch of the layers of material installed for impregnation is shown in Figure 5.15. The coil is vacuum impregnated with CTD-101k epoxy using a fixture similar to the one used for reaction. The nominal impregnated coil dimensions are shown in Table 5.7.

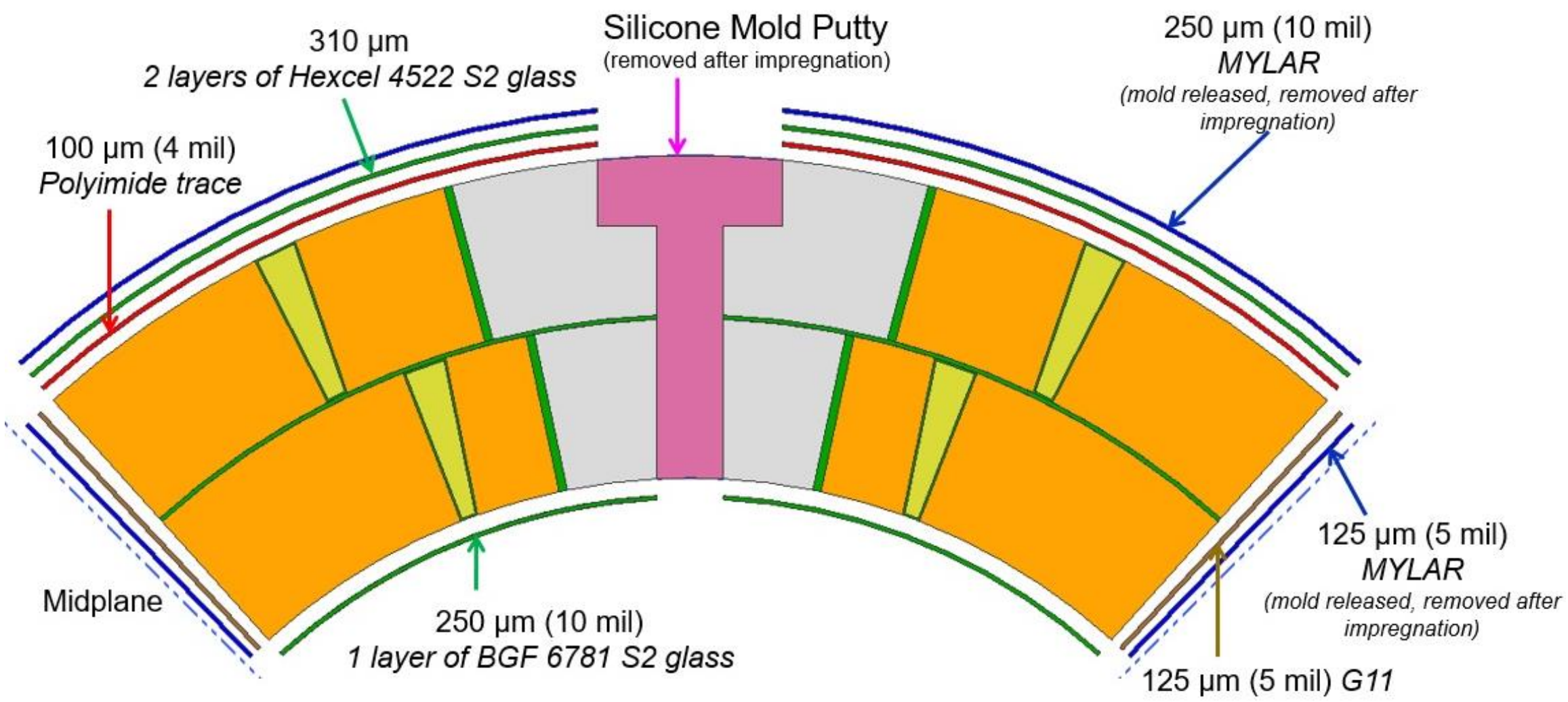

Figure 5.15: Coil cross section during Impregnation.

Materials:

- Trace – Polyimide + copper plated stainless steel.
- Fiberglass – Hexcel 4522.
- G11 – .250 mm thick.
- Mylar, midplanes – .125 mm thick (mold release applied).
- Mylar, coil OD – .250 mm thick (mold release applied).

Table 5.7: QXF Nominal Impregnation Coil Dimensions.

| **Description** | **Dimension** | **Tolerance** |
|---|---|---|
| Coil inner radius | 74.750 mm | +0<br>-0.05 mm |
| Coil outer radius | 113.376 mm | +0.08<br>-0.03 |
| Coil midplane offset | 0.125 mm | |

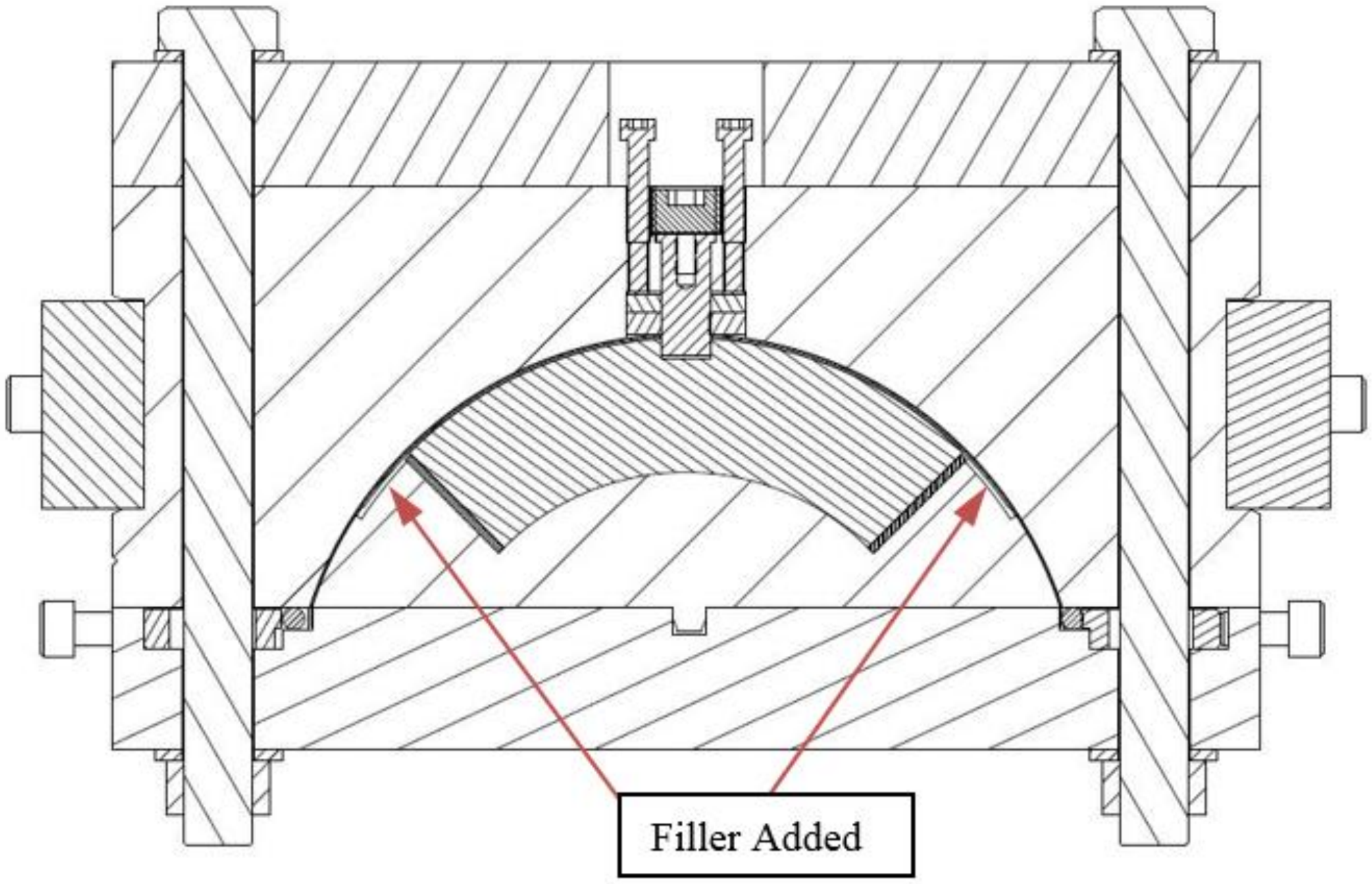

Figure 5.16: Impregnation fixture cross section.

The impregnation tooling is used to epoxy impregnate the coil in a vacuum oven. Fillers are installed to fill the gaps on each side of the mandrel blocks, as shown in Figure 5. This is done to help force the epoxy to flow through the coil instead of through the gaps. The coil is positioned at an incline within the vacuum furnace with the lead end of the coil elevated above the return end. External heaters mounted to the tooling are used for all steps of the impregnation cycle. While under vacuum, the coil is baked at 110º C then cooled back down to 55ºC before the impregnation is started. The epoxy is mixed, warmed to 55º C, and degassed to the same vacuum level as the coil. Once degassed, the vacuum level in the fixture and in the vacuum tank is equalized. The epoxy tank is let up to atmospheric pressure and the pressure differential forces epoxy into the coil. The epoxy is introduced into the return end of the impregnation fixture. The fill rate is controlled by a peristaltic pump, with pump speed set to result in a fill time of about 4 hours. An exit line from the top of the impregnation fixture is connected to an overflow reservoir within the vacuum oven. Once the epoxy flows into the overflow reservoir from the lead end of the coil, the epoxy input hose is clamped off at the return end and a dwell period at atmospheric pressure begins. During the dwell period, small amounts of epoxy back-fills into the coil from the overflow reservoir. After the dwell, the coil is rotated to the horizontal position and the cure cycle in Table 5.8 is used.

Table 5.8: QXF Coil Impregnation and Cure Cycle.

| Step | Description | Time / Rate |
|---|---|---|
| Ramp | From 20°C to 110°C | 10 C/hour |
| Soak | 110°C | 8 hours |
| Cool | From 110°C to 55°C | n/a |
| Resin transfer | Coil epoxy impregnation | 8 - 22 ml/min |
| Ramp | from 55°C to 110°C | 10 C/hour |
| Soak | 110°C | 5 hours |
| Ramp | from 110°C to 125°C | 10 C/hour |
| Soak | 125°C | 16 hours |

## 5.4 Coil Fabrication at BNL

### 5.4.1 Winding and Curing

Coil winding is done on the long shuttle winder (Fig. 5.17) using an automated winding program. Insulated cable and coil parts are procured by FNAL and supplied to BNL. The insulated cable is re-spooled, without any reverse bending of the cable, and split onto two spools for the inner and outer coil layers. The inner layer is wound first. Then it is painted with CTD-1202 binder, packaged for curing and transferred to the curing form block.

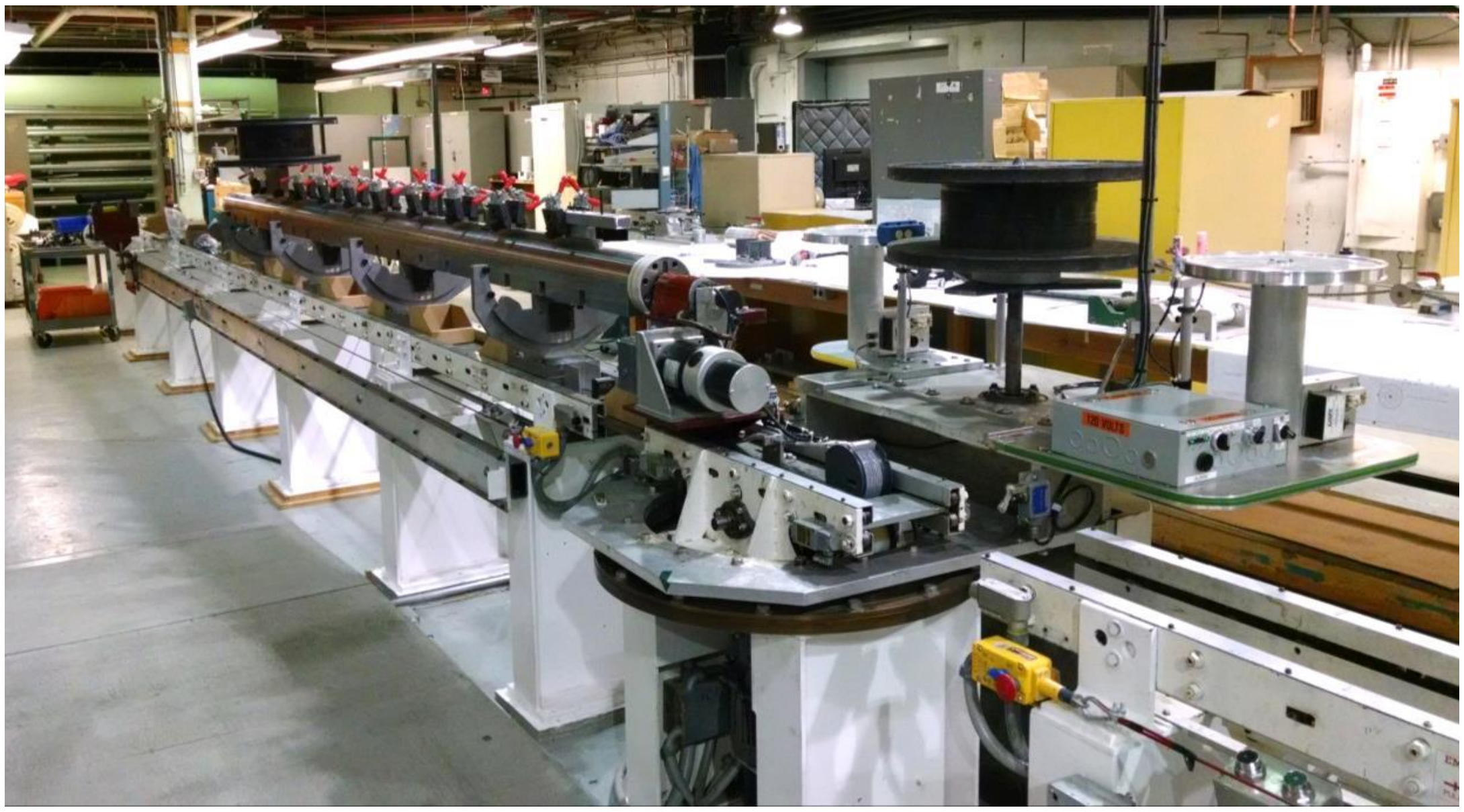

Figure 5.17: Winding Machine.

Coil curing is done utilizing the long curing press displayed in Fig. 5.18. The coil mandrel assembly is bolted down into the form-block to set the radial cavity (Figure 5.19). Azimuthal pressure is applied to the coil using 40 six inch diameter hydraulic cylinders. Pressure is applied to close the mold to the nominal coil size. The tooling is then heated with electric cartridge heaters installed in both the mandrel and the form-block to cure the binder. Cross sections with tooling dimensions are shown in Figure 20.

After the inner layer has been cured, the coil is transferred back to the winding machine and the process is repeated for the outer layer. The winding machine is configured so that both layers are wound without any reverse bending of the cable.

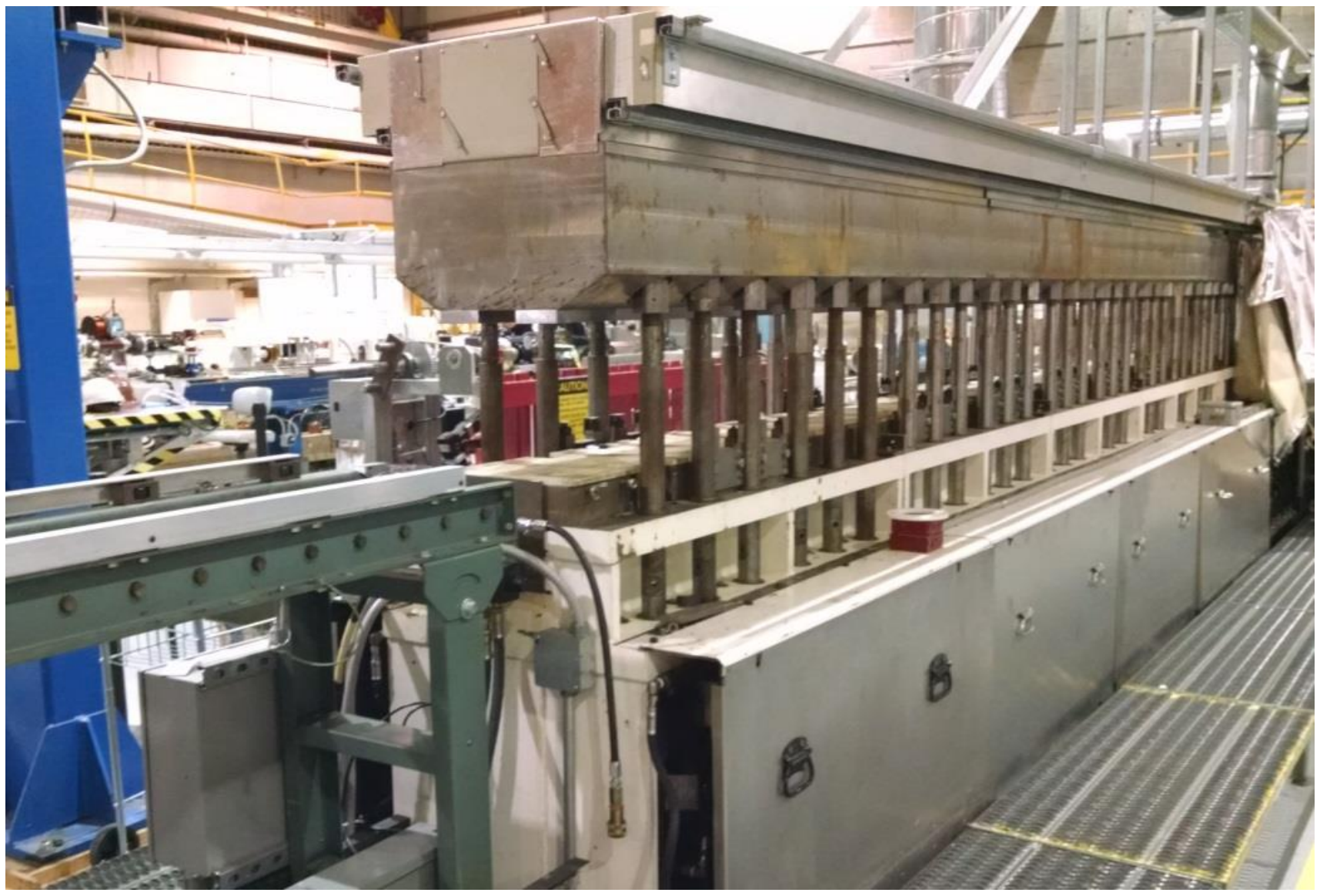

Figure 5.18: Curing Press

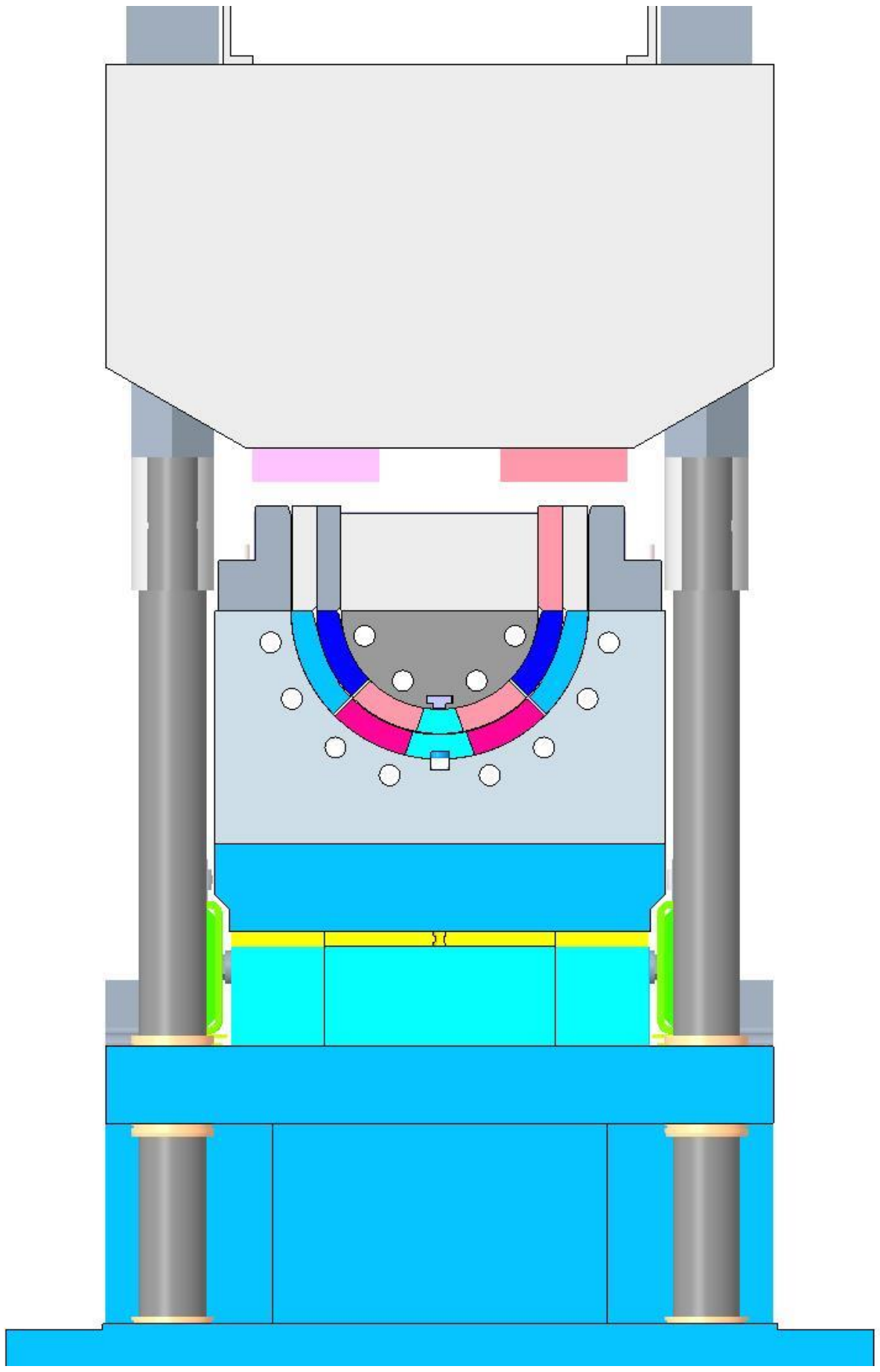

Figure 5.19. Curing tooling in the curing press

| Tooling Dimensions | in | mm |
|---|---|---|
| | | |
| Mandrel OD | 5.906 | 150.00 |
| | | |
| L1 Quad Pusher ID | 5.906 | 150.00 |
| L1 Quad Pusher OD | 7.374 | 187.30 |
| | | |
| L2 Quad Pusher ID | 7.382 | 187.50 |
| L2 Quad Pusher OD | 8.895 | 225.93 |
| | | |
| Formblock ID | 8.903 | 226.14 |
| | | |
| L1 Cure Only | | |
| Spacer ID | 7.382 | 187.50 |
| Spacer OD | 8.900 | 226.06 |
| | | |
| (Teflon wrap ) | 0.008 | 0.20 |

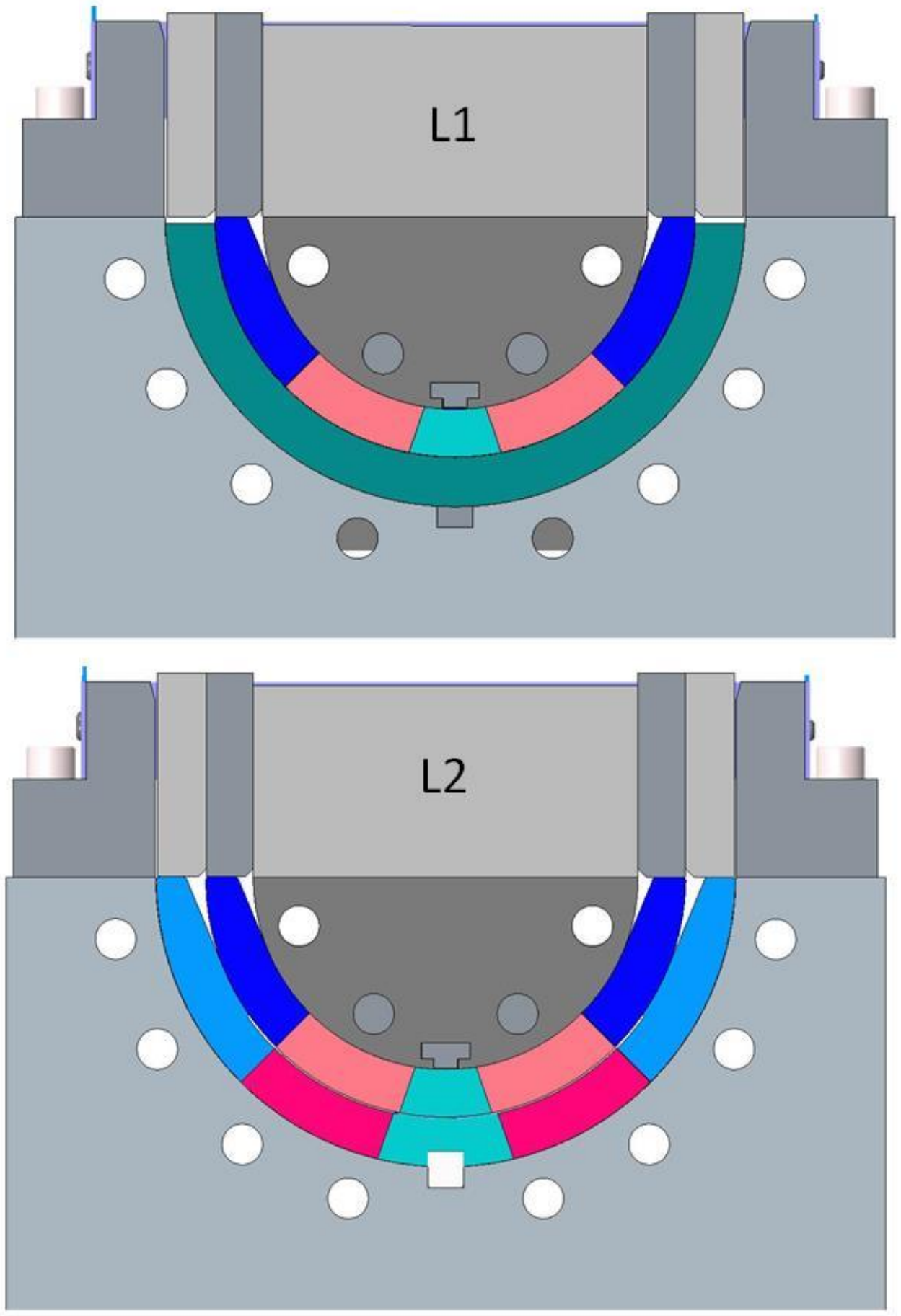

Figure 5.20: Curing tooling dimensions

### 5.4.2 Reaction

The QXF coil reaction/impregnation tooling design is based on the designs used for HQ and LHQ coils. A cross section of the reaction fixture is shown in Figure 5.11. The closed cavity mold defines the coil size precisely and alignment pins are used to position the pole pieces during reaction and impregnation. The nominal fixture dimensions are shown in Table 5.9. Slender links connect the end saddles to the coil pole to keep the saddles in contact with the cable turns during reaction. The fixture temperature during reaction is monitored continuously by thermocouples bolted to the outside of the reaction fixture.

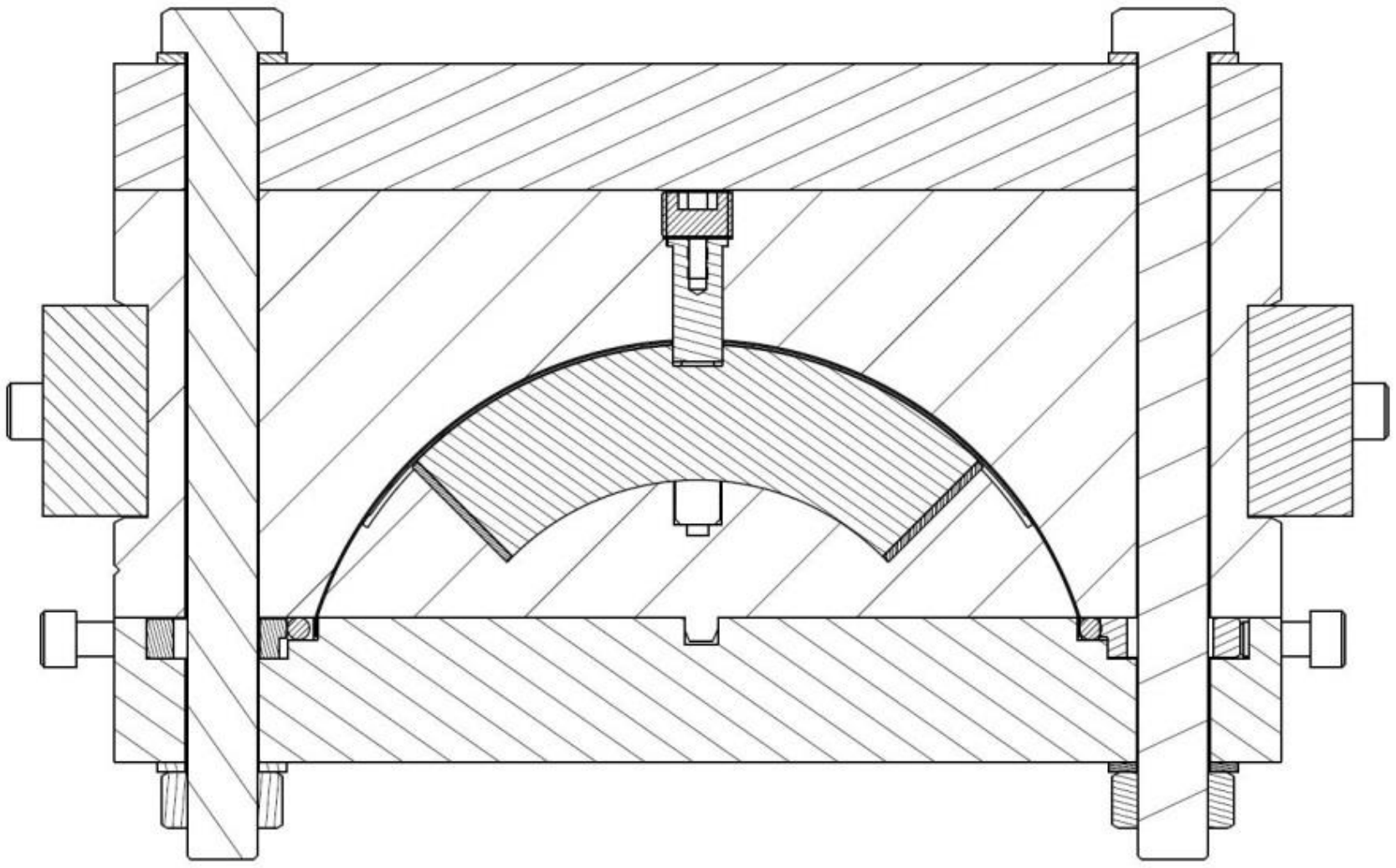

Figure 5.11: Reaction fixture cross section.

Table 5.9: QXF Reaction/Impregnation Fixture Dimensions.

| **Description** | **Dimension** |
|---|---|
| Inner radius | 74.750 mm |
| Outer radius | 113.630 mm |
| Midplane offset | 0.125 mm |

Note: Dimensions are with liner, radial filler and midplane shims installed.

Reaction takes place in a gas tight 5 meter long oven using an automatic program. The target cycle for the reaction fixture is defined in the series coil specification [1]. The durations and ramp rates for the oven control may be varied in order to achieve the target temperature for the reaction fixture. The reaction fixture is sealed and a continuously flowing argon atmosphere is used to carry away any contaminates that are released during the reaction cycle. Flow rates are sufficient to maintain the oven at a small positive pressure relative to the external atmosphere and the reaction fixture at a positive pressure relative to the internal oven atmosphere. Argon flow is maintained throughout the entire cycle, until the fixture has cooled to a temperature below 100°C.

Cogebi Cogemica Hi-Temp 710-1 mica sheets are set around the coil in preparation for the reaction in order to reduce friction between materials with different thermal expansions

### 5.4.3 Impregnation

After reaction, coil lead splices are made. The $Nb_3Sn$ coil leads are soldered to a pair of NbTi cables, using MOB39 flux and 96/4 tin/silver solder. Next the protection heater trace circuits are installed. The trace circuits consist of copper plated stainless steel foil glued to a layer of polyimide film. Fiberglass cloth and G11 shims are then added to the coil. The fixture is closed and rotated end-over-end to access the coil ID after work on the OD is complete. The coil is vacuum impregnated with CTD-101k epoxy using a fixture similar to the one used for reaction.

The impregnation fixture is closed and sealed. The fixture is hung vertically, lead end up, in a vacuum tank. Strip heaters mounted to the outside of the fixture are used for heating. The fixture and the vacuum tank are evacuated, baked and degassed. The coil is then cooled back down to 55ºC before the impregnation is started. The epoxy is mixed, warmed and degassed. Once degassed, the epoxy pot is let up to atmospheric pressure and the pressure differential forces epoxy into the coil. The epoxy is introduced into the bottom of the impregnation fixture. The fill rate is controlled by a peristaltic pump, with pump speed set to result in a fill time of about 4 hours. An exit line from the top of the impregnation fixture is connected to a resin trap outside the vacuum tank. When epoxy reaches the resin trap, the exit line is closed. The supply line remains open, with atmospheric pressure continuing to force resin into the coil. The coil is held overnight, for approximately 16 hours, at 55ºC while allowing epoxy to continuously draw into the coil. After the overnight soak, the cure cycle is initiated. The epoxy is cured using an automatic cycle. Vacuum is maintained in the vacuum tank until the cure is complete.

## 5.5 Coil Handling and Shipment

The MQXFA coil handling and shipping requirements are specified in [6]. We provide here a summary for information only. MQXFA coil shipping plans are presented in [7].

MQXFA coils are made of brittle $Nb_3Sn$ conductor. In order to prevent conductor degradation, all handling and shipment operations shall meet the following requirement:

**Coil H&S Requirement #1: The conductor strain shall never exceed 500 microstrain in any part of the coils during any handling or shipping operation.**

A coil shipping fixture is used to support the coils for shipment between labs. The fixture consists of an aluminum support tube mounted in an aluminum channel using rubber shock mounts. Side rails support the full length of the coil midplane. A series of clamps are applied over the coil OD. Longitudinal restraint is provided by bolts contacting the ends of the coil saddles. The fixture is shown in Figure 5.24.

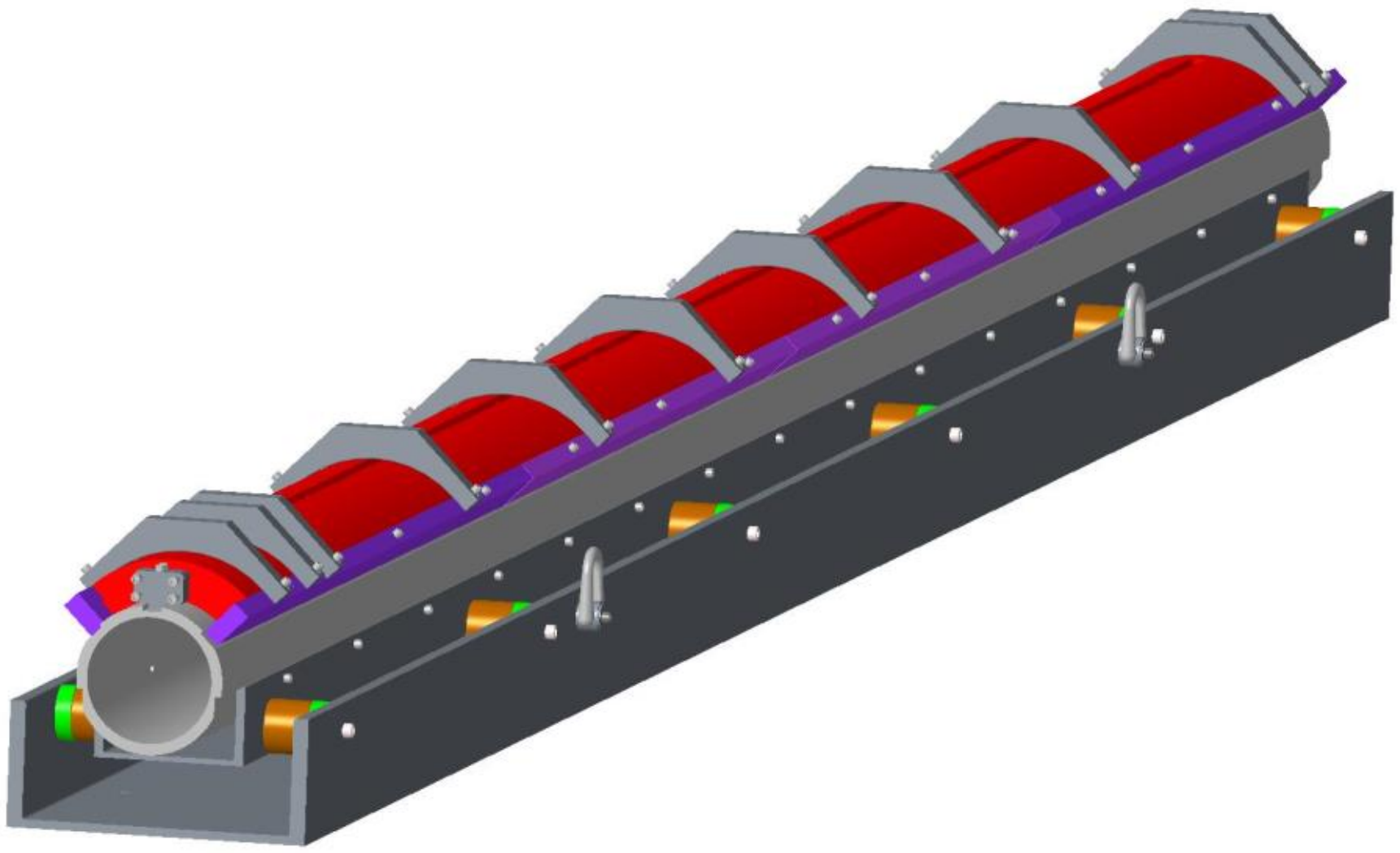

Figure 5.24: Coil shipping fixture.

The shipping fixture is installed inside a wooden crate. Shipment is made using a dedicated truck such that the once the crate is loaded onto the truck at the origin lab, it remains on the truck until it arrives at the destination lab. There is no loading and unloading permitted while on route. The shipping fixture is instrumented with accelerometers and shock watches are placed on the crate. If there are no shocks above 5 g the standard QC for coil reception can be performed. This threshold was determined thought analysis of coil shipments and magnet test results. FEM analysis confirmed that shocks up to 5 g are acceptable using the AUP MQXFA coil shipping fixture [6].

# 6 MQXFA Structure and Magnet Assembly

## 6.1 Magnet Structures Features

As described in the section 4.3 (Structural Design), the main features of the support structure design as shown in Fig. 6.1 are:

1. Shell-based support structure relying on the “bladder and key” technology to perform azimuthal pre-loads, which allows a reversible assembly process and tunable preload;
2. The aluminum shell is segmented into sections to minimize axial tension in the shell and to ensure a uniform azimuthal load on the coil;
3. The use of master key packages between the yokes and the pads allow for assembly clearances prior to preload operations;
4. G11 alignment keys inserted into the pole pieces provide the coil azimuthal alignment;
5. Axial pre-load is applied by four stretched axial stainless-steel rods connected to endplates that transfer the forces to the coil ends.

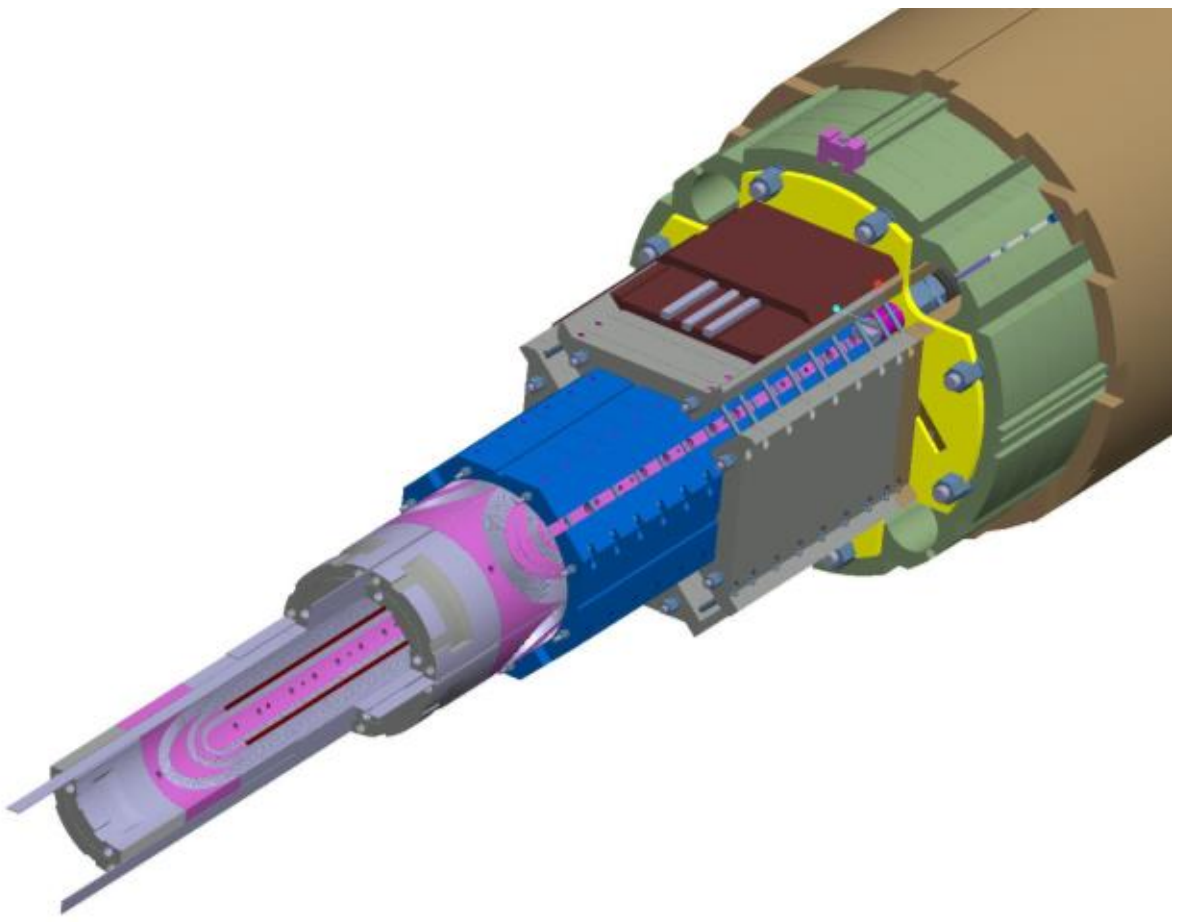

Fig. 6.1: 3D exploded view of the end of an MQXF magnet (axial end loading and splice connection components not shown).

## 6.2 Magnet Assembly Requirements

The Functional Requirements Specification, presented in Section 2, describe the functional requirements for the MQXFA magnets. Specifically, for the MQXFA magnet structures Table 6.1. lists the relevant requirements from the complete list, and how they are satisfied in this structure design.

Table 6.1. Relevant Functional Requirements for the MQXFA magnet structures

| Requirement | Description | Criteria Met |
|---|---|---|
| | **Threshold Requirements** | |
| R-T-01: | The MQXFA Coil aperture at R.T. after preload shall be 146.7 mm, guaranteeing an annulus free of HeII of at least 1.2 mm | Measure bore after magnet preload using a bore clearance gauge. |
| R-T-02: | Physical outer diameter must not exceed 614 mm. | O.D of shells are 614 mm. Verified after magnet has been preloaded. |
| R-T-06: | The MQXFA cooling channels must be capable of accommodating two (2) heat exchanger tubes running along the length of the magnet in the yoke cooling channels. The minimum diameter of the MQXFA yoke cooling channels that will provide an adequate gap around the heat exchanger tubes is 77 mm. | Yoke cooling hole dimensions are designed to 77 mm, measured after magnet preload has been completed. |
| R-T-08: | The MQXFA structure must have provisions for the following cooling passages: (1) Free passage through the coil pole and subsequent G-11 alignment key equivalent of 8 mm diameter holes repeated every 50 mm; (2) free helium paths interconnecting the yoke cooling channels holes; and (3) a free cross sectional area of at least 150 $cm^2$ | (1) G11 alignment keys are provided with holes for through passage of helium.<br>(2) Yoke laminations at shell junctions provide a total of 15.5 $cm^2$ cross talk between each cooling holes.<br>(3) To be verified at the cold mass. |
| R-T-14 | Splices are to be soldered with CERN approved materials. | Splice joints are made using CERN-approved Sn96Ag4 solder. |
| R-T-15 | Voltage Taps: the MQXFA magnet shall be delivered with (a) three redundant (3x2) quench detection voltage taps located on each magnet lead and at the electrical midpoint of the magnet circuit; (b) two (2) voltage taps for each quench strip heater; and (c) two (2) voltage taps for each internal MQXFA Nb3Sn-NbTi splice. Each voltage tap used for critical quench detection must have a redundant voltage tap. | The presence of these Voltage Taps are verified after the connectors have been installed on a preloaded magnet prior to vertical magnet testing. |
| R-T-16 | MQXFA magnet coils and quench protections heaters shall pass the hipots to specified voltages | The coils and magnets have electrical QC performed at key steps during the assembly; final hipot of preloaded magnet is verified passing before vertical magnet testing |
| R-T-20 | All MQXFA components must withstand a radiation dose of 35 MGy, or shall be approved by CERN for use in a specific location as shown in [6]” | Materials chosen for the magnet structures have been approved by CERN. |
| | **Objective Requirements** | |
| R-O-01 | Variation of local position of magnetic center must be within ±0.5 mm; variation of local position of magnetic axis within ±2 mrad. Local positions are measured with a 500 mm long probe every 500 mm. | Measurements to be performed and analyzed after each magnet assembly. |
| R-O-04 | Splice resistance target is less than 1.0 nΩ at 1.9K. | Magnet splices from MQXFS and MQXFAP1 have been measured to have <1.0 nΩ. |

### 6.2.1 MQXFA Magnet Assembly Specification

The magnet assembly must meet two primary specifications:

- Ensure coil pre-load is uniform to ±10 MPa at the pole
- R-O-01: Variation of local position of magnetic center must be within ±0.5 mm; variation of local position of magnetic axis within ±2 mrad. Local positions are measured with a 500 mm long probe every 500 mm.

#### 6.2.1.1 Uniformity of coil preload

The uniformity of the preload will be determined by the tolerances of the parts when fabricated. Figure 6.2 shows a quadrant of the MQXFA magnet cross-section showing the interfaces of the various components when assembled. Table 6.2 lists those parts and their respective allowed manufacturing tolerances. A worst-case scenario (though unlikely) would be that the all parts would vary at the maximum extent of the tolerance band and add up (not including the coil variances), which would represent a total of ±325 µm of variance, translating to ±43 MPa. However, a statistical spread would better represent how all these tolerances stack up. The RMS tolerance stack up is then calculated to be ±31 µm, which translates to ±9 MPa. It should be noted that although the variances of the impregnated coil can be ±250 µm, the radial shims used to build up the coil pack should be able to accommodate gross variances.

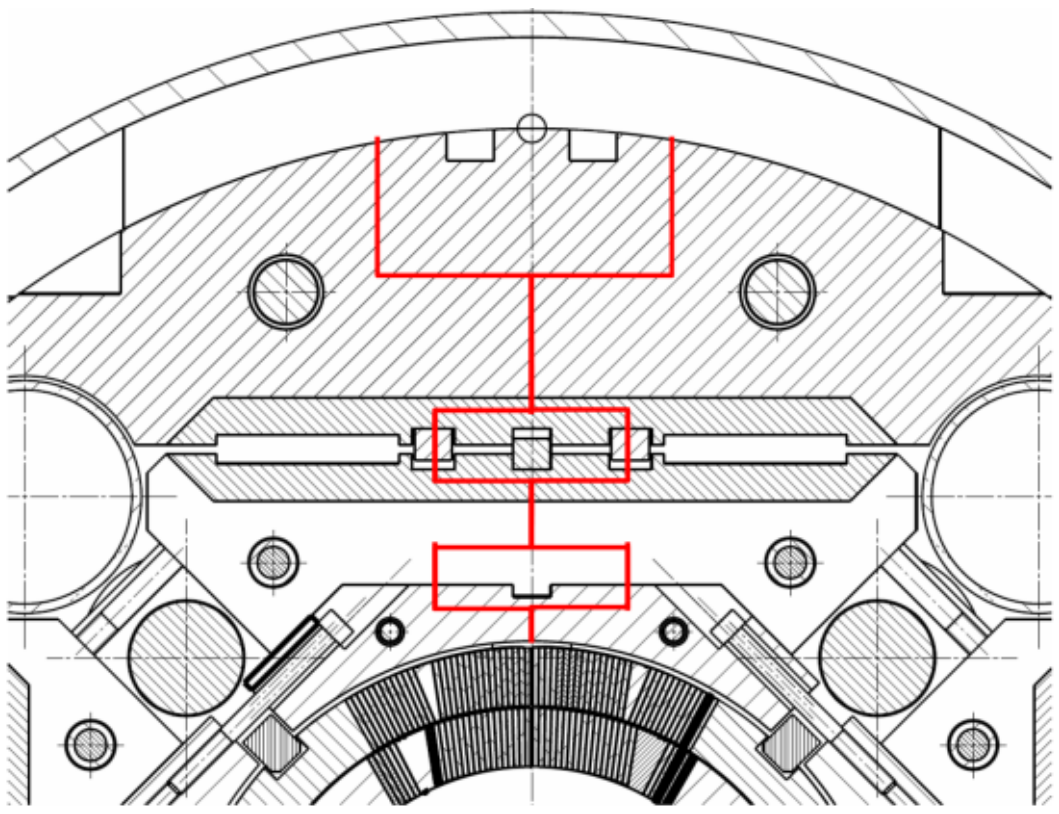

Figure 6.2: Magnet cross-section, showing areas of interfaces.

Table 6.2. Magnet Components Dimensional Specifications.
Different colors represent subassembly interfaces.

| Comp | Description | Stack tol. [μm] |
|---|---|---|
| Shell | Inner radius | ± 25 |
| Yoke | Outer radius | ± 25 |
| | Main master contact | ± 25 |
| Master 1&2 | Main yoke/pad contact x2 | ± 50 |
| | Load key x2 | ± 25 |
| Pad | Main master contact | ± 25 |
| | Main collar contact | ± 25 |
| Collar | Main pad contact | ± 25 |
| | Inner radius | ± 25 |
| Coil (see Section 5.1) | Outer Radius | ± 250 |

### 6.2.2 Magnet QA/QC Plan

- **Magnet Structures Components Mechanical Measurements**

The magnet structure parts will be inspected by the respective vendors where the parts will be fabricated. Components that consist of large quantities of parts (ie. Yoke laminations, etc.) will have only 10% of the batch measured, with an incoming inspection of a random sample performed at LBNL.

- **Magnet Electrical Measurements**

At various stages of the magnet assembly process electrical QC measurements will be taken. This will ensure that throughout the entire assembly process the electrical integrity of the magnet is maintained. US-Hilumi Doc 956 describes the details, but the various measurement hold points are listed in Table 6.3.

Table 6.3. Magnet Assembly Electrical Measurement hold points.

| Hold Point | Measurement |
|---|---|
| *Coil Reception* | Hipot, Impulse, Seq. R, Inductance |
| *Coil Pack Assembled* | Basic R checks |
| *Magnet Azimuthally loaded* | Basic R checks, Impulse, Hipot (optional) |
| *Magnet Axially loaded* | Hipot, impulse (individual coils) |
| *Magnet spliced and connectorized* | Seq. R, Inductance, Impulse (magnet), PH R, Hipot |
| *Magnet Pre-shipment* | Basic R checks to ground |

## 6.3 Magnet Assembly Breakdown Structure

Fig. 6.3 shows the magnet assembly breakdown structure (ABS) flowchart of the long magnet assembly. The ABS shows the subassemblies that make up the assembled magnet and helps to organize the required processes and captured data. The primary assembly drawings are listed in Table 6.4, although the full complement of approximately 130 drawings for the magnet structures components are listed elsewhere. Uncontrolled versions of the drawing package can be found in US-HiLumi Doc 953; all current version-controlled drawings are maintained on the LBNL Windchill document database.

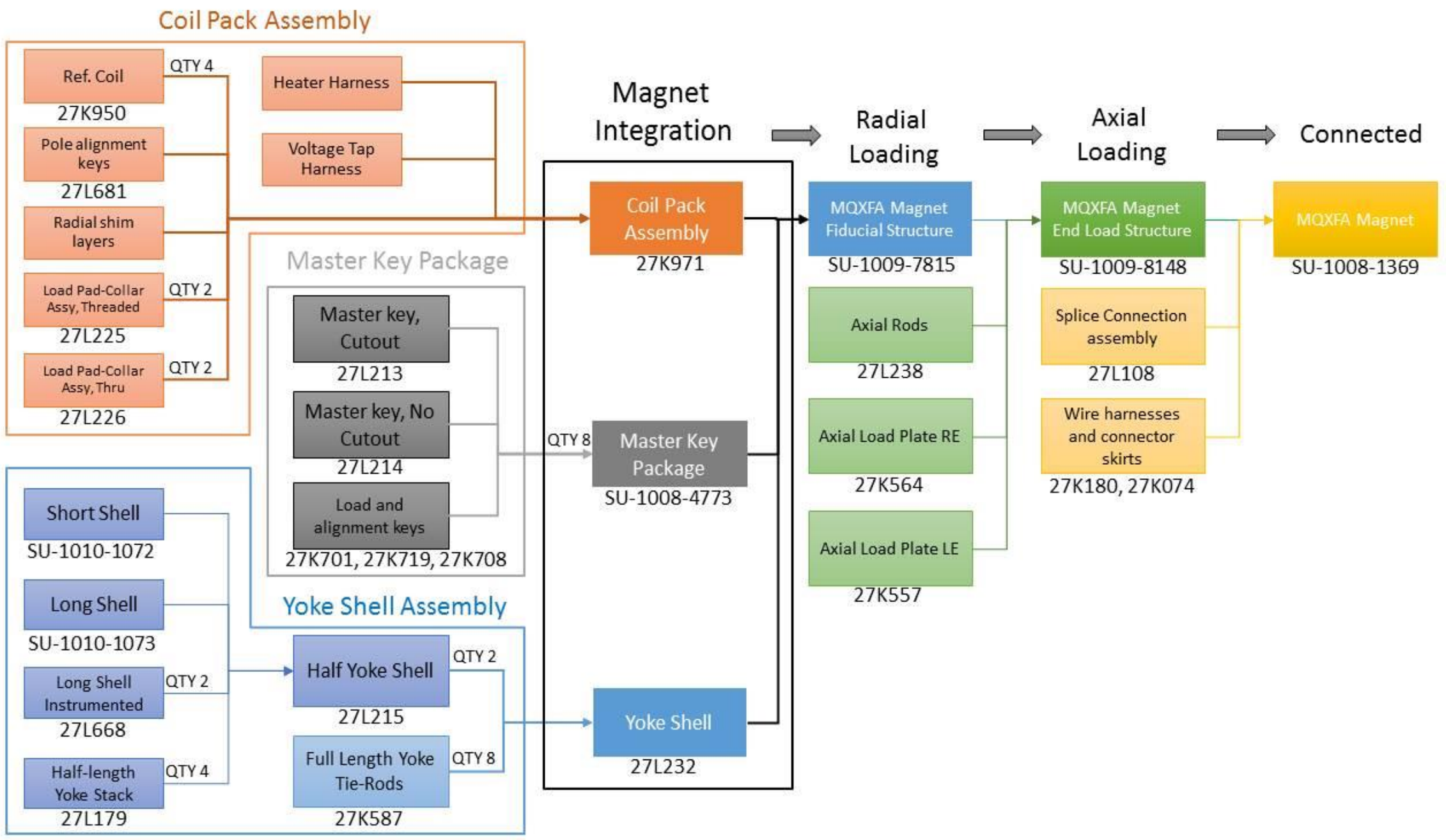

Fig. 6.3: MQXFA magnet assembly breakdown structure flowchart with associated drawings.

Table 6.4. Magnet Assembly drawing packages.

| Description | Drawing # | Notes |
|---|---|---|
| MQXFA Magnet Assembly | SU-1011-0518 | 26 additional dwgs in this package |
| MQXFA Axial loaded structure | SU-1009-8148 | 11 additional dwgs in this package |
| MQXFA Magnet fiducial structure | SU-1009-7815 | |
| Coil Pack Assembly | 27K971 | 47 additional dwgs in this package |
| Master Key Package | SU-1008-4773 | 10 additional dwgs in this package. |
| Yoke-shell Subassembly | 27L232 | 31 additional dwgs in this package |
| | | **131 Total drawings** |

### 6.3.1 **Shell-Yoke Subassembly**

### 6.3.2 **Shells**

The shells are 614 mm O.D. by 556 mm I.D. The MQXF magnet adopts segmented shells to minimize the axial tension in the shell segments and to ensure a uniform azimuthal load on the coils. There are two lengths of shell segments used in the assembly: a 325.65 mm long segment (two ea.) used at each extremity; and six 651.29 mm long segments stacked in between the short ones.

As shown in the Fig 6.4 each shell segment has various cutouts on each end, for alignment, tooling, and supports. There are pin slots machined into each quadrant of the ID on each end of the shells, though pins will be used only in one quadrant during magnet assembly operations.

The shells are fabricated and inspected by the vendor, then each individual shell will then be measured again upon receipt. The average I.D. will also be measured at five axial locations along the length of each shell segment: approximately 25 mm from each end, and three equally spaced mid-length locations. These measurements are used to determine an optimal arrangement and order that will be used when assembling these shells in the magnet structure. Once the order of shells is determined, three of the six long shells will be instrumented with strain gauges so that preload operations can be measured against the analytical targets.

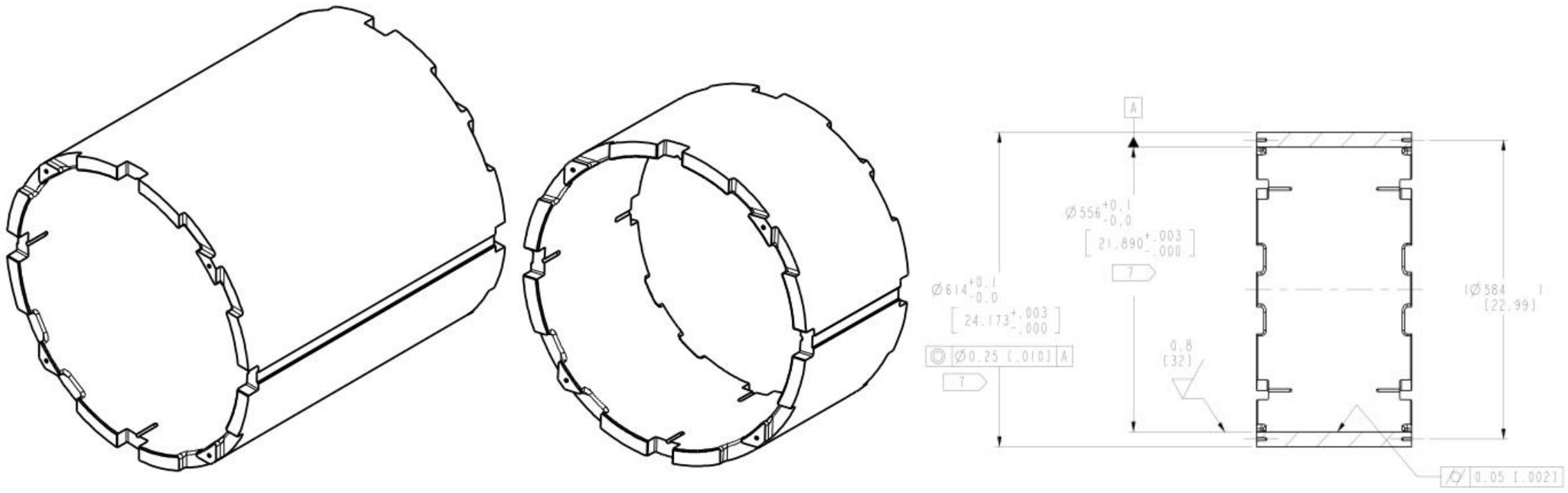

Fig. 6.4: The long and short shell segments, and dimensions of the cross section of the short shells.

### 6.3.3 Yokes

The iron yoke laminations are typically 49 mm (~1.93 in.) thick and made of ARMCO Pure Iron (Grade 4). Each quadrant of the yoke stacks is joined using two 19.05 mm (¾ in.) diameter stainless steel tie rods, and aligned by close-fitting aluminum bronze bushing sleeves. Each tie rod is pre-tensioned to 40,000 N (9000 lbs). Since the cooling holes in the MQXFA magnet are found in each quadrant, the required 77 mm clearance bore is actually formed by two adjacent yoke profiles.

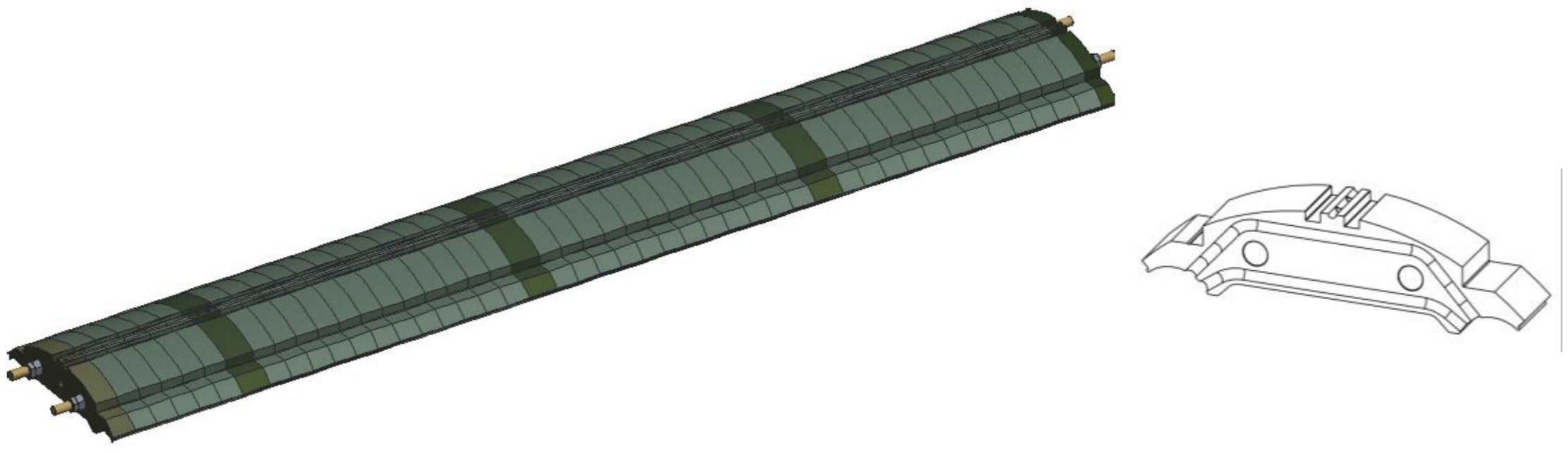

Fig. 6.5. (L) 3D model of a half-length yoke stack, and (R) a typical yoke lamination used at the junction between shells.

The yoke laminations are initially stacked into “sub-stacks” corresponding to the junction of the shell segments. These sub-stacks are then assembled into eight 2281.5 mm (89.82”) long sub-assemblies, which is half the length of the magnet (see Fig. 6.5). Four of these subassemblies will be assembled per half-length shell-yoke subassembly prior to the final integration of the full-length structure. The fact that these laminations bridge between shell segments increases the longitudinal stiffness of the shell assembly. However, when joined together the half-length shell-yoke subassemblies are not bridged at the center: only the yoke bushings and the coil pack assembly components bridge the full-length structure. It is for this reason that the yoke tie rods must be tensioned properly, since the tensioned yoke stacks provide the stiffness for the structure; the next section describes this process.

Special intermediate yoke laminations separate these pre-stacks in a half-length assembly. Though keeping the same basic geometry, the laminations located at the junction between two shell segments have additional features. First, there are two grooves cut into the mating faces of these yoke laminations to satisfy the cooling requirements (R-T-08(3)) when the magnet is in service, with each groove providing about 0.48 $cm^2$ of clear passage. A full-length yoke stack therefore provides approximately 15.5 $cm^2$ passage between each quadrant's cooling holes. Secondly, there are tapped holes at the crown, which are provisions for the welding block that will be attached when the SST LHe cryostat shell is welded over the magnet structure after a successful cold test.

### 6.3.4 Shell-Yoke Subassembly Operations

When assembling the shell-yoke subassembly, four shells (three long, one short) are initially stacked vertically on the assembly stand. Pins are inserted only in the "Top" quadrant slots initially to align the shells prior to the insertion of the half-length yoke stacks.

The four yoke stacks are then inserted vertically inside the stack of shells (Fig 6.6). A bladder operation is performed to insert the gap keys and constrain the yokes against the shells. The cooling holes in the yokes are utilized as the bladder locations to open up the gaps, which is more efficient than the operating the bladders at the internal cross support geometry. The nominal yoke gap is 12 mm, but the gap keys are shimmed to approximately 12.1 mm total to achieve a shell strain of approximately 200 μm/m.

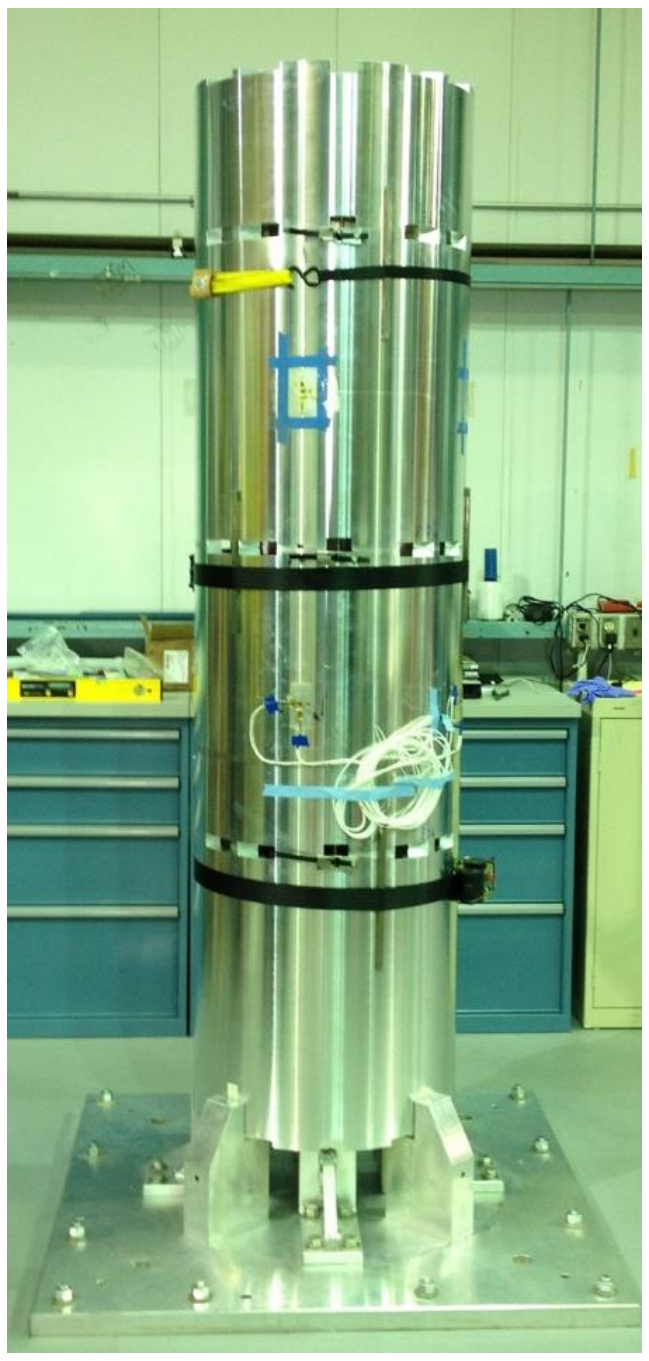

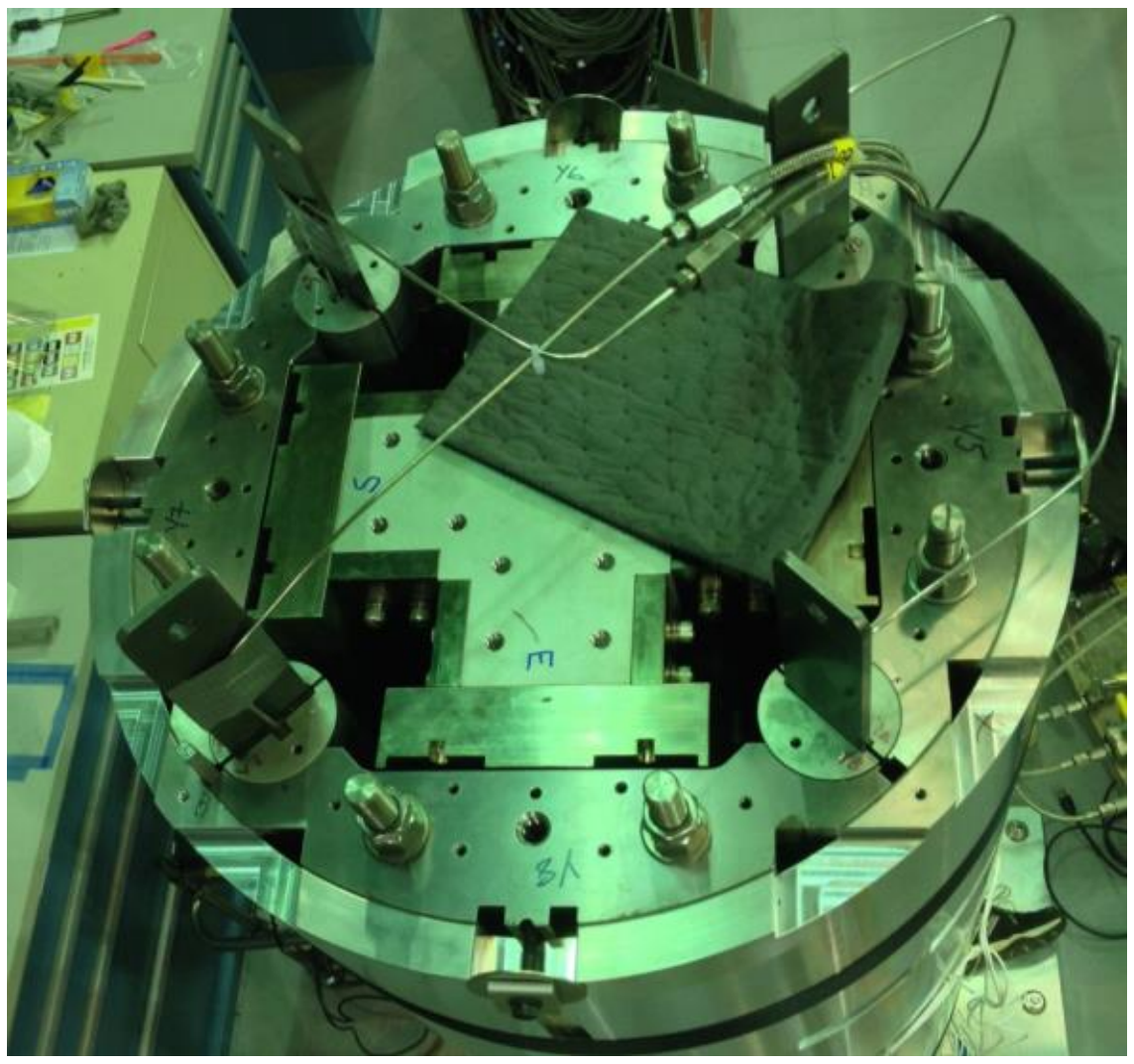

Fig. 6.6. (L) Half-length shell stack on the assembly stand. (R) View from the top, with the yoke quadrants ready to be preloaded.

Once the tooling is removed and the assembly has been moved, the process is repeated for the second half-length subassembly, after which they are both placed on the magnet integration table (see Figure 6.7). The dimensional size of the yoke opening is measured, and these values are used to calculate the master key initial shim package. When both half-length assemblies are joined together the centermost bronze bushing sleeves are replaced to span the halves, in addition to the short yoke tension rods being replaced with full-length ones. A hydraulic rig is used to pre-tension these full-length yoke tension rods to carry 9000 lbs. each (see Fig. 6.8). This completes the shell-yoke subassembly.

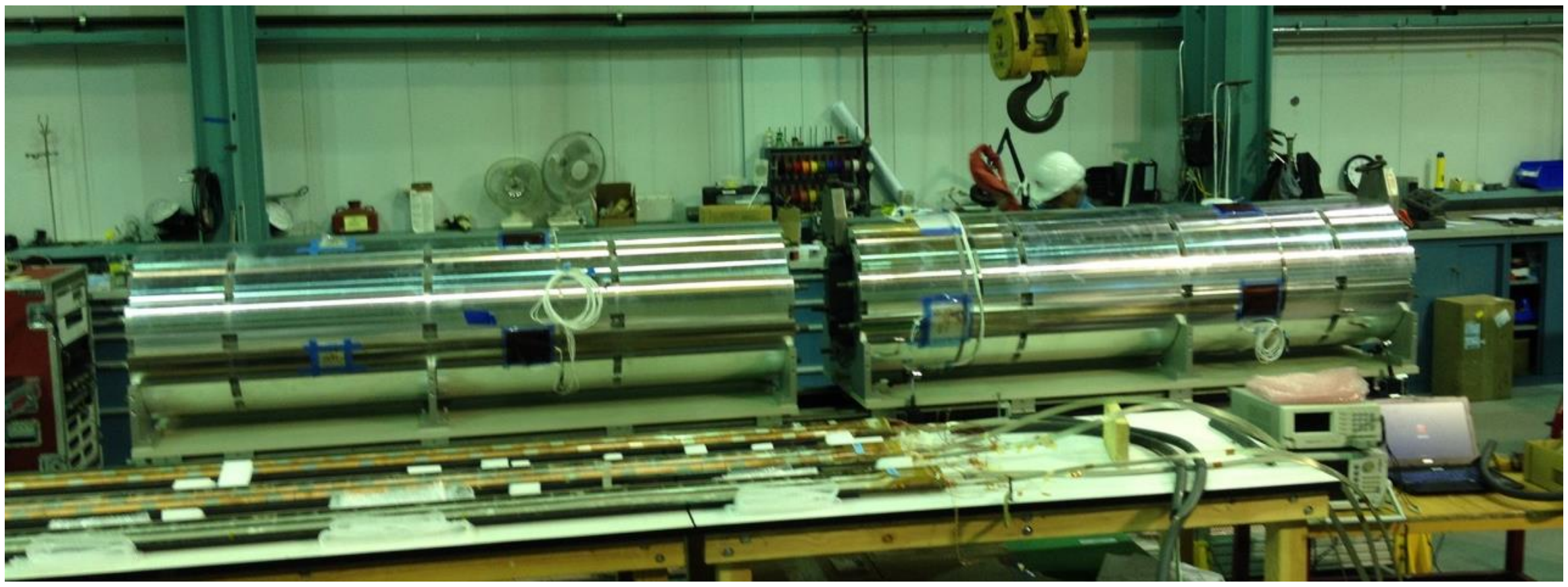

Fig. 6.7. Two half-length yoke-shell subassemblies, ready to be joined.

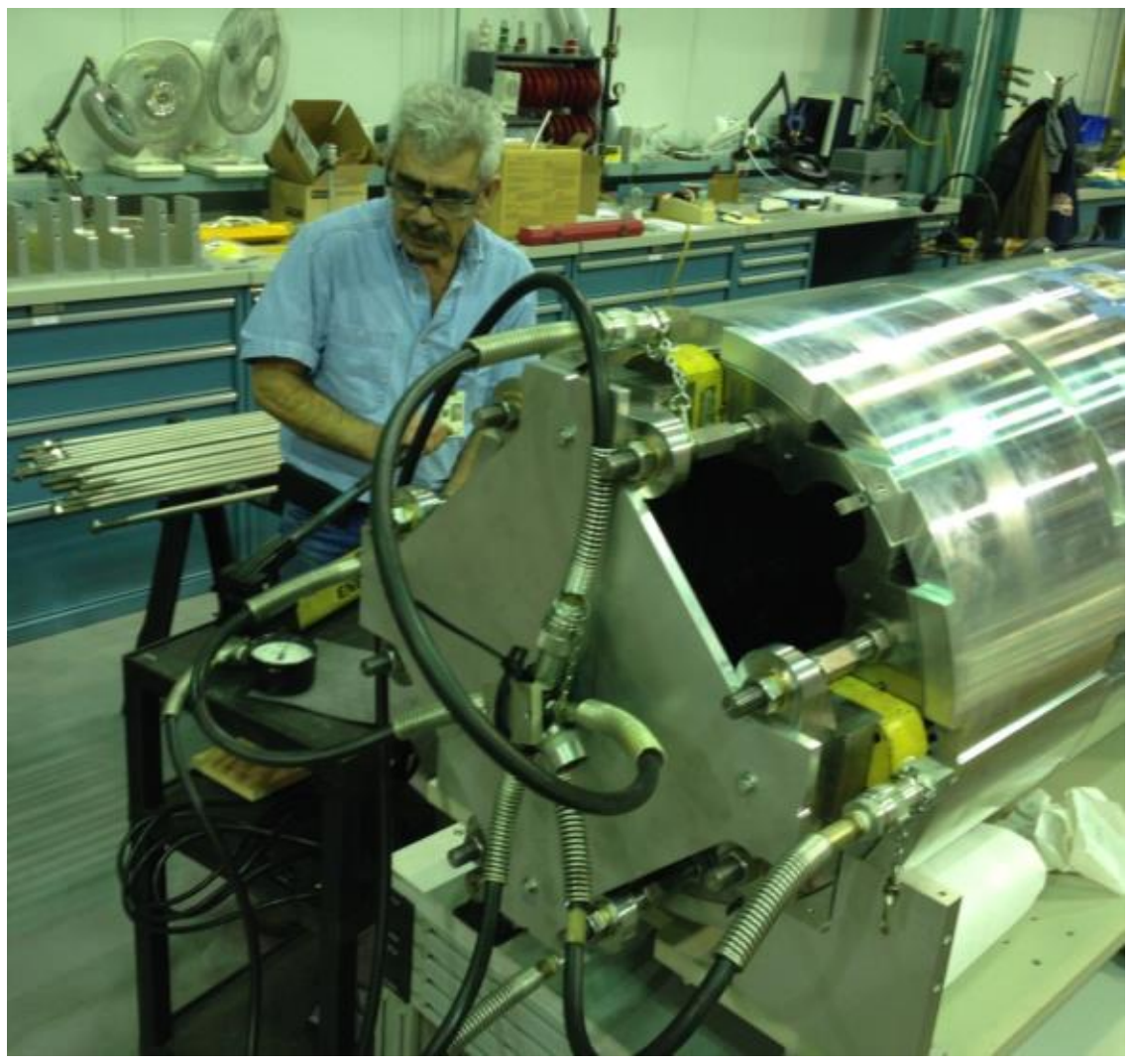

Fig. 6.8. Hydraulic rig for pre-tensioning the yoke tie rods.

## 6.4 Coil Pack Subassembly

### 6.4.1 Dressed Coils

When received, each impregnated coil undergoes a series of electrical and mechanical CMM measurements. These measurements will be used to determine the order and placement of each coil in the coil pack assembly. Once complete, a ground plane insulation (GPI) layer, made up of 75 μm polyimide and a 37 μm B-stage epoxy layer, is attached to each coil around the OD and midplanes. As described in [1], the GPI layer is bonded to the entire OD except for the areas over the end spacers and end shoe transition (LE and RE). In addition, active temperature is monitored during the GPI application.

The CMM measurements may also define whether an additional shim layer at the midplane, or additional radial shims, are required in later assembly operations. Support pads, called pions (LBNL Drawing SU-1010-1713), for the eventual beam tube will also be attached to the coils at this point while the coil poles are accessible before the coil pack is assembled. These will be nominally spaced 800 mm apart, but the bottom pair of coils will have theirs offset by 400 mm with respect to those found on the upper pair of coils.

In the prototype phase, coils were instrumented with strain gauges on the pole islands in three axial locations allowing for strain measurements in the azimuthal and axial directions. The final design utilizes only a single axial station on the coil at the return end of the magnet, which is then be removed after a successful magnet test, before the insertion of the beam tube as part of the cryostat and cold mass assembly. Finally, the G-11 pole keys are installed (but not shimmed to size yet) at this time. Note again, that these G11 pole keys also have 50 mm spacing of 8 mm holes to provide for the cooling required through the coil poles.

### 6.4.2 Pad-Collar Stacks

The aluminum collar laminations and the load pad laminations will be assembled together in four full-length stacks, one for each quadrant. Both lamination types are typically 50 mm (1.969 in) thick. While the collars are all 6061 aluminum, the load pads are made of ARMCO Pure Iron (Grade 4) in the straight section; each end must be stainless steel (304CO, low-cobalt) on the ends, in order to reduce the peak magnetic field.

The load pad is assembled first using short "sub-stacks" that correspond to the length between particular collar laminations. These sub-stacks are assembled into a full-length load pad assembly stack, aligned by aluminum bronze bushing sleeves, and joined using two 11.11 mm (7/16") pre-tensioned stainless steel tie rods.

Collar laminations are then stacked upon each assembled load pad. The collars are also aligned by aluminum bronze bushing sleeves, as well as by the key feature of the load pad geometry. Two 6.35 mm (¼ in.) stainless steel tie rods are tensioned to 1000 lbs. each. The collars assembly is held to the load pads via bolts through five load pad and collar lamination pairs along the length. These bolts are removed prior to the final assembly (see Fig 6.9).

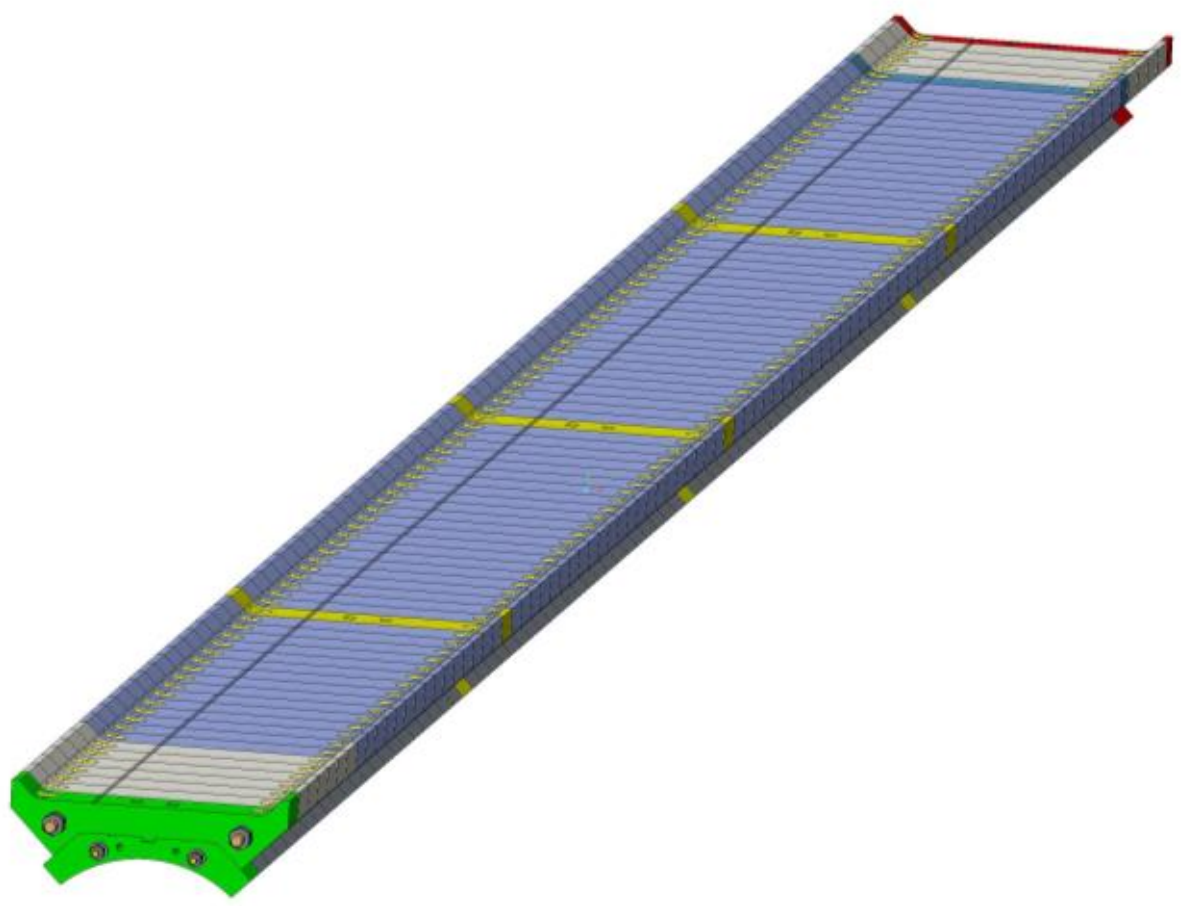

Fig. 6.9. Full-length load pad-collar assembly. The lighter colored ends of the load pads are SST instead of ARMCO Pure Iron.

### 6.4.3 Coil Pack Assembly Operations

The bottom pad-collar assembly is placed on the coil pack assembly table, and the two side pad-collar assemblies are assembled on the two pivot tables beside the primary assembly table (see Fig. 6.10). Based on the earlier CMM measurements performed on the coils, the inner curved surface of the collar stacks are lined with layers of G11 and/or polyimide as radial shims. While the nominal radial shim stack is 0.5 mm, including the coil GPI layer, the actual stack may vary depending on these CMM measurements. The first build also incorporates the use of pressure sensitive film as part of the radial stack up. These layers are eventually replaced by either polyimide or G11 in the final build (see Fig. 6.11).

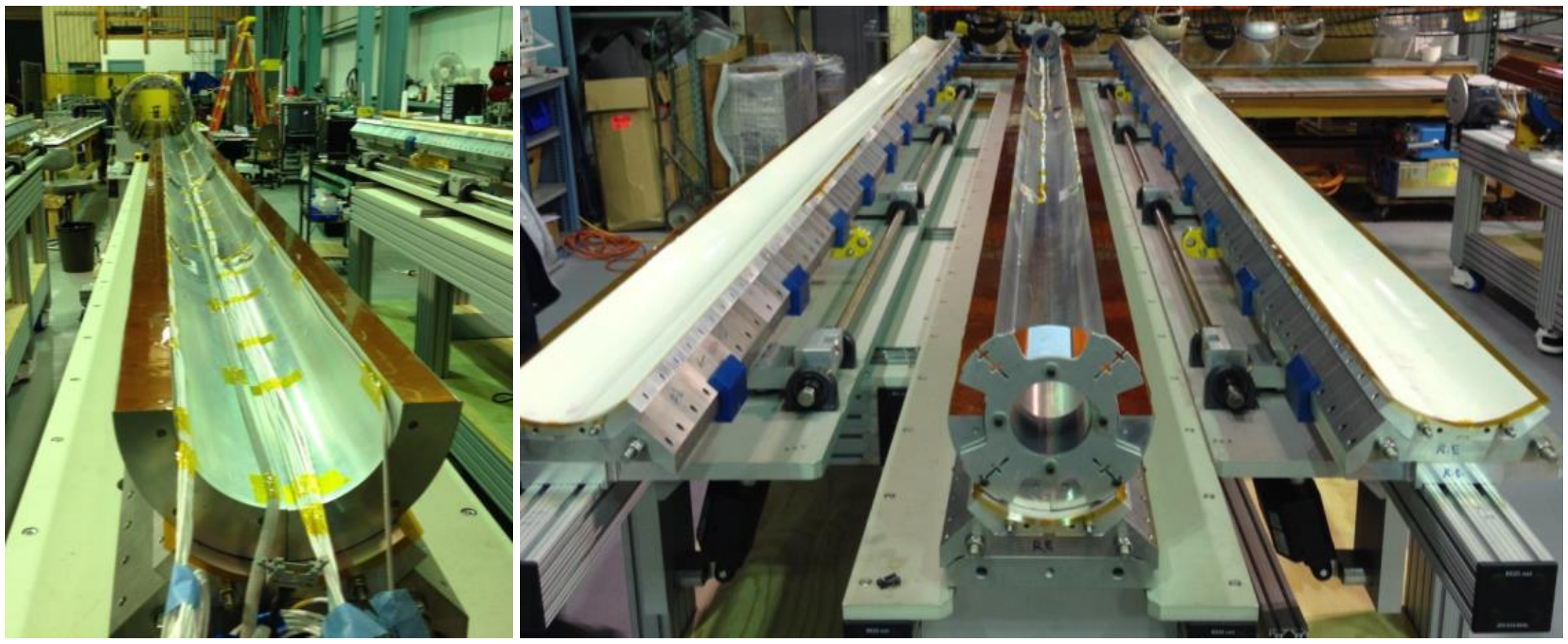

Fig. 6.10. (L) The bottom pair of (dummy) coils placed on the bottom pad-collar assembly. (R) The side pivot tables shown.

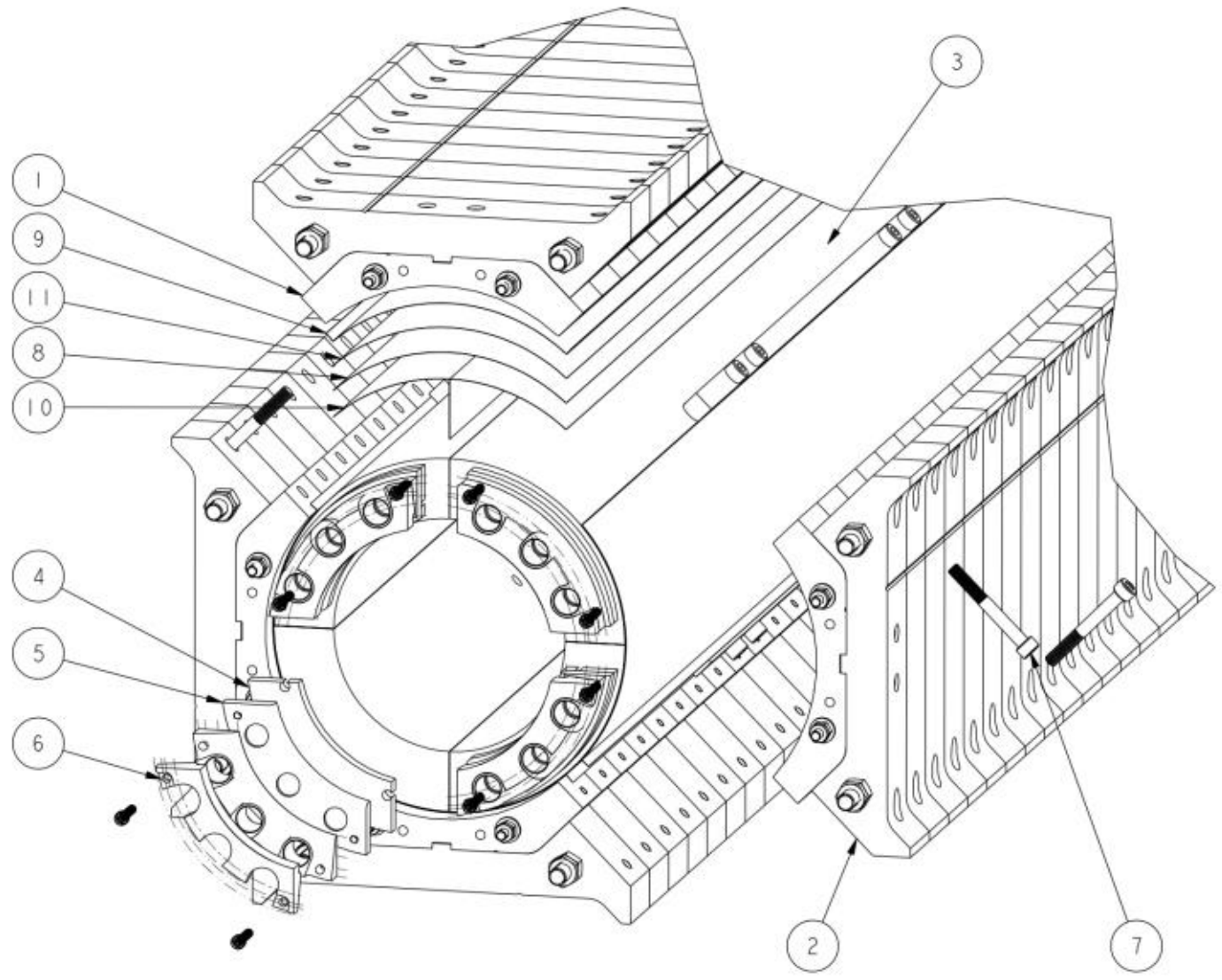

| QTY | ITEM | PART NO. | DESCRIPTION |
|---|---|---|---|
| 4 | 11 | | MQXFA COIL PACK SHIM, RADIAL, .003" |
| 8 | 10 | | MQXFA COIL PACK GPI, COIL |
| 4 | 9 | | MQXFA COIL PACK GPI, COLLAR |
| 4 | 8 | | MQXFA COIL PACK SHIM, RADIAL, .005" |
| 8 | 7 | | 91292a210_mc-m8x65shcs.prt |
| 8 | 6 | | su-1004-1403.asm |
| 8 | 5 | | lhcmqxfm0039.prt |
| 8 | 4 | | lhcmqxfm0038.prt |
| 1 | 3 | | MQXFA-I COIL SET |
| 2 | 2 | 2TK607 | MQXFA-I LOADPAD-COLLAR SUBASSEMBLY, THRU |
| 2 | 1 | 2TK606 | MQXFA-I LOADPAD COLLAR SUBASSEMBLY THREADED |

Fig. 6.11. The coil pack assembly shim layers.

The bottom pair of dressed coils is then moved onto the bottom pad-collar assembly. The upper pair of coils follows this operation, being secured on the assembly before the side pad-collar assemblies are rotated into place. Finally, the upper pad-collar assembly is placed on top, and all the pads are bolted together. At this stage, with unshimmed alignment keys inserted into the coils the gaps between these keys and the collars are measured.

This first assembly with pressure-sensitive paper is disassembled so that the layers can be replaced with the appropriate radial shims, and the alignment keys can be shimmed accordingly to ensure the collar-key contact conditions after cool-down. According to FEM computations, and as described in [2], this condition is achieved by assembly the coil pack at room temperature with a collar to pole-key gap of 0.200 mm per side This will ensure both proper alignment and minimum interception of the force from the shell at 1.9 K. The final coil pack is measured; these measurements are used to calculate the initial shim package of the master keys described in the next section (see Fig. 6.12).

A warm magnetic measurement of the coil pack will be performed at this time to ensure that there is nothing grossly out of alignment, and this information will be compared with the later final fiducialization measurement performed after the magnet assembly is complete.

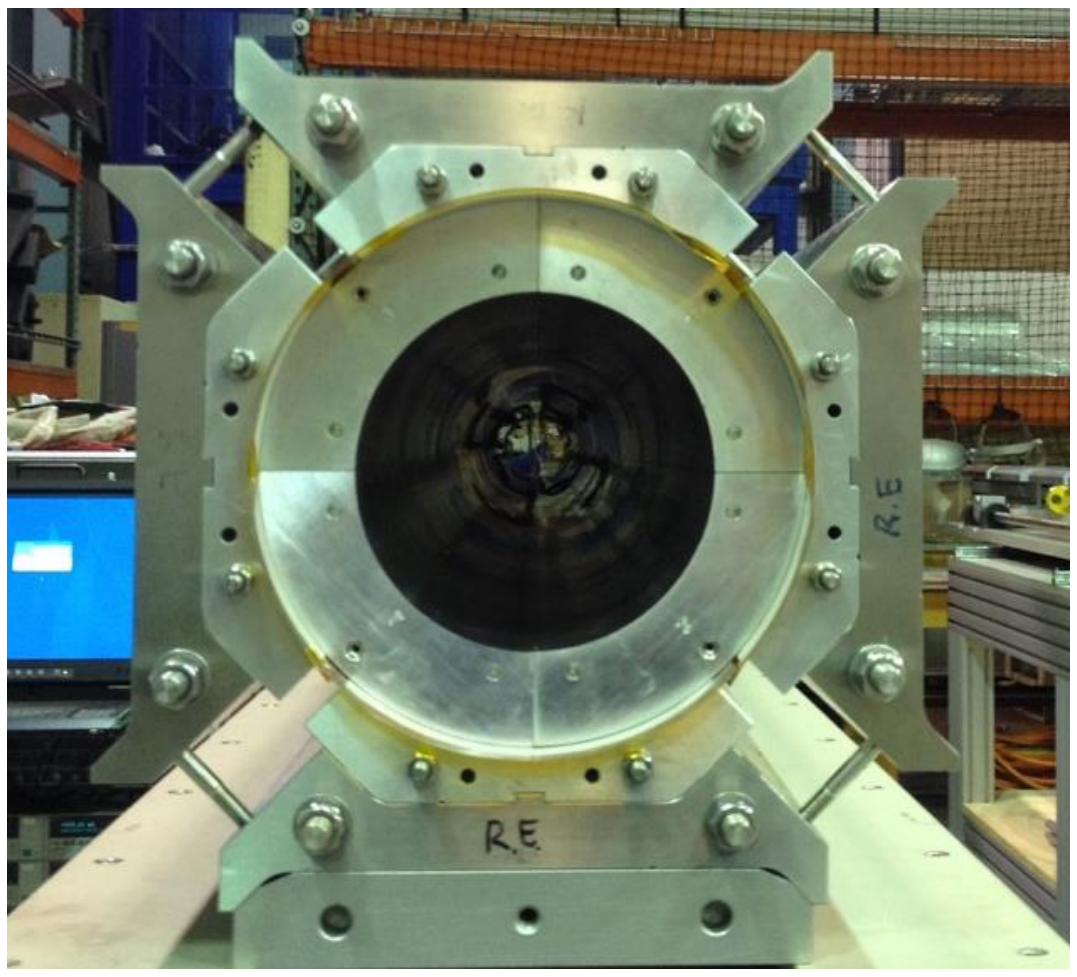

Fig. 6.12. The (dummy) coil pack assembly, with shim layers visible between the coils and collars.

### 6.4.4 Magnet Integration

#### 6.4.4.1 Master Key Packages

The mechanical measurements of the yoke opening and the coil pack are used to calculate the initial master key package shims. The master assembly consists of two master key plates, each with slots for bladders, load keys, and an alignment key. The master assemblies will be assembled ("kitted") with a uniform amount of shim in all the quadrants initially (see Fig 6.13).

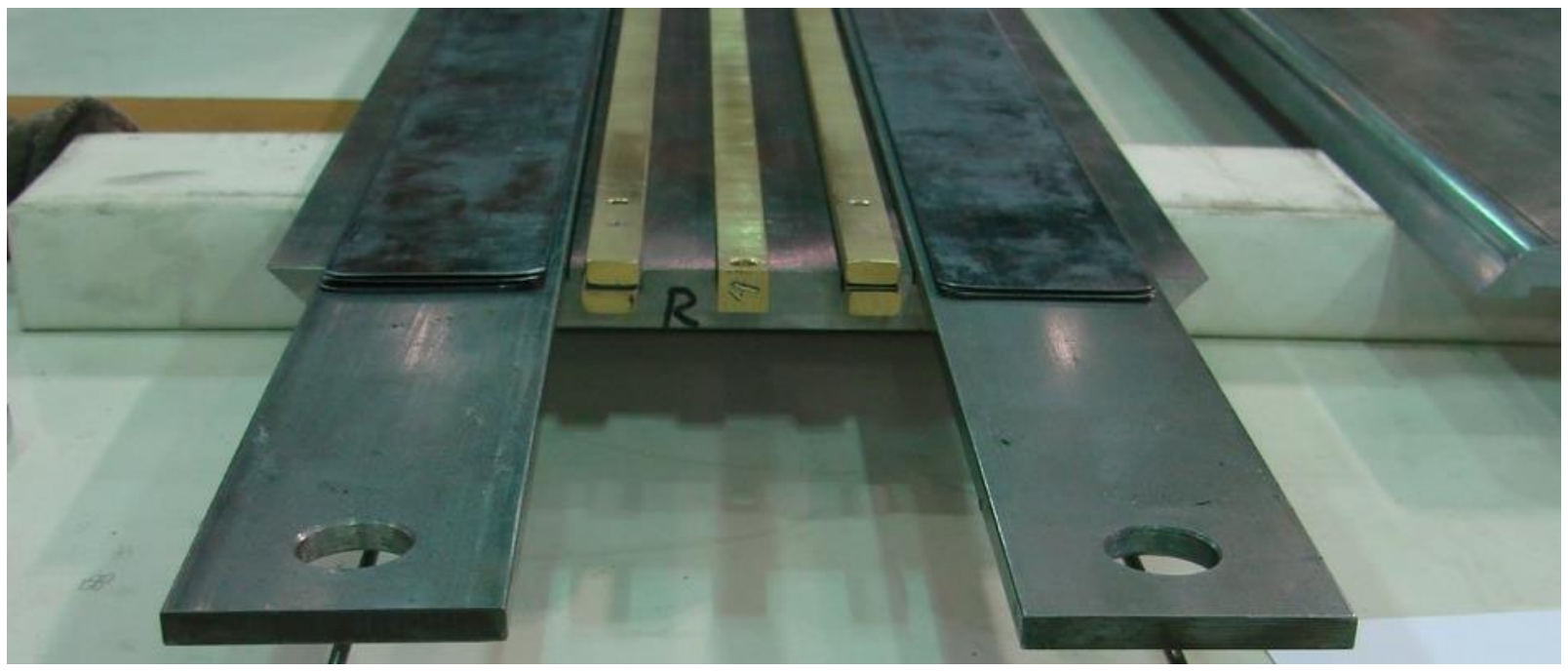

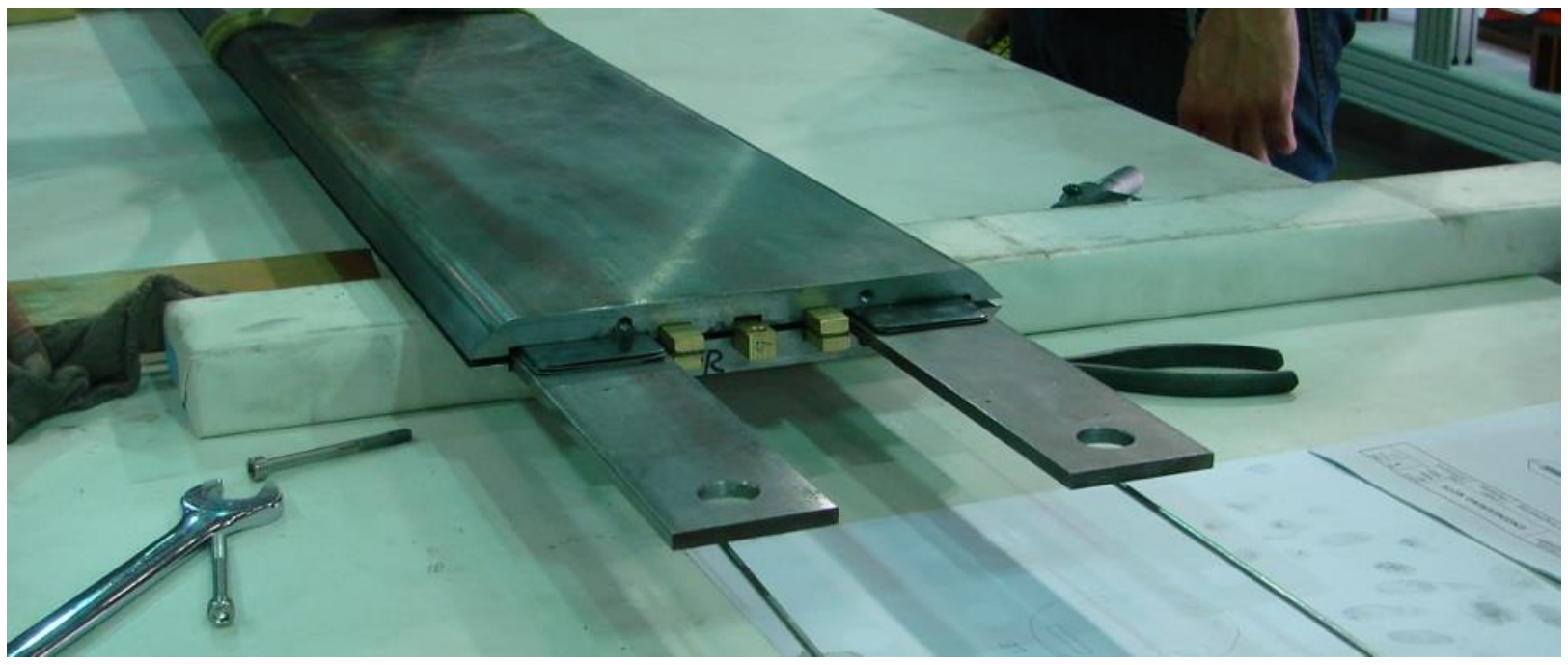

Fig. 6.13. (Top) Kitting the master key packages; (bot) assembled master key package.

The coil pack is inserted into the yoke opening using a set of rollers in the yoke cooling hole slots. Only the bottom quadrant master keys are present in this stage (Fig 6.13). Once the coil pack inserted, however, the rest of the master keys can be inserted into each quadrant.

#### 6.4.4.2 Magnet Preload Operations

The high-pressure pump is attached to the bladders on both ends of the magnet (keys and bladders are half-length), and the axial loading rig is attached to the axial rods to perform the axial preload operations (Fig 6.14). The load keys will be shimmed to obtain the target strain on the shells and coils, as measured by the strain gauges, determined by the FEA model, and the axial rod strain gauges are used to monitor the axial load prestress.

As described in [2], the loading operation begins with the axial rods snugged but not prestressed with the loading rig. The azimuthal preload is applied to a target of half the planned azimuthal load and shimmed so the bladders are deflated. At that point the axial loading rig is applied to yield half the planned axial preload, and the axial rod nuts tightened to maintain that pressure. The azimuthal preload is then reapplied, this time to the target load, and final shims installed. The last loading step is to apply the full axial preload, again using the axial rig. As can be seen in Fig. 6.15, where the red square indicates the target values and the markers show the measured data taken after the pre-loading operation of MQXFA03 and MQXFA04, the target pre-stress for coil (winding pole) and shell and the target strain for rods, with acceptance ranges, are:

- Shell stress: +58 ± 6 MPa
- Coil (winding pole) stress: -80 ± 8 MPa
- Rod strain: +950 με ± 95 με

A final magnet (mechanical and magnetic) fiducial survey will be performed at this stage.

This completes the magnet preload operations. Electrical QC measurements are taken at this stage to ensure that none of these operations have damage coils, or insulation layers.

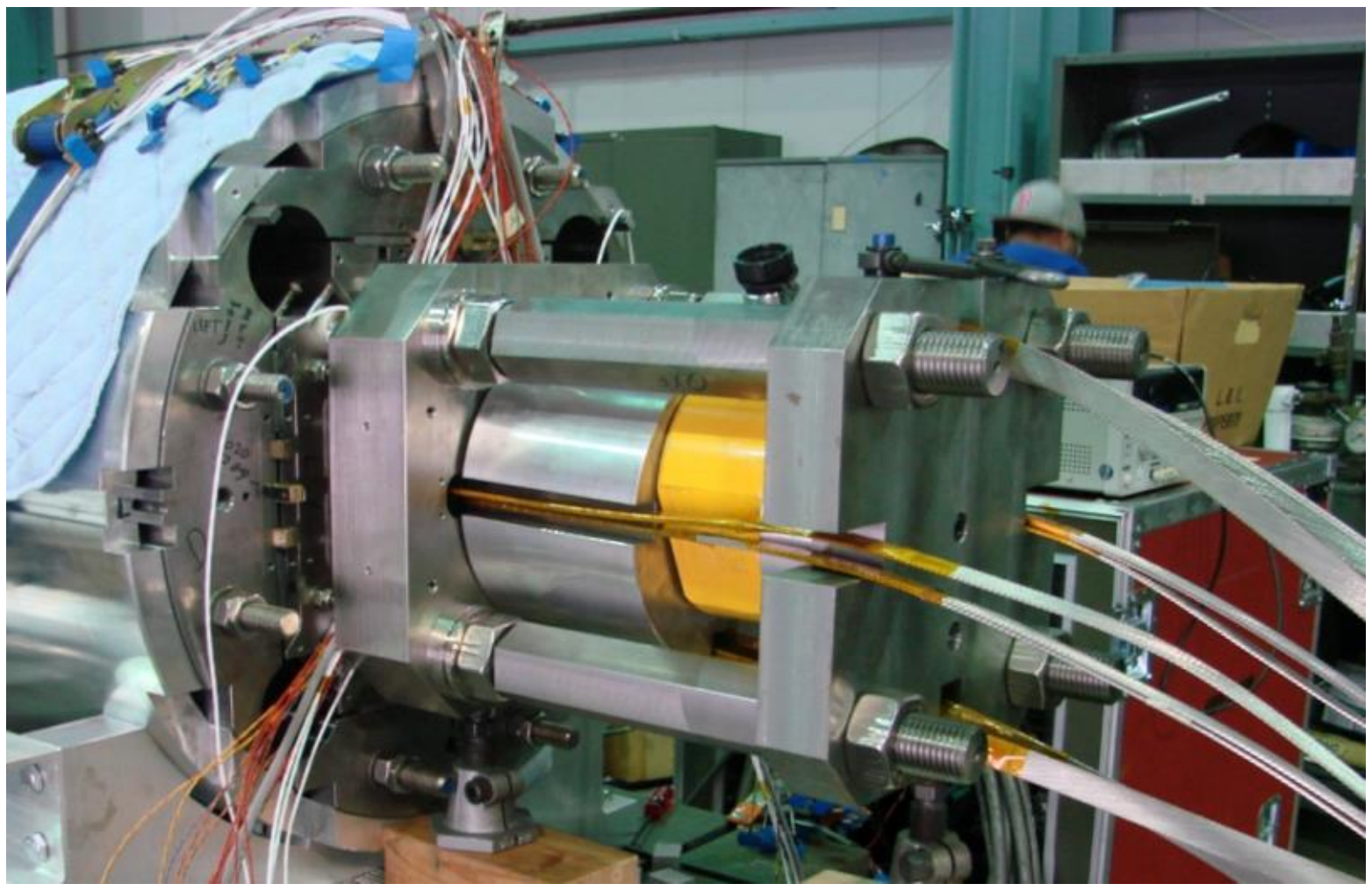

Fig. 6.14. Axial preloading operation with the axial loading rig.

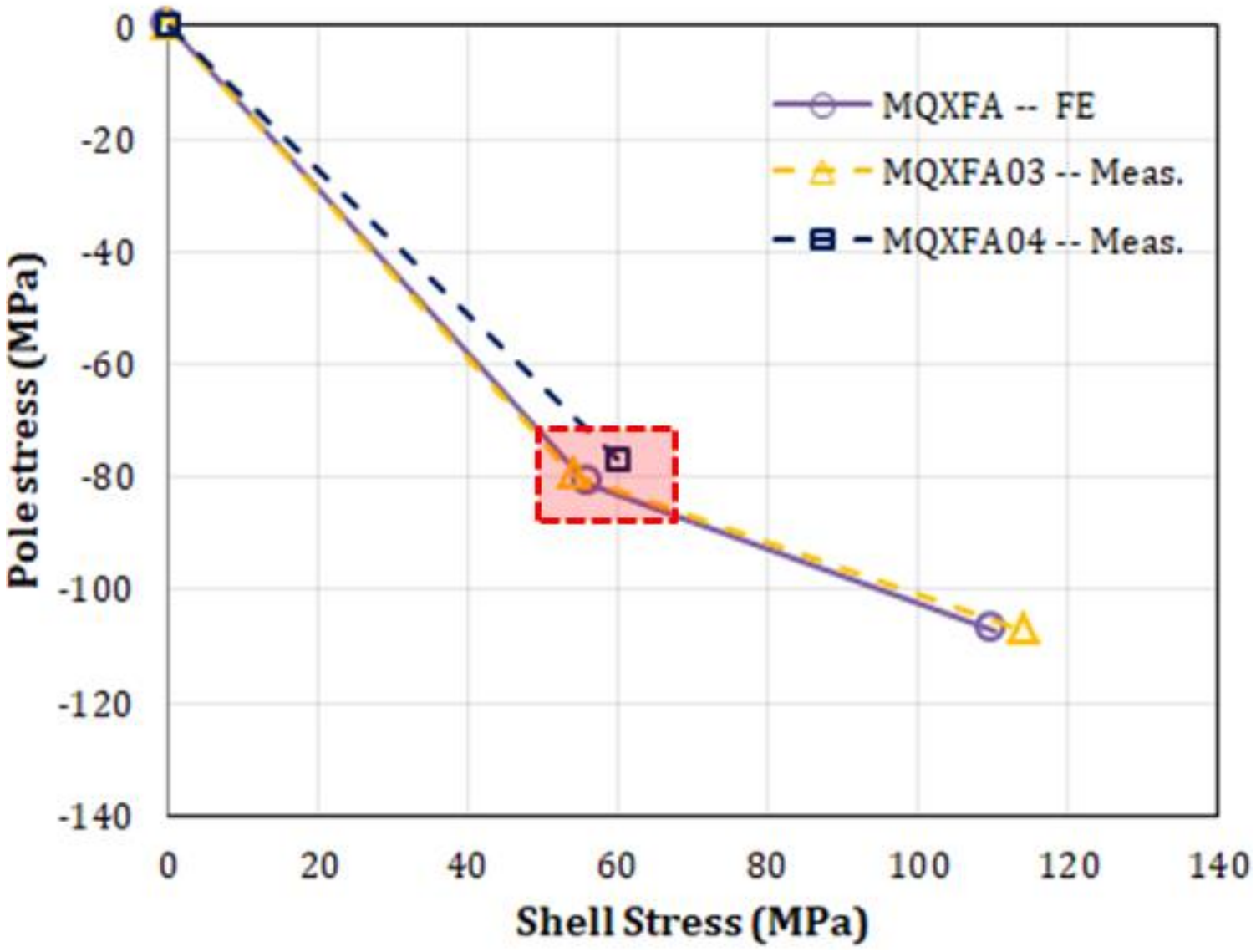

Fig. 6.15. Coil stress vs shell stress at 293 K and 1.9 K: measured data, FEM computations, and acceptance ranges (red box).

### 6.4.5 **Magnet Finishing Operations**

#### **6.4.5.1 Connector Skirts**

Before the splice connections are completed, all the instrumentation connectors are mounted to the skirts. This includes protection heaters, voltage taps, strain gauges, etc. for the entire magnet (see

Fig 6.16). A detailed list of MQXFA Magnet Connectors and Finishing Work Instructions (WI) can be found in [3].

It is important to point out that the instrumentation described here is related to the magnet for the vertical test. A different configuration is used for the magnet in the cold mass.

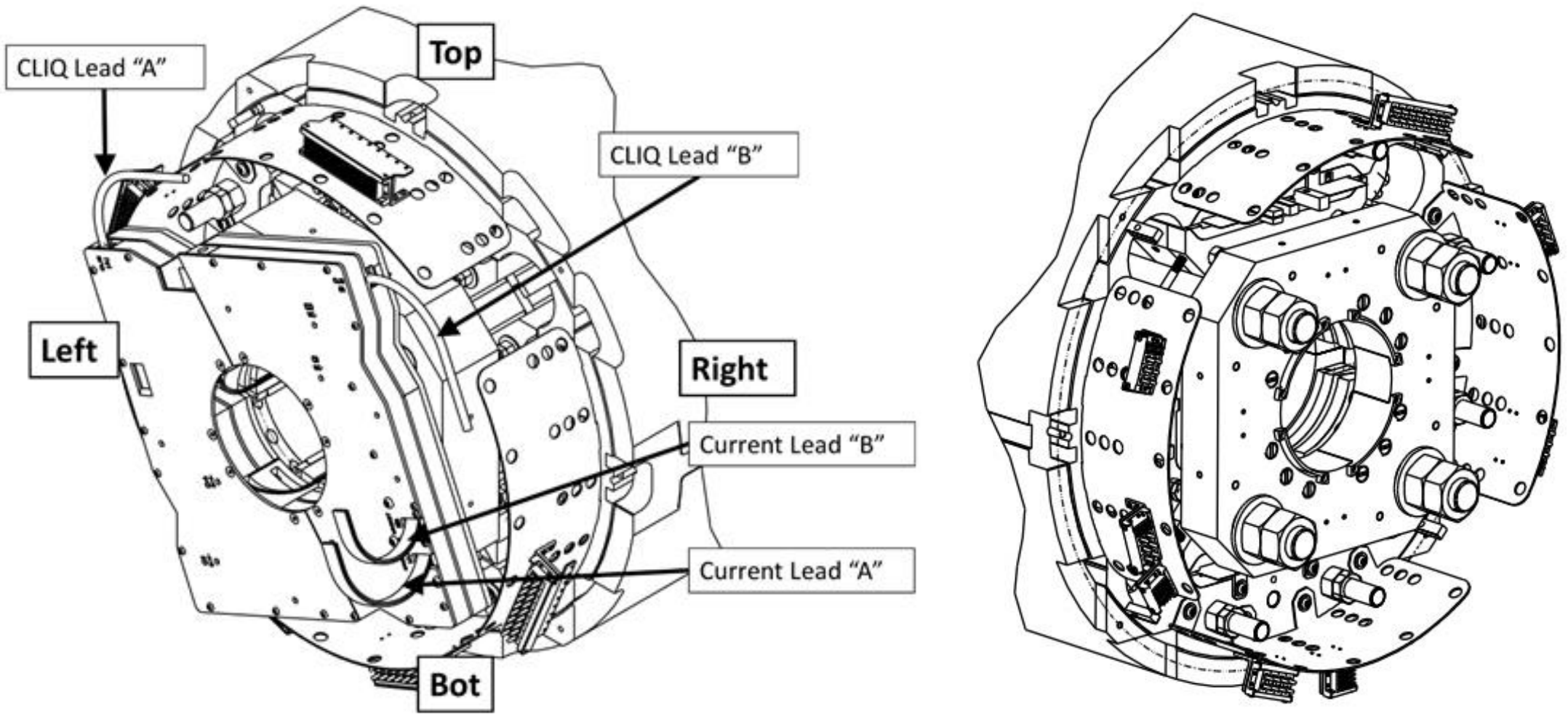

Fig. 6.16. Location of the connector skirts for the LE (L), and RE (R).

#### 6.4.5.2 Axial restraining plates

The End-plates on both Lead End (LE) and Return End (RE) are designed to keep the coil compressed axially at all stages. Off-center movements of any side End-plate will cause unbalanced axial force in the four rods, potentially affecting the axial load status of the coils. To prevent any motion of the end-plates during magnet handling and shipping, restraining plates are inserted between the loading keys and the end-plates (see Fig. 6.17).

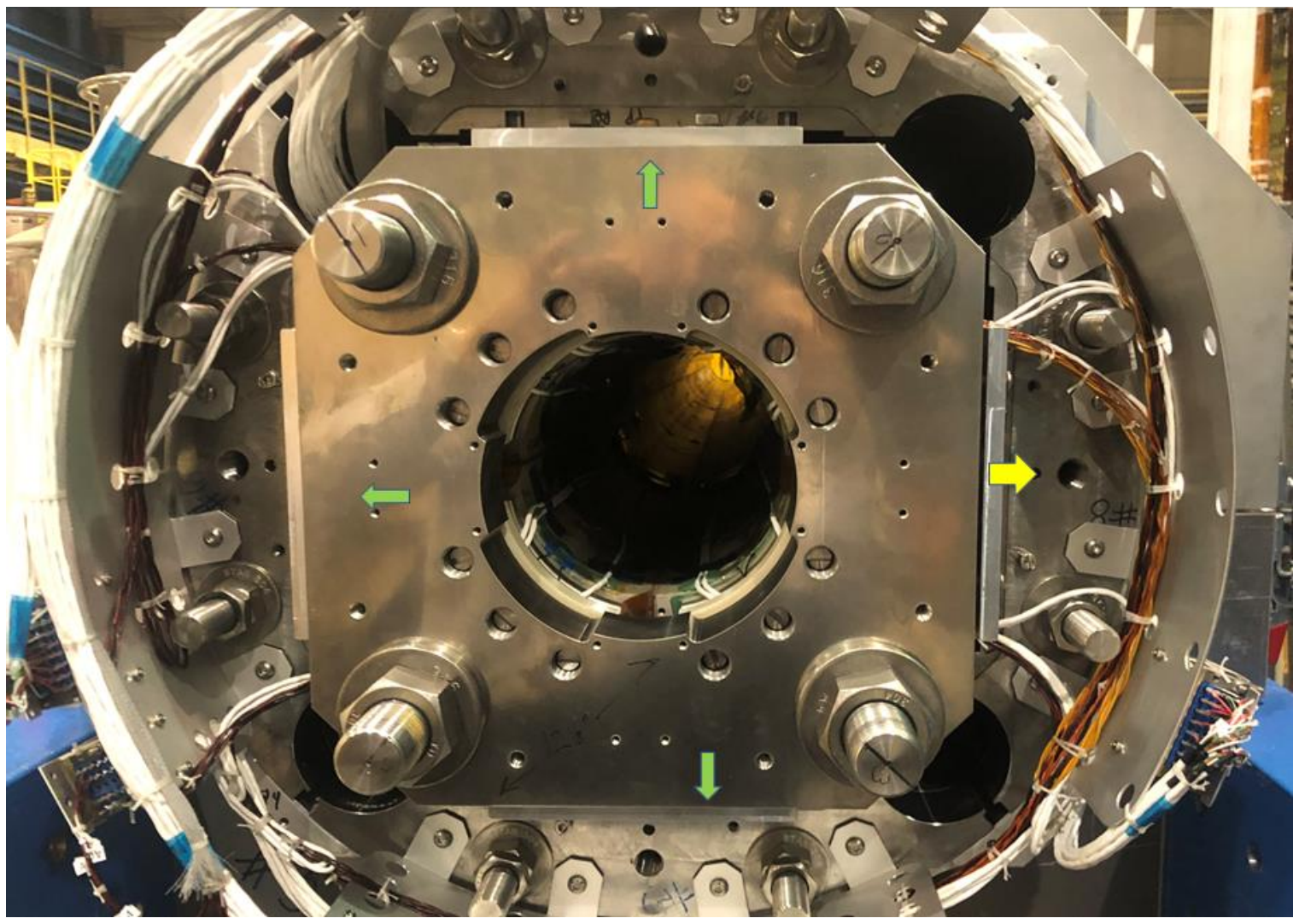

Fig. 6.17. Axial restraining plates (indicated by the arrows).

#### 6.4.5.3 Splice Connections

The magnet splice connections are made after the magnet preload operations are complete and the instrumentation and skirts are assembled. Fig. 6.18 shows the physical layout of the magnet splice connections box, which is made up of two layers due to space constraints and the bend radius of the cables [4].

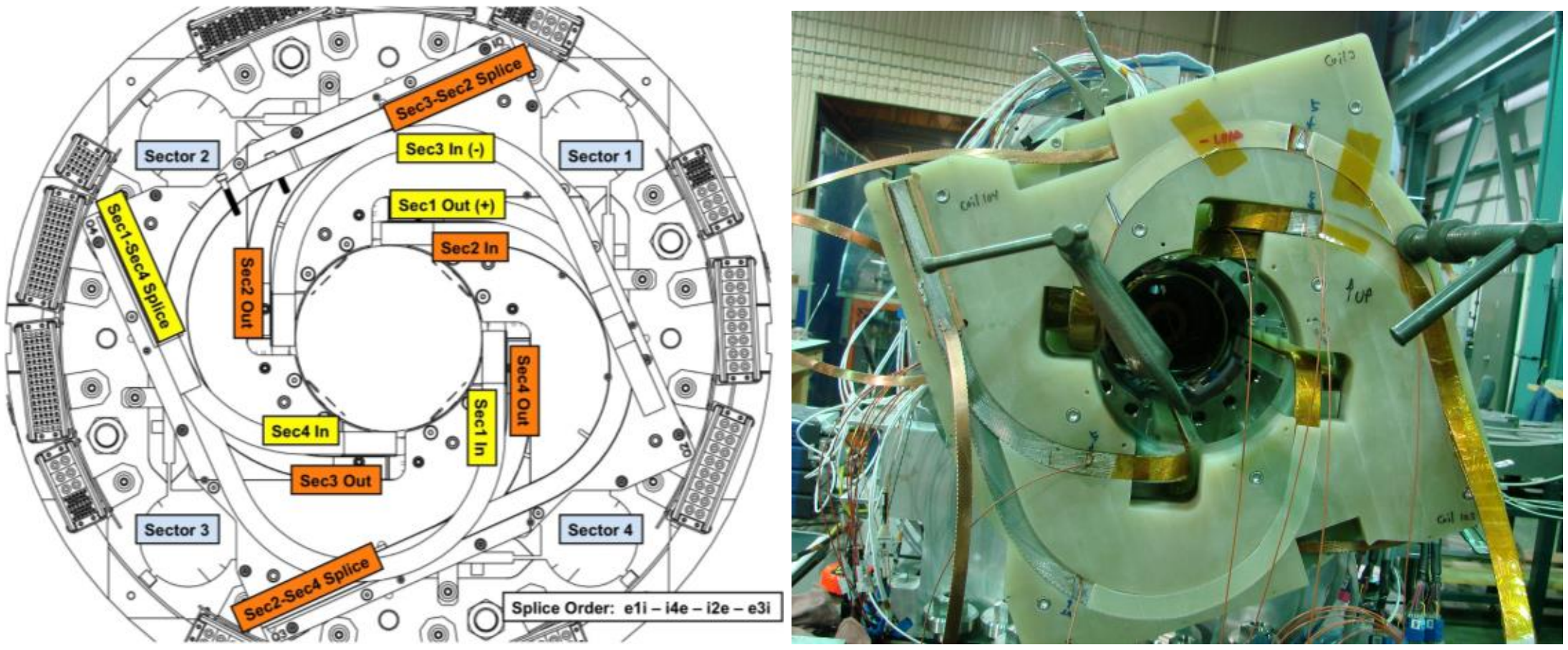

Fig. 6.18. Splice connections for the MQXFA.

The “A” and “B” magnet leads are uncut leads from two separate coils in the magnet: outer layer (OL) leads correspond to the “A” magnet lead, and inner layer (IL) leads correspond to the “B” magnet lead. The “A” lead requires 455 mm of length from the end of the coil to the exit of the splice connection box. The “B” lead requires 670 mm from the end of the coil to the exit of the splice connection box (see Fig. 6.19).

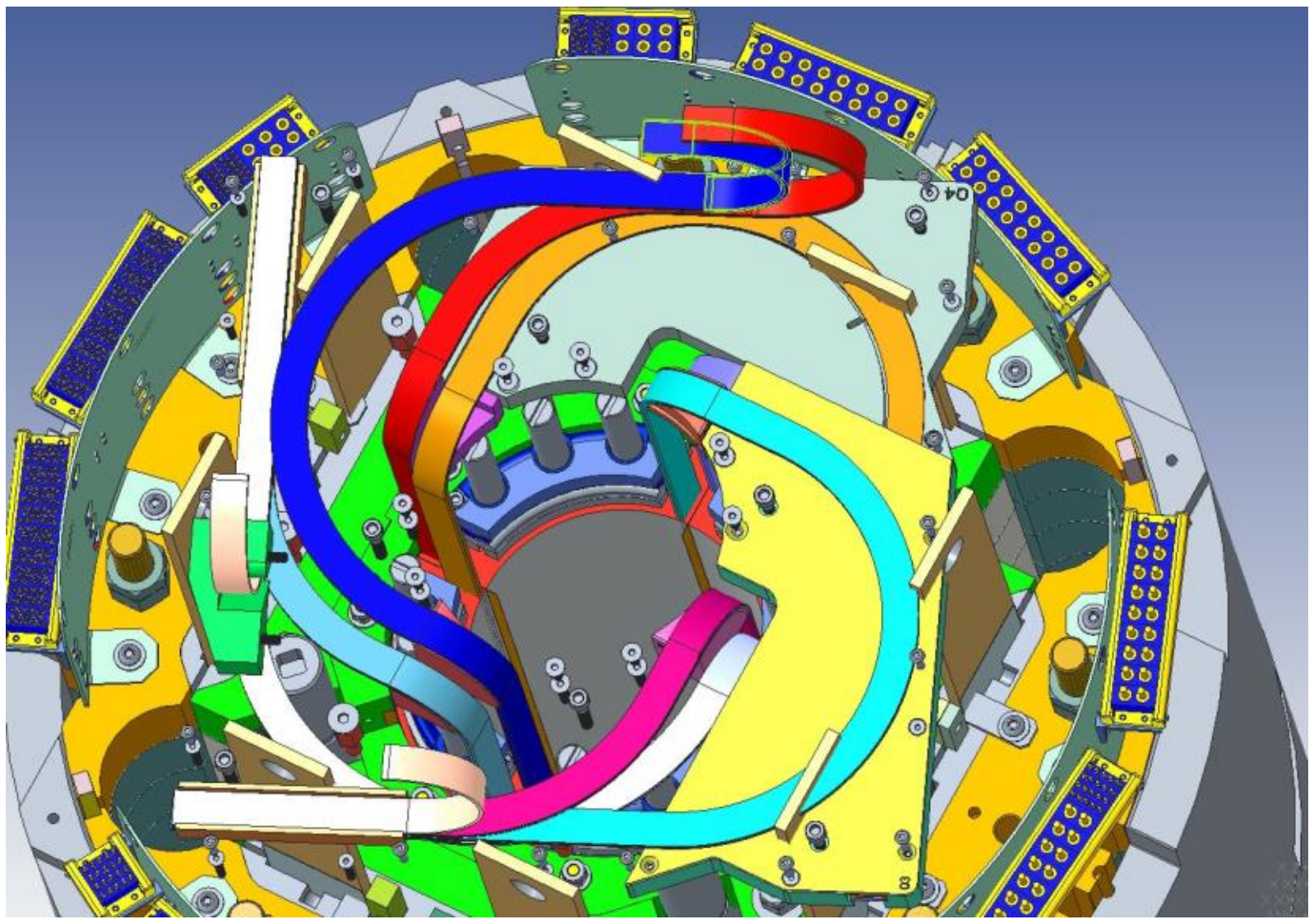

Fig. 6.19. MQXFA coil leads: The red lead is the OL lead (A) and the blue one is the IL lead (B),

#### 6.4.5.4 Voltage taps for quench detection

The quench detection instrumentation includes three redundant (3x2) quench detection voltage taps located on each magnet lead and at the electrical midpoint of the magnet circuit, two (2) voltage taps for each internal MQXFA $Nb_3Sn$-NbTi splice, OL heater current leads.

#### 6.4.5.5 Magnet handling and shipment

MQXFA magnets handling and shipping requirement are specified in ref. [5].

MQXFA magnets have four coils made of strain sensitive $Nb_3Sn$ conductor. To prevent conductor degradation, handling and shipping operations shall meet the following requirement for each MQXFA magnet:

**MQXFA H&S Requirement #1: The conductor strain shall never exceed 500 microstrain in any part of the coils during any MQXFA handling (for instance lifting, lowering, rotation and up-righting) and shipping operation.**

The following requirement preserves MQXFA longitudinal rigidity, and prevents structure damage:

**MQXFA H&S Requirement #2: Where the tie-rods go through the iron laminations, the laminations shall always remain in contact during any MQXFA handling and shipping operation.**

MQXFA magnets will be shipped from LBNL (assembly site) to BNL (vertical test site), and then to FNAL (cold-mass assembly site) or back to LBNL (repair site) using a shipping fixture designed to minimize shocks during transportation, and truck loading/unloading. Two accelerometers will be placed on the fixture and a dedicated truck equipped with air-raide suspension will be used for transportation.

If there are no shock above 4.5 g, the standard QC for magnet reception can be performed

A more detailed description of those requirements together with the shipping procedure and plans are report in ref. [5,6]

## 6.5 Magnet Fabrication Infrastructure at LBNL

### 6.5.1 Yoke-Shell Subassembly Tooling

A single assembly pad is designed to assemble the half-length shell-yoke assemblies vertically. This can be seen in both Fig 6.6, and Fig 6.20. Originally designed to use the same bladder configuration as in a real magnet (eight bladders, two per quadrant), it has since been configured to employ bladders inserted into the cooling holes instead. This configuration improves efficiency by reducing the required amount of bladders (eight => four), and the forces are more directly applied to the gap.

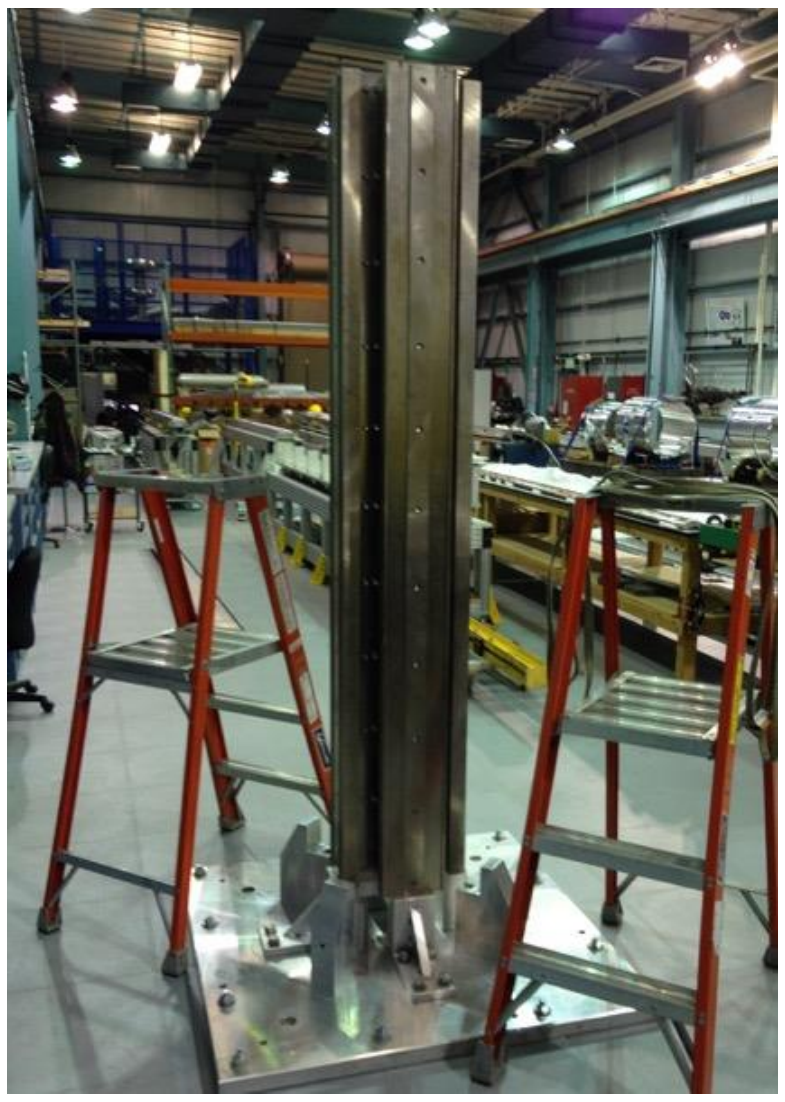

Fig. 6.20. Half shell-yoke assembly stand.

The integration table is where the majority of the magnet assembly activities take place. It consists of two support plates on Thompson linear rails so that each half-length shell-yoke assembly can be independent handled, then joined (see Fig 6.21). This table is also where the final magnet integration takes place, and all the finishing and magnetic measurements take place.

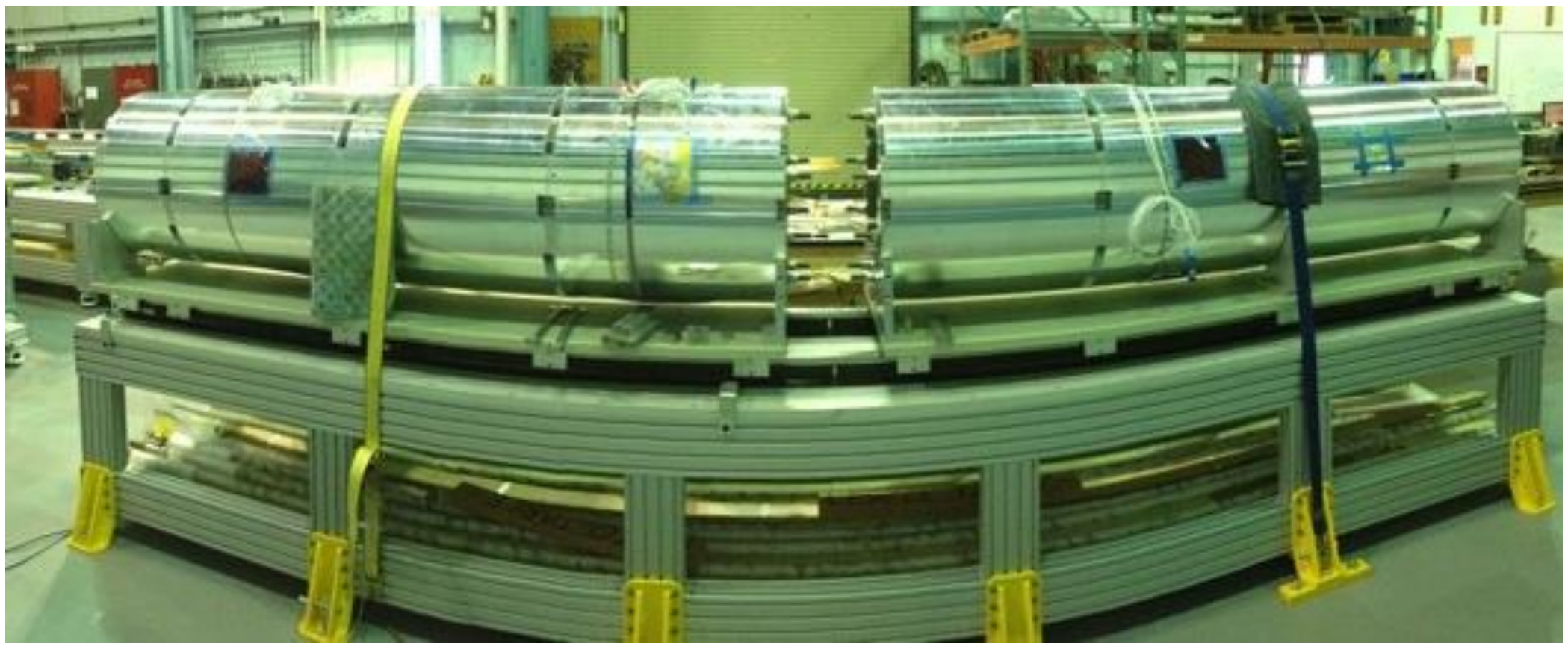

Fig. 6.21. “Panoramic” view of magnet integration table, shown with two half-length shell-yoke assemblies prior to joining.

### 6.5.2 Coil Pack Assembly Tooling

The coil pack assembly table incorporates a platen that a pad-collar assembly would be resting on. The platen can be raised via hydraulic pressure for when an assembled coil pack is ready to have its insertion rails attached. It is mounted on wheels so that the coil pack can be moved to the integration table when ready, however it would be moved to location where clear access to the ends is available during coil pack assembly steps. Two auxiliary tables are also part of this table, which facilitate the handling and assembly of the full length pad-collar assemblies. These tables are visible in Fig 6.10. This set of coil pack assembly tooling tables will be duplicated during the production run in order to allow for two technician teams to work on parallel activities of magnet assembly.

### 6.5.3 Coil Rollover Tooling

Tooling for safely handling coil rotation operations are shown in Fig 6.22. During coil preparations it is required that access to a coil's inner bore and outer surfaces be maintained. Coils are captured in "wheels" are driven by a gear assembly to rotates single coils or coil pairs. A second rollover tooling table will be procured for the project, allowing for coils in varying stages of processing to be handled at once when magnet production ramps up.

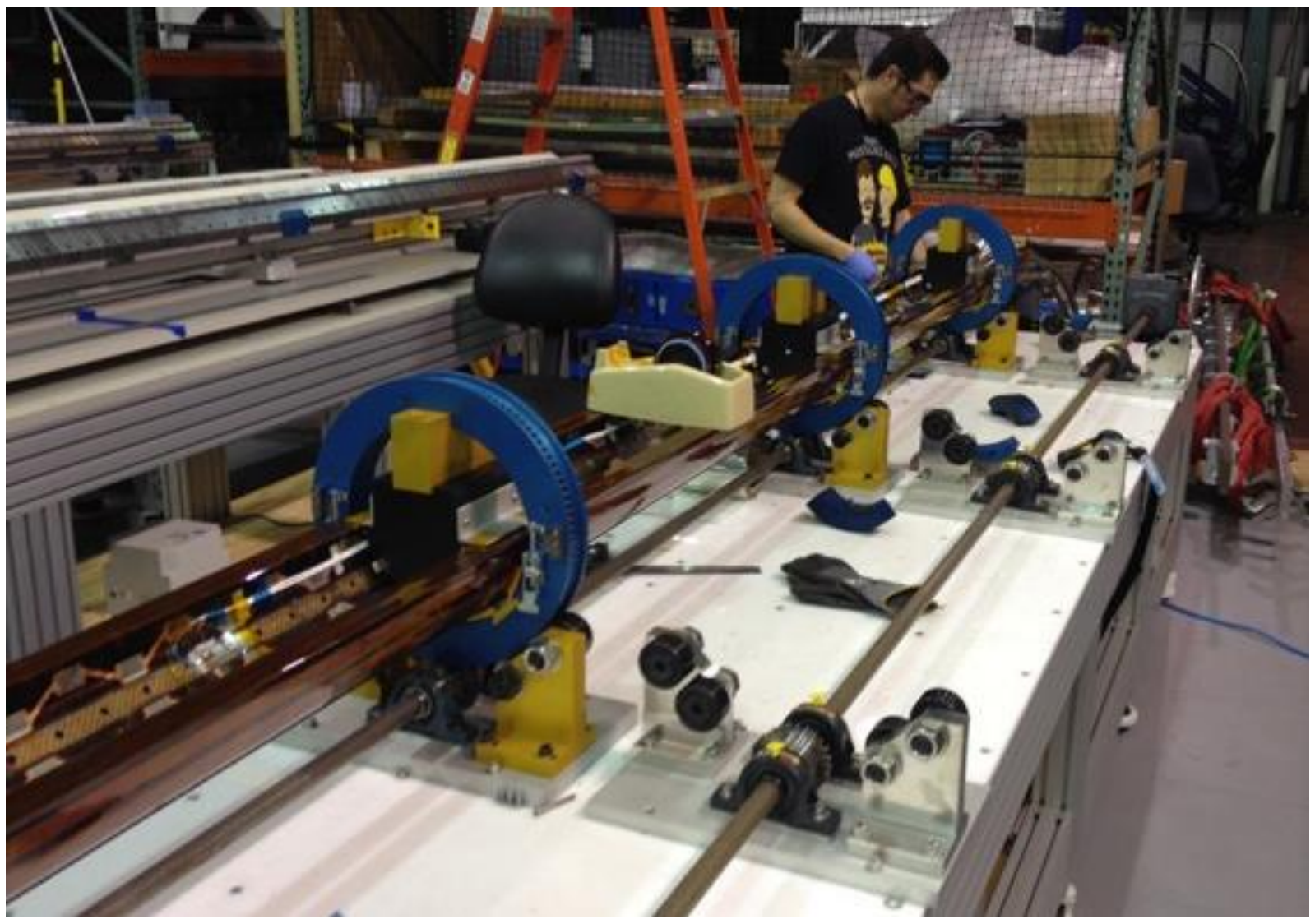

Fig. 6.22. Coil rollover tooling table.

# 7 MQXFA Vertical Test

The MQXFA test requirements are specified in the Functional Test Requirements document for MQXFA magnets [1]. Summary information is provided here for information only.

All prototype, pre-series and most series MQXFA magnets will be fully trained and tested in the vertical test facility (Figure 7.1) at Brookhaven National Laboratory (BNL), and on acceptance will be shipped to FNAL for assembly into LQXFA cold masses to be tested there. Comprehensive magnetic field measurements will also be performed at BNL. After prototype tests there will be 5 pre-series magnet tests and 15 series magnet tests at the BNL vertical test facility. The cold masses at FNAL will be assembled into the cryo-assemblies that will become the Q1 and Q3 elements of the HL-LHC inner triplets.

Vertical test stand 2 (Figure 7.2), to be used for the MQXFA testing, is one of the vertical test cryostats previously used for over 30 years to test a wide variety of superconducting magnets, including RHIC and SSC arc dipole and quadrupole short prototypes, all the RHIC corrector magnets, and many others. This particular test stand has been upgraded to accommodate the size of the MQXFA magnets and also to provide testing at 1.9 K, and less, and at a pressure of 1 bar, or more, and to also provide powering up to 24 kA with fast switching capabilities for quench protection. In parallel with this, the cryogenics facility has also seen extensive upgrades in 2016, most of which have been funded by BNL. In the fall of 2016, the new facility was successfully commissioned, and in late 2016 and Jan of 2017, the first ever long coil of MQXFA design was tested in a mirror configuration (MQXFPM1) and it reached 17.890 kA (ultimate current plus margin) in 11 quenches, and after 19 quenches exceeded this current by 8% (Figure 7.3). Along with this, quench protection heater tests and other R&D type measurements were done.

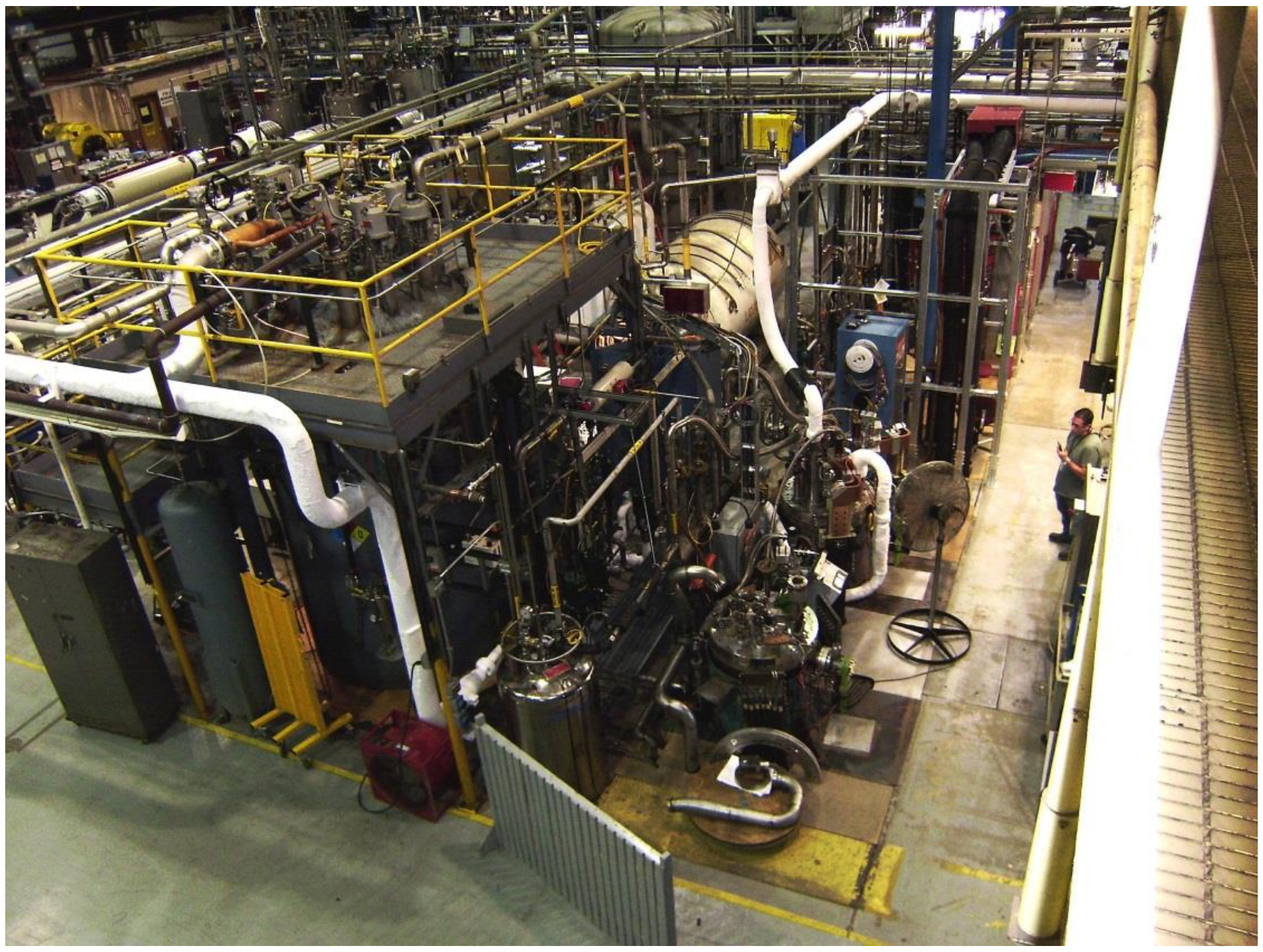

Figure 7.1: Vertical Test Facility at BNL, showing two of the five test cryostats.

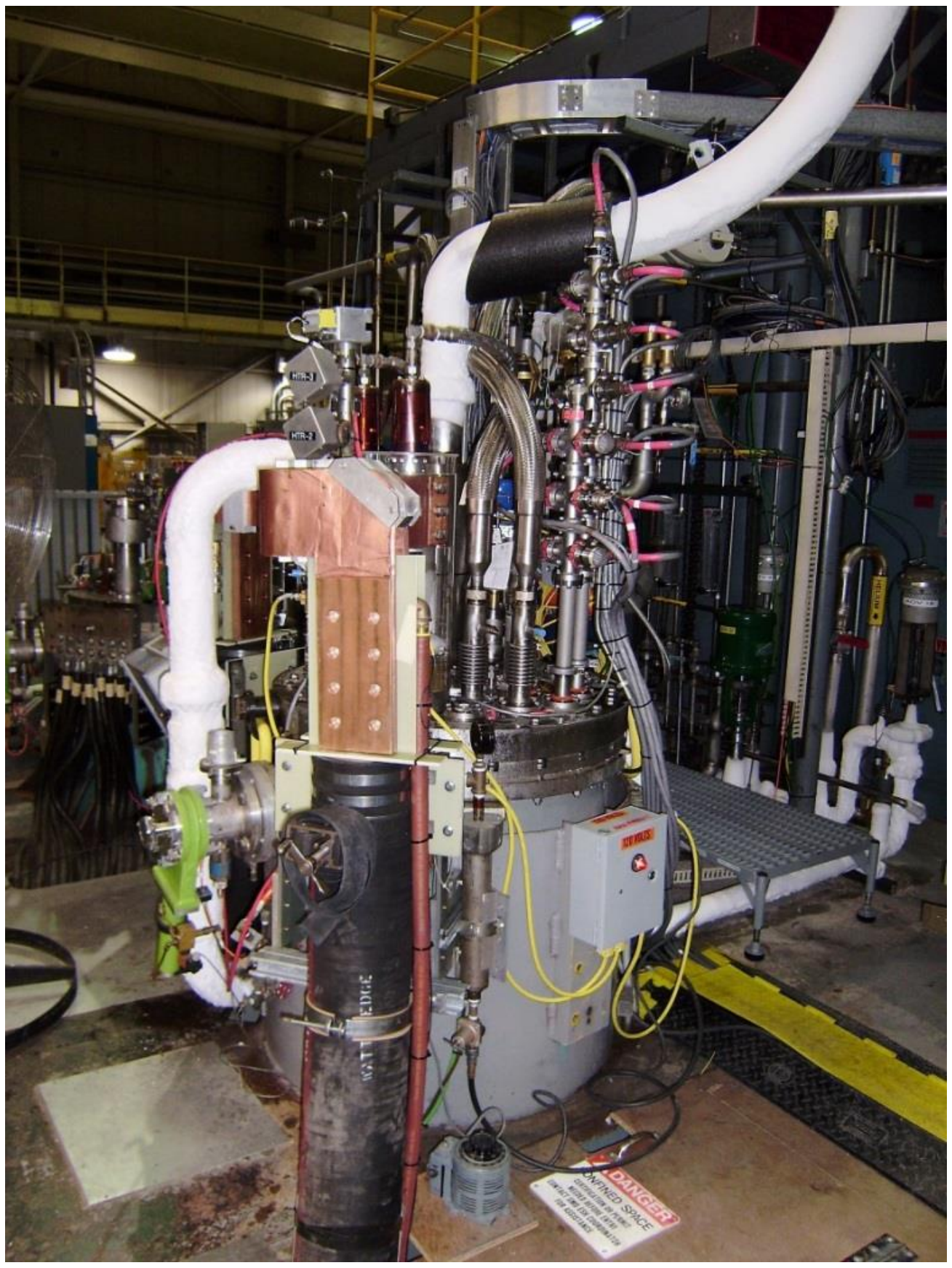

Figure 7.2: Vertical Test Stand 2, upgraded to provide 1.9 K at 1 bar (nom) and 24 kA, for testing of the 4.2 m-long MQXFA quadrupoles. The picture shows the test stand with the mirror MQXFPM1 under test.

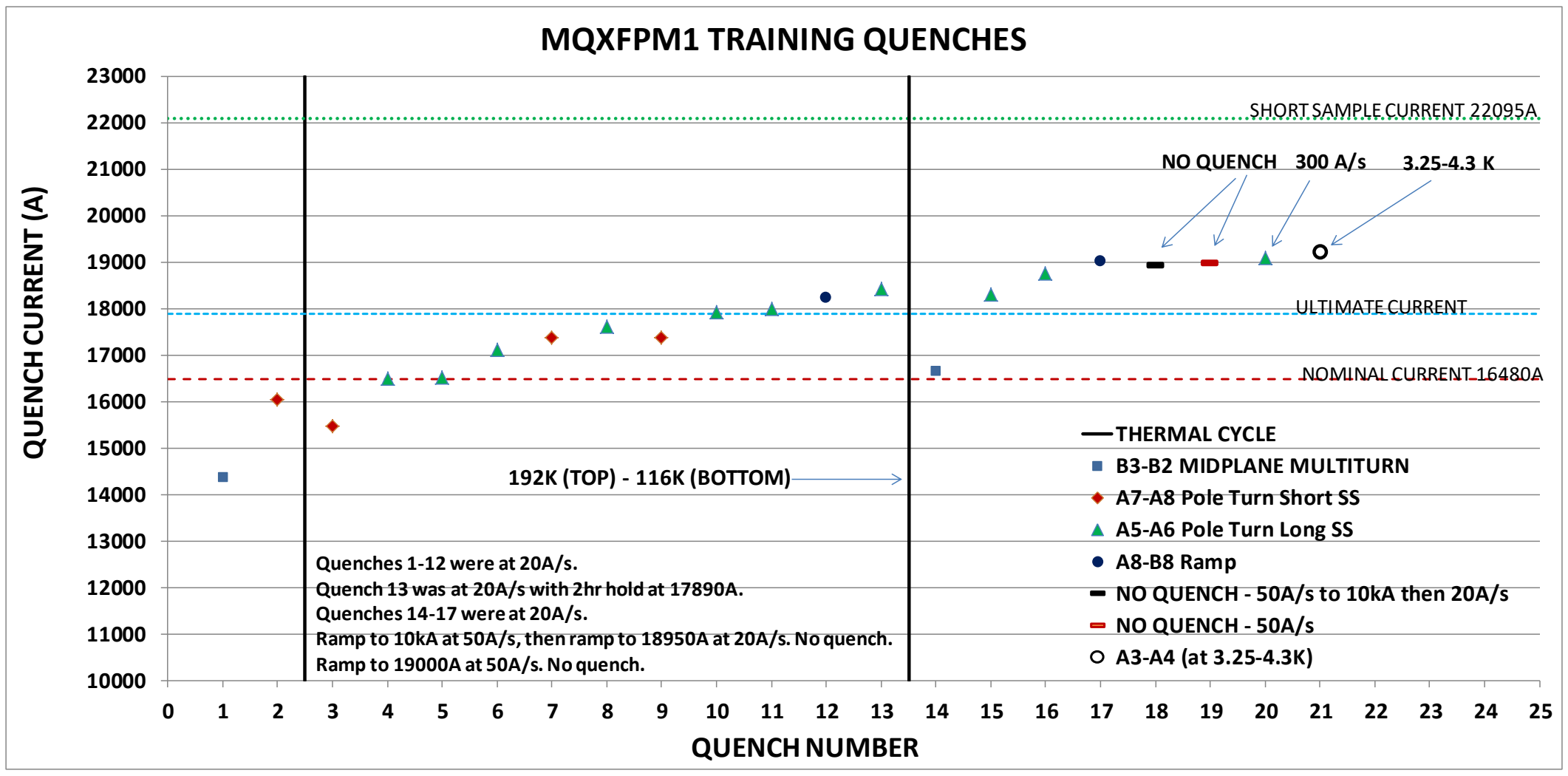

Figure 7.3: Plot of training quenches for the MQXFPM1 mirror at the BNL Vertical Test Facility. The red dotted line shows nominal current plus margin (16.48 kA), and the blue dotted line shows ultimate current plus margin (17.89 kA).

In late 2017 and early 2018, until 20-Feb, the first long prototype quadrupole MQXFAP1 was tested. A total of 18 training quenches were performed with the magnet reaching 17.426 A, 99.6% of the ultimate operating current, at Quench 18. Due to the development of a coil short to ground, testing was stopped to disassemble the magnet to investigate the short and re-assemble with a spare coil for later retest. During the training of MQXFAP1, the quench protection heaters satisfied the LHC requirement or better, and CLIQ was used successfully for the first time on a long magnet. MQXFAP1 quench history is shown in Fig. 7.4 below

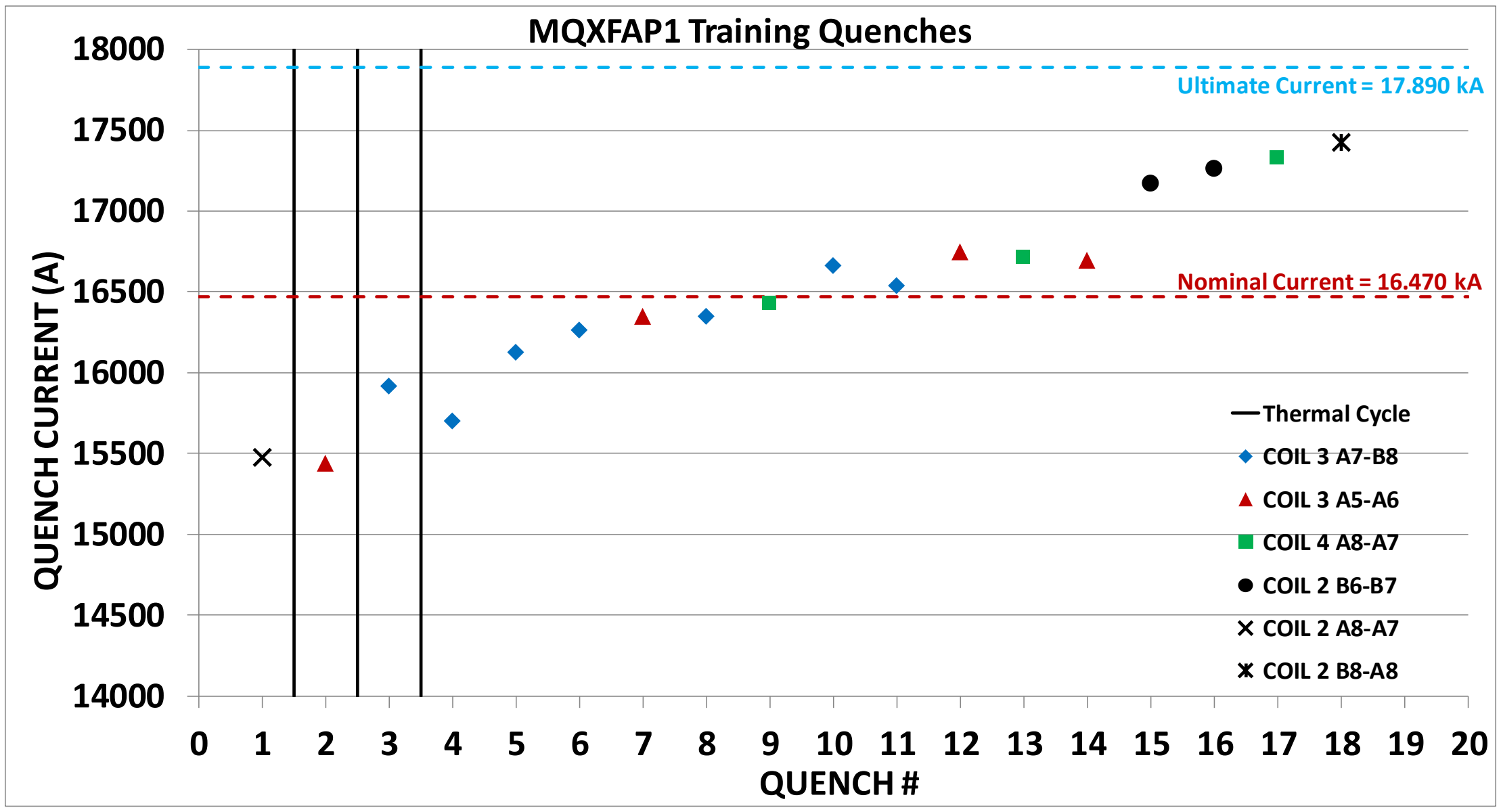

Fig. 7.4: Plot of training quenches for the MQXFAP1 prototype at the BNL Test Facility. The red dotted line shows nominal current plus margin (16.48 kA), and the blue dotted line shows ultimate current plus margin (17.89 kA).

As a result of He gas recovery issues during the early training quenches of MQXFAP1, additional upgrades were made to the cryogenics facility He gas recovery system in order to mitigate the loss of He gas after a quench caused by the opening of relief valves. This was due to fast and large increases in boiloff pressure after a quench (stored energies in the magnet can be as high as 6 MJ). With these upgrades, the test stand has now been able to operate without loss of He gas during quenches.

In addition, three more tests of full-length magnets have since been completed: the second prototype MQXFAP2, a rebuild of MQXFAP1 with a new coil replacing the coil shorted in the first test, and in Fall 2019, MQXFA03, the first pre-series MQXFA magnet, was successfully tested above nominal current. Fig. 7.5 shows the quench test plot for MQXFA03. This is the fifth full length MQXFA test at the BNL Vertical Test Facility, and the first that can operate in the HL-LHC.

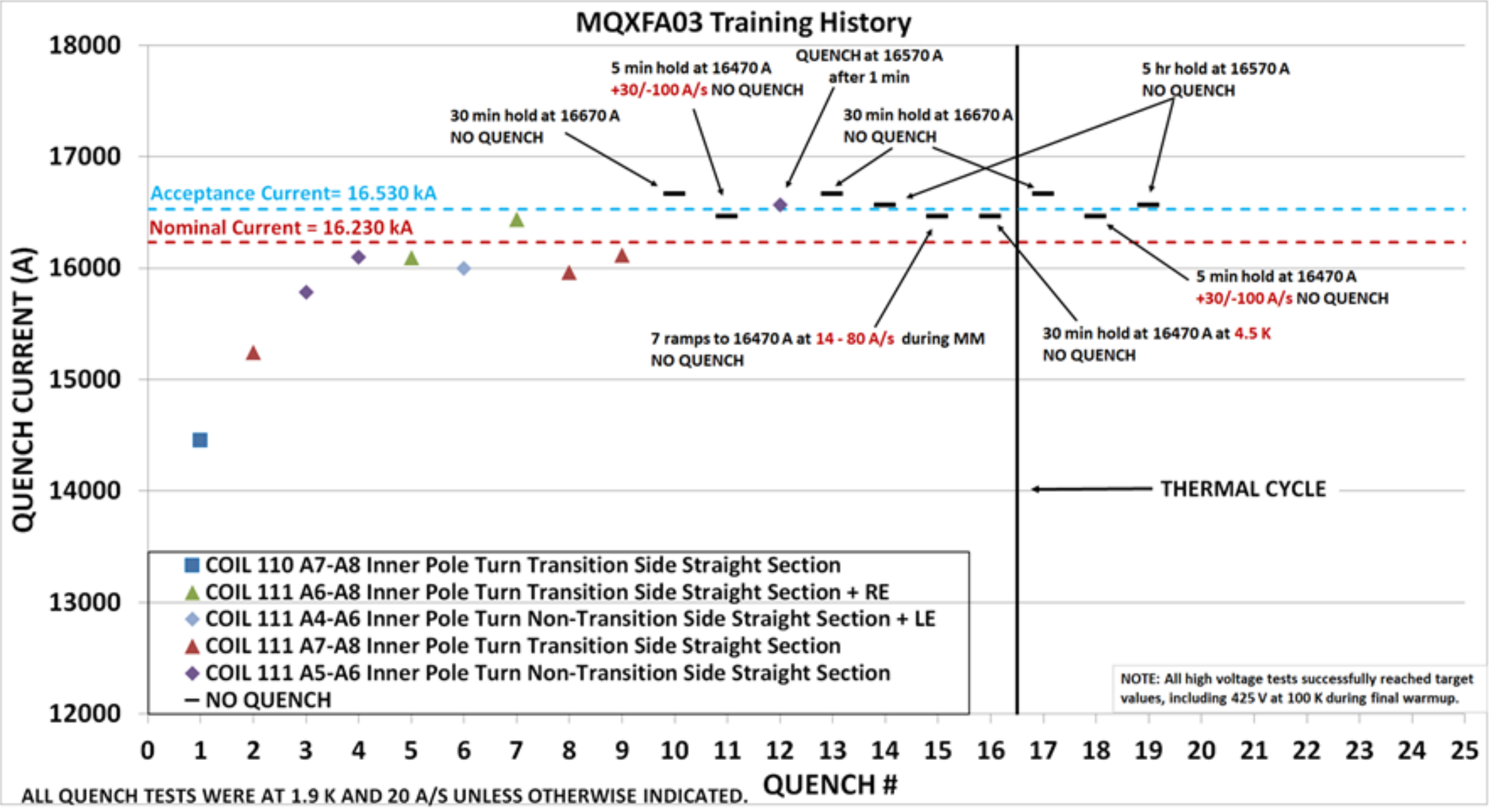

Fig.7.5. Plot of MQXFA03 quench test results.

## 7.1 Vertical Test Scope and Objectives - Prototypes

The primary objective of the prototype testing was to provide information to magnet designers to validate design choices, verify performance, and reduce risk of MQXFA production. The scope of these tests included quench performance and training of magnets, room temperature (warm) and 1.9 K (cold) magnetic field quality and integral field measurements, and tests designed to determine the characteristics and parameters of the quench protection systems, among others. As already noted, three MQXFA prototypes have already been tested at the BNL Vertical Test Facility, with the objectives having been met.

MQXFA prototype magnets were quench-tested to train to the maximum required test current of 19.8 kA, and to verify stable mechanical and electrical operation. Though the LHC will not use external energy extraction, magnet protection during training made use of energy extraction to an external dump resistor as well as copper clad stainless-steel quench protection heater strips, and CLIQ. In addition, quench protection studies with the protection heaters were performed. Training and other spontaneous quench tests were done with the external energy extraction system included in order to save liquid helium and reduce the quench recovery time during training.

The resistances of all NbTi-NbTi and NbTi-$Nb_3Sn$ splices will be monitored during the cold testing and measured during powering to operating current. Major checkouts and measurements to be conducted for the prototypes at the vertical test stand included:

At room temperature before cool down

- Electrical checkouts, and tests ensuring the integrity of the insulation (high voltage, or hipot)
- Magnetic measurements: integral field strength and field harmonics (using rotating probe)
- Strain gauge measurements

Measurements during cooldown and warmup

- Temperature sensor (RTD) measurements
- Strain gauge measurements
- Residual Resistivity Ratio (RRR) measurements

After cool down to 1.9 K

- Hipot measurements
- Magnetic Measurements
  - **Using rotating probe**
- Quench training up to 19.8 kA
- Quench characterization
  - **Voltage taps measurements**
  - **Quench Antenna measurements**
- Strain gauge measurements while cold and during powering
- Ramp Rate Dependence measurements
- Holding ultimate operation current (2-8 hours)
- Quench protection studies with heaters and/or CLIQ
- Measurement of the magnet inductance versus current
- Quench propagation studies
- Voltage spike Measurements
- Energy Loss measurements
- Splice measurements (at 1.9 or 4.5 K)

Measurements at 4.5 K

- Assess short sample limit percentage at 4.5 K after 1.9 K training

- Ramp Rate Dependence at 4.5 K in case of issues at 1.9 K

At room temperature after warm-up

- Electrical checkouts and tests ensuring the integrity of the insulation (high voltage, or hipot)
- Magnetic measurements: integral field strength and field harmonics (using rotating probe)

Thermal cycles were included to verify that the magnet retains its training.

For planning purpose, we assumed 50 training quenches.

A magnet re-test after pre-stress adjustment or coil exchange was done for the rebuild of MQXFAP1 to MQXFAP1b. The test plan for this prototype re-test should be and was shorter than the plan for a new magnet, aiming at assessing quench performance improvements/changes and completing the test plan. For planning purpose, we assumed 30 training quenches and ½ effort for protection and other studies. A thermal cycle was planned also for MQXFA re-tests.

## 7.2 Vertical Test Scope and Objectives – Production (Pre-Series and Series) Magnets

Production magnets, which have the final design for operation in the HL-LHC, include 5 pre-series magnets, which will have most of the instrumentation of the prototypes, and 15 series magnets, which will have limited instrumentation.

The primary objective of the production testing is quench training, magnet qualification, and quality assurance. Each magnet will be tested and trained prior to cold mass (LMQXFA) assembly, in order to check its quality and verify that it meets all requirements. Production magnets will not be tested above 17.5 kA (ultimate operating current).

The test cycle of production magnets is expected to be shorter than for prototypes: series magnets in particular will not have prototype instrumentation (for instance only voltage taps for splice measurement and quench detection and not for quench localization). Presently, the following numbers of quenches are suggested for planning purpose: 25 quenches at 1.9 K and, 2 quenches at 1.9 K after the thermal cycle (if needed).

The resistances of all NbTi-NbTi and NbTi-Nb3Sn splices will be monitored during the cold testing and measured during powering to operating current.

Major checkouts and measurements to be conducted for the production models at the vertical test stand include:

At room temperature before cool down

- Electrical checkouts and tests ensuring the integrity of the insulation (high voltage, or hipot)
- Magnetic measurements: integral field strength and field harmonics (using rotating probe)

After cool down to 1.9 K

- Electrical checkouts (including high voltage)
- Quench test training – to 16.53 kA (nominal current plus 300 A margin as specified in the Acceptance Criteria Part A: MQXFA Magnet [5]); with the exception of minimum two magnets, which will be trained to 17.5 kA (ultimate current)
  - Magnetic measurements: integral field strength and field harmonics (using rotating probe)
- Holding the nominal current plus 300 A margin for an extended period of time
- Quench Protection Heater tests to verify nominal operation
- Splice measurements

At room temperature after warm-up

  - Magnetic measurements: integral field strength and field harmonics (using rotating probe)
  - Electrical checkouts before shipping to FNAL

## 7.3 Field Quality Measurements

Magnetic field measurements are performed at room temperature before and after the cold test, and at 1.9 K. The field harmonics are measured with a rotating probe based on the multi-layer printed circuit board design [2]. Integral field strength is determined with z-scans using the rotating probe.

At room temperature, field harmonics will be measured at low currents.

Field measurements will be done after cool down to 1.9 K. Integral field strength and field harmonics will be measured at a series of currents including injection (960 A) and collision (16480 A) currents.

A standard list of magnetic measurements with rotating probe for prototype magnets developed jointly with CERN include and already being performed on short models and the latest full-scale models are:

- Z-scan
- Integral field measurements (as much as possible within test facility constraints)
- Loops (14, 20 and 40 A/s)
- Stair steps (with large steps)
- Accelerator cycle

## 7.4 Acceptance Tests

Requirements for qualification (acceptance) of the MQXFA magnets must be met before shipment to FNAL and assembly into LQXFA cold masses. The magnets will be accepted for shipment if all requirements are satisfied, including quench performance at 1.9 K and specified tolerances for field quality, for the integral magnetic field and the field harmonics before and after and during testing at 1.9 K.

## 7.5 High-Voltage Withstand Levels (Hipot Tests)

The MQXFA electrical design requirements and test values are specified in [3, 4]. We provide here a summary table for information only. The official documents can be accessed through the links provided below.

During operation the MQXFA magnets may experience high voltages in case of magnet quench. An extensive study of peak voltages during operation of the HL-LHC Inner Triplet magnets has been performed taking into account the possible impact of different conductor/coil parameters, different quench-start locations, and in case of heater failure scenarios. The results are summarized in [3] for three possible quench protection systems: (i) only Outer Layer Quench Heaters (O-QH), (ii) Outer and Inner Layer Quench Heaters (O-QH + I-QH), and (iii) Outer Layer Quench Heaters and Coupling-Loss Induced Quench system (O-QH + CLIQ). The third option is the baseline quench protection system for the Inner Triplet of HL-LHC.

Based on these studies, electrical test criteria have been defined to qualify the magnets for machine operation. Two different voltage levels are defined for: (1) Nominal Operating Conditions (NOC); and (2) tests at warm (room temperature in dry air with T = 20±3 °C and humidity lower than 60%).

The MQXFA electrical test values are summarized in Table 7.1. Once again, this table is provided for information only while the official reference is provided in [3,4,5]. It should also be noted that once the components have been exposed in a previous stage to helium, initial values at magnet reception cannot be longer applied as the presence of helium may weaken insulation by creating creepage paths.

Table 7.1 MQXFA required hi-pot test voltages and leakage current [5].

| **Circuit Element** | **Expected Vmax [V]** | **V hi-pot [V]** | **I hi-pot [μA]** | **Minimum time duration [s]** |
|---|---|---|---|---|
| Coil to Ground at RT before helium exposure | n.a. | 3680 | 10 | 30 |
| Quench heater to Coil at RT before helium exposure | n.a. | 3680 | 10 | 30 |
| Coil to Ground at cold | 353 | 1840 | 10 | 30 |
| Quench Heater to Coil at cold | 900 | 2300 | 10 | 30 |
| Coil to Ground at RT after helium exposure | n.a. | 368 | 10 | 30 |
| Quench heater to Coil at RT after helium exposure | n.a. | 460 | 10 | 30 |
| Quench heater to Coil at 100 K helium gas | 353 | 425 | 10 | 30 |